\documentclass[a4paper,12pt]{article}
\pdfoutput=1
\usepackage{epsfig}
\usepackage{amssymb}
\usepackage{amsfonts}
\usepackage{amsmath}
\usepackage{euscript}
\usepackage{verbatim}
\usepackage{latexsym}
\usepackage{graphicx}
\usepackage{caption}
\usepackage{float}
\usepackage{subcaption}
\usepackage[colorlinks=true, linkcolor=black, citecolor=black, urlcolor=blue]{hyperref}
\usepackage[T1]{fontenc}
\usepackage[utf8]{inputenc}

\usepackage{listings}

\usepackage{booktabs}
\usepackage{appendix}
\newif\ifdtup

\catcode`\@=11

\@addtoreset{equation}{section}

\def\@normalsize{\@setsize\normalsize{15pt}\xiipt\@xiipt
\abovedisplayskip 14pt plus3pt minus3pt%
\belowdisplayskip \abovedisplayskip
\abovedisplayshortskip \z@ plus3pt%
\belowdisplayshortskip 7pt plus3.5pt minus0pt}

\def\small{\@setsize\small{13.6pt}\xipt\@xipt
\abovedisplayskip 13pt plus3pt minus3pt%
\belowdisplayskip \abovedisplayskip
\abovedisplayshortskip \z@ plus3pt%
\belowdisplayshortskip 7pt plus3.5pt minus0pt
\def\@listi{\parsep 4.5pt plus 2pt minus 1pt
     \itemsep \parsep
     \topsep 9pt plus 3pt minus 3pt}}

\relax

\catcode`@=12

\catcode`\@=11

\def\section{\@startsection{section}{1}{\z@}{3.5ex plus 1ex minus
   .2ex}{2.3ex plus .2ex}{\large\bf}}

\def\SymBoxes#1#2#3#4{\newdimen\un@t \un@t#3%
\raisebox{#1}{\rule{#2\un@t}{#4}\hskip-#2\un@t% lower horizontal
\@tempdimb\un@t \advance\@tempdimb by-#4\@tempcntb#2\relax%
\@whilenum{\@tempcntb>0}\do{%                         % #2 vertical lines
\rule{#4}{\un@t}\hskip\@tempdimb \advance\@tempcntb by\m@ne}%
\hskip-#2\un@t \rule[\un@t]{#2\un@t}{#4}%
\rule[\un@t]{#4}{#4}\hskip-#4%             % upper horizontal line
\rule{#4}{\un@t}}\hskip-#4}                % rightest vertical line
\begin{document}
%\begin{letter}{~}

%%%%%%Define some new commands and  macros
\newcommand{\beq}{\begin{equation}}
\newcommand{\eeq}{\end{equation}}
\newcommand{\bea}{\begin{eqnarray}}
\newcommand{\eea}{\end{eqnarray}}
\newcommand{\beas}{\begin{eqnarray*}}
\newcommand{\eeas}{\end{eqnarray*}}
\newcommand{\defi}{\stackrel{\rm def}{=}}
\newcommand{\non}{\nonumber}
\newcommand{\bquo}{\begin{quote}}
\newcommand{\enqu}{\end{quote}}
%%%%%%%%%%%%%%%%
\renewcommand{\(}{\beq}
\renewcommand{\)}{\eeq}
%%%%%%%%%%%%%%%%%%%%%%%%%%%%%%%%%% definitions
\def \eqn#1#2{\beq#2\label{#1}\eeq}

\def\e{\epsilon}
\def\IZ{{\mathbb Z}}
\def\IR{{\mathbb R}}
\def\IC{{\mathbb C}}
\def\IQ{{\mathbb Q}}
\def\de{\partial}
\def\Tr{ \hbox{\rm Tr}}
\def\H{ \hbox{\rm H}}
\def\HE{ \hbox{$\rm H^{even}$}}
\def\HO{ \hbox{$\rm H^{odd}$}}
\def\K{ \hbox{\rm K}}
\def\Im{ \hbox{\rm Im}}
\def\Ker{ \hbox{\rm Ker}}
\def\const{\hbox {\rm const.}}
\def\o{\over}
\def\im{\hbox{\rm Im}}
\def\re{\hbox{\rm Re}}
\def\bra{\langle}\def\ket{\rangle}
\def\Arg{\hbox {\rm Arg}}
\def\Re{\hbox {\rm Re}}
\def\Im{\hbox {\rm Im}}
\def\exo{\hbox {\rm exp}}
\def\diag{\hbox{\rm diag}}
\def\longvert{{\rule[-2mm]{0.1mm}{7mm}}\,}
\def\a{\alpha}
\def\dag{{}^{\dagger}}
\def\tq{{\widetilde q}}
\def\p{{}^{\prime}}
\def\W{W}
\def\N{{\cal N}}
\def\hsp{,\hspace{.7cm}}

\def\br{\nonumber}
\def\IZ{{\mathbb Z}}
\def\IR{{\mathbb R}}
\def\IC{{\mathbb C}}
\def\IQ{{\mathbb Q}}
\def\IP{{\mathbb P}}
\def \eqn#1#2{\beq#2\label{#1}\eeq}

\newcommand{\C}{\ensuremath{\mathbb C}}
\newcommand{\Z}{\ensuremath{\mathbb Z}}
\newcommand{\R}{\ensuremath{\mathbb R}}
\newcommand{\rp}{\ensuremath{\mathbb {RP}}}
\newcommand{\cp}{\ensuremath{\mathbb {CP}}}
\newcommand{\vac}{\ensuremath{|0\rangle}}
\newcommand{\vact}{\ensuremath{|00\rangle}                    }
\newcommand{\oc}{\ensuremath{\overline{c}}}
\newcommand{\psizero}{\psi_{0}}
\newcommand{\phizero}{\phi_{0}}
\newcommand{\hzero}{h_{0}}
\newcommand{\psiin}{\psi_{\rh}}
\newcommand{\phiin}{\phi_{\rh}}
\newcommand{\hin}{h_{\rh}}
\newcommand{\rh}{r_{h}}
\newcommand{\rb}{r_{b}}
\newcommand{\psibnd}{\psi_{0}^{b}}
\newcommand{\psibndp}{\psi_{1}^{b}}
\newcommand{\phibnd}{\phi_{0}^{b}}
\newcommand{\phibndp}{\phi_{1}^{b}}
\newcommand{\gbnd}{g_{0}^{b}}
\newcommand{\hbnd}{h_{0}^{b}}
\newcommand{\zh}{z_{h}}
\newcommand{\zb}{z_{b}}
\newcommand{\man}{\mathcal{M}}
\newcommand{\hbr}{\bar{h}}
\newcommand{\tbr}{\bar{t}}

\begin{titlepage}
%\begin{flushright} CHEP XXXXX
%ULB-TH/09-10\\
%hep-th/yymmnnn\\ \end{flushright}
%\bigskip

\def\thefootnote{\fnsymbol{footnote}}

\begin{center}
{\large
{\bf Geodesics from Quantum Field Theory: \\ 
A Case Study in AdS

%Localized States, Geodesics, and Position Operators in Curved Spacetime

}}
\end{center}

%\bigskip
\begin{center}
Vaibhav Burman$^a$\footnote{\texttt{vaibhav2021@iisc.ac.in}}, \ Chethan Krishnan$^a$\footnote{\texttt{chethan.krishnan.physics@gmail.com}}, \ Livesh Parajuli$^a$\footnote{\texttt{liveshparajuli@gmail.com}} 
%\vspace{0.1in}

\end{center}

\renewcommand{\thefootnote}{\arabic{footnote}}

\begin{center}
%\vspace{0.2cm}

  $^a$ {Center for High Energy Physics,\\
Indian Institute of Science, Bangalore 560012, India}\\

\end{center}
\vspace{-0.15in}
\noindent

\begin{center} {\bf Abstract} \end{center}

Localized one-particle states of a quantum field theory--whether in flat space or on a curved background--are expected to exhibit geodesic motion in an appropriate semiclassical regime. This expectation is often invoked heuristically: in this work we develop two precise implementations and test them in detail in global AdS$_3$. First, we define a covariant ``center-of-mass'' trajectory from the expectation value of the stress tensor operator and show, using only $\nabla_\mu\langle T^{\mu\nu}\rangle=0$, that it obeys the geodesic equation in the monopole (sufficiently localized) approximation in a general spacetime. This provides a QFT-in-curved-spacetime generalization of the Mathisson-Papapetrou-Dixon framework in classical general relativity. Second, we construct position operators from the Klein--Gordon inner product and mode completeness, and compute their expectation values in generic single-particle wave packet states. We then build explicit normalizable wave packets of a free scalar field in empty AdS$_3$ with tunable energy and angular momentum, and demonstrate analytically and numerically that both prescriptions reproduce the expected radial, circular, and elliptical-like timelike and null geodesics. Our discussion also isolates a natural ultra-relativistic regime in which the wave packet trajectory exhibits a controlled crossover from timelike to null geodesic behavior. We identify precise limits where the localized geodesic interpretation of the wave packet breaks down. On the CFT side, we show that bulk localization--specifically the radial data--is captured by how the state is distributed over global descendants of the dual primary.

\vspace{1.6 cm}
\vfill

\end{titlepage}

\setcounter{footnote}{0}

\tableofcontents

\section{Introduction and Outline}

One of the basic expectations in quantum field theory is that sufficiently localized one-particle states should propagate along classical geodesics in an appropriate semiclassical regime. In flat space this expectation is so familiar that it is often left at the level of intuition. In curved spacetime, however, and especially in anti-de Sitter space where one would like to understand bulk localization from the boundary CFT, the question becomes more interesting and has got some attention \cite{Berenstein:2019,Goto:2017,Terashima:2023WavePackets, Kevin1, Kevin2}: how does a quantum CFT state manage to encode the data of an approximately bulk localized classical trajectory? 

At the same time, relativistic localization is famously subtle. The classic analyses of Newton and Wigner, Wightman and Hegerfeldt make clear that sharply localized position eigenstates are not innocuous physical objects in relativistic quantum theory \cite{Newton:1949,Wightman:1962,Hegerfeldt1974,Hegerfeldt1998}. Any attempt to extract geodesic motion from quantum states must therefore be formulated with some care.

Two broad motivations sit in the background of the present work. The first is the origin of the bulk radial coordinate in holography, as we mentioned above. The second is the emergence of the black hole interior: our proximate cause being the calculations in \cite{Burman1,Burman2}. A controlled understanding of how an approximately localized bulk trajectory emerges from quantum data is a natural prerequisite for both problems. Before one can ask how a boundary description organizes an entire bulk region, or more ambitiously an interior region, one should first understand how it organizes the kinematics of an ordinary semiclassical particle in the bulk.

One aim of this paper is to make the emergence of geodesic motion from quantum field theory precise in two complementary ways. The first is intrinsically covariant and makes no reference to a position operator. Starting from the expectation value of the stress tensor, we define a center-of-mass trajectory by taking the energy-weighted centroid on a constant-time slice, and show that in the monopole approximation it obeys the geodesic equation in a general background spacetime. In this sense, the construction may be viewed as a quantum-field-theoretic analogue of the Mathisson--Papapetrou--Dixon framework familiar from classical general relativity \cite{Mathisson1937,Papapetrou:1951,Dixon1970I,Dixon1970II,Dixon1974III}. The second approach is more directly operatorial. In the concrete setting of AdS$_3$ (but it allows generalizations to broad classes of spacetimes), we use the Klein--Gordon inner product together with the completeness of the modes to construct position operators whose expectation values can be evaluated in arbitrary one-particle states.

A key conceptual point is that we are not trying to rehabilitate sharply localized relativistic position eigenstates as physical states. That is precisely where the standard no-go results become relevant. Instead, we use smooth, normalizable wave packets and ask a different question: when do their suitably defined peaks behave classically? The answer is nontrivial already in flat space, and more so in AdS where the spectrum is discrete and the natural observables are adapted to curved-space mode functions. The AdS/CFT language provides a useful interpretation of several of the results.

After setting up the general formalism, we test both prescriptions in detail in global AdS$_3$. We construct explicit normalizable wave packets of a free scalar field with tunable initial mean position, (a proxy for the) initial radial momentum and angular momentum, and show analytically and numerically that both the stress-tensor centroid and the position-operator expectation values reproduce the expected radial, circular and elliptical-like timelike geodesics, as well as the corresponding null trajectories, when the packet is sufficiently localized.  There is a controlled crossover in which the trajectory interpolates from timelike to null behavior in an ultra-relativistic regime. We isolate a breakdown of the localized geodesic interpretation when the packet becomes too narrow compared to the scale set by its energy. 
We also explain why the conserved charges of the wave packet need not coincide exactly with the conserved charges of the point-particle geodesic that best fits its trajectory: finite width effects shift this identification, even when the curve itself matches very well.

Even though we work with AdS in most of this paper, we use coordinates that only manifest time translations and rotations. We do {\em not} exploit the underlying maximal symmetry. If we work with (say, embedding) coordinates and methods that exploit the special properties of the AdS vacuum, we can make some of the discussions simpler. But this defeats our purpose in two ways: Firstly, the mechanisms that are valid in more general spacetimes than AdS become less transparent. Secondly, our underlying motivation is to connect bulk locality in general with the properties of states in the boundary CFT, and exploiting the enhanced symmetries of the vacuum does not help our agenda\footnote{A further comment is that even though we work with AdS$_3$, much of our discussion should be adaptable (with more angles) to higher dimensional AdS spaces as well: the special features of 2+1 dimensional gravity and Virasoro symmetry are not of much relevance to our QFT-in-curved-space discussion.}.

A special feature of empty AdS is that the evenly spaced scalar spectrum and the associated selection rules imply exact statements on one-particle states. For suitable observables in AdS$_3$ that we identify, their expectation values in \emph{arbitrary} one-particle states obey the same equations as the corresponding classical geodesic combinations. These exact relations do not by themselves imply that an arbitrary state is localized or semiclassical: rather, they cleanly separate the exact kinematical part of the evolution from the variance corrections that control the actual quality of the geodesic approximation. This is one of the places where AdS exhibits extra structure beyond the generic curved-space discussion.

A final section of the paper relates things to the CFT side. The one-particle Hilbert space of a bulk scalar in global AdS$_3$ is naturally identified with the global conformal module of the dual scalar primary, and Euclidean-regularized operator insertions provide a standard class of CFT states to compare with our bulk Gaussian packets \cite{Berenstein:2019,Goto:2017}. By rewriting those CFT states in the same $(n,m)$ basis used in the bulk analysis, we make the comparison fully explicit. This also gives some useful intuition about the first broad motivation mentioned above. Section~\ref{sec:CFT_interpretation} can be viewed as a step in the direction of deriving the bulk radial coordinate from the CFT: starting from CFT data, one can infer the mean conserved charges, reconstruct the associated semiclassical orbit, and thereby reverse-engineer a notion of radial location and radial motion for the corresponding bulk state. In that concrete sense, the discussion there illuminates how radial information is encoded.

The organization of the paper is as follows. In Section~\ref{sec:geodesic} we review the relevant timelike and null geodesics in global AdS$_3$, including a pair of exact classical relations that later reappear at the quantum level. Section~\ref{sec:relativistic_localization} briefly reviews the relativistic localization problem and explains how our use of position operators differs from the Newton--Wigner question. In Section~\ref{sec:com} we derive the geodesic equation from the stress tensor centroid in a general background, and in Section~\ref{sec:operators} we construct position operators in AdS$_3$. Section~\ref{sec:EV_of_rho_phi_Ttt} develops the relevant expectation values, while Section~\ref{sec:AdS3_implementation} presents explicit wave packets and their numerical evolution. Section~\ref{sec:geo-vs-wave-charges} discusses the relation between geodesic charges and wave-packet charges, and Section~\ref{sec:breakdown} analyzes the breakdown of the classical description. In Section~\ref{sec:Q-C_corres} we show that certain operator combinations satisfy the classical geodesic equations exactly at the level of expectation values, and in Section~\ref{sec:CFT_interpretation} we give the CFT interpretation of the bulk wave packets and compare our construction to Euclidean-regularized CFT states. 

The appendices collect a number of technical derivations and supplementary checks. Appendix \ref{appendix_geo_from_stress_tensor} presents the general derivation of the geodesic equation from the stress tensor, together with the exact moment relations used in the main text. Appendix \ref{appendix_recipe_for_positiooperators_static} discusses the construction of position operators in a broader class of static spacetimes, while Appendix \ref{appendix_scalar_field_ads3} summarizes the scalar field, conserved charges, and stress tensor in global AdS$_3$. Appendices \ref{appendix_why_is_the_expval_m_0?} and \ref{appendix_why_is_n_0_initial_radial_mom?} justify two of the parameter identifications used in our wave-packet construction, namely $\langle \hat L \rangle = - \langle \hat m \rangle = m_0$ and the interpretation of $n_0$ as a proxy for the initial radial momentum. Appendices \ref{appendix_2d_plots_com} and \ref{appendix_2d_plots_pos_ope} collect additional two-dimensional plots for the stress-tensor and position-operator approaches respectively. Appendix \ref{appendix where we show that the definitions give straight line} provides the flat-space analogue in both Cartesian and plane polar coordinates, Appendix \ref{appendix_Q-C_corresp_com} discusses the operator-level quantum-classical correspondence in the stress-tensor framework, and Appendix \ref{appendix_sec:symmetry_origin} explains the origin of the selection rules that underlie the exact AdS relations. In Appendix \ref{app:eikonal} we discuss the eikonal/WKB limit of a (classical) scalar field equation in a general curved spacetime. Appendix \ref{eknl_lim_get_geo_and_all} connects this to the one particle states and wave packets of the corresponding quantum field theory  -- we do the calculations in detail for our AdS$_3$ wave packets, but the lessons generalize.

\section{Geodesics in Global \texorpdfstring{AdS$_{3}$}{AdS3}}\label{sec:geodesic}

We begin by recalling the geometry of global AdS$_{3}$. In coordinates that cover the entire manifold, the metric with AdS radius $\mathcal{R}$ takes the form:
\begin{equation}\label{metric1}
ds^{2} = \mathcal{R}^{2}\sec^{2}\rho\left(dt^{2}-d\rho^{2}-\sin^{2}\rho\, d\phi^{2}\right)
\end{equation}
where $\rho\in[0,\pi/2)$, $t\in\mathbb{R}$, and $\phi\sim\phi+2\pi$. The conformal boundary is located at $\rho=\pi/2$, while $\rho=0$ represents the center in these coordinates. The metric is static and rotationally symmetric, being independent of both $t$ and $\phi$. Consequently, there are two manifest Killing vectors, $\partial_{t}$ and $\partial_{\phi}$, whose associated conserved quantities along a geodesic are given by (see \cite{Carroll:2004st}):
\begin{equation}\label{constantEandL}
E = g_{\mu\nu}(\partial_{t})^{\mu}u^{\nu}= \mathcal{R}^{2}\sec^{2}(\rho)\frac{dt}{d\lambda},\qquad
L = -g_{\mu\nu}(\partial_{\phi})^{\mu}u^{\nu}= \mathcal{R}^{2}\tan^{2}(\rho)\frac{d\phi}{d\lambda}.
\end{equation}
Here, $x^{\mu}(\lambda)$ is a geodesic parametrized by an affine parameter $\lambda$, with tangent vector $u^{\mu}=dx^{\mu}/d\lambda$. The sign convention for $L$ is chosen so that $L$ is positive for motion in the direction of increasing $\phi$. In addition, the norm of the tangent vector is itself constant along the geodesic:
\begin{equation}
\epsilon = g_{\mu\nu}\frac{dx^{\mu}}{d\lambda}\frac{dx^{\nu}}{d\lambda},
\end{equation}
with $\epsilon=+1$ for timelike geodesics (here $\lambda$ may be taken as proper time $\tau$), $\epsilon=0$ for null geodesics, and $\epsilon=-1$ for spacelike geodesics. We will focus on the timelike and null cases and will demonstrate the emergence of corresponding classical trajectories from the quantum evolutions in subsequent sections.

By eliminating $dt/d\lambda$ and $d\phi/d\lambda$ in favor of $E$ and $L$, one can integrate the $\rho$ component of the geodesic once to yield a first-order radial equation. A convenient way to obtain it is to substitute the expressions for $\dot{t}$ and $\dot{\phi}$ from \eqref{constantEandL} into the norm condition $\epsilon = g_{\mu\nu}\dot{x}^{\mu}\dot{x}^{\nu}$. After a straightforward rearrangement, one finds:
\begin{equation}\label{radial-eq}
\left(\frac{d\rho}{dt}\right)^{2}=1-\frac{\epsilon\mathcal{R}^{2}}{E^{2}\cos^{2}\rho}-\frac{L^{2}}{E^{2}\sin^{2}\rho}.
\end{equation}
This equation governs all geodesic motion in global AdS$_{3}$. The turning points of the radial motion occur when $d\rho/dt=0$, which determines the allowed range of $\rho$ for given $E$, $L$ and $\epsilon$.

\subsection{Timelike Geodesics \texorpdfstring{($\epsilon=+1$)}{epsilon}}\label{sec:geodesic:timelike}

For a massive particle, $\epsilon=+1$, and the radial equation becomes:
\begin{equation}\label{timelike}
\left(\frac{d\rho}{dt}\right)^{2}=1-\frac{\mathcal{R}^{2}}{E^{2}\cos^{2}\rho}-\frac{L^{2}}{E^{2}\sin^{2}\rho}.
\end{equation}
The motion is confined to the region where the right-hand side is non‑negative, which defines an interval $\rho_{\text{min}}<\rho<\rho_{\text{max}}$. Now let us see how much time advances when $\rho$ goes from $\rho_{min}$ to $\rho_{max}$. Substituting $u=\sin^{2}(\rho)$ in the above equation, we get the expression for time advanced to be:
\begin{equation}
t =  \frac{1}{2}\int_{u_{1}}^{u_{2}}\frac{du}{\sqrt{(u_{2}-u)(u-u_{1})}}    
\end{equation}
where $u_{2}> u_{1}$. This integral evaluates to $t = \frac{\pi}{2}$. When the particle moves from $\rho_{\max}$ to $\rho_{\min}$, the time increases by another $\frac{\pi}{2}$. This is half a period. Hence, the time period of the  complete $\rho $-oscillations is $T_{\rho} = 2 \pi$.

\subsubsection{Radial Infall \texorpdfstring{($L=0$)}{L}}\label{sec:geodesic:timelike:radial}
When the angular momentum vanishes, \eqref{timelike} simplifies to:
\begin{equation}\label{radial}
\frac{d\rho}{dt}= \pm\sqrt{1-\frac{1}{\xi^{2}\cos^{2}\rho}},\qquad \xi\equiv\frac{E}{\mathcal{R}}.
\end{equation}
The $\pm$ signs correspond to outgoing and ingoing radial geodesics respectively. Integrating this equation gives:
\begin{equation}
t = \pm\left[\sin^{-1}\left(\frac{\sin\rho}{\sqrt{1-\xi^{-2}}}\right)-\sin^{-1}\left(\frac{\sin\rho_{0}}{\sqrt{1-\xi^{-2}}}\right)\right],
\end{equation}
where $\rho_{0}$ is the radial coordinate at $t=0$. From the condition that the argument of the square root, in \eqref{radial}, be non‑negative we obtain $\xi^{2}\cos^{2}\rho\geq1$ throughout the motion, which implies $\rho\leq\rho_{\text{max}}$ with:
\begin{equation}
\sin\rho_{\text{max}}=\sqrt{1-\xi^{-2}}.
\end{equation}
Thus, a massive particle on a radial geodesic can never reach the boundary $\rho = \pi/2$ unless $E \to \infty$; instead, it oscillates between $\rho = 0$ and a maximum radius $\rho_{\max}$, passing through the origin and re-emerging on the opposite side each cycle, with $\rho_{\max}$ determined by its energy. In the limit $E\to\infty$, we have $\rho_{\text{max}}\to\pi/2$ and the motion reduces to the null radial case $t=\pm(\rho-\rho_{0})$.

\subsubsection{Orbital Motion \texorpdfstring{($L\neq0$)}{L neq 0}}\label{sec:geodesic:timelike:orbital}
For non‑zero angular momentum, it is useful to derive the shape of the orbit. Dividing $d\rho/dt$ by $d\phi/dt$ eliminates time and yields:
\begin{equation}\label{d_rho_by_dt_mmaassiivvee_case}
\frac{d\rho}{d\phi}= \pm\frac{\mathcal{R}\sin^{2}\rho}{L}\sqrt{\frac{E^{2}}{\mathcal{R}^{2}}-\frac{L^{2}}{\mathcal{R}^{2}\sin^{2}\rho}-\frac{1}{\cos^{2}\rho}}.
\end{equation}
The quantity under the square root must be non‑negative, which defines two turning radii $\rho_{1}\leq\rho_{2}$ (the roots of the quadratic obtained after setting $u=\sin^{2}\rho$). The integral can be evaluated explicitly, leading to the implicit orbit equation:
\begin{equation}\label{orbit-massive}
\sin(\phi-\phi_{0})\sqrt{1-\frac{\sin^{2}\rho_{1}}{\sin^{2}\rho_{2}}}
= \sqrt{1-\frac{\sin^{2}\rho_{1}}{\sin^{2}\rho}}.
\end{equation}
This describes a closed curve without precession. Let us call this curve an ellipse like curve. As $\rho$ increases from $\rho_{1}$ to $\rho_{2}$, $\phi$ advances by $\pi/2$, and after a full cycle when $\rho$ returns back to its initial value, $\Delta\phi=\pi$. The full azimuthal period becomes $T_{\phi}=2\pi$.

The two extrema of the orbit are obtained by solving the following quadratic equation for given $E$ and $L$:
\begin{equation}\label{the quadratic equation to find the two extreme rho of themassive orbit}
u^{2}-u\left( 1+\frac{L^{2}-\mathcal{R}^{2}}{E^{2}}\right)+\frac{L^{2}}{E^{2}}=0    
\end{equation}

\subsubsection*{Circular Orbits} 
The geodesic equation can be written as:
\begin{equation}
\frac{\mathcal{R}^{4}}{\cos^{4}(\rho)}\dot{\rho}^{2} = E^{2}-V_{eff}   
\end{equation}
where $\dot{\rho}=\frac{d\rho}{d\lambda}$ and the effective potential is given by $V_{\text{eff}}=L^{2}/\sin^{2}\rho+\mathcal{R}^{2}/\cos^{2}\rho$. Circular geodesics correspond to constant $\rho=\rho_{0}$, which requires $\dot{\rho}=0$ and $\partial V_{\text{eff}}/\partial\rho=0$. These conditions give:
\begin{equation}
\frac{\sin^{4}\rho_{*}}{\cos^{4}\rho_{*}}=\frac{L^{2}}{\mathcal{R}^{2}},\qquad E=\pm\frac{L}{\sin^{2}\rho_{*}},\qquad \frac{d\phi}{dt}= \pm1
\end{equation}
The second derivative of the effective potential is positive for these orbits:
\begin{equation}
\frac{\partial^{2}V_{\text{eff}}}{\partial\rho^{2}}\Big|_{\rho_{*}}=\frac{8\mathcal{R}^{2}}{\cos^{4}\rho_{*}}>0
\end{equation}
so all massive circular orbits in global AdS$_{3}$ are stable.

\subsection{Null Geodesics \texorpdfstring{($\epsilon=0$)}{epsilon0}}\label{sec:geodesic:null}

For massless particles, setting $\epsilon=0$ in \eqref{radial-eq} gives:
\begin{equation}
\frac{d\rho}{dt}= \pm\sqrt{1-\frac{L^{2}}{E^{2}\sin^{2}\rho}}
\end{equation}

\subsubsection{Null Geodesics \texorpdfstring{($L=0$)}{L0}}\label{sec:geodesic:null:radial}

When $L=0$, the above equation reduces to the simple form $d\rho/dt=\pm1$, whose solution is:
\begin{equation}\label{the classical geodesic eqn for massless case radially infalling}
t = \pm(\rho-\rho_{0}).
\end{equation}
Thus radial null geodesics reach the boundary in finite coordinate time and reflect back.

\subsubsection{Null Geodesics \texorpdfstring{($L\neq0$)}{L neq 0}}\label{sec:geodesic:null:orbital}
For $L\neq0$, the motion is confined to $\sin^{2}\rho\geq L^{2}/E^{2}$. Writing $\alpha=L/E$, the radial equation becomes:
\begin{equation}
\frac{d\rho}{dt}= \pm\sqrt{1-\frac{\alpha^{2}}{\sin^{2}\rho}}.
\end{equation}
Integration yields:
\begin{equation}
t = t_{0}\pm\left[\sin^{-1}\left(\frac{\cos\rho_{0}}{\sqrt{1-\alpha^{2}}}\right)-\sin^{-1}\left(\frac{\cos\rho}{\sqrt{1-\alpha^{2}}}\right)\right].
\end{equation}
Here, $\rho_{0}$ is the launching radius. The equation of orbit is obtained as before from $d\rho/d\phi$, which after integration gives:
\begin{equation}\label{non_radial_orbit_for_null_case}
\rho = \sin^{-1}\left[\frac{\sin\rho_{0}}{\sqrt{1-\cos^{2}\rho_{1}\sin^{2}(\phi-\phi_{0})}}\right].
\end{equation}
This is precisely the $E\to\infty$ limit of the massive orbit \eqref{orbit-massive}. The azimuthal advance from the minimum radius $\rho_{1}$ to the boundary and back is exactly $\pi$. Again, the full period is $2 \pi$.

\subsubsection*{Null Circular Orbits }
In null case the potential becomes $V_{eff}= \frac{L^{2}}{\sin^{2}(\rho)}$. Demanding $\partial V_{\text{eff}}/\partial\rho=0$ gives:
\begin{equation}
\rho_{*}=\frac{\pi}{2},\qquad E=\pm L,\qquad \frac{d\phi}{dt}= \pm1
\end{equation}
The second derivative is:
\begin{equation}
\frac{\partial^{2}V_{\text{eff}}}{\partial\rho^{2}}\Big|_{\rho_{*}=\frac{\pi}{2}}=2\,L^{2}>0
\end{equation}
So this means we get stable null orbits only at the boundary of AdS.

\subsection{Two Exact Relations}\label{sec:geodesic:exact_relations}

For later reference we note two exact differential equations satisfied by any geodesic. First, introducing $u=\cos(2\rho)$, a short calculation starting from \eqref{radial-eq} leads to:
\begin{equation}\label{exact relation 1}
\frac{d^{2}u}{dt^{2}}+4u = \frac{4(\epsilon\mathcal{R}^{2}-L^{2})}{E^{2}}
\implies \cos(2\rho)=\,A\,+\,B\cos(2t+\delta).
\end{equation}
Here $A=\frac{\epsilon \mathcal{R}^{2}-L^{2}}{E^{2}}$, $B=\frac{\sqrt{(E^{2}+L^{2}-\mathcal{R}^{2})^{2}-4L^{2}E^{2}}}{E^{2}}$ and $\delta$ is a phase controlled by initial conditions\footnote{The constant $A$ is straightforward to determine, so we comment here about $B$. Solving $\cos(2\rho_{\text{max}}) = A - B$ and $\cos(2\rho_{\text{min}}) = A + B$, we identify the amplitude as $B = u_2 - u_1$, where $u = \sin^2\rho$ and $u_2 > u_1$. By writing $|u_2 - u_1| = \sqrt{(u_1 + u_2)^2 - 4u_1 u_2}$ and utilizing the fact that the roots of Eq.~\eqref{the quadratic equation to find the two extreme rho of themassive orbit} satisfy the relations $u_1 + u_2 = 1 + \frac{L^2 - \mathcal{R}^2}{E^2}$ and $u_1 u_2 = \frac{L^2}{E^2}$, we obtain the given form of $B$.}. Thus $\cos(2\rho)$ oscillates harmonically with frequency $2$ with amplitude determined by the conserved quantities. Second, consider the complex combination $\mathcal{Z}=\sin(\rho)\,e^{i\phi}$. Using the geodesic equations one finds:
\begin{equation}\label{exact relation 2}
\frac{d^{2}\mathcal{Z}}{dt^{2}} = -\mathcal{Z} \implies \mathcal{Z}(t)= \tilde A\,e^{it}\,+\,B\,e^{-it}
\end{equation}
so that $\mathcal{Z}$ executes simple harmonic motion of unit frequency. From this, we can write the expression for $\arg(e^{i\phi})$ as:
\begin{equation}
\arg(e^{i\phi})= \arg(\tilde A\,e^{it}\,+\,\tilde B\,e^{-it})    
\end{equation}

The relations \eqref{exact relation 1} and \eqref{exact relation 2} are the classical counterparts of the quantum exact results that will appear later in the paper.

\section{An Aside on Relativistic Localization}\label{sec:relativistic_localization}

In non-relativistic quantum mechanics, position is an observable on the same footing as momentum: the operator $\hat{x}$ acts by multiplication in the coordinate representation, its eigenstates $|x\rangle$ form a complete orthonormal basis (in the distributional sense), and a particle can in principle be prepared in a state of arbitrarily sharp spatial localization whose subsequent time evolution is entirely consistent. In a relativistic setting, localization is subtler. The familiar non-relativistic ``package'' of sharp localization, positive energy, and causal propagation cannot be maintained simultaneously in the same naive way. In particular, localization schemes built from positive-frequency one-particle states generically exhibit instantaneous spreading under time evolution. This is the sense in which relativistic localization becomes problematic \cite{Wightman:1962,Fleming:1965_covariant,Fleming1965,Steinmann1968,Schweber:1961,Kalnay1971,Pavsic:2017}.

\medskip
\noindent
\textbf{The Newton--Wigner construction.}
The most systematic attempt to define a position operator for relativistic particles is due to Newton and Wigner~\cite{Newton:1949}. Their construction starts from the observation that the single-particle Hilbert space of a massive field carries a unitary representation of the Poincar\'e group, with basis states $|\mathbf{k}\rangle$ labeled by spatial momentum and an invariant inner product weighted by $d^d k/(2\omega_{\mathbf{k}})$, where $\omega_{\mathbf{k}} = \sqrt{\mathbf{k}^2 + M^2}$. Newton and Wigner sought a position operator $\hat{\mathbf{x}}_{\mathrm{NW}}$ whose eigenstates $|\mathbf{x}\rangle_{\mathrm{NW}}$ satisfy: (i) orthonormality, ${}_{\mathrm{NW}}\!\langle \mathbf{x}|\mathbf{x}'\rangle_{\mathrm{NW}} = \delta^{(d)}(\mathbf{x} - \mathbf{x}')$; (ii) correct transformation under spatial rotations and translations; and (iii) construction from positive-frequency modes only. These requirements uniquely fix the momentum-space wavefunction of the localized state to be
\bea\label{NW-state}
    \langle \mathbf{k} | \mathbf{x}\rangle_{\mathrm{NW}} \propto \sqrt{\omega_{\mathbf{k}}}\; e^{-i\mathbf{k}\cdot\mathbf{x}}\,.
\eea
The factor $\sqrt{\omega_{\mathbf{k}}}$ compensates for the Lorentz-invariant measure and ensures delta-function normalization with respect to the standard $d^d k$ integration. The NW position operator is the canonical conjugate of momentum in this re-weighted basis. It is self-adjoint, has a complete set of (distributional) eigenstates, and reduces to the standard non-relativistic position operator in the limit $|\mathbf{k}| \ll M$.

However, the $\sqrt{\omega_{\mathbf{k}}}$ factor in~\eqref{NW-state} means that, when an NW eigenstate is represented as an ordinary equal-time positive-frequency Klein--Gordon wavefunction, its spatial profile is not compactly supported. It is sharply peaked, but it has Compton-scale tails; more precisely, for large $r$ one finds an asymptotic falloff of the form $e^{-Mr}$ up to dimension-dependent power-law factors. In this sense, the state that the Newton--Wigner framework identifies as ``localized at $\mathbf{x}$'' already has nonzero amplitude arbitrarily far away at $t=0$. Thus the Newton--Wigner construction does not furnish compactly supported physical states in the usual sense.

\medskip
\noindent
\textbf{Why this is inevitable.}
The acausal spreading is not a defect specific to the Newton--Wigner construction. Hegerfeldt's theorem~\cite{Hegerfeldt1974,Hegerfeldt1998} shows that, in a theory with a Hamiltonian bounded below, strict localization and relativistic causal propagation are in tension: if a state is strictly localized in a bounded region at one time, then under time evolution the associated localization probability immediately develops nonzero tails arbitrarily far away. The underlying mechanism is tied to the spectral condition on $H$ and can be understood through Paley--Wiener type analyticity arguments. For our purposes, the important point is simply that sharply localized relativistic one-particle states are not stable under time evolution. We will therefore not rely on any claim of exact compact support within the positive-frequency one-particle Hilbert space.

\medskip
\noindent
\textbf{What we do instead.}
The localization pathologies reviewed above arise when one attempts to use position eigenstates--or any sharply localized states--as physical states. The Newton--Wigner program, and much of the subsequent literature, was motivated by the question ``what does it mean for a relativistic particle to be at a definite position?'' 

We ask a different question: ``When does a smooth wave packet behave like a geodesic?'' For this purpose, position operators serve as tools for computing expectation values, but not as a localization scheme. The eigenstates are simply not our focus.
Concretely, we define the ``position'' of a wave packet via two complementary prescriptions:
\begin{itemize}
    \item A \emph{center-of-mass trajectory} $\bar{x}^\sigma(t)$ defined as the energy-weighted centroid of $\langle T^{tt}\rangle$ (Section~\ref{sec:com}). Here $x^\sigma$ is a classical integration variable---no position operator is needed, and the definition makes sense in any spacetime.
    \item \emph{Position operators} $\hat{\rho}, \hat{\phi}$ constructed from the Sturm--Liouville
completeness of the radial mode functions (Section~\ref{sec:operators}). These operators act on the
positive-frequency one-particle Hilbert space appropriate to the AdS scalar field, in a role
analogous to that of the Newton--Wigner operator in flat space. Their distributional
eigenstates should therefore be viewed with the same general caution familiar from
relativistic localization: within a positive-frequency one-particle framework, exact sharp
localization is not expected to define stable physical states under time evolution. The
difference is entirely in how we use these operators: we compute expectation values in
smooth, normalizable wave packets and never require the eigenstates themselves to serve as
physical states.
\end{itemize}
For sufficiently localized smooth wave packets, the expectation values under both prescriptions are well-defined and do not rely on acausal sharply localized states. As we demonstrate in detail below, they track classical geodesics up to variance corrections that are suppressed by the wave packet width.

In our framework, the tension between localization and relativistic dynamics shows up as a
restriction on how sharply a wave packet can behave semiclassically over time. In the AdS
examples we study, once the packet becomes too narrow relative to the scale set by its
energy, the variance grows, the semiclassical approximation deteriorates, and the centroid
trajectory begins to deviate appreciably from the corresponding geodesic. The scale $1/E$
provides a useful diagnostic for this crossover in the examples below, but it should not be
interpreted as a universal sharp bound. We investigate this breakdown quantitatively in
Section~\ref{sec:breakdown}. Thus the tension between localization and relativity is not ``resolved'' but
controlled: it sets the regime of validity of our construction rather than producing any
paradox.

\section{Geodesic Equation from the Stress Tensor Operator}\label{sec:com}

In this and the subsequent section, we develop two distinct methodologies for defining and tracking the macroscopic trajectory of a quantum wave packet in curved spacetime. We first construct a framework based on the stress-energy tensor operator of the quantum field, to define a natural notion of position {\em expectation value}. In the next section we will present an approach that more directly involves the definition of a position {\em operator}\footnote{Even though the stress tensor operator approach leads to a well-defined position expectation value, it does not involve the definition of a position operator.}, but it should be emphasized that we are {\em not} following the Newton-Wigner path: we are not trying to use eigenstates of the position operator as physical states representing a particle at a definite location. In subsequent sections, we will perform explicit theoretical calculations and numerical simulations of both approaches for single particle states of a free scalar field in AdS$_{3}$, identifying the state profiles that correspond to various classical geodesics. 

For a general spacetime metric written in the ADM formalism,
\bea\label{ADM metric}
ds^{2} = -N_{\Sigma}^{2}\,dt^{2}+\sigma_{ab}(dx^{a}+N_{\Sigma}^{a}\,dt)(dx^{b}+N_{\Sigma}^{b}\,dt)
\eea
we can define the mass/energy ``operator'' of the field as:
\bea\label{mass}
M = \int_{\Sigma} dV \sqrt{\sigma}N_{\Sigma}n_{\mu}n_{\nu}T^{\mu\nu}.
\eea
Here $T^{\mu\nu}$ is the stress tensor of the field theory living on the above background geometry. We will work with free scalar field theory in this paper, but more general discussions are possible where the quantized particle of the field carries spin or charge labels. The spatial volume element is $dV = d^{d-1}x$,\,$\Sigma$ is a constant time slice hypersurface, $\sigma$ is the determinant of the induced metric on $\Sigma$ and $n^{\mu}$ is the timelike unit normal to $\Sigma$. Note that the derivation below does not require the spacetime to have a timelike Killing vector. If spacetime has a timelike Killing vector, then the above definition of $M$ coincides with the conserved Noether energy associated to the (scalar) field.

Given a state in the Hilbert space of the scalar field, we define the center of mass of the stress tensor expectation value distribution as\footnote{These definitions connect naturally with the definition of the center of mass in flat Minkowski spacetime -- they are related to boost charges in that limit. We thank Nirmal Raj for emphasizing this to us.}:
\bea\label{COM def}
\bar{x}^{\sigma} = \frac{\int_{\Sigma}dV \sqrt{\sigma}N_{\Sigma}\,x^{\sigma}\,n_{\mu}\langle T^{\mu\nu}\rangle n_{\nu}}{\int_{\Sigma}dV\sqrt{\sigma}N_{\Sigma}n_{\mu}\langle T^{\mu\nu}\rangle n_{\nu}}
\eea
Using the ADM metric \eqref{ADM metric}, \eqref{COM def} reduces to:
\bea\label{xbar T00}
\bar{x}^{\sigma} = \frac{\int_{\Sigma}dV x^{\sigma}\langle \mathcal{T}^{00}(x)\rangle N_{\Sigma}^{2}(x)}{\int_{\Sigma}dV\langle \mathcal{T}^{00}(x)\rangle N_{\Sigma}^{2}(x)}
\eea
where we have defined $\mathcal{T}^{\mu\nu} = \sqrt{|g|}\,\,T^{\mu\nu}$. It is worth pointing out here that an expectation value evaluation in the state, is present in the denominator as well. In other words, the LHS {\em cannot} be viewed simply as the expectation value of some position {\em operator}: we are instead using the ``center of mass'' of the state to define its location via $\langle T^{\mu\nu} \rangle$. We are of course also dropping the demand for strict localization. Note that there is nothing that forces us to require that the wave packet should be an eigenstate of a position operator. Rather, we track the physical distribution of its energy and momentum.

To show that $\bar{x}^{\sigma}$ satisfies the geodesic equation, we use the following property:
\bea
\nabla_{\mu}\langle\mathcal{T}^{\mu\nu}(x)\rangle = 0
\eea
This gives us the following identity:
\begin{align}
&\,\,\,\,\,\,\,\,\,\,\,\,\,\,\,\partial_{\mu}(\langle\mathcal{T}^{\mu\nu}\rangle)+\Gamma^{\nu}_{\alpha\beta}\langle\mathcal{T}^{\alpha\beta}\rangle=0 \label{identity1}\\
&\implies\partial_{\mu}\langle\tilde{\tau}^{\mu\nu}\rangle-\partial_{\mu}(N_{\Sigma}^2)\langle\mathcal{T}^{\mu\nu}\rangle+\Gamma^{\nu}_{\alpha\beta}\langle\tilde{\tau}^{\alpha\beta}\rangle=0\label{identity2}
\end{align}
where $\tilde{\tau}^{\mu\nu}=N_{\Sigma}^2\mathcal{T^{\mu\nu}}$.
By employing this identity and performing the calculations detailed in Appendix \ref{appendix_geo_from_stress_tensor}, we obtain:
\begin{align}\label{gd eqn}
\ddot{\bar{x}}^i + \bar{\Gamma}^{i}_{\alpha\beta} \dot{\bar{x}}^\alpha \dot{\bar{x}}^\beta - \bar{\Gamma}^{0}_{\alpha\beta} \dot{\bar{x}}^\alpha \dot{\bar{x}}^\beta \dot{\bar{x}}^i = \frac{1}{M^{00}} \left[ \dot{\bar{x}}^i \left( \bar{\Gamma}^{0}_{\alpha\beta} \epsilon^{\alpha\beta} + \delta\Gamma^0 \right) - \dot{\epsilon}^i - \bar{\Gamma}^{i}_{\alpha\beta} \epsilon^{\alpha\beta} - \delta\Gamma^i \right].
\end{align}
The result requires some explanation. Here $M^{00}$ is the $00$-component of $M^{\mu\nu}$, the macroscopic moment defined as $M^{\mu\nu} = \int \langle \mathcal{T}^{\mu\nu}\rangle d^3x$ and $\bar{\Gamma}^{\alpha}_{\mu\nu}=\Gamma^{\alpha}_{\mu\nu}(\bar{x})$. The inner moments of the deviations (which act as error terms on the right-hand side of the above equation) are defined as:
\begin{itemize}
    \item $\epsilon^{\alpha \beta} = \dot{\bar{x}}^\beta \epsilon^{\alpha} + \dot{D}^{\beta \alpha 0} + \bar{\Gamma}^{\alpha}_{\mu\nu} D^{\beta\mu\nu} + \delta\Gamma^{\beta \alpha}$
    \item $\epsilon^{0}=0\,\, \text{and}\,\,\epsilon^i \equiv\epsilon^{0i} = \dot{D}^{i00} + \bar{\Gamma}^{0}_{\alpha\beta} D^{i\alpha\beta} + \delta\Gamma^{i0}$
    \item $D^{\mu \alpha\beta} \equiv \int X^\mu \, \langle\mathcal{T}^{\alpha\beta}\rangle d^3x$ \quad (Dipole moments)
    \item $\delta\Gamma^\mu \equiv \int \delta\Gamma^{\mu}_{\alpha\beta} \, \langle\mathcal{T}^{\alpha\beta}\rangle d^3x$ \quad (Connection deviation integral)
    \item $\delta\Gamma^{\lambda\mu} \equiv \int X^\lambda \, \delta\Gamma^{\mu}_{\alpha\beta} \, \langle\mathcal{T}^{\alpha\beta}\rangle d^3x$ \quad (Dipole connection deviation)
\end{itemize}
where $X^{\mu}\,=\,x^{\mu} -\bar{x}^{\mu}$ and $\delta\Gamma^{\mu}_{\alpha\beta}(x)=\Gamma^{\mu}_{\alpha\beta}(x) -\bar{\Gamma}^{\mu}_{\alpha\beta}$. To see how these terms arise, see Appendix \ref{appendix_geo_from_stress_tensor}. Note that $X^{0}=0$ because of the definition \eqref{xbar T00} where the integration happens on a constant-time hypersurface.

The left-hand side of Eq. \eqref{gd eqn} is precisely the geodesic equation when coordinate time is chosen as the parameter, and the right-hand side contains the error terms. Note that this result is valid for very general background metrics, and not just (say) stationary spacetimes. In the derivation above, we made no assumption about the state with respect to which the expectation values are being calculated. So, the result holds for both single-particle as well as multi-particle states. The only criterion for the geodesic equation to emerge for the center of mass is the suppression of the additional terms on the RHS. For a background with generic Christoffel symbols (like curved spacetime or even flat space in curvilinear coordinates) we expect the terms in the RHS to be suppressed, only if the state is spatially localized\footnote{It should be clear that flat space in Cartesian coordinates where all $\Gamma_{\alpha\beta}^{\mu}$ are 0, satisfies the geodesic equation for the center of mass irrespective of the localization properties of the state.}. This is loosely the content of the ``monopole approximation'' in the classical analogue of this problem -- the so-called Mathisson-Papapetrou-Dixon framework in classical general relativity.  We will formulate a systematic discussion of the variance of the state after the introduction of the position operator language of the next section. 

\section{Position Operators in \texorpdfstring{AdS$_{3}$}{AdS3}}\label{sec:operators}

Complementing the stress-energy approach established in the previous section, we now introduce our second method: the explicit construction of position operators acting directly on the Hilbert space. 

At first glance, construction of position operators acting on the Hilbert space might appear to go against the spirit of the no-go theorem of Wightman \cite{Wightman:1962}, Hegerfeldt \cite{Hegerfeldt1974,Hegerfeldt1998}, and others \cite{Steinmann1968,Schweber:1961,Kalnay1971,Gerlach1968,Gerlach1969,Gromes1970} regarding relativistic localization. However, the point is that we do not aim to view eigenstates of these position operators as localized physical states -- the latter is what leads to conceptual problems. We use position operators only to compute expectation values in smooth wave packets, which is a  different enterprise. We outline our approach in the concrete setting of AdS$_3$, noting that it can be naturally generalized to higher-dimensional AdS spacetimes. 

We work with the massive scalar field $\Phi(t,\rho,\phi) \sim \frac{1}{\sqrt{2\pi}}e^{-i\omega t}e^{im\phi}R_{nm}(\rho)$ in the AdS$_3$ background \eqref{metric1}. The differential equation for the radial part of the field becomes:
\bea\label{Diff equation}
-\frac{d}{d\rho}\Big[\tan(\rho)\frac{dR_{nm}(\rho)}{d\rho}\Big]+\tan(\rho)\Big[\frac{m^{2}}{\sin^{2}(\rho)}+\frac{M^{2}\mathcal{R}^{2}}{\cos^{2}(\rho)}\Big]R_{nm}(\rho) = \omega^{2}\tan(\rho)R_{nm}(\rho).
\eea
This is a standard Sturm-Liouville problem with orthonormality 
\bea\label{orthogonality relation}
\int_{0}^{\frac{\pi}{2}}d\rho\,\tan(\rho)R_{nm}^{*}(\rho)R_{n'm}(\rho) = \delta_{nn'},
\eea
and completeness
\bea\label{completeness relation}
\sum_{n=0}^{\infty}\tan(\rho)R_{nm}^{*}(\rho)R_{nm}(\rho') = \delta(\rho-\rho').
\eea
These relations can be drived directly from the Klein-Gordon inner product (see Appendix \ref{appendix_recipe_for_positiooperators_static}).
The explicit form of the field $\Phi(\rho,\phi,t)$ (that is consistent with the canonical commutation relation) and $R_{nm}(\rho)$ is given in \eqref{mode expansion1} and \eqref{Form of Rnmrho that is used} respectively. 

With these, we can construct position operators, $\hat{\rho}$ and $\hat{\phi}$:
\bea\label{demand}
\hat{\rho}|\rho,\phi\rangle = \rho|\rho,\phi\rangle, \qquad \hat{\phi}|\rho,\phi\rangle = \phi|\rho,\phi\rangle
\eea
The position eigenstates $|\rho,\phi\rangle=a^{\dagger}(\rho,\phi)|0\rangle$, with $a(\rho,\phi)$ and its Fourier transform $a_{nm}$ defined as:
\begin{align}
&a(\rho,\phi) = \frac{1}{\sqrt{2\pi}}\sum_{n,m}R_{nm}(\rho)e^{im\phi}a_{nm}\label{annihilation pos}\\  &a_{nm} = \frac{1}{\sqrt{2\pi}}\int d\rho\, d\phi\, \tan(\rho)\,R_{nm}^{*}(\rho)e^{-im\phi}a(\rho,\phi)\label{annihilation mom}
\end{align}
The $a_{nm}$'s are the Fourier coefficients in the mode expansion of $\Phi(\rho,\phi,t)$ (see eqn. \eqref{mode expansion1}) which satisfy $[a_{nm},a^{\dagger}_{n'm'}]=\delta_{nn'}\delta_{mm'}$ .
With this commutation relation in hand, one can show that $a(\rho,\phi)$'s satisfy:
\bea\label{commutation relation}
[a(\rho,\phi),a^{\dagger}(\rho',\phi')]=\frac{1}{\tan(\rho)}\delta(\rho-\rho')\delta(\phi-\phi')
\eea
It is easy to see that if we define $\hat{\rho}$ and $\hat{\phi}$ as:
\begin{align}
\hat{\rho} &= \int d\rho\,d\phi\,\tan(\rho)\,a^{\dagger}(\rho,\phi)\rho\,a(\rho,\phi)\label{rho operator} \\
\hat{\phi} &= \int d\rho\,d\phi\,\tan(\rho)\,a^{\dagger}(\rho,\phi)\phi\,a(\rho,\phi)\label{phi operator}
\end{align}
then the equation \eqref{demand} is satisfied. 

The above procedure for defining position operators can be generalized to more general spacetimes. We discuss this in Appendix \ref{appendix_recipe_for_positiooperators_static} for a class of static spacetimes. Note that our earlier center-of-mass definition in eqn.\eqref{COM def} is valid for any spacetime\footnote{More precisely, we expect the approach to work in any spacetime in which quantum field theory in curved spacetime and the associated definition of the stress tensor operator make sense.}.

Our definition of $\hat{\phi}$ operator is a bit misleading, because it does not take into account the periodicity of $\phi$. In the next section, we will see how to define $\hat{\phi}$ so that the periodicity of $\phi$ becomes manifest.

\section{Expectation Values of \texorpdfstring{$\hat{\rho}$}{rho-hat}, \texorpdfstring{$\hat{\phi}$}{phi-hat} and \texorpdfstring{$\hat{T}^{00}$}{phi-hat}}\label{sec:EV_of_rho_phi_Ttt}

In this section, we explicitly compute the expectation values of $\hat{\rho}$ and $\hat{\phi}$, defined in \eqref{rho operator} and \eqref{phi operator}, respectively, for generic single-particle states. We also compute the expectation value of $\hat{T}^{tt}$ in order to evaluate $\bar{x}^{\sigma}$ as defined in \eqref{xbar T00}. To proceed, we need to specify the single particle state at $t=0$ which in the momentum basis is of the form:
\bea\label{generic single particle state at t=0}
|\Psi(0)\rangle=\sum_{n,m}g(n,m)a^{\dagger}_{nm}|0\rangle
\eea
This state is normalized i.e $\langle\Psi(0)|\Psi(0)\rangle=1$. The evolution of a state $|\Psi\rangle$ is given by the equation:
\bea
i\frac{\partial}{\partial t}|\Psi(t)\rangle\,=\,\hat{H}\,|\Psi(t)\rangle
\eea
In our case, the Hamiltonian operator is given by \eqref{Ham_op_in_AdS3}. From the above equation, we get the time evolved packet profile to be $g(n,m,t)=g(n,m)e^{-i\omega_{nm}t}$. This gives our time evolved state as:
\begin{align}\label{generic state}
|\Psi(t)\rangle = \sum_{n,m}g(n,m)e^{-i\omega_{nm}t}a^{\dagger}_{nm}|0\rangle = \int d\rho\,d\phi\,\tan\rho \,f(\rho,\phi,t)a^{\dagger}(\rho,\phi)|0\rangle
\end{align}
where $f(\rho,\phi,t)$ is the time-dependent position-space profile whose relation with $g(n,m,t)$ should be derived. Using \eqref{annihilation pos} in \eqref{generic state} we obtain the following relation:
\begin{align}\label{gnm}
g(n,m)e^{-i\omega_{nm}t} = \frac{1}{\sqrt{2\pi}}\int d\rho\,d\phi\,\tan(\rho)\,R^{*}_{nm}(\rho)f(\rho,\phi,t)e^{-im\phi}
\end{align}
The orthornomality and completeness relations \eqref{orthogonality relation} and \eqref{completeness relation} yield \footnote{The equation below means $f(\rho,\phi,t)=\langle\rho,\phi|\Psi(t)\rangle$. One can also show that $\langle n,m|\Psi(t)\rangle\,=\,g(n,m,t)$, where $|n,m\rangle=a^{\dagger}_{nm}|0\rangle$.}:
\begin{align}\label{f}
f(\rho,\phi,t) = \frac{1}{\sqrt{2\pi}}\sum_{n,m}g(n,m)e^{-i\omega_{nm}t}R_{nm}(\rho)e^{im\phi}
\end{align}
We put $t=0$ in \eqref{gnm} and substitute the expression of $g(n,m)$ into \eqref{f}, which finally leads to:
\begin{align}\label{f(rho,phi,t)}
f(\rho,\phi,t) = \frac{1}{2\pi}\sum_{n,m}\int d\rho'\,d\phi'\,\tan(\rho')\,R_{nm}(\rho)R^{*}_{nm}(\rho')e^{im(\phi-\phi')}e^{-i\omega_{nm}t}f(\rho',\phi',0)
\end{align}
Using the commutation relation \eqref{commutation relation}, it is easy to see that the expectation value of $\hat{\rho}$ in the state $|\Psi(t)\rangle$ becomes:
\begin{align}\label{EV eqn}
\hspace{-0.5cm}\langle\Psi|\hat{\rho}|\Psi\rangle = \sum_{n,m,n',m'}\int d\rho\,d\phi\,\tan(\rho)\rho\, g^{*}(n',m',t)g(n,m,t)\langle0|a_{n'm'}a^{\dagger}(\rho,\phi)a(\rho,\phi)a^{\dagger}_{nm}|0\rangle
\end{align}

Substituting \eqref{annihilation pos} into \eqref{EV eqn} and using the commutation relation $[a_{nm},a^{\dagger}_{n'm'}]=\delta_{nn'}\delta_{mm'}$, we find:
\begin{align}\label{EV rho}
\langle\hat{\rho}\rangle = \int d\rho\,d\phi\,\tan(\rho)\,\rho\,|f(\rho,\phi,t)|^{2}
\end{align}
where $f(\rho,\phi,t)$ is defined in \eqref{f(rho,phi,t)}. A similar calculation applies to the expectation value of $\hat{\phi}$ in the state $|\Psi(t)\rangle$. We will simply state the result here:
\begin{align}\label{EV phi}
\langle\hat{\phi}\rangle = \int d\rho\,d\phi\,\tan(\rho)\,\phi\,|f(\rho,\phi,t)|^{2}
\end{align}
As noted in the previous section, the definition of $\hat{\phi}$ does not account for the $2\pi$-periodicity of the angular coordinate. Consequently, direct evaluation of $\langle\hat{\phi}\rangle$ does not give the numerically correct value. Instead, we use the following periodic definition:
\bea\label{EV phi new}
\langle \widehat{e^{i\phi}}\rangle = \int d\rho\,d\phi\,\tan(\rho) e^{i\phi}|f(\rho,\phi,t)|^{2}
\eea
We get this expression by using the Taylor expansion of $e^{i\phi}$ and noting that $\langle\hat{\phi^{n}}\rangle$ has the same form as \eqref{EV phi} with $\phi$ replaced by $\phi^{n}$ inside the integral.\footnote{Equations \eqref{EV rho}, \eqref{EV phi}, and \eqref{EV phi new} can also be derived using the fact that $ \hat{A} = \int d\rho\, d\phi\, \tan(\rho)\, |\rho,\phi\rangle \langle \rho,\phi|$ acts as the identity operator, i.e., $\hat{A}|\rho',\phi'\rangle = |\rho',\phi'\rangle$. One can insert this identity into expressions such as $ \langle \Psi | h(\rho,\phi) | \Psi \rangle = \langle \Psi | \hat{\mathbb{I}}\, h(\rho,\phi) | \Psi \rangle,$ where $h(\rho,\phi)$ is an arbitrary function that has a Taylor expansion. Using Eq.~\eqref{demand} and the relation $f(\rho,\phi,t) = \langle \rho,\phi | \Psi(t) \rangle$, the above equations follow.} The expectation value of $\hat{\phi}$ is obtained by taking the argument of \eqref{EV phi new}. We will use \eqref{EV rho} and \eqref{EV phi new} to do numerics with the operator formalism.

Now let us go to the stress-tensor approach. We are interested in calculating the expectation value of $\hat{T}^{00}$ in a generic single particle state whose initial profile is $g(n,m)$. We will work in the Heisenberg picture where the operator $\hat{T}^{00}$ evolves not the state $|\Psi\rangle$.

For the scalar field $\Phi$ the stress-energy tensor is given by:
\bea\label{Stress Tensor for free scalar field}
\hat{T}^{\mu\nu} = \partial^{\mu}\Phi\partial^{\nu}\Phi - \frac{1}{2}g^{\mu\nu}(\partial^{\lambda}\Phi\partial_{\lambda}\Phi - M^{2}\Phi^{2} )
\eea
Here $M$ is the mass of the scalar field. The state at $t=0$ is (put $t=0$ in equation \eqref{generic state}) given by:
\bea\label{state}
|\Psi\rangle\equiv|\Psi(t=0)\rangle = \sum_{n,m}g(n,m)  a^{\dagger}_{nm}|0\rangle = \int d\rho' d\phi' \tan(\rho') f_{0}(\rho',\phi') a^{\dagger}(\rho',\phi')|0\rangle
\eea
The expectation value of $\hat{T}^{00}$ in $|\Psi\rangle$ becomes:
\bea
\hspace{-2mm}\langle\hat{T}^{00}\rangle = \hspace{-7mm}\sum_{n_{1},m_{1},n_{2},m_{2}}\hspace{-3mm}\frac{A(n_{1},m_{1};n_{2},m_{2}) }{4\pi\sqrt{\omega_{n_{1}m_{1}}\omega_{n_{2}m_{2}}}} e^{-i(\omega_{n_{1}m_{1}}-\omega_{n_{2}m_{2}})t}e^{i(m_{1}-m_{2})\phi}\Big(\hspace{-1mm}\langle a_{n_{1}m_{1}}a^{\dagger}_{n_{2}m_{2}}\rangle\hspace{-1mm} + \hspace{-1mm}\langle a^{\dagger}_{n_{2}m_{2}}a_{n_{1}m_{1}}\rangle\Big)
\eea
Here $\langle... \rangle$ denotes the expectation value. See Appendix \ref{appendix_scalar_field_ads3} for explicit calculation. Using the following identities:
\begin{align}
\langle a_{n_{1}m_{1}}a^{\dagger}_{n_{2}m_{2}}\rangle &= \langle a^{\dagger}_{n_{2}m_{2}}a_{n_{1}m_{1}}\rangle + \langle\delta_{n_{1}n_{2}}\delta_{m_{1}m_{2}}\rangle \nonumber\\
\langle a^{\dagger}_{n_{2}m_{2}}a_{n_{1}m_{1}}\rangle &= g^{*}(n_{2},m_{2})g(n_{1},m_{1}) \nonumber \\
\langle\delta_{n_{1}n_{2}}\delta_{m_{1}m_{2}}\rangle &= \delta_{n_{1}n_{2}}\delta_{m_{1}m_{2}}\sum_{n_{3},m_{3}}|g(n_{3},m_{3})|^{2} = \delta_{n_{1}n_{2}}\delta_{m_{1}m_{2}}
\end{align}
where in the last equation, we used the identity $\langle\Psi|\Psi\rangle =1,$ the expression for $\langle\hat{T}^{00}\rangle$ becomes:
\begin{align}
\hspace{-3mm}\langle\hat{T}^{00}\rangle &= \sum_{n_{1},m_{1},n_{2},m_{2}}\frac{A(n_{1},m_{1};n_{2},m_{2})}{2\pi\sqrt{\omega_{n_{1}m_{1}}\omega_{n_{2}m_{2}}}}g^{*}(n_{2},m_{2})g(n_{1},m_{1}) e^{-i(\omega_{n_{1}m_{1}}-\omega_{n_{2}m_{2}})t}e^{i(m_{1}-m_{2})\phi} + \nonumber\\
&\hspace{9cm}+ \sum_{n_{1},m_{1}}\frac{A(n_{1},m_{1};n_{1},m_{1})}{4\pi\omega_{n_{1}m_{1}}}
\end{align}
where $A(n_{1},m_{1};n_{2},m_{2})$ is defined in Appendix \ref{appendix_scalar_field_ads3}. The second term in the above expression is time-independent and therefore does not influence the temporal behavior of $\langle \hat{T}^{00} \rangle$. We may thus subtract this constant contribution. The rationale is analogous to subtracting the infinite zero-point energy in the Hamiltonian of a free scalar field. In the same way, this constant term does not arise when working with the normal-ordered stress-energy tensor $\hat{T}^{\mu\nu}$. We can write the expression of $\langle\hat{T}^{00}\rangle$ as:
\bea\label{T00 expectation value}
\langle\hat{T}^{00}\rangle = I + II + III + IV
\eea
where $I, II, III$ and $IV$ are again defined in Appendix \ref{appendix_scalar_field_ads3}. We put this expression of $\langle\hat{T}^{00}\rangle$ in equation \eqref{xbar T00},
\bea\label{xbar-AdS3}
\bar{x}^{\sigma} = \frac{\int d\rho\,d\phi\,x^{\sigma}\langle\mathcal{T}^{00}\rangle \sec^{2}(\rho)}{\int d\rho\,d\phi\,\langle\mathcal{T}^{00}\rangle \sec^{2}(\rho)}
\eea
We will use this equation with $x^{\sigma}=\rho$ and $e^{i\phi}$ to do numerics with the stress tensor formalism. As before, the angle is extracted by taking the argument of the centroid of $e^{i\phi}$.

For a single-particle state, the spatial probability density $|f(\rho, \phi, t)|^2 \tan(\rho)$ in the position operator formalism takes the same significance as the normalized energy density $\langle \mathcal{T}^{00}(x) \rangle$ in the stress tensor approach. 

\section{Implementation in \texorpdfstring{AdS$_{3}$}{AdS3}}\label{sec:AdS3_implementation}

Having constructed two formalisms for tracking quantum wave packets, we now turn to their explicit realization in a scalar field theory on AdS$_{3}$ background. Our primary objective is to numerically evaluate and directly compare the macroscopic trajectories produced by our two parallel frameworks: the  stress-energy center of mass and the position operators. We begin by defining a highly tunable spatial wave packet capable of modeling purely radial, circular, and elliptical-like motion. By systematically varying the scalar mass $M$, the initial conditions and the initial localization widths, we show that a properly tuned wave packet follows the expected classical trajectory with remarkable fidelity, while states lacking the necessary coherence (e.g., a delocalized null packet) fail to exhibit classical motion. Through a series of 3D visual simulations, we demonstrate that both methodologies successfully and consistently recover classical geodesic motion for well-localized states. Furthermore, we explicitly capture the physical breakdown of this classical behavior manifesting as wave packet delocalization and splitting -- when fundamental localization bounds are violated.

\subsection{Choice of the Wave Packet}\label{sec:wp_choice}

To make our framework from previous sections explicit, we must specify the spatial profile $f(\rho, \phi)$ that encodes the initial state of the wave packet. By tailoring its functional form we can prepare wave packets that mimic radial, circular, or elliptical-like motion in the AdS$_3$ background. The general form of the wave packet that we will work with is\footnote{A useful way to motivate the ansatz is to recall the standard Gaussian wave packet in ordinary quantum mechanics,
\[
\psi(x)\propto \exp\!\left[-\frac{(x-x_0)^2}{4\sigma^2}\right]\exp\!\big(i p_0 (x-x_0)\big),
\]
whose modulus is peaked at $x_0$ while the linear phase makes its momentum-space wavefunction peak at $p_0$. Equation~(7.1) is simply the natural AdS$_3$ analogue of this idea in the $(\rho,\phi)$ variables: we choose a Gaussian envelope localized near $(\rho_0,\phi_0)$, and multiply it by phases $e^{-i n_0(\rho-\rho_0)}$ and $e^{-i m_0(\phi-\phi_0)}$ so that the packet is simultaneously localized in position and biased toward definite radial and angular motion. In this sense $n_0$ plays the role of a radial momentum label (more precisely, a proxy for the initial radial momentum), while $m_0$ plays the role of the angular-momentum label. Because $\phi$ is a compact coordinate, the angular Gaussian is made periodic by summing over images $\phi\to \phi+2\pi n$, which yields the explicitly periodic form in~(7.1). In the regime where the packet is well localized away from the identification region near $\phi=0\sim 2\pi$, the image sum is dominated by a single term, and one may use the simpler approximate expression~(7.2).}:
\begin{equation}\label{general choice of packet}
f(\rho,\phi) = \mathcal{N}_{\rho}\,e^{-\frac{(\rho-\rho_{0})^{2}}{4\sigma_{1}^{2}}}e^{-in_0(\rho-\rho_0)}\mathcal{N}_{\phi}\sum_{n=-\infty}^{\infty} e^{-\frac{(\phi-\phi_{0}+2n\pi)^{2}}{4\sigma_{2}^{2}}}e^{-im_0(\phi-\phi_0+2n\pi)}.
\end{equation}
In this expression, $n_{0}$ and $m_{0}$ are real parameters that correspond to the initial radial momenta (as a proxy) and angular momenta respectively (see Appendix \ref{appendix_why_is_the_expval_m_0?} and \ref{appendix_why_is_n_0_initial_radial_mom?}). The radial and angular variances we will denote as $\sigma_1\equiv\sigma_\rho$ and $\sigma_2 \equiv \sigma_\phi$. The state is clearly periodic in $\phi$, but for all practical purposes, as long as we are far from the $(0,2\pi)$ region, this can be approximated to:
\begin{equation}\label{approximate choice of packet}
f(\rho,\phi) = \mathcal{N}_{\rho}\,e^{-\frac{(\rho-\rho_{0})^{2}}{4\sigma_{1}^{2}}}e^{-in_0(\rho-\rho_0)}\mathcal{N}_{\phi}\,e^{-\frac{(\phi-\phi_{0})^{2}}{4\sigma_{2}^{2}}}e^{-im_0(\phi-\phi_0)}
\end{equation}

\subsubsection{Radial Geodesic}\label{sec:wp_choice:radial}
For a purely radial trajectory, the conserved angular momentum $m_{0}$ must vanish. For simplicity, we further assume a vanishing initial radial momentum, $n_{0} = 0$. Under these conditions, the spatial wave packet profile reduces to a decoupled Gaussian form:
\begin{equation}\label{wave packet for radial infall}
f(\rho,\phi) = \mathcal{N}_{\rho}\,e^{-\frac{(\rho-\rho_{0})^{2}}{4\sigma_{1}^{2}}}\mathcal{N}_{\phi}\,e^{-\frac{(\phi-\phi_{0})^{2}}{4\sigma_{2}^{2}}}.
\end{equation}
In this expression, $\mathcal{N}_{\rho}$ and $\mathcal{N}_{\phi}$ denote normalization constants, while $(\rho_{0}, \phi_{0})$ represent parameters specifying the approximate\footnote{For the radial coordinate, the presence of ``$\tan \rho$'' in the KG measure means that $\rho_0$ is slightly different from the initial Gaussian wave packet peak. But this is a tiny effect for $\rho_0$ not too close to $\pi/2$, and we usually launch our states far enough from the boundary. We have also checked the evolution with other shapes of the wave packets, and the results are robust as long as the peaks and spreads are comparable.} initial spatial coordinates at $t = 0$. Under sufficient localization, these parameters closely correspond to the initial expectation values (or the center-of-mass coordinates) of the system. The parameters $\sigma_{1}$ and $\sigma_{2}$ are suitably defined standard deviations that dictate the radial and angular widths of the wave packet, respectively.

\subsubsection{Elliptical-like and Circular Geodesics}\label{sec:wp_choice:ellip_and_cir}
To model orbital motion, we introduce a non-zero angular momentum parameter $m_{0}$ and for convenience keep $n_{0} = 0$. The corresponding wave packet profile is given by:
\begin{equation}\label{wave packet for circular motion}
f(\rho,\phi) = \mathcal{N}_{\rho}\,e^{-\frac{(\rho-\rho_{0})^{2}}{4\sigma_{1}^{2}}}\mathcal{N}_{\phi}\,e^{-\frac{(\phi-\phi_{0})^{2}}{4\sigma_{2}^{2}}}e^{-im_{0}(\phi-\phi_{0})}.
\end{equation}

We will now compute the expectation values of the stress tensor and position operator approaches in these wave packet states. 

\subsection{Stress Tensor Approach}\label{sec:AdS3_implementation_com}

We start with the stress tensor approach and discuss the massive and massless field cases separately. 

Within the AdS$_{3}$ geometry, the specific coordinate expression for $\bar{x}^{\sigma}$ is given by ($\mathcal{T}^{00} = \sqrt{|g|}\,T^{00}$):
\bea\label{rho bar}
\bar{\rho} = \frac{\int d\rho\,d\phi\,\rho\,\langle\mathcal{T}^{00}\rangle \sec^{2}(\rho)}{\int d\rho\,d\phi\,\langle\mathcal{T}^{00}\rangle \sec^{2}(\rho)}
\eea
and
\bea\label{phi bar}
\overline{e^{i\phi}} = \frac{\int d\rho\,d\phi\,e^{i\phi}\,\langle\mathcal{T}^{00}\rangle \sec^{2}(\rho)}{\int d\rho\,d\phi\,\langle\mathcal{T}^{00}\rangle \sec^{2}(\rho)}.
\eea
To evaluate $\langle \mathcal{T}^{00} \rangle(t)$, we need to compute the propagators appearing in Eq.~\eqref{equation for the 4 propagators to calculate T^00}. 

Let us make a comment about the numerical implementation. The truncation parameters $n_{\max}$ and $m_{\max}$ in the Fourier sums (see e.g., Appendix \ref{appendix_why_is_the_expval_m_0?}) are chosen such that the norm of the state remains approximately unity, while the expectation value of the angular momentum operator remains close to $m_0$ throughout the evolution. The wave packets we work with are (often) Gaussians and have excellent convergence properties in Fourier sums.

\subsubsection{Massive Case: Elliptical-like}\label{sec:com:massive_ellip}

\begin{figure}[H]
    \centering

    \begin{subfigure}[b]{0.48\textwidth}
        \includegraphics[width=\linewidth]{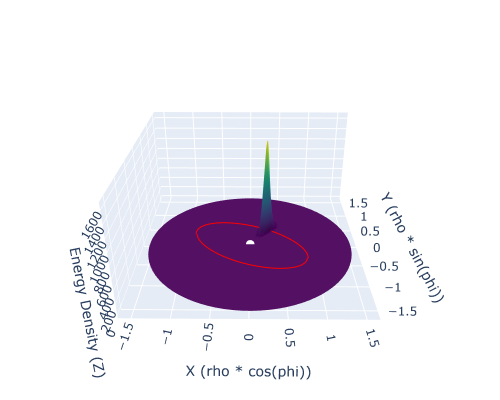}
        \caption{ The packet profile in the beginning. $(t=0.0)$}
    \end{subfigure}\hfill
    \begin{subfigure}[b]{0.48\textwidth}
        \includegraphics[width=\linewidth]{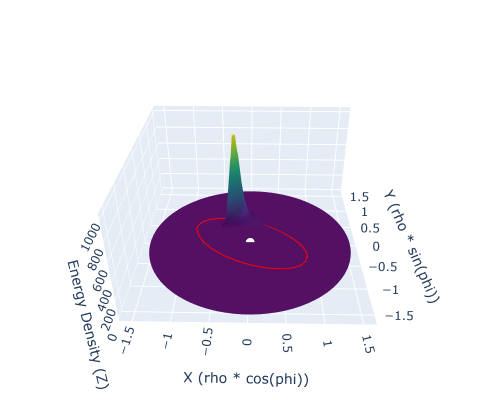}
        \caption{ Packet evolving. $(t=0.6)$}
    \end{subfigure}\hfill

    \begin{subfigure}[b]{0.48\textwidth}
        \includegraphics[width=\linewidth]{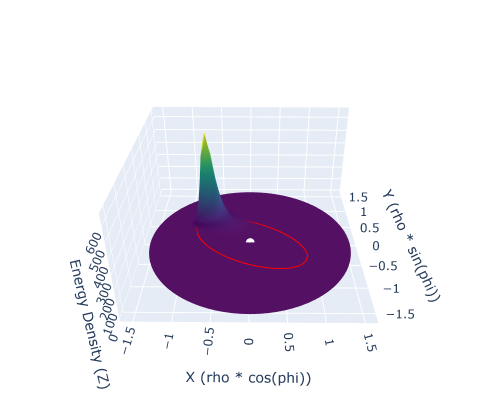}
        \caption{ Packet evolving further. $(t=1.1)$}
    \end{subfigure}\hfill
    \begin{subfigure}[b]{0.48\textwidth}
        \includegraphics[width=\linewidth]{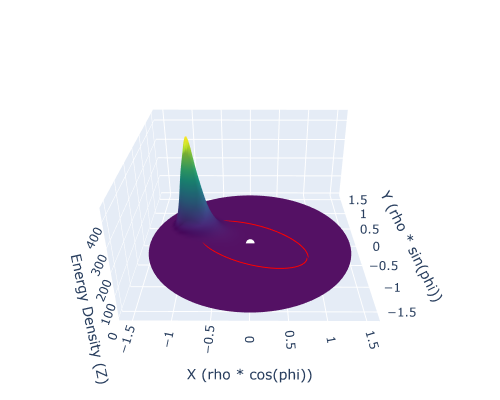}
        \caption{ Profile when the packet is at the maximum radial distance. $(t=\pi/2)$}
    \end{subfigure}\hfill
    
    \caption{Parameters: $M=25, m_{0}=-20,\rho_{0}=0.5,\phi_{0}=\pi/3,\sigma_{1}=0.09,\sigma_{2}=0.05,n_{max}=m_{max}=75$}
    \label{fig:fig1}

\end{figure}

For the elliptical-like geodesics, we choose the profile defined in \eqref{wave packet for circular motion}:
\bea
f(\rho,\phi) = \mathcal{N}_{\rho}\,e^{-\frac{(\rho-\rho_{0})^{2}}{4\sigma_{1}^{2}}}\mathcal{N}_{\phi}\,e^{-\frac{(\phi-\phi_{0})^{2}}{4\sigma_{2}^{2}}}e^{-im_{0}(\phi-\phi_{0})}
\eea
With this choice of profile, we perform numerical evaluations of the trajectories as defined in \eqref{rho bar} and \eqref{phi bar} for various configurations of the parameters $\rho_{0}$, $\phi_{0}$, $\sigma_{1}$, $\sigma_{2}$, and $m_{0}$. 
\begin{itemize}
    \item $M=25, m_{0}=-20$
\end{itemize}
To visualize the evolution of the center of mass, we present the 3D plots in Fig. \ref{fig:fig1}.  The $z$ axis shows the energy/mass distribution. These plots illustrate the dynamics for first quarter of the orbital period, spanning the interval from $t=0$ to $t=\pi/2$ with the choice of parameters stated in the caption. The remaining three-quarters of the evolution exhibit similar behavior. In these plots, the exact classical geodesic is superimposed as a solid red line to facilitate a direct comparison with the numerical evolution of the wave packet.

The plot of $\bar{\rho}(t)$ versus $\bar{\phi}(t)$ for the above case is presented in Appendix~\ref{appendix_2d_plots_com_massive_elliptical}, along with additional examples.

\subsubsection{Massive Case: Radial Infall}\label{sec:com:massive_radial}

For purely radial infall, we choose the profile defined in \eqref{wave packet for radial infall}:
\begin{equation}
f(\rho,\phi) = \mathcal{N}_{\rho}\,e^{-\frac{(\rho-\rho_{0})^{2}}{4\sigma_{1}^{2}}}\mathcal{N}_{\phi}\,e^{-\frac{(\phi-\phi{0})^{2}}{4\sigma_{2}^{2}}}.
\end{equation}

\begin{figure}[H]
    \centering

    \begin{subfigure}[b]{0.48\textwidth}
        \includegraphics[width=0.9\linewidth]{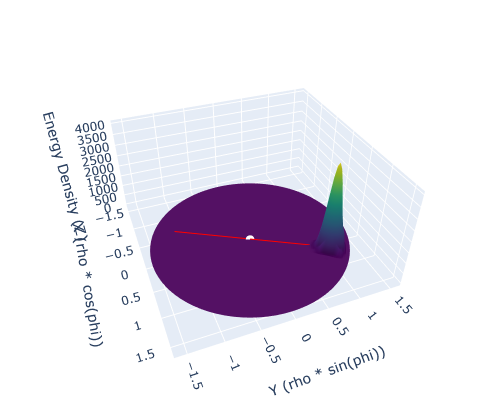}
        \caption{ The packet profile in the beginning. $(t=0.0)$}
    \end{subfigure}\hfill
    \begin{subfigure}{0.48\textwidth}
        \includegraphics[width=0.9\linewidth]{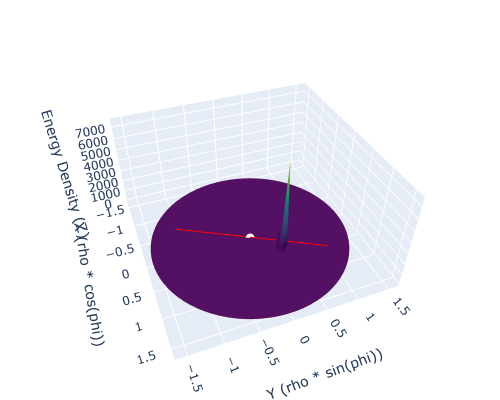}
        \caption{ The packet moves closer to the origin. $(t=1.0)$}
    \end{subfigure}\hfill
    
    \begin{subfigure}[b]{0.48\textwidth}
        \centering
        \includegraphics[width=0.9\linewidth]{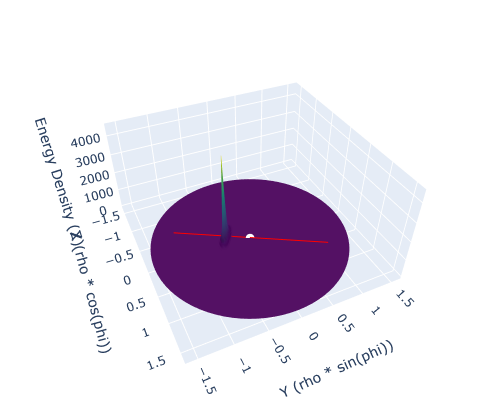}
        \caption{ The profile when packet crosses the origin. $(t=2.0)$}
    \end{subfigure}\hfill
    \begin{subfigure}{0.48\textwidth}
        \centering
        \includegraphics[width=0.9\linewidth]{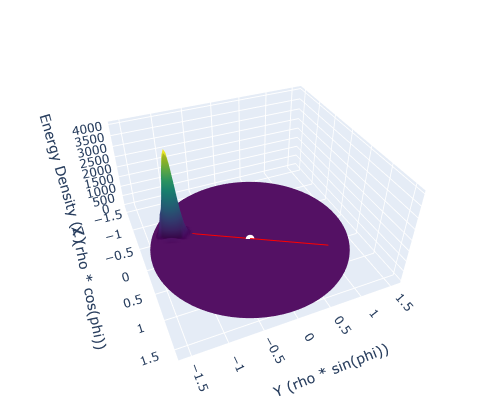}
        \caption{ The packet is now at the antipodal point. $(t=\pi)$}
    \end{subfigure}\hfill
    \caption{Parameters: $\rho_{0}=1.2,\phi_{0}=\pi/3,\sigma_{1}=0.09,\sigma_{2}=0.05,n_{max}=m_{max}=200$, $M=40, m_{0}=0 \hspace{2mm}$}
    \label{fig:fig:com_mass_rad}

\end{figure}

To illustrate the evolution of a radially infalling state, we present 3D plots for a scalar of mass $M=40$. The wave packet is initialized at coordinate time $t = 0$ with $\rho_{0} = 1.2$ and $\phi_{0} = \pi/3$. Fig. \ref{fig:fig:com_mass_rad} illustrate its evolution as the packet propagates through the center of the geometry to the antipodal point at $\phi = 4\pi/3$, which is reached at $t = \pi$. For this specific configuration, the numerically evaluated energy of the wave packet is $E \approx 127.9$.

The plots of $\bar{\rho}(t)$ vs $\bar{\phi}(t)$ corresponding to the above case can be found in Appendix \ref{appendix_2d_plots_com_massive_radial_infall}. Also see Appendix \ref{appendix_2d_plots_com_rho_vs_t} for the $\bar{\rho}(t)$ vs $t$ plots for this and other examples.

\subsubsection{Null Case: Non-radially Infalling Localized Wave Packet}\label{sec:com:non_radial_null_loc}

For the null case we choose $M=0$ with:
\begin{equation}
f(\rho,\phi) = \mathcal{N}_{\rho}e^{-\frac{(\rho-\rho_{0})^{2}}{4\sigma_{1}^{2}}}\mathcal{N}_{\phi}e^{-\frac{(\phi-\phi_{0})^{2}}{4\sigma_{2}^{2}}}e^{-in_{0}(\rho-\rho_{0})}e^{-im_{0}(\phi-\phi_{0})}.
\end{equation}
As before, for convenience, we set $n_{0} = 0$. If we had set say positive value for the radial momentum parameter $n_{0}$, the wave packet would have been initialized to propagate inward from its starting coordinates point.

\begin{figure}[H]
    \centering

    \begin{subfigure}[b]{0.48\textwidth}
        \includegraphics[width=0.9\linewidth]{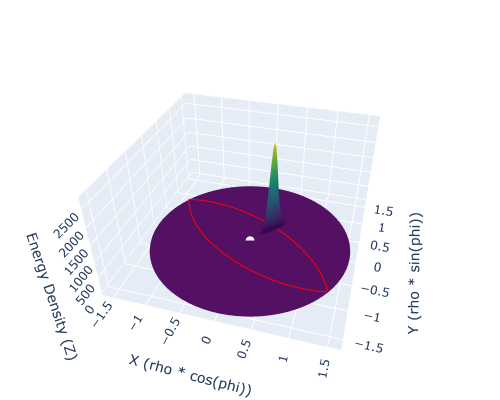}
        \caption{ The massless packet profile in the beginning. $(t=0.0)$}
    \end{subfigure}\hfill
    \begin{subfigure}[b]{0.48\textwidth}
        \includegraphics[width=0.9\linewidth]{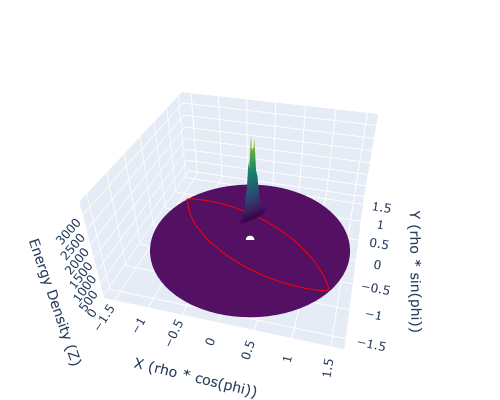}
        \caption{ The packet moving towards the boundary. $(t=0.4)$}
    \end{subfigure}\hfill

    \begin{subfigure}[b]{0.48\textwidth}
        \includegraphics[width=0.9\linewidth]{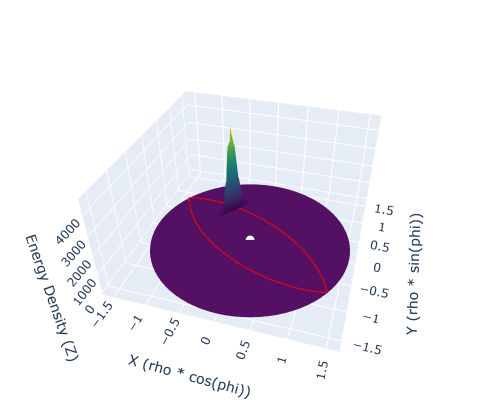}
        \caption{ The packet moving towards the boundary. $(t=0.8)$}
    \end{subfigure}\hfill
    \begin{subfigure}[b]{0.48\textwidth}
        \includegraphics[width=0.9\linewidth]{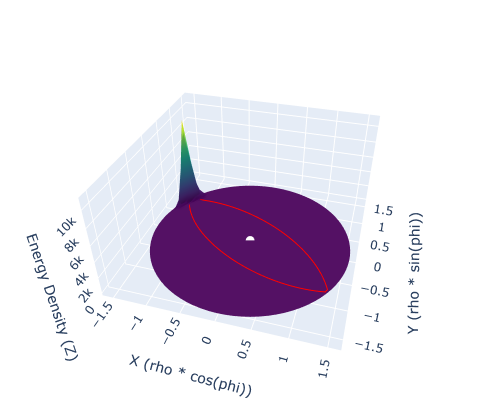}
        \caption{ The packet starts bouncing off from the boundary. $(t=1.60)$}
    \end{subfigure}\hfill

    \caption{Parameters: $\rho_{0}=0.5,\phi_{0}=\pi/3,\sigma_{1}=0.09,\sigma_{2}=0.05,n_{max}=m_{max}=300$,  $M=0, m_{0}=-40 \hspace{2mm}$. The packet reaches the boundary at $t=\frac{\pi}{2}$.}
    \label{fig:fig:com_null_non-rad}
\end{figure}

Fig. \ref{fig:fig:com_null_non-rad} illustrates the 3D trajectory of the wave packet's evolution with the choice of parameters given in the caption. The null scalar goes to the boundary, see \eqref{non_radial_orbit_for_null_case}, and bounces back because of the Dirichlet boundary conditions. The numerically evaluated energy of the wave packet is $E \approx 83$.

The plots of $\bar{\rho}(t)$ vs $\bar{\phi}(t)$ corresponding to the above case can be found in Appendix \ref{appendix_2d_plots_com_null_localized}.

\subsubsection{Null Case: De-localized Wave Packet}\label{sec:com:null_deloc}

\begin{figure}[H]
    \centering
    \begin{subfigure}[b]{0.45\textwidth}
        \includegraphics[width=\linewidth]{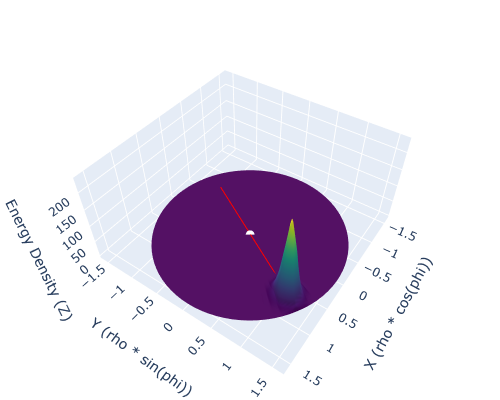}
        \caption{ The massless packet profile in the beginning. $(t = 0.00)$}
    \end{subfigure}\hfill
    \begin{subfigure}[b]{0.45\textwidth}
        \includegraphics[width=\linewidth]{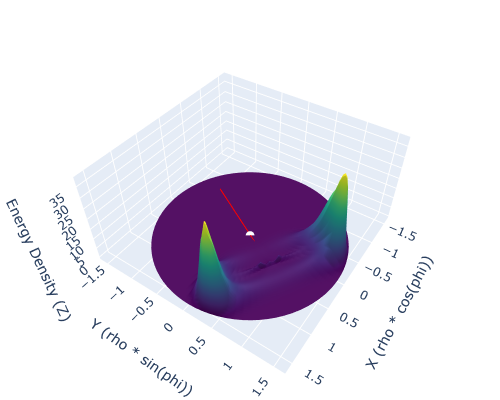}
        \caption{ Onset of delocalization. $(t = 0.90)$}
    \end{subfigure}
    
%\vspace{-1em}

    \begin{subfigure}[b]{0.45\textwidth}
        \includegraphics[width=\linewidth]{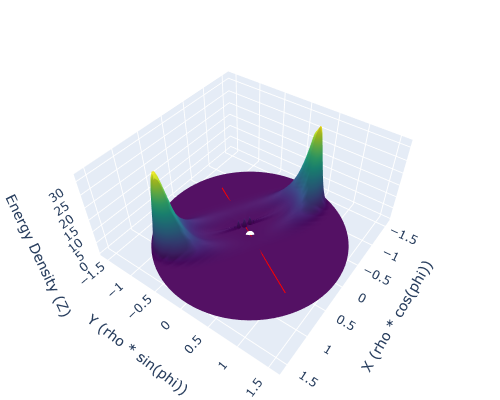}
        \caption{ Packet on the other side. $(t = 1.80)$}
    \end{subfigure}\hfill
    \begin{subfigure}[b]{0.45\textwidth}
        \includegraphics[width=\linewidth]{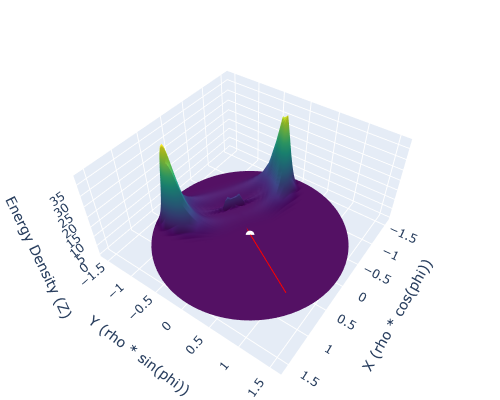}
        \caption{ The packet starts relocalizing. $(t = 2.25)$}
    \end{subfigure}

%\vspace{-1em}

    \begin{subfigure}[b]{0.45\textwidth}
        \includegraphics[width=\linewidth]{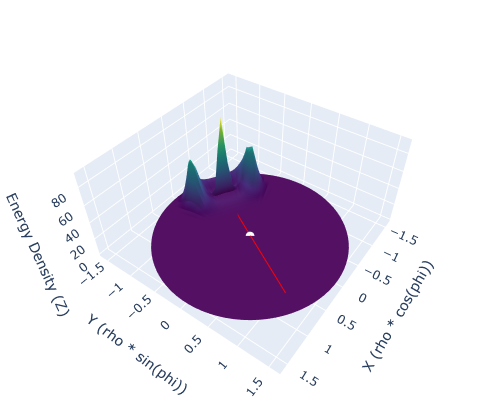}
        \caption{ The packet starts relocalizing. $(t = 2.70)$}
    \end{subfigure}\hfill
    \begin{subfigure}[b]{0.45\textwidth}
        \includegraphics[width=\linewidth]{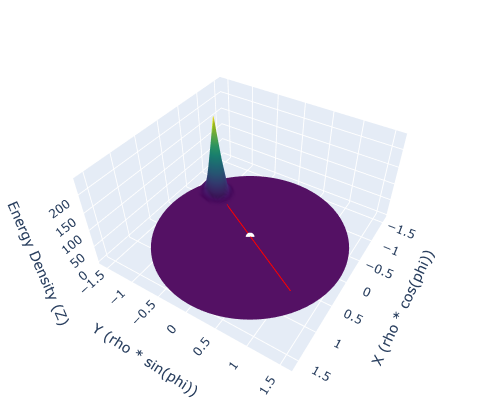}
        \caption{ The massless packet is near the antipodal point. $(t = 3.15)$}
    \end{subfigure}

    \caption{Time evolution of the delocalized wave packet ($M=0$, $m_0=0$, $n_0=0$) 
    with parameters $\rho_0=1.2$, $\phi_0=\pi/3$, $\sigma_1=0.09$, $\sigma_2=0.05$,  
    $n_{\max}=m_{\max}=200$. Panels (a)--(f) show snapshots at increasing times; 
    at $t=\pi$, the packet reaches the antipodal point.}
    \label{fig:delocalized}
\end{figure}

Using the same choice of wave packet as used for radially infalling massive case (see Eq. \eqref{wave packet for radial infall}) results in a highly delocalized evolution for the massless case. Fig. \ref{fig:delocalized} present the 3D plots for the first half of this evolution with the choice of parameters stated in the Figure below.

It is evident from these plots that this specific choice of parameters does not maintain a well-localized state. For this configuration, the numerically evaluated energy of the state is $E \approx 11$. In section \ref{sec:breakdown}, we will see the fundamental reason for this delocalization is the inherent length scale $1/E$. Increasing $n_{0}$ enhances the initial radial momentum, and consequently the energy of the state. As a result, as $n_0$ is increased, the wave packet exhibits localized classical motion that follows the classical trajectory for $\rho(t)$, given by $\rho(t) = \rho_{0} \pm t$. See Appendix~\ref{appendix_2d_plots_com_rho_vs_t_only_radial} for a plot of $\bar{\rho}(t)$ as a function of $t$ in this case, where $n_{0} = 200$, corresponding to $E \approx 200$.

\subsubsection{Circular Geodesic at a Given Radius for Fixed \texorpdfstring{$m_{0}$}{m0}}\label{sec:com:cir}

We now hold the angular momentum parameter fixed at $m_{0}=-30$ to systematically determine the critical scalar mass $M$ required to sustain circular orbits at various initial radii. Our numerical evaluations reveal a clear inverse relationship between the orbital radius and the required mass. Beginning with an initial radial coordinate of $\rho_{0}=0.45$, a nearly exact circular trajectory emerges when the mass is tuned to $M \approx 120$. Extending this analysis to progressively larger radii, we observe that the wave packet trajectory circularizes at correspondingly reduced scalar masses: $M \approx 40$ for $\rho_{0}=0.70$, and $M \approx 17.3$ for $\rho_{0}=0.90$. Continuing this outward progression, the required mass drops to $M \approx 6.5$ at $\rho_{0}=1.10$, and finally to $M \approx 2.75$ at $\rho_{0}=1.20$.

This establishes a clear inverse relationship between the orbital radius and the mass required to maintain a circular trajectory. Extrapolating this trend to the spatial asymptotic limit ($\rho \to \pi/2$) implies that for strictly massless states ($M \to 0$), circular geodesics manifest exclusively at the boundary of the AdS$_{3}$ manifold.

A comprehensive set of these trajectories, along with their specific evolution and parameter configurations, is documented in Appendix \ref{appendix_2d_plots_com_circular_orbits}. From the overlaid plots of the geodesics, it is clear that the numerical state evolution matches the geodesic extremely well.

\subsection{Position operator Approach}\label{sec:pos}

Having analyzed the wave packet dynamics through the stress-energy tensor approach, we now evaluate the macroscopic trajectories using our explicitly constructed position operators. This facilitates a direct, one-to-one comparison between the two theoretical frameworks. In this section, we compute the total energy of the wave packet using the expectation value of the AdS$_{3}$ Hamiltonian in state \eqref{generic state}, which (as it should) matches the energy computed using the stress-tensor approach in the previous section for the same parameter choice (see Appendix \ref{appendix_writing_hamiltonian_and _other_operators_ads3}).

Recall from Section \ref{sec:EV_of_rho_phi_Ttt} that the expectation values for the radial and periodic angular position operators over a generic single-particle state are given by:
\begin{align}\label{expectation values}
\langle\hat{\rho}\rangle &= \int d\rho\, d\phi\, \tan(\rho)\,\rho\,|f(\rho,\phi,t)|^{2} \nonumber\\
\langle \widehat{e^{i\phi}}\rangle &= \int d\rho\, d\phi\, \tan(\rho)\,e^{i\phi}\,|f(\rho,\phi,t)|^{2}
\end{align}
where $f(\rho,\phi,t)$ is defined in \eqref{f(rho,phi,t)} in terms of $f(\rho,\phi)$ as:
\bea
f(\rho,\phi,t) = \frac{1}{2\pi}\sum_{n,m}\int d\rho'\,d\phi'\,\tan(\rho')R_{nm}(\rho)R_{nm}^{*}(\rho')e^{-im(\phi-\phi')}e^{-i\omega_{nm}t}f(\rho',\phi',0)
\eea
As we said before, the initial choice of the wave packet profile $f(\rho,\phi)$ is given by,
\bea
f(\rho,\phi) = \mathcal{N}_{\rho}e^{-\frac{(\rho-\rho_{0})^{2}}{4\sigma_{1}^{2}}}e^{-in_{0}(\rho-\rho_{0})}\mathcal{N}_{\phi}e^{-\frac{(\phi-\phi_{0})^{2}}{4\sigma_{2}^{2}}}e^{-im_{0}(\phi-\phi_{0})}
\eea

\subsubsection{Massive Case: Elliptical-like}\label{sec:pos:massive_ellip}
\begin{figure}[H]
    \centering

    \begin{subfigure}[b]{0.48\textwidth}
        \includegraphics[width=0.8\linewidth]{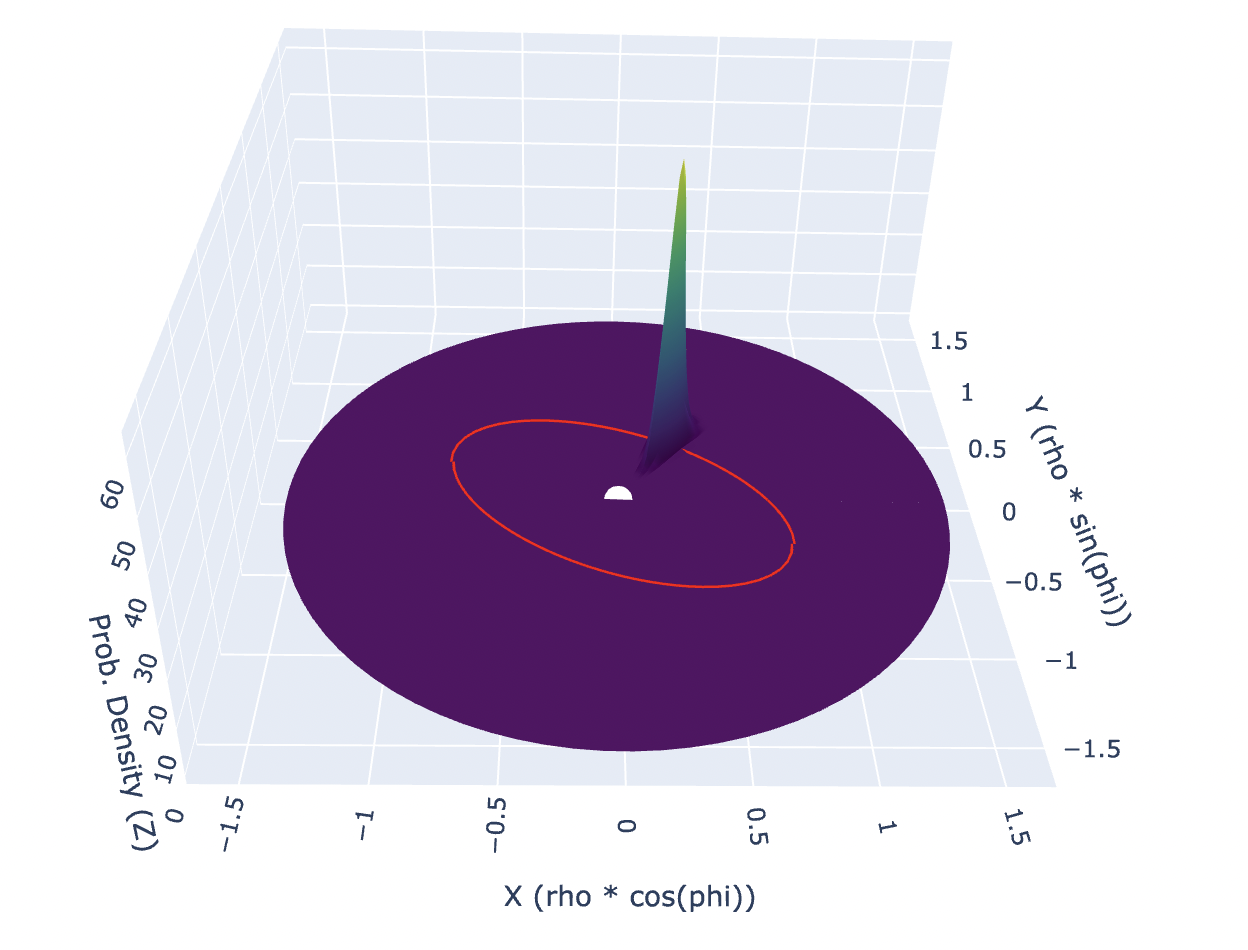}
        \caption{ The packet profile in the beginning. $(t=0.0)$}
    \end{subfigure}\hfill
    \begin{subfigure}[b]{0.48\textwidth}
        \includegraphics[width=0.8\linewidth]{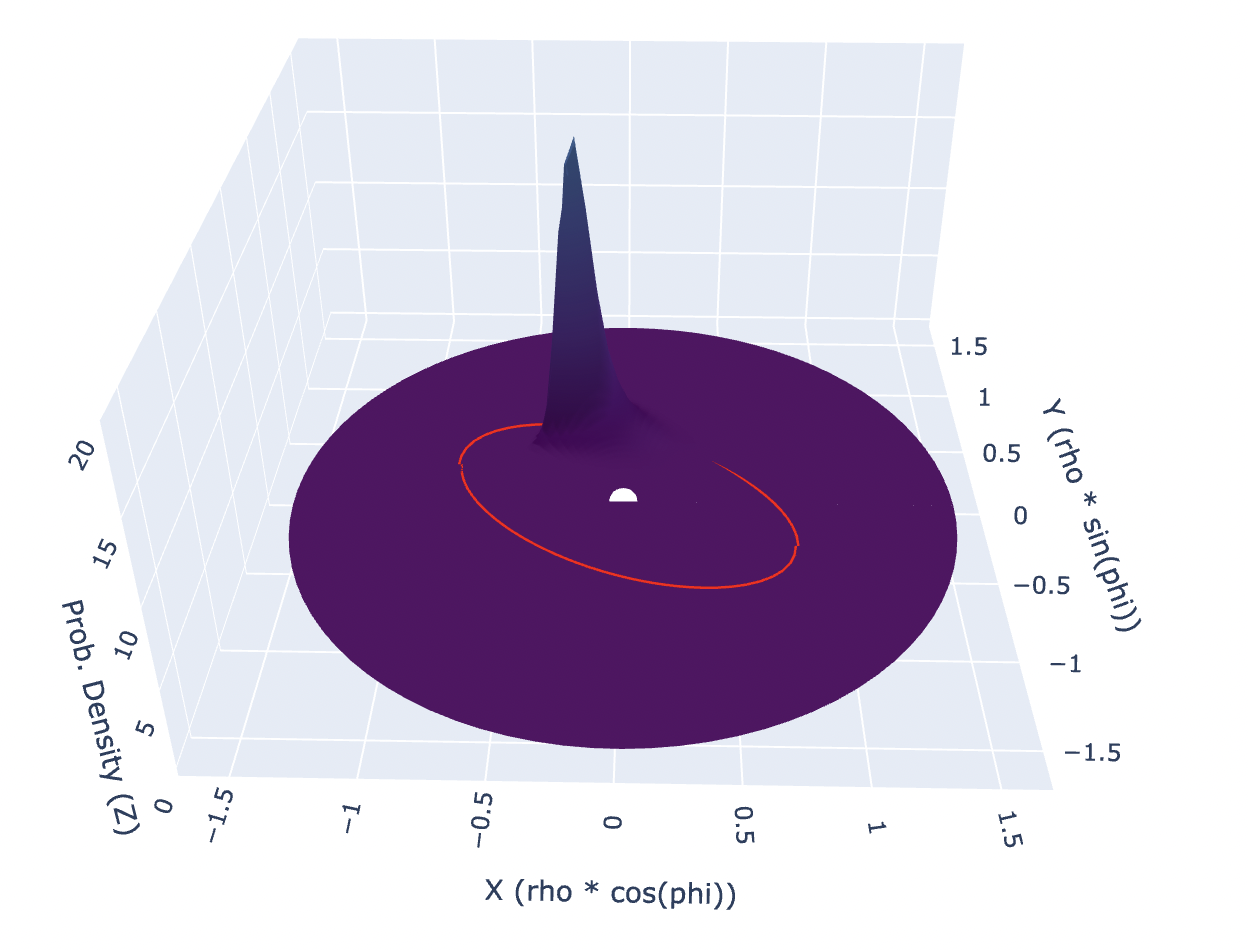}
        \caption{ Packet evolving. $(t=0.6)$}
    \end{subfigure}

    \begin{subfigure}[b]{0.48\textwidth}
        \includegraphics[width=0.8\linewidth]{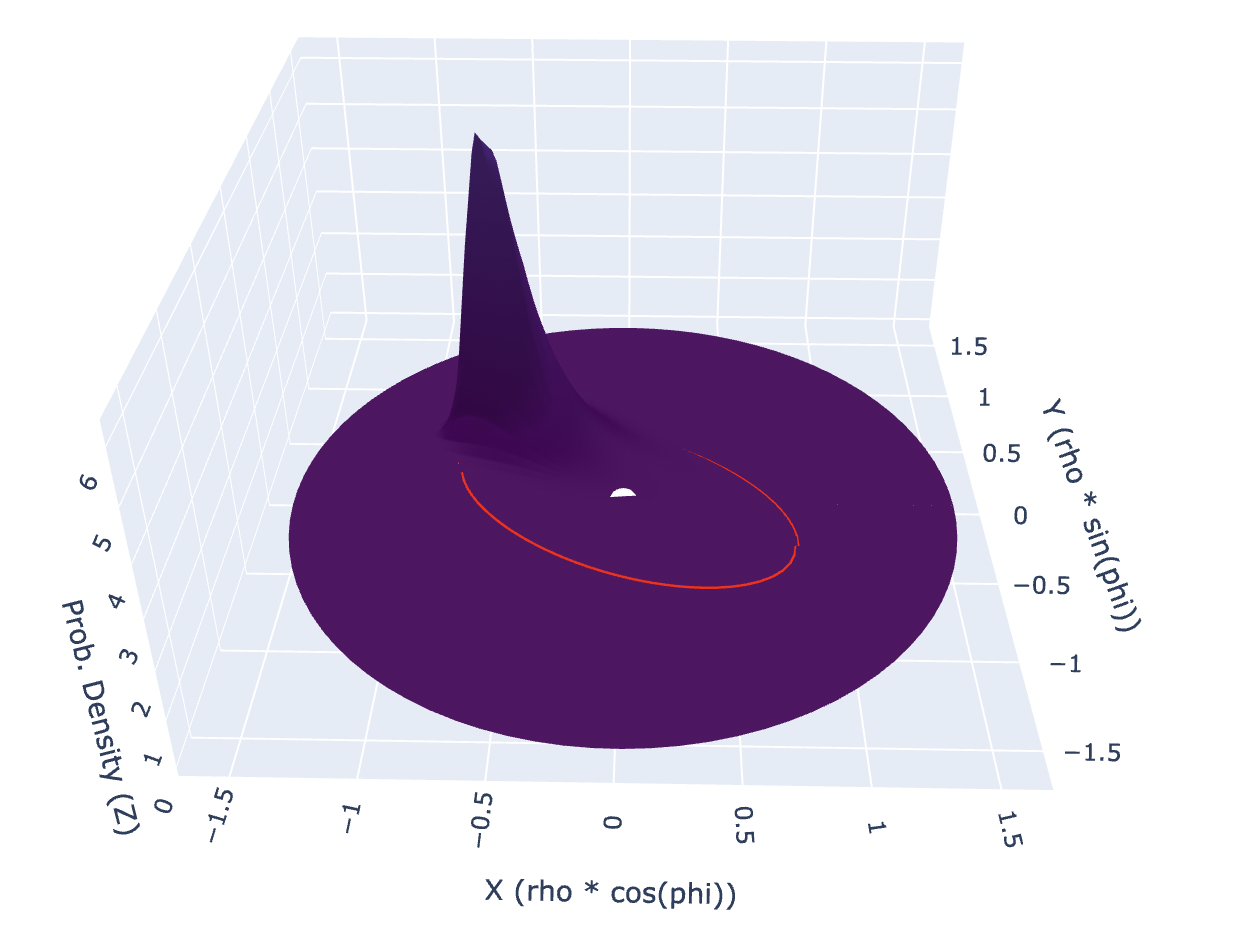}
        \caption{ Packet evolving further. $(t=1.1)$}
    \end{subfigure}\hfill
    \begin{subfigure}[b]{0.48\textwidth}
        \includegraphics[width=0.8\linewidth]{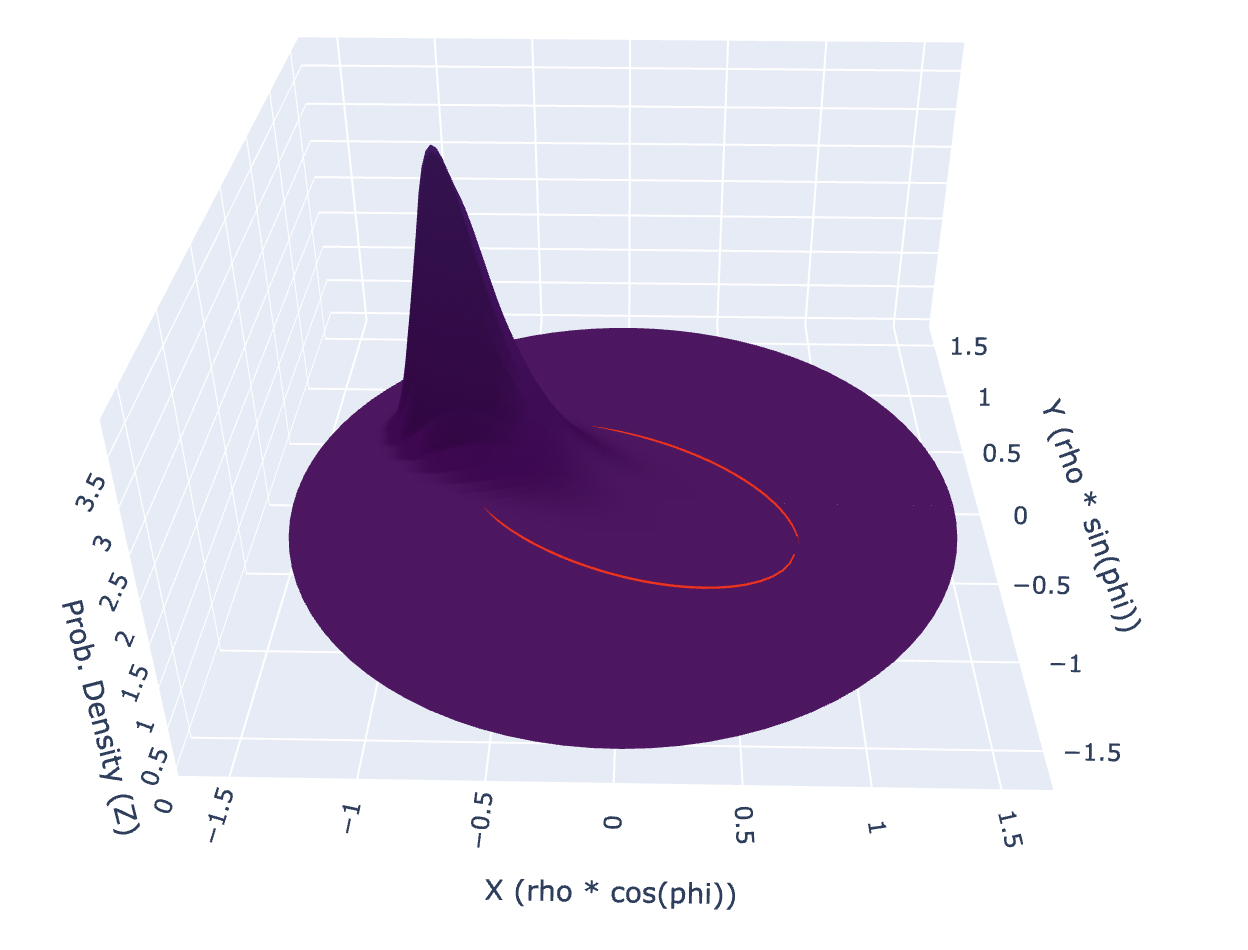}
        \caption{ Profile when the packet is at the maximum radial distance. $(t=\frac{\pi}{2})$}
    \end{subfigure}
    \caption{Parameters: $\rho_{0}=0.5,\phi_{0}=\pi/3,\sigma_{1}=0.09,\sigma_{2}=0.05, n_{max}=m_{max}=75 , M=25, m_{0}=-20,n_{0}=0$.}
    \label{fig:figMass_Ellip}

\end{figure}

For elliptical-type geodesics, we choose the profile defined in \eqref{wave packet for circular motion}:
\bea
f(\rho,\phi) = \mathcal{N}_{\rho}e^{-\frac{(\rho-\rho_{0})^{2}}{4\sigma_{1}^{2}}}\mathcal{N}_{\phi}e^{-\frac{(\phi-\phi_{0})^{2}}{4\sigma_{2}^{2}}}e^{-im_{0}(\phi-\phi_{0})}
\eea
We insert this profile into equations in \eqref{expectation values} and perform the numerical evolution by specifying the initial conditions $(\rho_{0},\phi_{0})$ and the parameters $\sigma_{1},\sigma_{2}, m_{0}$.
\begin{itemize}
    \item \textbf{$M=25, m_{0}=-20$}
\end{itemize}
To visualize the evolution of the probability density 
$\tan(\rho)\,|f(\rho,\phi,t)|^{2}$, we present 3D plots 
for the same case previously analyzed using the 
stress tensor approach:  $M = 25,m_0 = -20$. The subsequent 3D plots (Fig. \ref{fig:figMass_Ellip}) illustrate the first quarter of the orbital period, spanning the time interval from $t=0$ to $t=\pi/2$ with the choice of parameters stated in the figure below. The dynamics during the remaining three quarters of the evolution strictly mirror this behavior. For visual clarity, in the 3D plots below, the exact classical geodesic is overlaid as a solid red curve for direct comparison with the center of the probability density distribution. The peaks of the wave packets precisely coincide with the geodesic trajectories: to emphasize this, we present 2D plots in Appendix \ref{appendix_2d_plots_pos_ope}.  For this specific choice of initial conditions stated in Fig. \ref{fig:figMass_Ellip}, the numerically evaluated energy is $E \approx 53$. 

The plot of $\langle\rho(t)\rangle$ vs $\langle\phi(t)\rangle$ corresponding to this can be found in Appendix \ref{appendix_2d_plots_pos_ope_massive_elliptical} along with other examples.
%\noindent Now, let us look at the radial infall case.

\subsubsection{Massive Case: Radial Infall}\label{sec:pos:massive_radial}

\begin{figure}[H]
    \centering

    \begin{subfigure}[b]{0.48\textwidth}
        \includegraphics[width=0.8\linewidth]{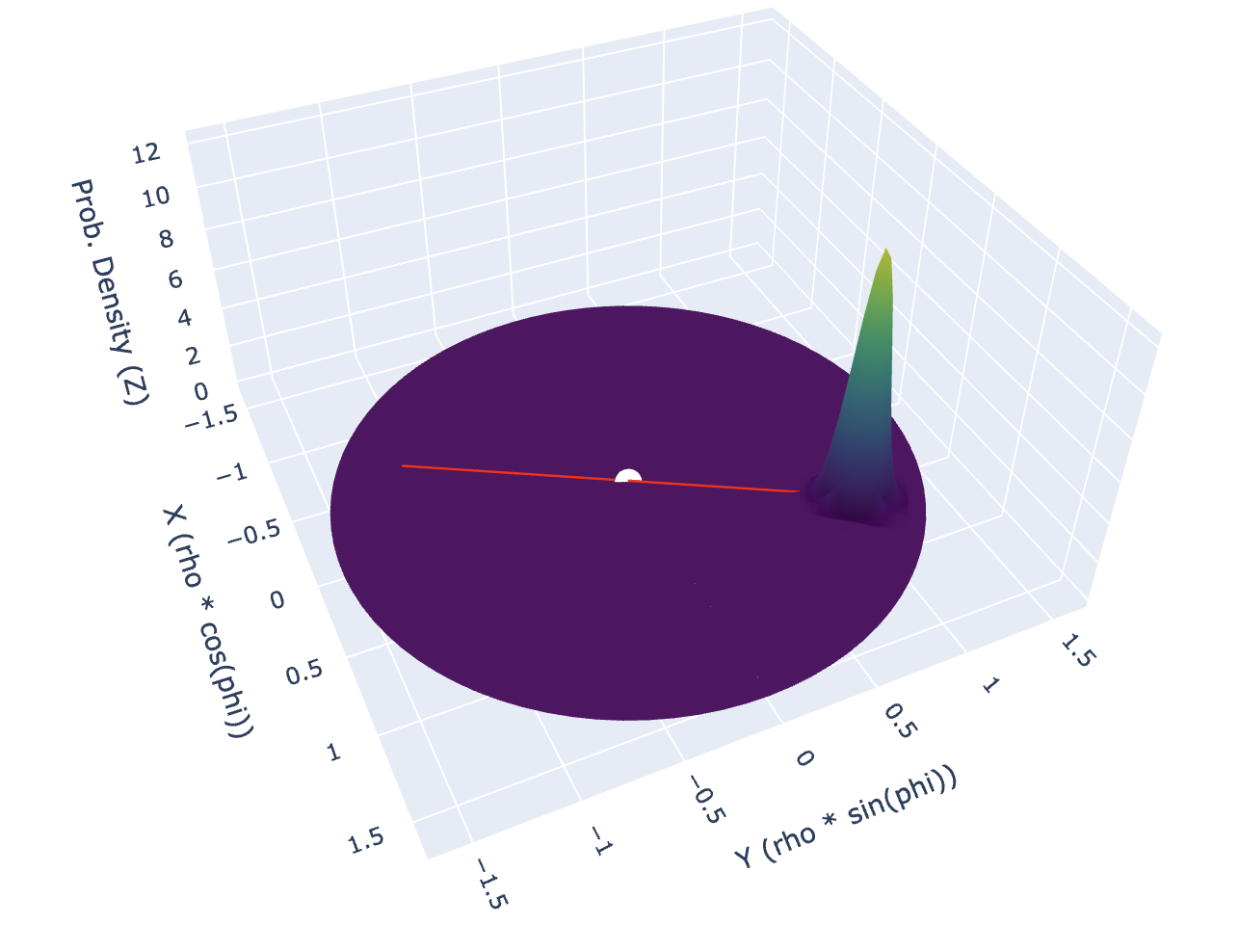}
        \caption{ The packet profile in the beginning. $(t=0.0)$}
    \end{subfigure}\hfill
    \begin{subfigure}[b]{0.48\textwidth}
        \includegraphics[width=0.8\linewidth]{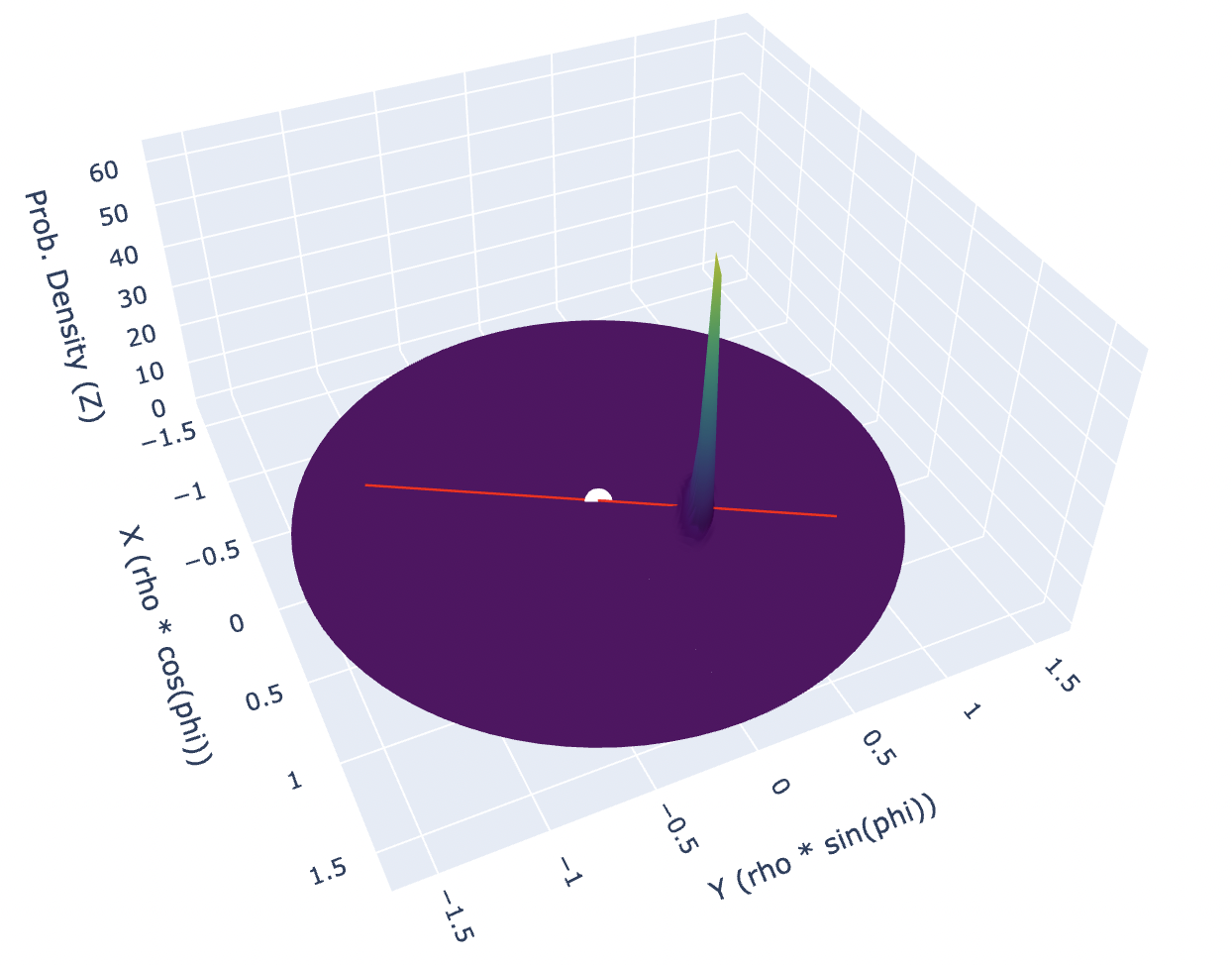}
        \caption{ The packet moves closer to the origin. $(t=1.0)$}
    \end{subfigure}\hfill
    
    \begin{subfigure}{0.48\textwidth}
        \includegraphics[width=0.8\linewidth]{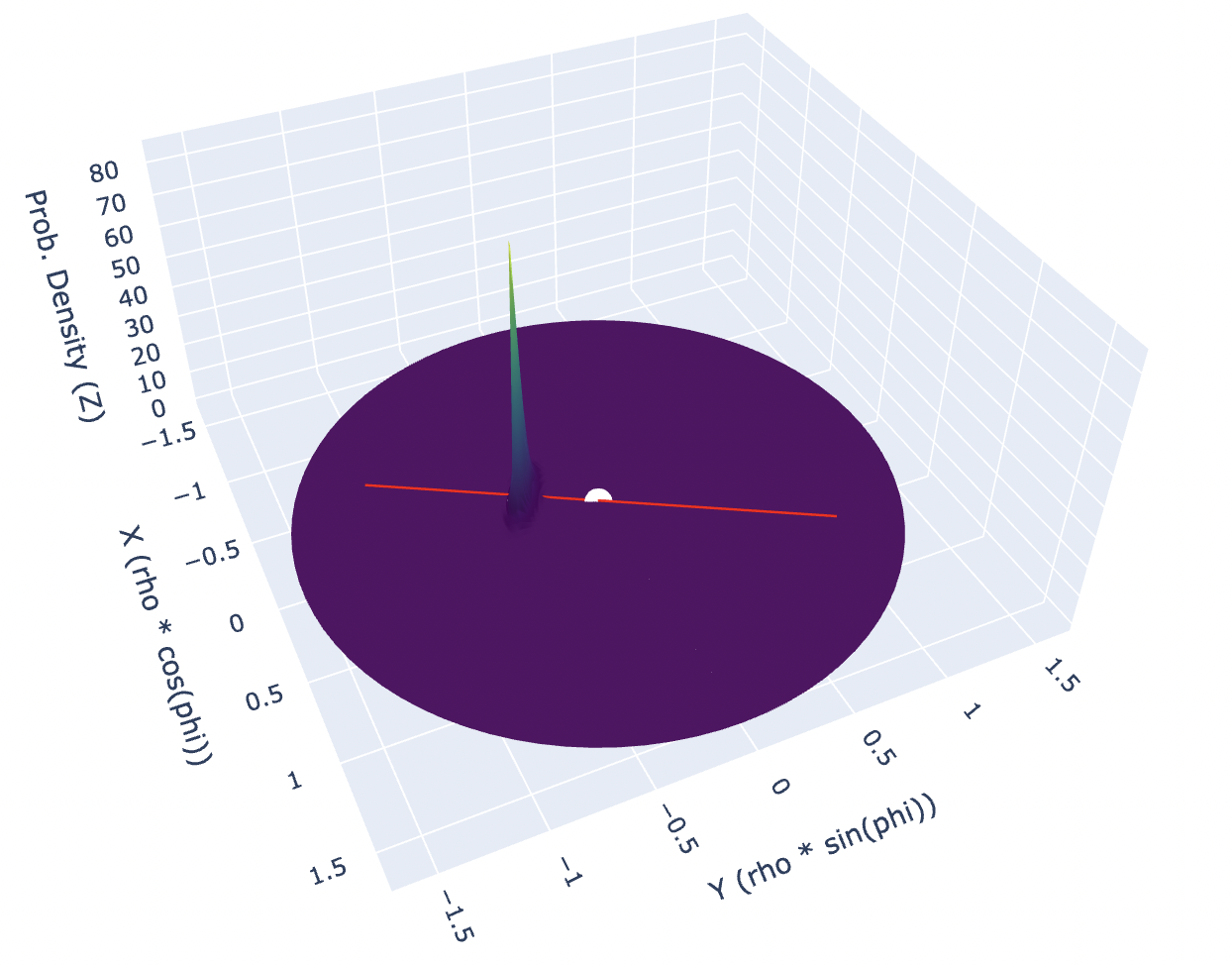}
        \caption{ The profile when packet crosses the origin. $(t=2.0)$}
    \end{subfigure}\hfill
    \begin{subfigure}{0.48\textwidth}
        \includegraphics[width=0.8\linewidth]{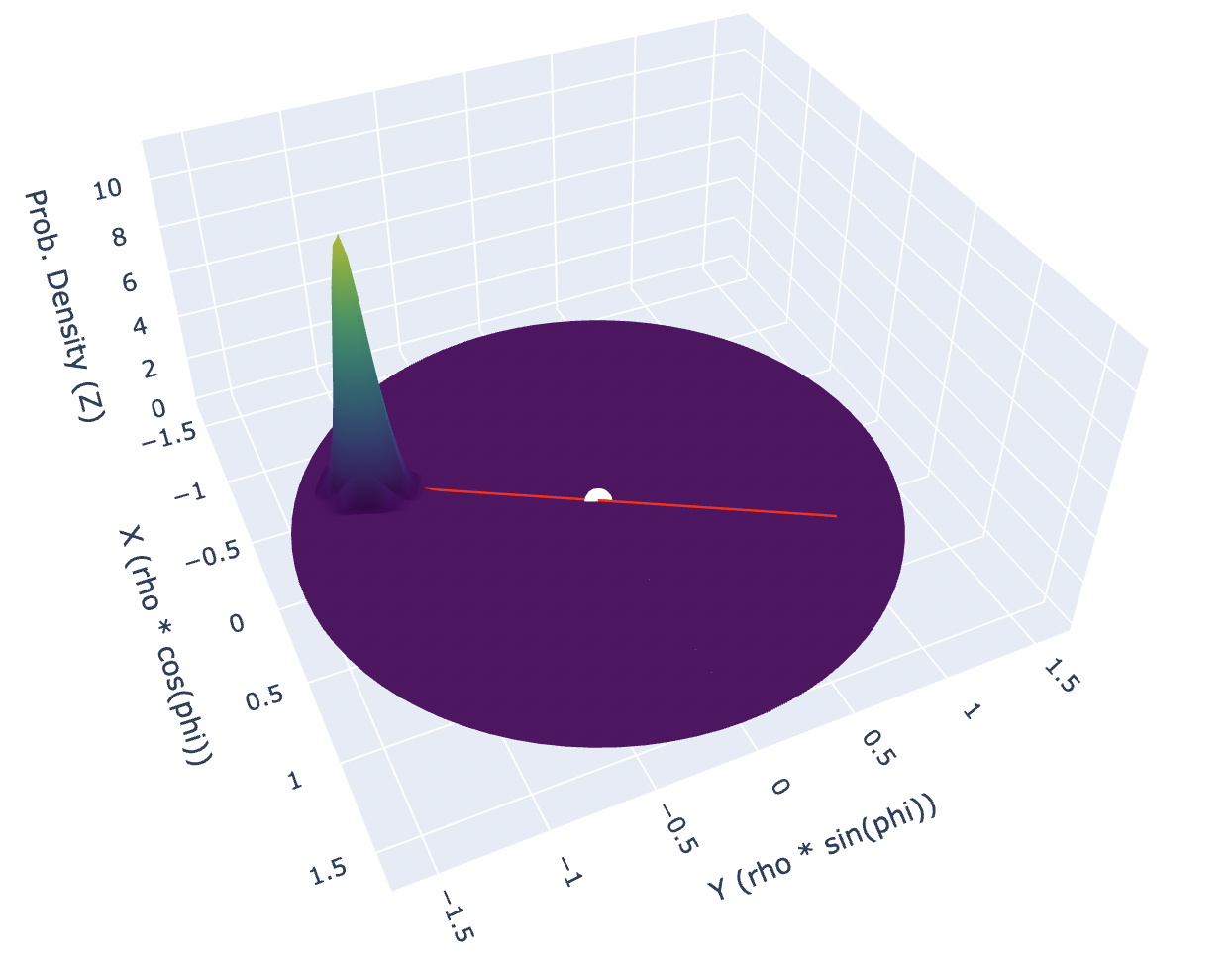}
        \caption{ The packet is now at the antipodal point. $(t=\pi)$}
    \end{subfigure}
    
    \caption{Parameters: $\rho_{0}=1.2,\phi_{0}=\pi/3,\sigma_{1}=0.09,\sigma_{2}=0.05, n_{max}=m_{max}=200 ,M=40, m_{0}=0, n_{0}=0$. The wave packet evolves through the center of the geometry to the antipodal point at $\phi = 4\pi/3$, which is reached at coordinate time $t=\pi$.}
    \label{fig:fig:pos_mass_rad}

\end{figure}

Radial infall we choose the profile defined in \eqref{wave packet for radial infall}:
\begin{equation}
f(\rho,\phi) = \mathcal{N}_{\rho}\,e^{-\frac{(\rho-\rho_{0})^{2}}{4\sigma_{1}^{2}}}\mathcal{N}_{\phi}\,e^{-\frac{(\phi-\phi_{0})^{2}}{4\sigma_{2}^{2}}}.
\end{equation}
Substituting this profile into the expectation value integrals \eqref{expectation values}, we numerically evaluate the probability density evolution by specifying the initial coordinate parameters ($\rho_{0}$, $\phi_{0}$) and the wave packet widths ($\sigma_{1}$, $\sigma_{2}$). To visualize the dynamics of radial infall, we present a 3D plot of the wave packet's trajectory in Fig. \ref{fig:fig:pos_mass_rad}. For the exact parameter configuration stated in Fig. \ref{fig:fig:pos_mass_rad}, the numerically evaluated energy of the wave packet is $E \approx 128$.

The plots of $\langle\rho(t)\rangle$ vs $\langle\phi(t)\rangle$ corresponding to the above case can be found in Appendix \ref{appendix_2d_plots_pos_ope_massive_rad_infall}. See Appendix \ref{appendix_2d_plots_pos_ope_massive_rad_infall_rho_vs_t} for the plots of $\langle\rho\rangle(t)$ vs $t$.

\subsubsection{Null Case: Non-radially Infalling Localized Wave Packet}\label{sec:pos:null:loc_non_radial}
\begin{figure}[H]
    \centering

    \begin{subfigure}[b]{0.48\textwidth}
        \includegraphics[width=0.8\linewidth]{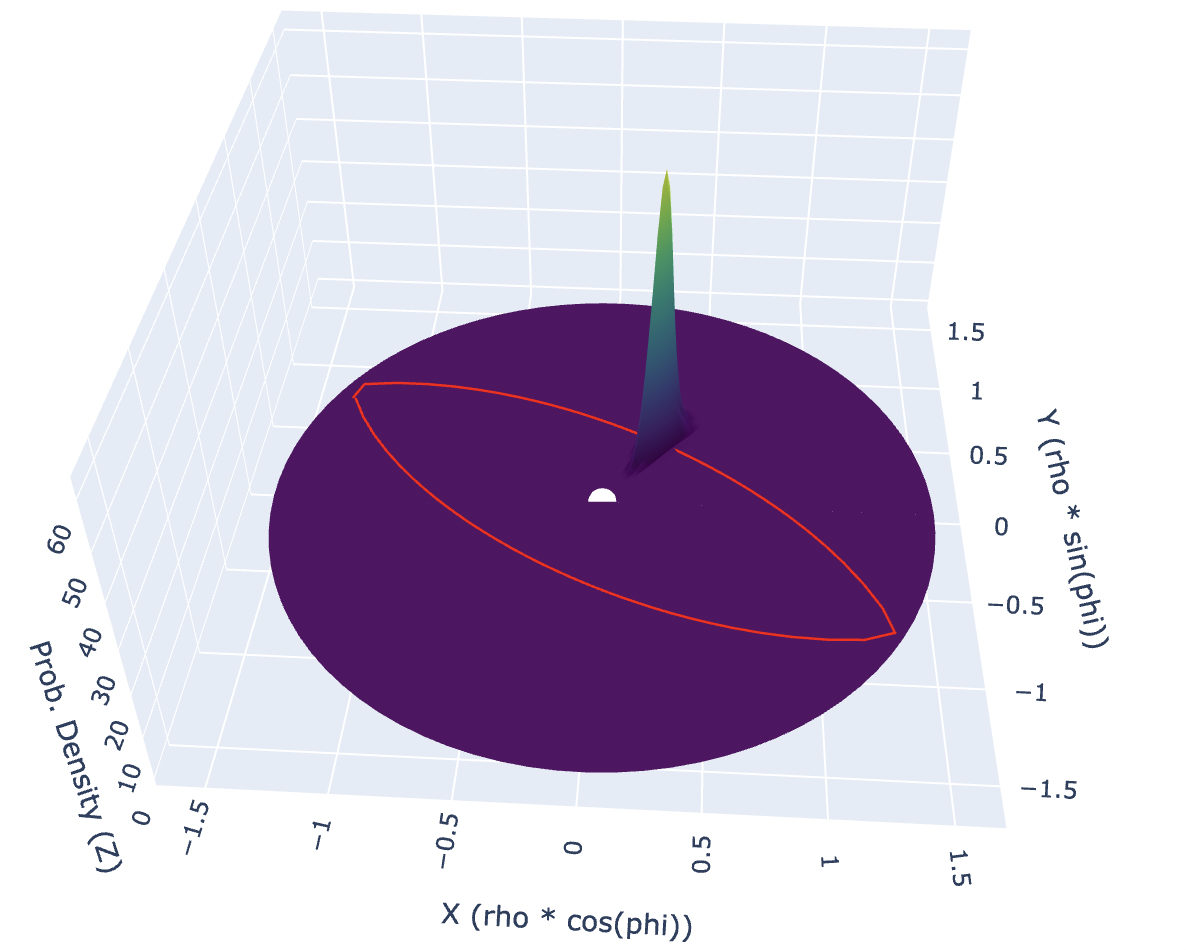}
        \caption{ The massless packet profile in the beginning. $(t=0.0)$}
    \end{subfigure}\hfill
    \begin{subfigure}{0.48\textwidth}
        \includegraphics[width=0.8\linewidth]{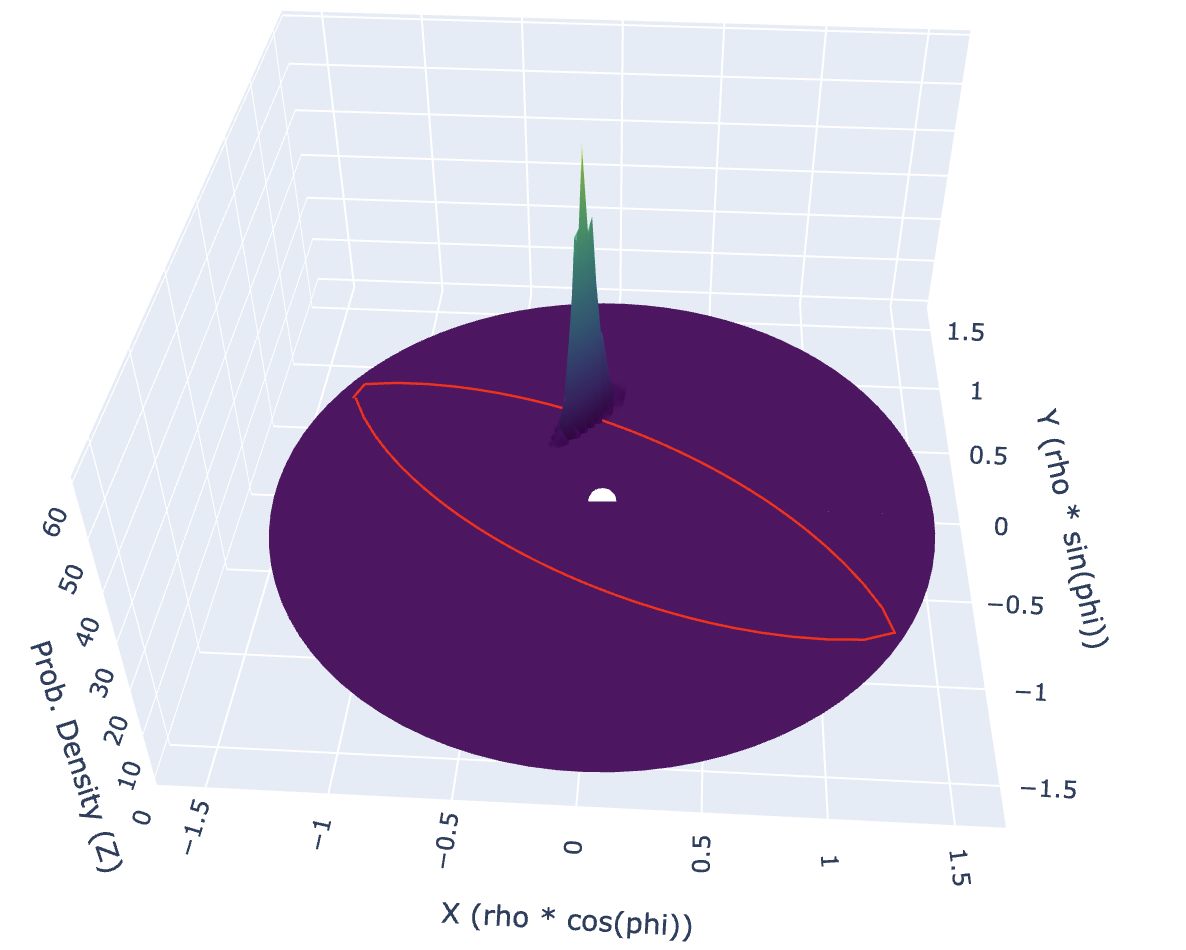}
        \caption{ The packet moving towards the boundary. $(t=0.4)$}
    \end{subfigure}\hfill

    \begin{subfigure}[b]{0.48\textwidth}
        \includegraphics[width=0.8\linewidth]{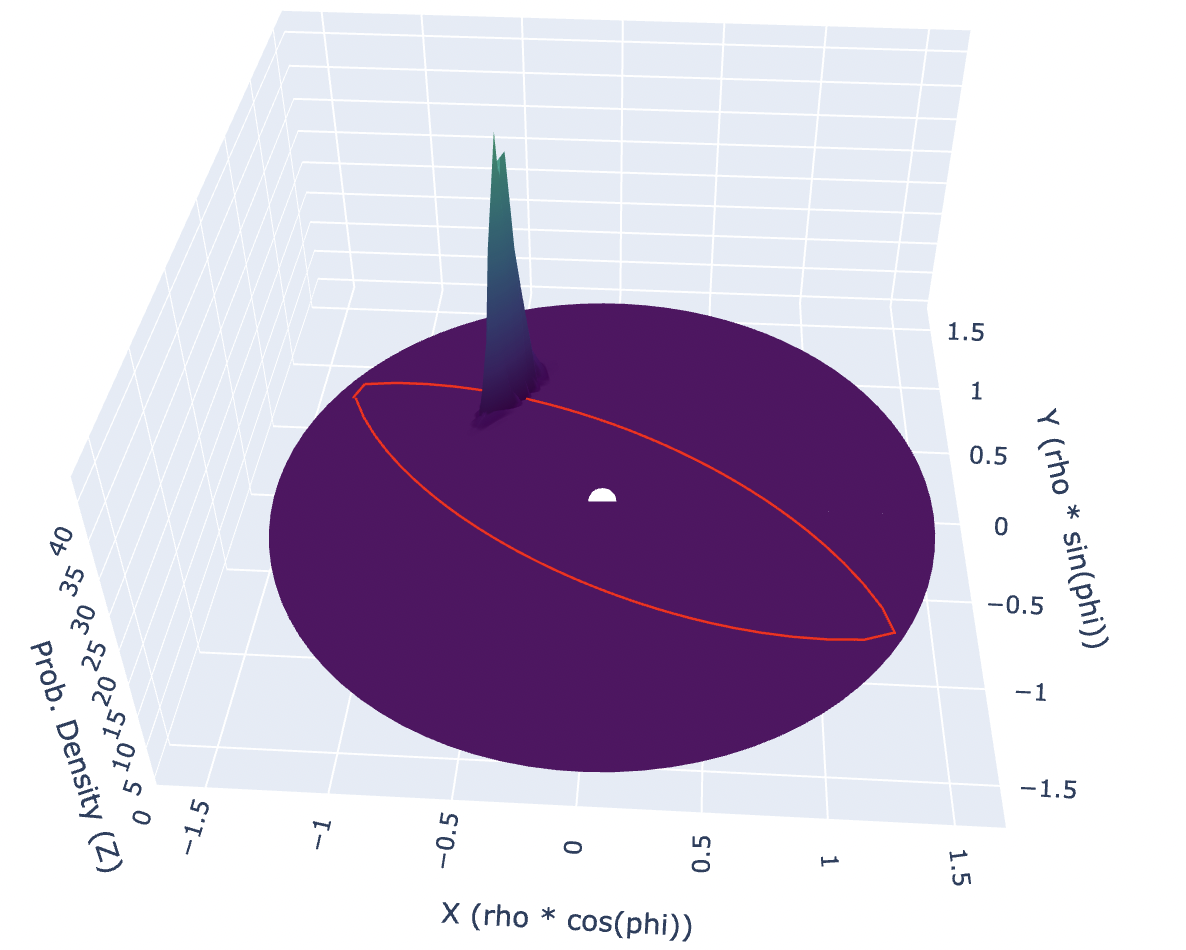}
        \caption{ The packet moving towards the boundary. $(t=0.8)$}
    \end{subfigure}\hfill
    \begin{subfigure}{0.48\textwidth}
        \includegraphics[width=0.8\linewidth]{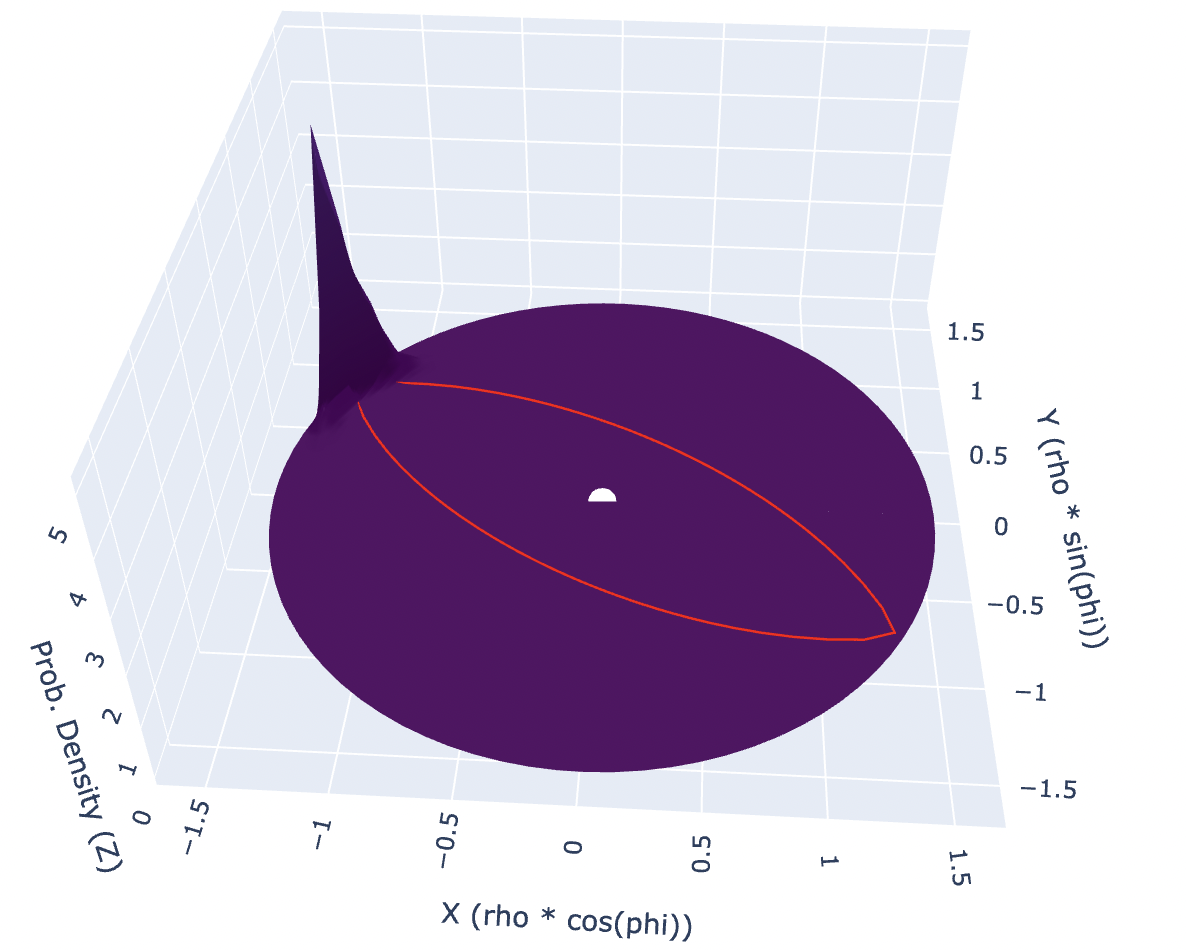}
        \caption{ The packet starts bouncing off from the boundary. $(t=1.60)$}
    \end{subfigure}\hfill
    
    \caption{Parameters: $\rho_{0}=0.5,\phi_{0}=\pi/3,\sigma_{1}=0.09,\sigma_{2}=0.05,n_{max}=m_{max}=300, M=0, m_{0}=-40, n_{0}=0$.}
    \label{fig:fig:pos_null_non-rad}

\end{figure}

Using the same wave packet as in Section \ref{sec:com:non_radial_null_loc}, we present below the 3D evolution of a non-radial null geodesic in Fig. \ref{fig:fig:pos_null_non-rad}. For the parameter choices specified in these figures, the numerically computed energy of the system is $E \approx 83.5$. 

The plot of $\langle\rho(t)\rangle$ vs $\langle\phi(t)\rangle$ corresponding to this can be found in Appendix \ref{appendix_2d_plots_pos_ope_null_localized_2d_plots}.

\subsubsection{Null Case: De-localized Wave Packet }\label{sec:pos:null_deloc}
We now consider the specific parameter choice $m_{0} = 0$ and $n_{0} = 0$ in the massless (null) limit, which yields a strictly real wave packet profile. One consequence of reality\footnote{This is a consequence of the fact that wave packet momentum is related to its phase gradient. Explicitly: the expanded expression  
\bea
\langle\hat{\rho}\rangle(t) = \frac{1}{2\pi}\sum_{n,m,n',m'}\int d\rho\,d\phi\,\tan(\rho)\,\rho\,g^{*}(n',m')g(n,m)e^{i\Delta\omega t}R^{*}_{n'm'}(\rho)R_{nm}(\rho)e^{i(m-m')\phi},
\eea
can be written using $\frac{1}{2\pi}\int d\phi\,e^{i(m-m')\phi}=\delta_{mm'}$ as
\begin{equation}\label{expression for rhohat's expectation value in terms of g and I}
\langle\hat{\rho}\rangle(t) = \sum_{n,m,n'}I(n,n',m)g^{*}(n',m)g(n,m)e^{2i(n'-n)t}
\end{equation}
where the integral $I$ is defined as:
$ I(n,n',m) = \int d\rho\,\rho\,\tan(\rho)\,R^{*}_{n'm}(\rho)R_{nm}(\rho).$
Taking the time derivative of $\langle\hat{\rho}\rangle$ gives:
$\frac{d\langle\hat{\rho}\rangle}{dt} = -2i\sum_{n,m,n'}(n-n')I(n,n',m)g^{*}(n',m)g(n,m)e^{2i(n'-n)t}.$
At $t=0$, this gives
$ \frac{d\langle\hat{\rho}\rangle}{dt}\bigg|_{t=0} = -2i\sum_{n,m,n'}n\,I(n,n',m)g^{*}(n',m)g(n,m) + 2i\sum_{n,m,n'}n'\,I(n,n',m)g^{*}(n',m)g(n,m).$
In the second term, we apply the substitution $m\rightarrow-m$, relabel $n\leftrightarrow n'$, and use the identities $I(n,n',-m)=I(n,n',m)$ and $g(n,-m)=g^{*}(n,m)$. Note that the second identity is valid only if the initial position-space profile $f(\rho,\phi,t=0)$ is purely real. Under these substitution, the two terms perfectly cancel, yielding:
$\frac{d\langle\hat{\rho}\rangle}{dt}\bigg|_{t=0} = 0.$
Note that till here we have made no reference to the mass $M$ of the scalar. So this equation holds in general if the initial packet choice is real. But this is immediately a problem for massless radial infall case because $\frac{d\langle\hat{\rho\rangle}}{dt}=1$ should hold at all times. This establishes a strict constraint on the choice of the initial wave packet profile for the massless case: a massless particle will not satisfy the radially infalling equation \eqref{the classical geodesic eqn for massless case radially infalling} for \emph{any} purely real choice of the initial wave packet profile $f(\rho,\phi,t=0)$.} is that $\frac{d\langle\hat{\rho}\rangle}{dt}\bigg|_{t=0} = 0$. This means that such a state completely fails to track the radial classical {\em null} geodesic equation, which in this case would be $d\langle \rho(t) \rangle/dt = \pm 1$. The physical mechanism driving this failure becomes visually evident in the subsequent 3D plots: rather than maintaining spatial coherence, the wave packet undergoes severe delocalization and splitting during its evolution.

\begin{figure}[H]
    \centering

    \begin{subfigure}[b]{0.48\textwidth}
        \includegraphics[width=0.8\linewidth]{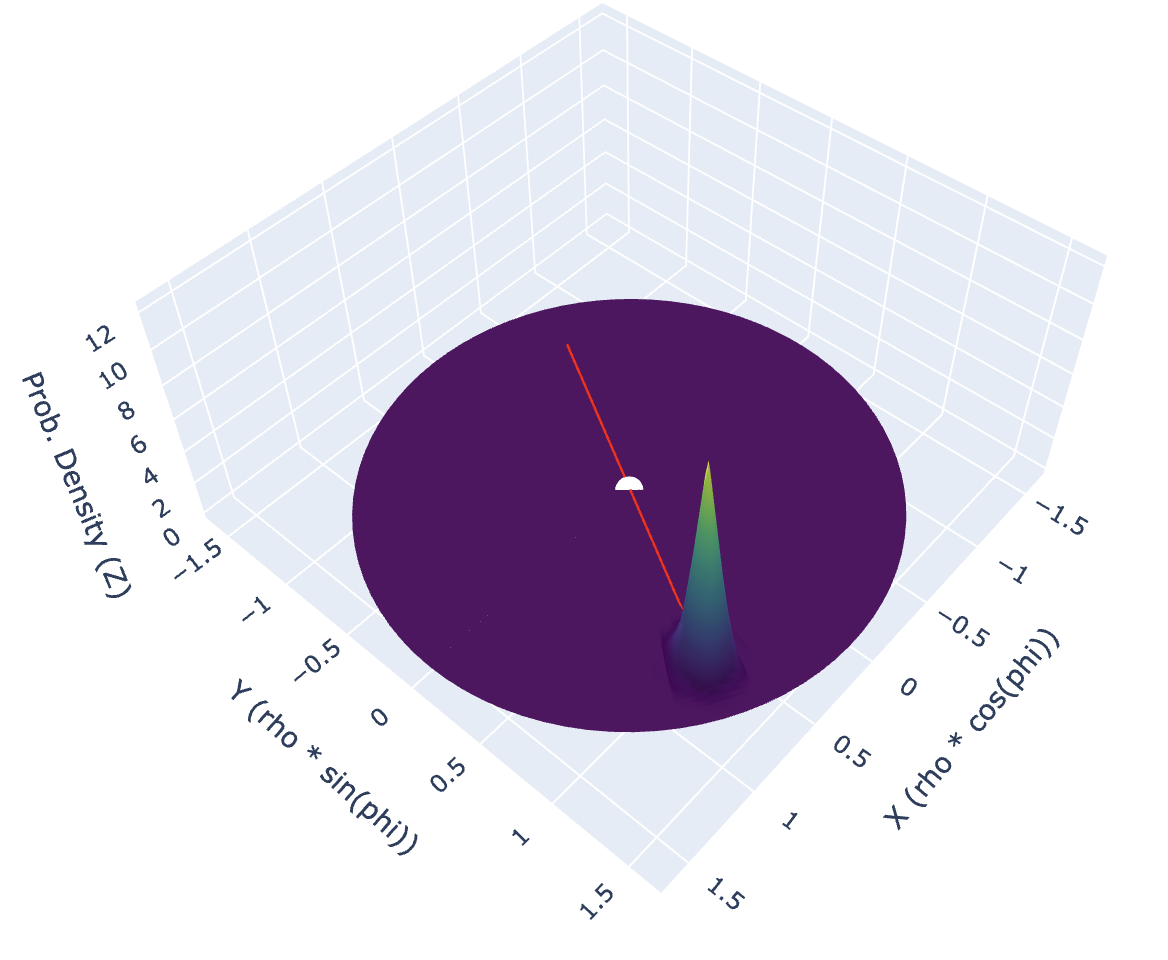}
        \caption{ The massless packet profile in the beginning. $(t = 0.00)$}
    \end{subfigure}\hfill
    \begin{subfigure}[b]{0.48\textwidth}
        \includegraphics[width=0.8\linewidth]{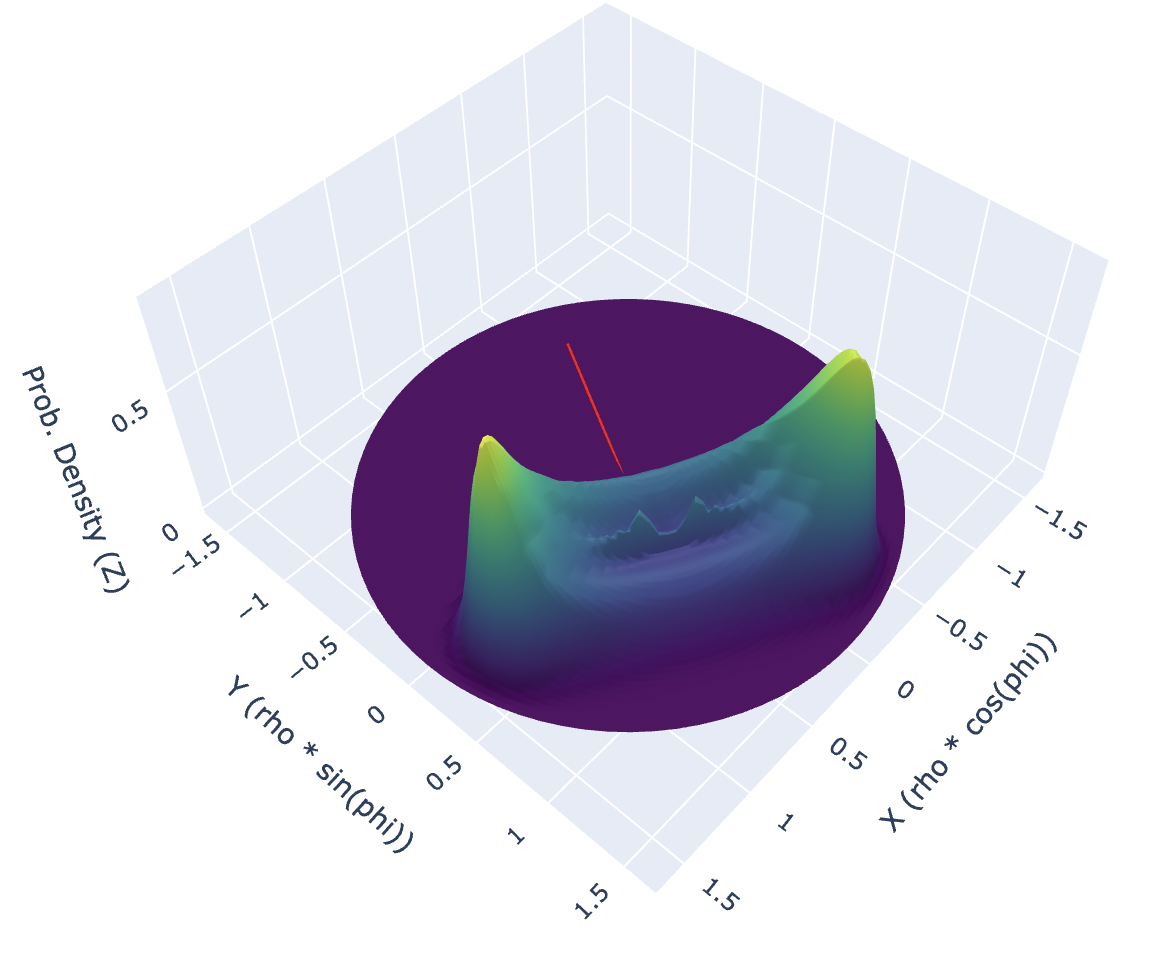}
        \caption{ Onset of delocalization. $(t = 0.90)$}
    \end{subfigure}\hfill

    \begin{subfigure}[b]{0.48\textwidth}
        \includegraphics[width=0.8\linewidth]{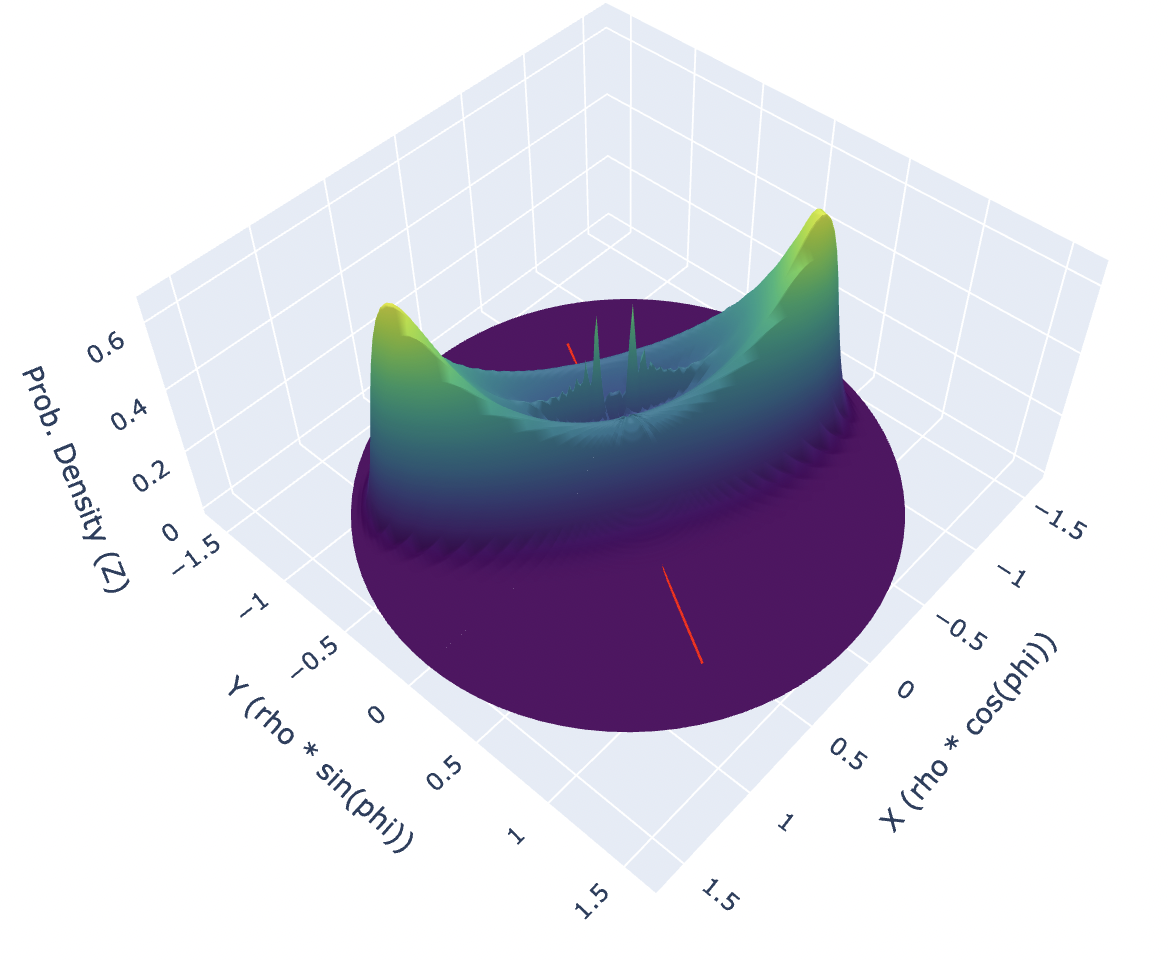}
        \caption{ Packet on the other side. $(t=1.8)$}
    \end{subfigure}\hfill
    \begin{subfigure}[b]{0.48\textwidth}
        \includegraphics[width=0.8\linewidth]{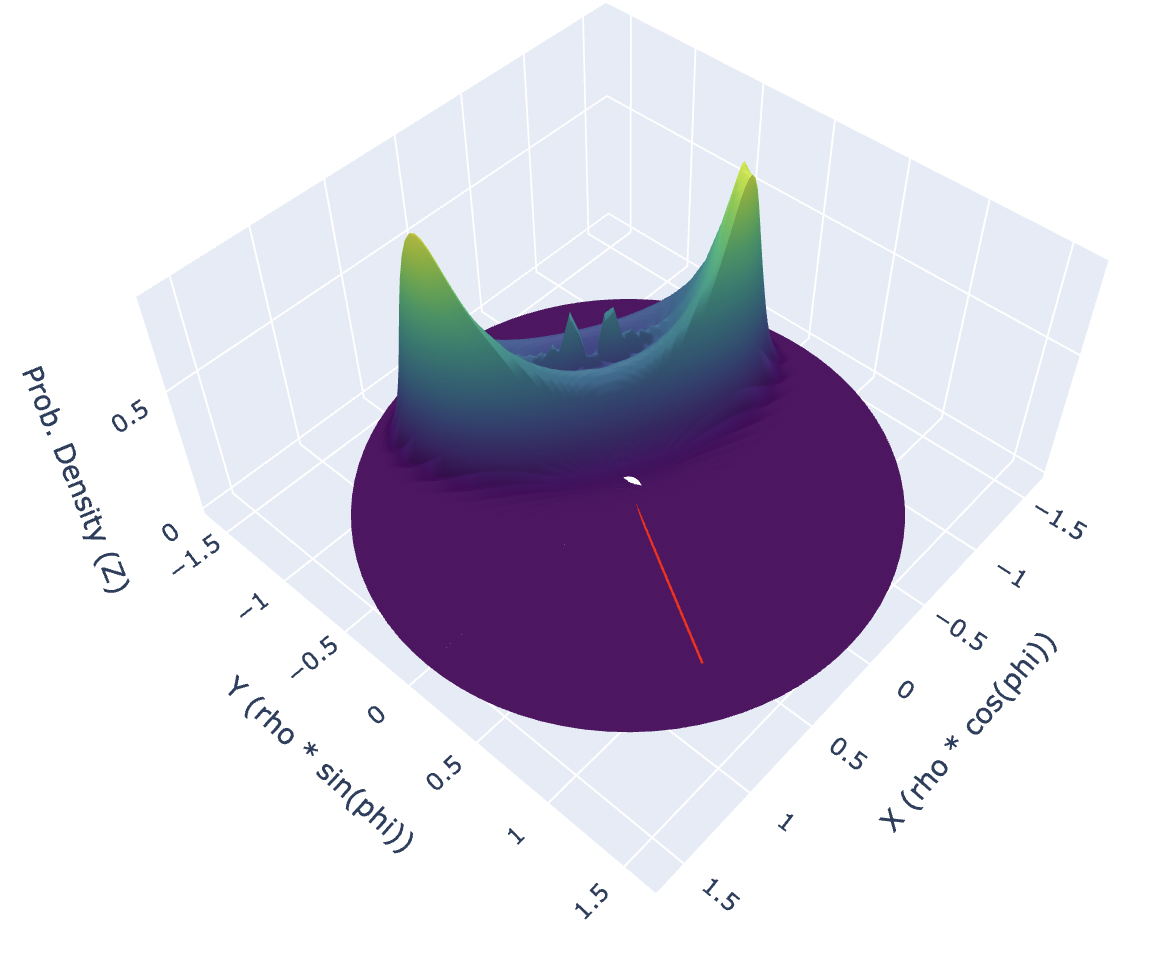}
        \caption{ The packet starts relocalizing. $(t = 2.25)$}
    \end{subfigure}\hfill

    \begin{subfigure}[b]{0.48\textwidth}
        \includegraphics[width=0.8\linewidth]{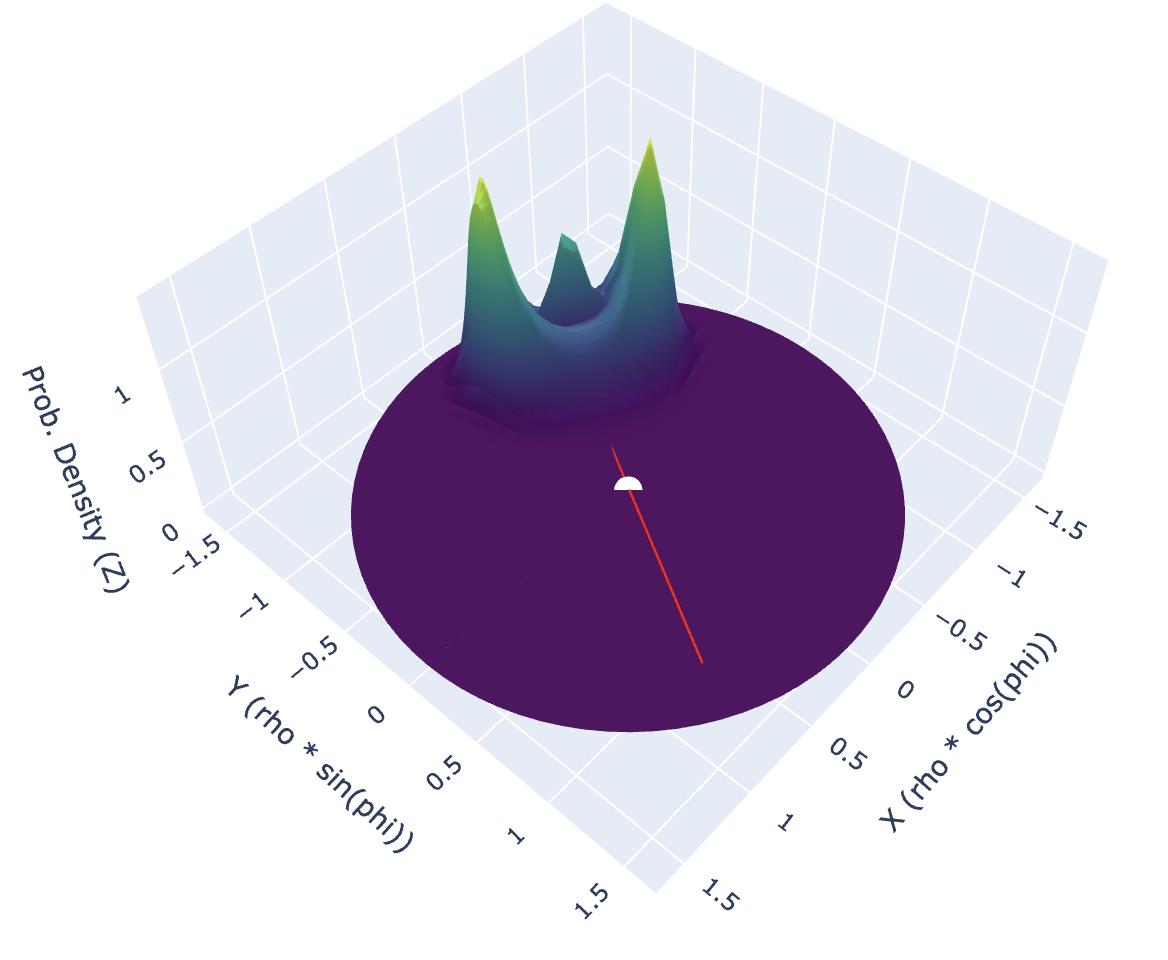}
        \caption{ The packet starts relocalizing. $(t = 2.70)$}
    \end{subfigure}\hfill
    \begin{subfigure}[b]{0.48\textwidth}
        \includegraphics[width=0.8\linewidth]{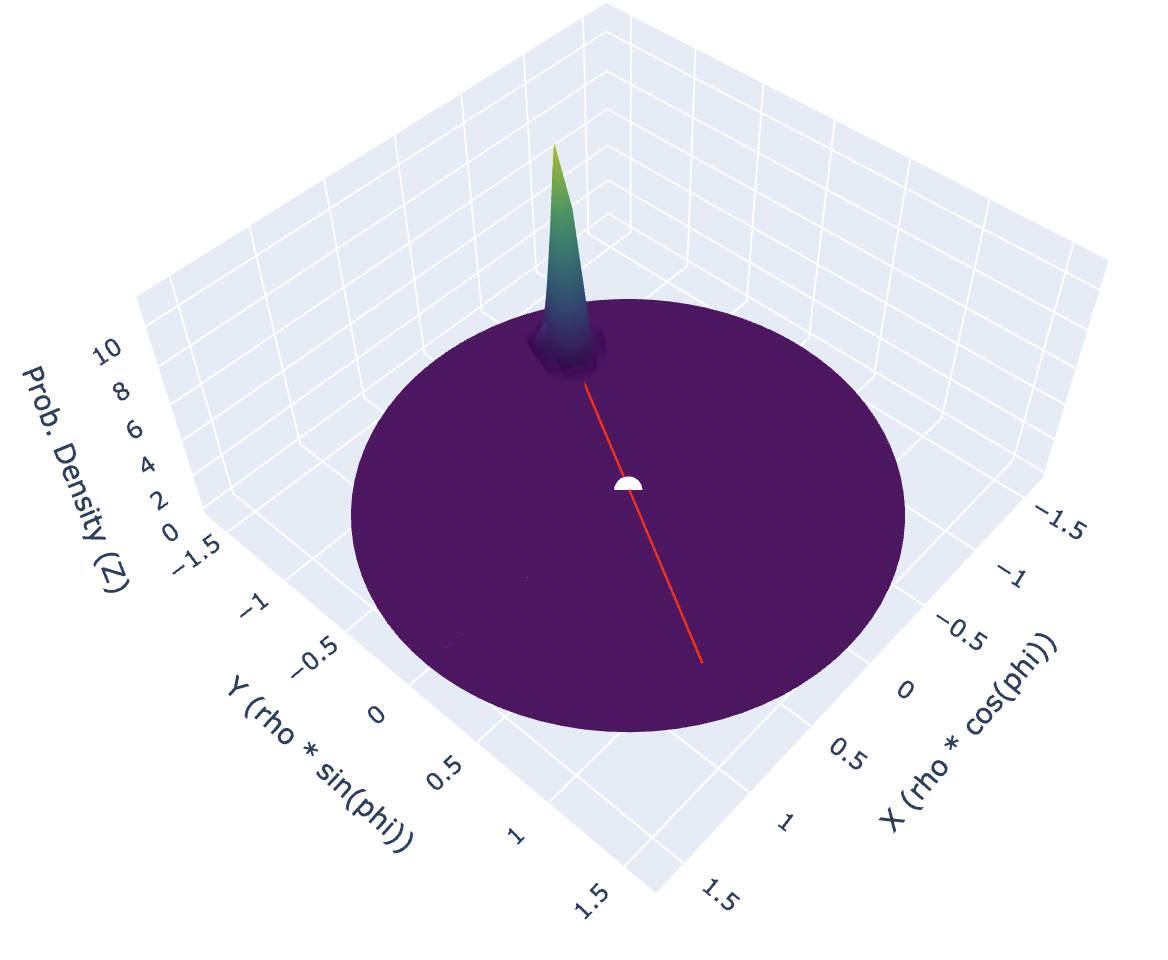}
        \caption{ The massless packet is at the antipodal point. $(t = 3.14)$}
    \end{subfigure}\hfill
    
    \caption{Parameters: $\rho_{0}=1.2,\phi_{0}=\pi/3,\sigma_{1}=0.09,\sigma_{2}=0.05,n_{max}=m_{max}=200, M=0, m_{0}=0, n_{0}=0$}
    \label{fig:fig:pos_null_de-loc}

\end{figure}

The 3D plots below (Fig. \ref{fig:fig:pos_null_de-loc}) capture the first half of the temporal evolution for this real, null wave packet. For the specific initial conditions detailed in the figure below, the numerically evaluated energy is $E \approx 11$.
While a rigorous physical justification for this phenomenon is provided in subsequent sections, it fundamentally arises from the existence of a characteristic minimum length scale $(1/E)$ below which coherent, localized macroscopic motion cannot be sustained. This splitting of the real null wave packet is a physical consequence of the localization limit we talked about earlier in section \ref{sec:relativistic_localization} and will establish  in more detail in section \ref{sec:breakdown}. 

In stark contrast, for the massive scalar case, increasing the mass systematically suppresses this delocalization, yielding a tightly confined wave packet even when $m_{0} = n_{0} = 0$. This progressive localization is confirmed by $\rho$ vs $t$ plots discussed in Appendix \ref{appendix_2d_plots_pos_ope_massive_rad_infall_rho_vs_t}, where the finite-width offset near the origin monotonically decreases at higher masses. Even in the massless case, as before, if a large initial radial momentum $n_0$ (corresponding to high energy $E$) is imparted, the wave packet remains well localized throughout the evolution and accurately follows the classical trajectory $\rho(t)$ (see \ref{appendix_2d_plots_pos_ope_null_localized_radial_infall}).

\subsubsection{Circular Geodesic at a Given Radius for Fixed \texorpdfstring{$m_{0}$}{m0}}\label{sec:pos:cir}

Given that we want to compare the two approaches, we set the angular momentum fixed at $m_{0}=-30$ as done before. As before, we see a clear inverse relationship between the orbit's radius and the mass of the particle. For $\rho_{0}=0.45$ we get circular motion with $M \approx 117$, $M \approx 38$ for $\rho_{0}=0.70$, $M \approx 16$ for $\rho_{0}=0.90$, $M \approx 5.8$ at $\rho_{0}=1.10$ and finally $M \approx 2.5$ at $\rho_{0}=1.20$. The trend is the same as the one we had before.
Extrapolating this trend to the spatial asymptotic limit ($\rho \to \pi/2$) implies that for strictly massless states ($M \to 0$), circular geodesics manifest exclusively at the boundary of the AdS$_{3}$ manifold.

A comprehensive set of these trajectories, along with their specific evolution and parameter configurations, is documented in Appendix \ref{appendix_2d_plots_pos_ope_circular_orbits}.

\section{Geodesic Charges vs Wave Packet Charges}
\label{sec:geo-vs-wave-charges}

Before proceeding, we note the following. In our numerical evolutions in the previous section, we noted that the peaks of the wave packets follow classical geodesic trajectories. But in order to make this correspondence, we need to find a map between the numerically obtained curves, and the geodesics. The latter can be characterized either by their conserved charges, or the sizes of their major and minor axes\footnote{We use these terms loosely, even though the curves are not exactly ellipses.}, which we will refer to as extrema ($\rho_{max}$ and $\rho_{min}$). It turns out however, that the numerically determined radial extrema and those obtained by putting the wave packet's energy and angular momentum into the classical geodesic formulae, are not the same. This may seem disconcerting, but it is physical, and (as we will see) is a direct result of the wave packet's finite spatial extent. As a result of the finite width, even though the wave packet's numerically evaluated energy and angular momentum are conserved charges for the wave packet, they cannot be directly associated to the classical geodesic.\footnote{\label{foot}To find the conserved charges associated to a geodesic for a point particle, we define the conserved current $J^{\mu} = T^{\mu\nu}\xi_{\nu}$ associated to the (massive) point particle stress-tensor $T^{\mu\nu}$, given by $T^{\mu\nu}(x) = \frac{M}{\sqrt{|g|}}\int u^{\mu}(\tau)u^{\nu}(\tau)\delta^{D+1}(x-z(\tau))d\tau$ \cite{Weinberg1972}. Here $M$ is the mass of the particle, $z(\tau)$ is the location of the particle in spacetime, $u^{\mu} = \frac{dz^{\mu}(\tau)}{d\tau}$ and $\xi_{\nu}$ is the Killing vector. The expression for conserved charge is given by $Q = \int J^{0}\sqrt{g}d^{D}\vec{x}$. By doing the integration over $d^{D}\vec{x},$ and using properties of delta functions, we get the conserved charge $Q = M \xi_{\nu}(z(\tau))u^{\nu}(\tau)$. We rewrite it in the form $q = g_{\mu\nu}\xi^{\mu}\frac{dx^{\nu}}{d\tau}$, where $q$ is the conserved quantity per unit mass. For spacetimes having timelike and rotational killing vectors this gives us the conserved energy and conserved angular momentum per unit mass, respectively, which are defined as $e = g_{tt}\frac{dt}{d\tau} ; \hspace{0.2cm}  l = -g_{\phi\phi}\frac{d\phi}{d\tau}$. This derivation allows us to relate the conserved charges of the point particle stress tensor to those of the geodesic.} Instead, what we do is to determine the extrema of the wave packet trajectories and then identify them with the extrema of the geodesics. This leads to a precise match between the curves -- but the conserved charges of the geodesic are slightly shifted from those of the wave packet. 

One way to understand this phenomenon is to realize that the functional dependence of the extrema of the wave packet on the charges of the wave packet need not be the same as the functional dependence of the extrema of the geodesic on the charges of the geodesic.

In the next subsection, to provide some intuition for  this phenomenon, we describe a toy model: we will identify an energy shift in the 0+1D wave packets of an anharmonic potential. This has qualitatively similar origins to the shifts we note in our geodesics.

\subsection{Toy Model: Gaussian Wave Packet in Anharmonic Potential}\label{anharmonic_toy_model}

From our numerical results and the relationship between the extrema and the geodesic charges, the expectation is that the energy associated to the wave packet is larger than the energy that can be associated to the geodesic. Below we will consider the 1D wave packet in an anharmonic potential to motivate our expectation.

We consider the potential,
\bea
V(x) = a\,x^{4}+b\,x^{2}+c
\eea
where $a,b,c\in\mathbb{R}$ with $a>0$ and $b>0$. The total classical energy of the wave packet can be written as:
\bea
E_{cl.} = \frac{p^{2}}{2M} + V(x)
\eea
The expectation value of the wave packet Hamiltonian $\hat{H}$ in the state $|\Psi(t)\rangle$ is given as
\bea\label{1D Hamiltonian}
E_{qn.}\equiv\langle\hat{H}\rangle=\frac{\langle\hat{p}^{2}\rangle}{2M}+\langle V(\hat{x})\rangle 
\eea
For a gaussian wave packet of standard deviation $\sigma_{x}$ in position and $\sigma_{p}$ in momentum, the following relations hold:
\bea
\langle\hat{x}^{2}\rangle = \langle\hat{x}\rangle^{2} + \sigma_{x}^{2};\hspace{5mm} \langle\hat{p}^{2}\rangle = \langle\hat{p}\rangle^{2} + \sigma_{p}^{2};\hspace{5mm} \langle\hat{x}^{4}\rangle = \langle\hat{x}\rangle^{4} + 6\langle\hat{x}\rangle^{2}\sigma_{x}^{2}+3\sigma_{x}^{4};
\eea
With this we can write \eqref{1D Hamiltonian} as:
\bea
E_{qn.} = E_{cl.} + \Big( \frac{\sigma_{p}^{2}}{2M}+6\,a\,\langle\hat{x}\rangle^{2}\sigma_{x}^{2}+3\,a\,\sigma_{x}^{4}+b\,\sigma_{x}^{2} \Big)
\eea
Clearly, since $a>0$ and $b>0$, the variance terms are strictly positive, yielding
\bea\label{inequality}
E_{qn.}>E_{cl.}
\eea
Here $E_{cl.}$ means the energy obtained by putting expectation value of $\hat{x}$ operator in the classical energy equation. Clearly, the quantum energy of the wave packet is larger than the classical energy. 

This observation can be thought of as explaining why $\rho_{max}$ and $\rho_{min}$ of the geodesic cannot be obtained, if we plug in the wave packet energy and angular momentum into the geodesic formulas for $\rho_{max}$ and $\rho_{min}$. Note that we have chosen the Gaussian wave packet and the shape of the potential for simplicity/concreteness. Some of the choices can affect the specific inequality \eqref{inequality}, but the fact that the two expressions are different is robust for general potentials.

\section{On the Breakdown of Classical Behavior}\label{sec:breakdown}

The classical single-particle interpretation of a wave packet traversing a curved spacetime relies fundamentally on the spatial extent of the packet. In our framework, the initial packet profile is governed by the spatial width parameters $\sigma_{\rho}$ and $\sigma_{\phi}$, which control the extent of localization at $t=0$. We find that the system possesses an intrinsic characteristic length scale given by $1/E$ (or $\frac{\hbar c}{E}$ when restoring standard constants), where $E$ is the conserved energy associated with the wave packet. 

The onset of classical versus quantum behavior is dictated by the ratio of the wave packet widths to this characteristic length (which can be viewed as a Compton wavelength):
\begin{itemize}
    \item \textbf{Classical Regime ($1 \gg \sigma_{\rho}, \sigma_{\phi} \gg 1/E$):} The spatial widths are larger than the intrinsic length scale. The wave packet remains localized throughout its evolution, and its trajectory follows the classical geodesic. For the packet to be localized in AdS the widths have to be smaller than the AdS length scale (which we set to unity).
    \item \textbf{Quantum Regime ($\sigma_{\rho}, \sigma_{\phi} \lesssim 1/E$):} The spatial widths are comparable to or smaller than the intrinsic length scale of the ``particle$"$. The classical single-particle notion breaks down, the packet becomes highly delocalized, and the variances in the expectation values become large. Often, the center of mass or the peak of the probability density itself gets deviations as we will see below\footnote{But this can sometimes be avoided by choosing position operators judiciously, as we will see in the next section.}.
\end{itemize}

\subsection{Numerical Examples: The Massless Case}\label{sec:breakdown_massless}

To illustrate this breakdown, consider a massless wave packet ($M=0$) with the following initial parameters: $\rho_{0}=0.5$, $\phi_{0}=\pi/3$, $m_{0}=-40$, and $n_{0}=0$. 

\begin{figure}[H]
    \centering
    \begin{subfigure}{0.48\textwidth}
        \includegraphics[width=1\linewidth]{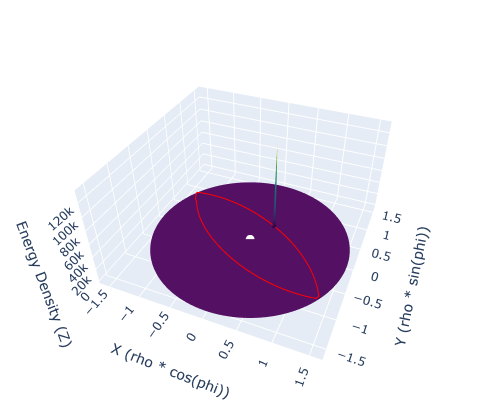}
        \caption{$t=0.0$}
    \end{subfigure}\hfill
    \begin{subfigure}{0.48\textwidth}
        \includegraphics[width=1\linewidth]{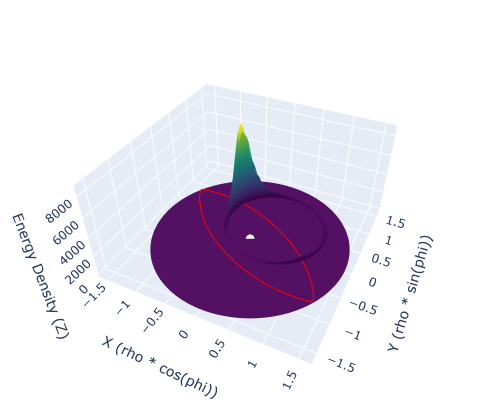}
        \caption{$t=0.8$}
    \end{subfigure}\hfill
    \begin{subfigure}{0.48\textwidth}
        \includegraphics[width=1\linewidth]{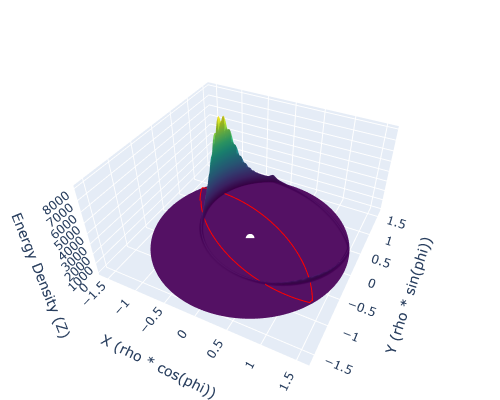}
        \caption{$t=1.2$}
    \end{subfigure}\hfill
    \begin{subfigure}{0.48\textwidth}
        \includegraphics[width=1\linewidth]{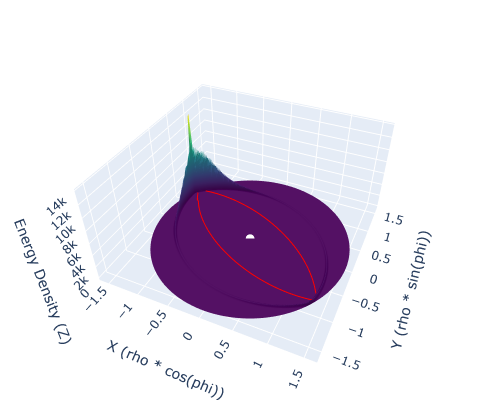}
        \caption{$t=1.6$}
    \end{subfigure}\hfill   
    
    \caption{Energy density distribution. Parameters: $\rho_{0}=0.50, \phi_{0}=\pi/3, \sigma_{\rho}=0.01, \sigma_{\phi}=0.01, n_{max}=m_{max}=300$}
    \label{fig:fig:breakdown_massless}
    
\end{figure}

\begin{figure}[H]
    \centering
    \begin{minipage}{0.48\textwidth}
        \centering
        \includegraphics[width=1\linewidth]{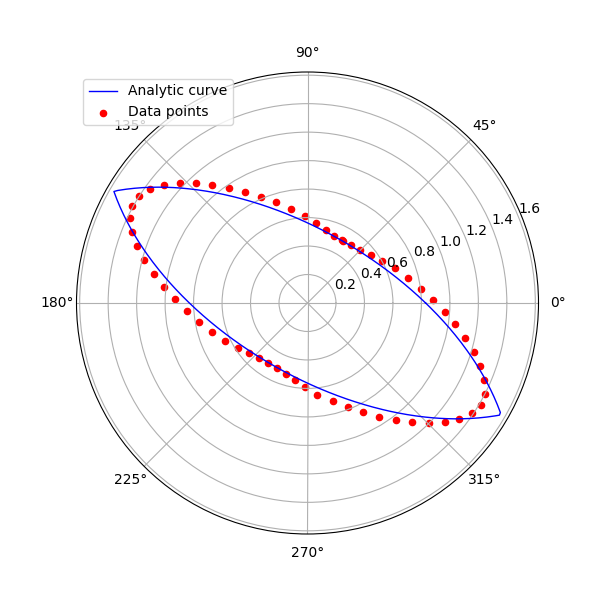}
    \end{minipage}
    \caption{Polar plot mapping the trajectory. Parameters: $\rho_{0}=0.50, \phi_{0}=\pi/3, \sigma_{\rho}=0.01, \sigma_{\phi}=0.01, M=0, m_{0}=-40, n_{0}=0$.}
    \label{fig:fig:breakdown_massless_2D}
\end{figure}

When we set $\sigma_{\rho}=0.01$ and $\sigma_{\phi}=0.01$, the conserved energy is evaluated to be $E \approx 122.65$. The characteristic length scale is therefore $1/E \approx 0.008$. Because our chosen sigmas ($0.01$) are comparable to this length scale, the classical nature of the packet breaks down. As shown in the energy density distribution plots below (Fig. \ref{fig:fig:breakdown_massless}), the packet delocalizes significantly and even fails to follow the classical geodesic curve. That the peak of the wave packet does not follow a geodesic is more clear from Figure \ref{fig:fig:breakdown_massless_2D}. This $\bar{\rho}$ vs $\bar{\phi}$ polar plot confirms the deviation from the predicted geodesic path.

In stark contrast, when we previously utilized broader initial widths ($\sigma_{\rho}=0.09, \sigma_{\phi}=0.05$) with the same parameters, the conserved energy was $E \approx 83$, yielding a characteristic length of $1/E \approx 0.012$. Because the spatial widths ($0.09$ and $0.05$) were significantly larger than $0.012$, the classical nature was preserved, and we observed localized motion tracing the exact null geodesic.

\subsection{Energy Dependence and Massive Radial Infall }\label{sec:length_scale}

This framework perfectly explains why classical behavior is restored when either the initial momentum or the mass is increased. For instance, for the packet considered earlier with parameters $M = 0$, $\rho_{0} = 1.2$, $\phi_{0} = \pi/3$, $m_{0} = 0$, and $n_{0} = 0$, and with widths ($\sigma_{\rho} = 0.09$, $\sigma_{\phi} = 0.05$), the energy was comparatively low ($E \approx 11$), corresponding to a characteristic length scale of $1/E \approx 0.091$. This length is comparable to $\sigma_{\rho}$ and larger than $\sigma_{\phi}$, resulting in delocalized motion. The corresponding behavior is illustrated in Fig. \ref{fig:delocalized} (stress tensor evolution) and Fig. \ref{fig:fig:pos_null_de-loc} (position operators).

However, introducing a large initial radial momentum ($n_{0}=200$) drives the energy up to $E \approx 200$, effectively shrinking the characteristic length to $1/E \approx 0.005$. Because $0.005 \ll \sigma_{\rho}, \sigma_{\phi}$, the classical behavior completely dominates, resulting in a tightly localized packet following a radially infalling geodesic (see Fig. \ref{fig:fig:com_null_rad_and_ellip_2D} in Appendix \ref{appendix_2d_plots_com} and Fig. \ref{fig:fig:pos_null_radial_and_ellip_2D} in Appendix \ref{appendix_2d_plots_pos_ope}). Similarly, increasing the mass parameter ($M=20, 40, 60, 80$) sufficiently raises the wave packet energy, pushing the $1/E$ scale well below the chosen spatial widths and stabilizing the localized classical trajectory (see Fig. \ref{fig:fig:com_massive_radial_2D} in Appendix \ref{appendix_2d_plots_com} and Fig. \ref{fig:fig:pos_massive_radial_2D} in Appendix \ref{appendix_2d_plots_pos_ope}).

\begin{figure}[H]
    \centering
    \begin{subfigure}{0.48\textwidth}
        \includegraphics[width=1\linewidth]{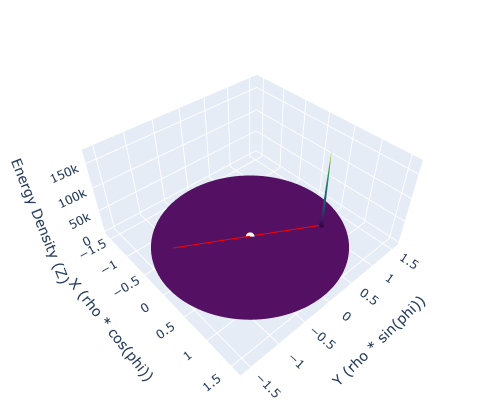}
        \caption{$t=0.0$}
    \end{subfigure}\hfill
    \begin{subfigure}{0.48\textwidth}
        \includegraphics[width=1\linewidth]{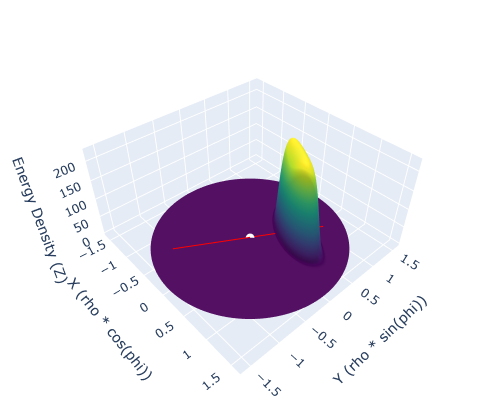}
        \caption{$t=0.8$}
    \end{subfigure}\hfill

    \begin{subfigure}{0.48\textwidth}
        \includegraphics[width=1\linewidth]{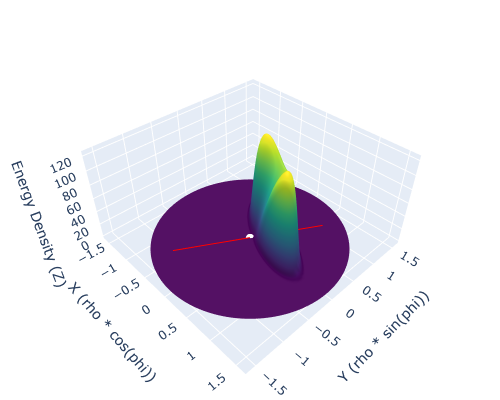}
        \caption{$t=1.2$}
    \end{subfigure}\hfill
    \begin{subfigure}{0.48\textwidth}
        \includegraphics[width=1\linewidth]{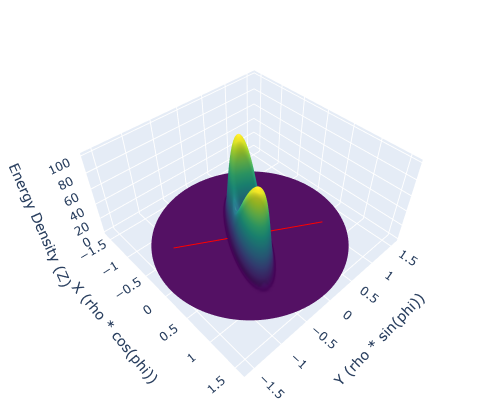}
        \caption{$t=1.6$}
    \end{subfigure}\hfill

    \caption{Radial infall delocalization. Parameters: $\rho_{0}=1.20, \phi_{0}=\pi/3, \sigma_{\rho}=0.01, \sigma_{\phi}=0.01, n_{max}=m_{max}=220, M=40, m_{0}=0, n_{0}=0$.}
    \label{fig:fig:breakdown_massive}
\end{figure}

Even with a large mass, if we reduce the spatial widths, we can trigger the onset of quantum behavior. Let us examine a massive radial infall case with $M=40$ and severely restricted widths ($\sigma_{\rho}=0.01, \sigma_{\phi}=0.01$). Previously, with wider sigmas, $E \approx 127.9$ gave $1/E \approx 0.008 \ll \sigma$, yielding localized motion. Now, with $\sigma=0.01$, the energy shifts slightly to $E=131.39$, making $1/E \approx 0.0076$. Since the characteristic length is again comparable to the spatial widths, the classical behavior should start breaking down. Fig. \ref{fig:fig:breakdown_massive} demonstrates this predicted delocalization.

\begin{figure}[H]
    \centering
    \begin{minipage}{0.48\textwidth}
        \centering
        \includegraphics[width=1\linewidth]{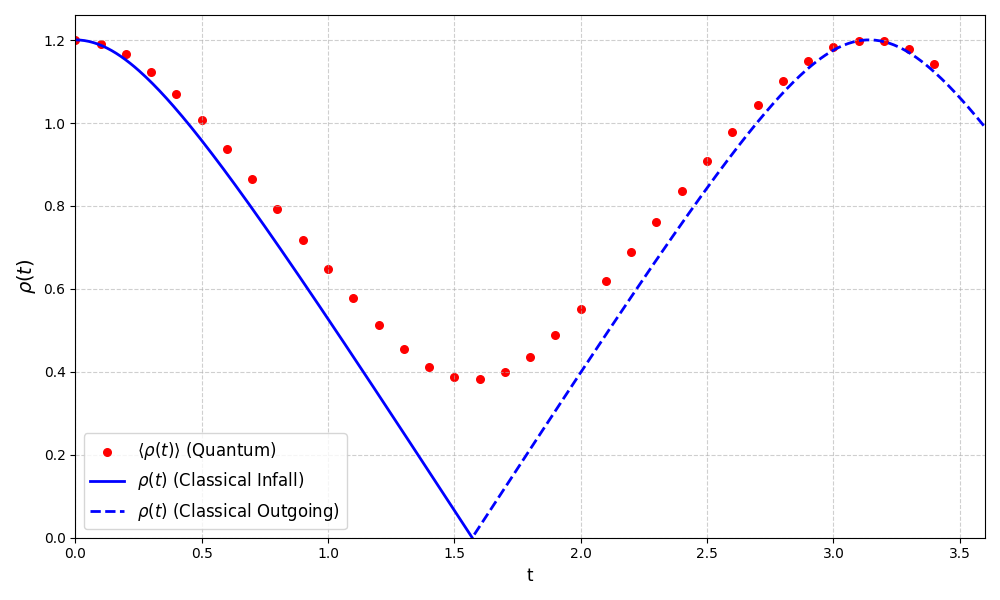}
    \end{minipage}
    \caption{$\rho$ vs $t$ plot for massive radial infall showing significant deviation from the classical geodesic.}
    \label{fig:fig:breakdown_massive_2D}
\end{figure}

Plotting the centroid of $\rho$ against $t$, Fig. \ref{fig:fig:breakdown_massive_2D} highlights a substantial offset compared to the classical trajectory, confirming that the spatial restrictions have allowed the error terms to cause large perturbation.

While the plots above reflect the stress tensor approach distribution, the behavior remains consistent for the position operator as well. Ultimately, these results confirm that wave packet dynamics in this spacetime are governed by the interplay between spatial width and the characteristic length scale $1/E$.

It is important to note that the parameter $1/E$ is not a sharp cutoff; the transition between localized classical motion and delocalized quantum behavior is continuous. Furthermore, delocalization  does not necessarily imply deviation of the packet's center  $\bar{\boldsymbol{x}}$ or $\langle\boldsymbol{\hat{x}}\rangle$ from its classical trajectory. An example of this is flat space in Minkowski coordinates. The packet spreads but still the cumulative effect is such that the centers $\bar{\boldsymbol{x}}$ and $\langle\boldsymbol{\hat{x}}\rangle$ follow a straight line (see \cite{Pavsic:2017} for diagrams and Appendix \ref{appendix where we show that the definitions give straight line} for the theoretical demonstration). In Anti-de Sitter (AdS) space, the error terms associated with the spatial spread actively contribute to a macroscopic deviation of the centroid values ($\bar{\rho}$ and $\bar{\phi}$) from the geodesic curve in the above cases we examined. But nonetheless we will see in the next Section, that special choices of position operators exist, such that the centroid and the expectation value follow the geodesic, even though the variances become large.

Finally, let us consider the case where the system is in the state $|\rho_{0}, \phi_{0}\rangle$ at $t=0$. This corresponds to the initial profile
\begin{equation}
f(\rho,\phi)=\frac{1}{\tan(\rho)}\,\delta(\rho-\rho_{0})\,\delta(\phi-\phi_{0}),
\end{equation}
which represents an idealized strictly localized (and therefore non-normalizable as a one-particle) state at the spatial point $(\rho_{0},\phi_{0})$.
As expected, the energy (or probability) density of such a state delocalizes over time, since one cannot anticipate classical motion for an extremely localized initial configuration. Numerical analysis confirms this behavior, in agreement with our previous results. This state may roughly be viewed as the limiting case $\sigma_{\rho}, \sigma_{\phi} \to 0$, for which delocalization persists irrespective of the mass.

The above statements are about (de)localization of the wave packets in the stress tensor or position operator language (with the position operators we had defined earlier). But it turns out that in AdS, with a particular definition of position operators, the evolutions of the expectation values in {\em any} state will follow the geodesic equation (modulo variance corrections). This is a consequence of certain selections rules that are in effect in AdS due to the integer spacings of the eigenmodes. We turn to this discussion next. 

\section{Operator-level Quantum-Classical Correspondence}\label{sec:Q-C_corres}

In this section, we will see (using selection rules imposed by the solutions of the Klein-Gordon equation) that the classical equations \eqref{exact relation 1} and \eqref{exact relation 2} are satisfied {\em exactly} at the level of expectation values on {\em arbitrary} single particle states. This can be viewed as a judicious choice of position operators\footnote{By operator-level quantum-classical correspondence, we mean operators projected on to the one-particle Hilbert space. The full Hilbert space of the quantum field theory in AdS$_3$ contains multi-particle states as well.} that exploits the special properties of  the AdS background.
Using these we can also provide an analytic demonstration that $\langle\hat{\rho}\rangle$ and $\langle\hat{\phi}\rangle$ satisfy the geodesic equations only up to variance corrections (as we have seen numerically).

Analogous to how we defined the $\hat{\rho}$ operator in \eqref{rho operator}, we define $\widehat{\cos(2\rho)}$ as:
\bea\label{cos(2rho) operator}
\widehat{\cos(2\rho)} = \int d\rho\,d\phi\,\tan(\rho)a^{\dagger}(\rho,\phi)\,\cos(2\rho)\,a(\rho,\phi)
\eea
When we take the expectation value of this over the generic time evolved state \eqref{generic state}, we get:
\begin{align}
\langle\widehat{\cos(2\rho)}\rangle_{t}= \sum_{{n_{1},m_{1},n_{2},m_{2}}} g(n_{1},m_{1})\, g^{*}(n_{2},m_{2})e^{i(\omega_{n_{2}m_{2}}-\omega_{n_{1}m_{1}})t} \int d\phi\, \frac{1}{2\pi} \,e^{-i(m_2-m_1)\phi}\times\nonumber\\ \times\int d\rho \tan(\rho) \cos(2\rho)\, R_{n_1 m_1}(\rho)\,R_{n_2 m_2}(\rho)
\end{align}
The integral over $\phi$ gives $\delta_{m_1,m_2}$. Summing over $m_{2}$ in the above expression, we get the integral over $\rho$ to be:
\bea\label{X}
X^{m_1}_{n_2,n_1}=\int d\rho \tan(\rho) \cos(2\rho)\, R_{n_1 m_1}(\rho)\,R_{n_2 m_1}(\rho)
\eea
This integral has the property that it is non zero only when either $n_1=n_2$ or $|n_1-n_2|=1$. Using this we can write the expression for 
$\langle\widehat{\cos(2\rho)}\rangle_{t}$ as:
\bea
\langle\widehat{\cos(2\rho)}\rangle_{t}=\sum_{n,m} \left(|g(n,m)|^{2} X^{m}_{n,n} +\left[g^{*}(n+1,m)\,g(n,m)\,e^{2it}X^{m}_{n+1,n}+hc\,\right]\right)
\eea
This can be written in the form:
\bea\label{final result for expectation value of cos2rho in the main body}
\langle\widehat{\cos(2\rho)}\rangle_{t}= A + B\cos(2 t+\delta)
\eea
where $A$, $B$ and $\delta$ are related to sums over terms involving $g(n,m)$'s\footnote{We will assume here that these sums are finite. This is related to the specific choice of the wave packet, loosely its normalizability. We have checked numerically for various choices that this is true.}.
This is exactly the classical equation \eqref{exact relation 1}. This means the operator in equation \eqref{cos(2rho) operator} follows its classical equation exactly when we take its expectation value over any generic normalizable single particle state \eqref{generic state}.

We may write \eqref{final result for expectation value of cos2rho in the main body} as:
\bea
\cos(2\langle\hat{\rho}\rangle)_{t}= A + B\cos(2 t+\delta)+\Delta_{1}(t)
\eea
where $\Delta_{1}(t)=\cos(2\langle\hat{\rho}\rangle)_{t}-\langle\widehat{\cos(2\rho)}\rangle_{t}$.
Using this we get:
\bea\label{expectation value of rho upto the variance term}
\langle\hat{\rho}\rangle_{t}=\frac{1}{2}\cos^{-1}\left(A + B\cos(2 t+\delta)+\Delta_{1}(t)\right)
\eea
This corresponds precisely to the classical equation for $\rho$, up to a variance-like correction term. It is therefore reasonable to assume that $\Delta_{1}(t)$ remains small, in which case $\langle \hat{\rho} \rangle_t$ closely follows its classical evolution as given by \eqref{exact relation 1}.

Now let us define a new operator%\footnote{Unlike much of the rest of our discussions in this paper, exact constructions that crucially rely on $e^{i \phi}$ will be non-trivial to generalize to higher dimensional spheres.}:
\bea\label{Z operator}
\widehat{\sin(\rho)e^{i\phi}}\equiv\widehat{\mathcal{Z}} = \int d\rho\,d\phi\,\tan(\rho)a^{\dagger}(\rho,\phi)\,\sin(\rho)e^{i\phi}\,a(\rho,\phi)
\eea
Going through the same steps as before, we get:
\begin{align}
\langle\widehat{\mathcal{Z}}\rangle_{t}= \sum_{{n_{1},m_{1},n_{2},m_{2}}} g(n_{1},m_{1})\, g^{*}(n_{2},m_{2})e^{i(\omega_{n_{2}m_{2}}-\omega_{n_{1}m_{1}})t} \int d\phi\, \frac{1}{2\pi} \,e^{-i(m_2-m_1-1)\phi}\times\nonumber\\ \times\int d\rho \tan(\rho) \sin(\rho)\, R_{n_1 m_1}(\rho)\,R_{n_2 m_2}(\rho)    
\end{align}
The integral over $\phi$ gives the selection rule $\delta_{m_2,m_{1}+1}$. The integral over $\rho$ then becomes:
\bea\label{Y}
Y^{m_1}_{n_2,n_1}=\int d\rho \tan(\rho) \sin(\rho)\, R_{n_1 m_1}(\rho)\,R_{n_2 m_{1}+1}(\rho)\equiv I
\eea
This integral has the following properties:
\[
I = \begin{cases} 
\text{non-zero} ;& \text{if } n_2 = n_1 \text{ or } n_2 = n_1 - 1\, (m\ge0) \\
\text{non-zero} ;& \text{if } n_2 = n_1 \text{ or } n_2 = n_1 + 1\, (m<  0) \\
0 ;& \text{otherwise}
\end{cases}
\]
To proceed, we split the sum for $\langle\widehat{\mathcal{Z}}\rangle_{t}$ into the $m \ge 0$ and $m < 0$ cases. 
For $m \ge 0$, we further partition the sum into:
\begin{itemize}
    \item The $n_2 = n_1 - 1$ terms,
    \item The $n_2 = n_1$ terms.
\end{itemize}
Similarly, for $m < 0$, we split the sum into:
\begin{itemize}
    \item The $n_2 = n_1 + 1$ terms,
    \item The $n_2 = n_1$ terms.
\end{itemize}
This grouping reveals two components with $e^{-it}$ time dependence (one from each $m$ case) and two components with $e^{it}$ time dependence.
So, the final expression for $\langle\widehat{\mathcal{Z}}\rangle_{t}$ looks like:
\bea\label{final expression for expectation value of sinrho expiphi}
\langle\widehat{\sin(\rho)e^{i\phi}}\rangle_{t}=\langle\widehat{\mathcal{Z}}\rangle_{t} = A\,e^{it}+B\,e^{-it}
\eea
Note that $A$ and $B$ here are not same as the ones in the expression for $\cos(2\langle\hat{\rho}\rangle)_{t}$. Differentiating this with respect to $t$ twice gives:
\bea
\frac{d^{2}\langle\widehat{\mathcal{Z}}\rangle_{t}}{dt^2}=-\langle\widehat{\mathcal{Z}}\rangle_{t}
\eea
This is exactly the classical equation for $\mathcal{Z}_{t}$ as in \eqref{exact relation 2} which is just a harmonic oscillator. Note that the linearity of the AdS$_{3}$ spectrum was crucial for the exact statements we found above. 
Now as done for $\rho$, we can write \eqref{final expression for expectation value of sinrho expiphi} in the following form
\begin{align}
\langle\widehat{\sin(\rho)}\rangle_{t}\langle\widehat{e^{i\phi}}\rangle_{t}= A\,e^{it}+B\,e^{-it}+\Delta_{2}(t)
\end{align}
where $\Delta_{2}(t)=\langle\widehat{\sin(\rho)}_{t}\rangle\langle\widehat{e^{i\phi}}\rangle_{t}-\langle\widehat{\sin(\rho)e^{i\phi}}\rangle_{t}$. This gives
\bea
\arg\left(\langle\widehat{e^{i\phi}}\rangle_{t}\right)= \arg\left(A\,e^{it}+B\,e^{-it}+\Delta_{2}(t)\right) 
\eea
This is precisely the classical equation for $\arg(e^{i\phi})$ up to the variance like error term on the RHS (see \eqref{exact relation 1}). It is reasonable to assume that $\Delta_{2}(t)$ is small in most of the situations causing $\arg\left(\langle\widehat{e^{i\phi}}\rangle_{t}\right)$ to closely follow its classical equation.

We get similar results from the stress tensor definition (i.e. $\overline{\cos(2\rho)}$ and $\overline{\sin(\rho)e^{i\phi}}$ follow their classical equation exactly). The details of the calculation can be found in the Appendix \ref{appendix_Q-C_corresp_com}. 

Let us see the time evolution of $\mathcal{Z}_{t}$ from both the operator formalism and the center of mass formalism. The plots below (Figures \ref{fig:fig31} and \ref{fig:fig32}) are for $M=30,m_{0}=-30,\rho_{0}=0.5,\phi_{0}=\pi,\sigma_{1}=0.09,\sigma_{2}=0.06,n_{max}=m_{max}=80$. $\mathcal{Z}$ can be written as:
\bea
\langle\mathcal{Z}\rangle_{t}= (A+B)\cos(t)+i(A-B)\sin(t)= C\cos(t)+i D \sin(t)
\eea
Note that C and D can be complex but for this example they happen to be real. The plot of real and imaginary part of $\langle\mathcal{Z}\rangle_{t}$ look as below:
\begin{figure}[H]
    \centering
    % First Image
    \begin{minipage}{1.0\textwidth}
        \centering
        \includegraphics[width=0.70\linewidth]{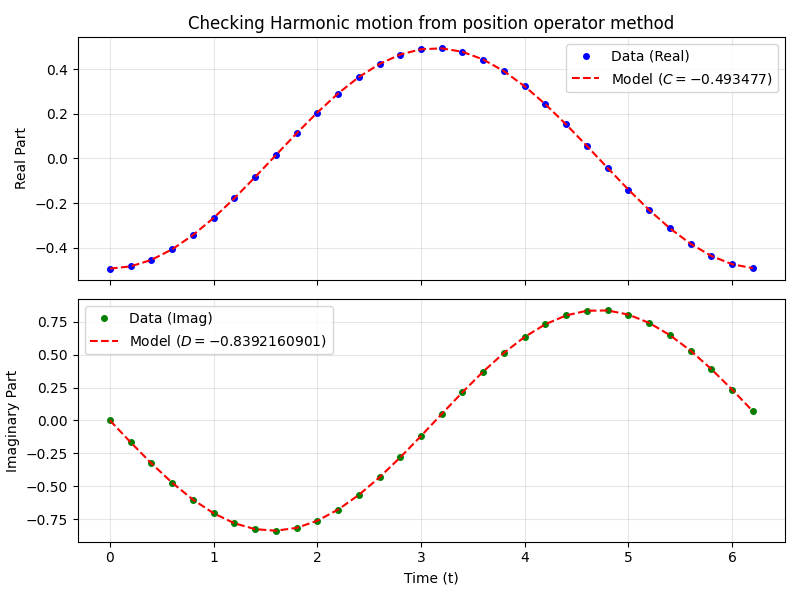}
        \caption{Operator formalism with parameters: $\rho_{0}=0.50, \phi_{0}=\pi, \sigma_{1}=0.09, \sigma_{2}=0.06, n_{max}=m_{max}=80, M=30, m_{0}=-30, n_{0}=0, C=-0.49347, D=-0.83922$}
        \label{fig:fig31}
    \end{minipage}

    % Second Image
    \begin{minipage}{1.0\textwidth}
        \centering
        \includegraphics[width=0.70\linewidth]{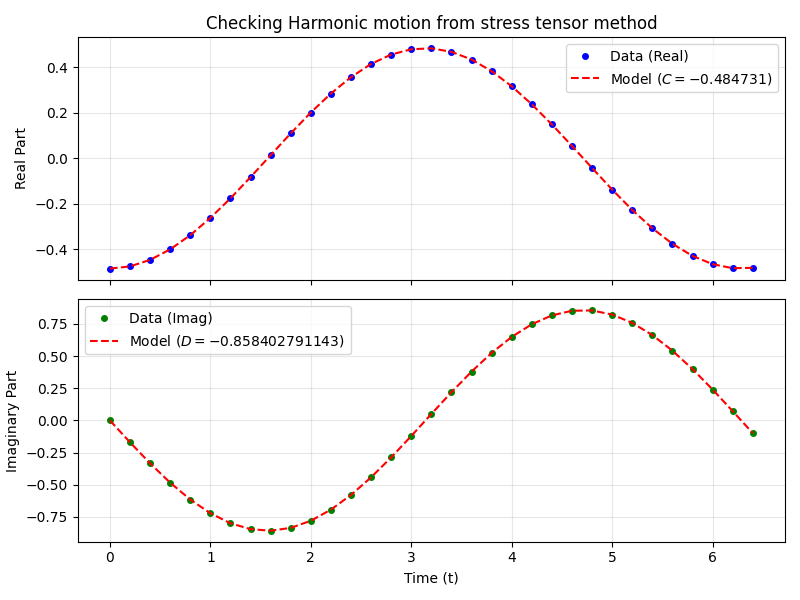}
        \caption{Center of mass formalism with parameters: $\rho_{0}=0.50, \phi_{0}=\pi, \sigma_{1}=0.09, \sigma_{2}=0.06, n_{max}=m_{max}=80, M=30, m_{0}=-30, n_{0}=0, C=-0.48473, D=-0.85840$}
        \label{fig:fig32}
    \end{minipage}
\end{figure} 
Clearly the quantum data from both approaches match with their corresponding classical curves. %As said before the constants $C$ and $D$ are different, but they are approximately the same.

Our discussion in this section was a direct consequence of the selection rules due to the integer-spacing in the AdS modes\footnote{Selection rules in AdS of a different kind have been noted before in \cite{Oleg}.}. We demonstrate this explicitly in terms of orthogonal polynomials in Appendix \ref{appendix_sec:symmetry_origin}. We emphasize however that while nice (and evidently a consequence of the underlying symmetries of AdS), these results should not be viewed as something mysterious: our general philosophy still holds. Note that Eq. \eqref{final result for expectation value of cos2rho in the main body} and \eqref{final expression for expectation value of sinrho expiphi} tell us nothing about the variance of the packet (defined as e.g., $\langle \cos^2 2 \rho\rangle - \langle \cos(2 \rho)\rangle^2$). This means these exact statements are not indicative of the spread the packet has, they simply reflect the propagation of the peak.

\section{CFT Interpretation of the Bulk Wave Packet}\label{sec:CFT_interpretation}

In this section we give a more systematic CFT interpretation of the bulk wave packet states used throughout the paper. The basic point is that the bulk one-particle Hilbert space of a free scalar in global AdS$_3$ can be identified with the global conformal module of a scalar primary operator in the dual CFT$_2$. The bulk Gaussian packet introduced earlier is then just one particular superposition of these one-particle states. On the CFT side, a natural and well-studied family of such superpositions is obtained from Euclidean-regularized local operator insertions \cite{Berenstein:2019,Goto:2017}\footnote{\cite{Terashima:2023WavePackets} develops a boundary description of a general high energy bulk wave packet, and diagnoses how that bulk-localized object appears to boundary observables, in particular the boundary stress tensor.}. Our aim here is to write the Euclidean-regularized states directly in the same $(n,m)$ basis used in the bulk analysis, and then compare them to the coefficients $g(n,m)$ that define our Gaussian packet.

We will work at leading order in the large-$c$/free-field limit, where only the global conformal descendants are needed. In particular, no Virasoro descendants beyond the global $\mathfrak{sl}(2,\mathbb{R})_{L}\oplus \mathfrak{sl}(2,\mathbb{R})_{R}$ module will play any role in the present discussion. This again aligns with our earlier comment that for the main conceptual discussions in this paper, AdS$_3$ is no more special than higher dimensional AdS spaces. The essential reason for this is that we are working with quantum field theory in curved space, and not directly with dynamical aspects of gravity.

\subsection{Bulk Modes as Boundary Modules}\label{subsec:CFT_bulk_modes_global_desc}

The scalar field $\Phi$ of bulk mass $M$ is dual to a scalar primary operator $\mathcal{O}_{\Delta}$ of dimension $\Delta$, related by
\begin{equation}\label{eq:CFT_bulk_mass_dimension_relation}
M^{2}\mathcal{R}^{2}=\Delta(\Delta-2),
\end{equation}
and by the extrapolate dictionary
\begin{equation}\label{eq:CFT_bulk_extrapolate}
\lim_{\rho\rightarrow \frac{\pi}{2}}(\cos\rho)^{-\Delta}\,\Phi(t,\rho,\phi)\;\propto\;\mathcal{O}_{\Delta}(t,\phi).
\end{equation}
Thus the same parameter $\Delta$ that appears in the bulk spectrum $\omega_{nm}=\Delta+2n+|m|$ is the CFT scaling dimension of the dual primary.

Let $|\Delta\rangle$ denote the highest-weight state created by $\mathcal{O}_{\Delta}$ under the state-operator correspondence. Since $\mathcal{O}_{\Delta}$ is a scalar, we have
\begin{equation}
h=\bar h=\frac{\Delta}{2},
\end{equation}
and the normalized global descendants are
\begin{equation}\label{eq:normalized_global_descendants}
|p,q;\Delta\rangle
\equiv
\frac{1}{\sqrt{p!\,(\Delta)_{p}}}\,
\frac{1}{\sqrt{q!\,(\Delta)_{q}}}\,
L_{-1}^{p}\bar L_{-1}^{q}\,|\Delta\rangle,
\qquad p,q=0,1,2,\dots,
\end{equation}
where $(a)_{r}\equiv \Gamma(a+r)/\Gamma(a)$ is the Pochhammer symbol. These states satisfy
\begin{equation}
H\,|p,q;\Delta\rangle=(\Delta+p+q)\,|p,q;\Delta\rangle,
\qquad
J\,|p,q;\Delta\rangle=(p-q)\,|p,q;\Delta\rangle,
\end{equation}
where, as usual, $H=L_{0}+\bar L_{0}$ and $J=L_{0}-\bar L_{0}$ (we suppress the vacuum Casimir shift, which is irrelevant for the one-particle problem).

To connect directly with the bulk mode labels $(n,m)$, define
\begin{equation}\label{eq:mplusmminus}
m_{+}\equiv \max(m,0),\qquad m_{-}\equiv \max(-m,0),
\end{equation}
and then set
\begin{equation}\label{eq:nm_to_pq_dictionary}
|n,m\rangle_{\rm CFT}\equiv |\,n+m_{+},\,n+m_{-};\Delta\rangle.
\end{equation}
By construction,
\begin{equation}\label{eq:CFT_n_m_eigenvalues}
H\,|n,m\rangle_{\rm CFT}=(\Delta+2n+|m|)\,|n,m\rangle_{\rm CFT},
\qquad
J\,|n,m\rangle_{\rm CFT}=m\,|n,m\rangle_{\rm CFT}.
\end{equation}
This exactly reproduces the bulk spectrum $\omega_{nm}$ and angular Fourier label $m$. Therefore the bulk one-particle states $|n,m\rangle=a^{\dagger}_{nm}|0\rangle$ can be identified with the CFT global descendants \eqref{eq:nm_to_pq_dictionary}. In other words, the coefficients $g(n,m)$ that appeared in Sections \ref{sec:EV_of_rho_phi_Ttt} and \ref{sec:Q-C_corres} may be interpreted equally well as bulk mode amplitudes or as amplitudes for normalized global descendants.

It is useful to compare this with the exact bulk-local state of Goto and Takayanagi \cite{Goto:2017}. At the center of global AdS, their construction gives
\begin{equation}
|\phi_{\Delta}(0)\rangle_{\rm GT}
=
\sum_{k=0}^{\infty}
(-1)^k\,
\frac{\Gamma(\Delta)}{k!\,\Gamma(k+\Delta)}\,
L_{-1}^k \bar L_{-1}^k |\Delta\rangle.
\end{equation}
This is exact in the sense of the leading-order bulk local-operator construction: it is the CFT state dual to a local bulk field acting on the vacuum, rather than a finite-energy normalizable wave packet. The Euclidean-regularized states we use below are different in spirit: they are finite-energy normalizable packets, built from the same global-descendant basis, and are therefore more directly comparable to our bulk Gaussian packets.

\subsection{States from Euclidean-Regularized Operators}\label{subsec:CFT_exact_coefficients}

Following \cite{Berenstein:2019}, consider the Euclidean-regularized operator
\begin{equation}\label{eq:CFT_epsilon_operator}
\mathcal{O}_{\Delta,\epsilon}(\phi,t)
=
e^{-\epsilon H}\,
\mathcal{O}_{\Delta}(\phi,t)\,
e^{+\epsilon H},
\qquad \epsilon>0.
\end{equation}
At $t=0$, the state created by a local insertion at angle $\phi$ is
\begin{equation}
|\phi;\epsilon\rangle
\equiv
\mathcal{O}_{\Delta,\epsilon}(\phi,0)|0\rangle. \label{def-g}
\end{equation}
To expand this state in the normalized descendant basis, we first use the local operator expansion on the cylinder:
\begin{equation}\label{eq:cylinder_local_expansion}
\mathcal{O}_{\Delta}(\phi,0)|0\rangle
=
\sum_{p,q\geq 0}
\frac{e^{\,i(p-q)\phi}}{p!\,q!}\,
\partial^{p}\bar\partial^{q}\mathcal{O}_{\Delta}(0)|0\rangle.
\end{equation}
Using the state-operator correspondence,
\begin{equation}
\partial^{p}\bar\partial^{q}\mathcal{O}_{\Delta}(0)|0\rangle
=
L_{-1}^{p}\bar L_{-1}^{q}\,|\Delta\rangle,
\end{equation}
and the normalization \eqref{eq:normalized_global_descendants}, we obtain
\begin{equation}\label{eq:exact_descendant_expansion_pq}
|\phi;\epsilon\rangle
=
\mathcal{N}_{\epsilon}
\sum_{p,q\geq 0}
e^{-\epsilon(\Delta+p+q)}
e^{\,i(p-q)\phi}
\sqrt{\frac{(\Delta)_{p}(\Delta)_{q}}{p!\,q!}}\,
|p,q;\Delta\rangle,
\end{equation}
where $\mathcal{N}_{\epsilon}$ is an overall normalization fixed by $\langle \phi;\epsilon|\phi;\epsilon\rangle=1$.

Rewriting this in the $(n,m)$ basis defined in \eqref{eq:nm_to_pq_dictionary} gives
\begin{equation}\label{eq:exact_descendant_expansion_nm}
|\phi;\epsilon\rangle
=
\sum_{n=0}^{\infty}\sum_{m\in \mathbb{Z}}
g_{\epsilon,\phi}(n,m)\,|n,m\rangle_{\rm CFT},
\end{equation}
with the exact coefficients\footnote{Note that this object is loosely analogous to the extrapolate limit of the bulk expectation value $\langle 0 | \Phi |n,m\rangle$, see \eqref{def-g}.}
\begin{equation}\label{eq:g_epsilon_theta_exact}
g_{\epsilon,\phi}(n,m)
=
\mathcal{N}_{\epsilon}\,
e^{\,im\phi}\,
e^{-\epsilon(\Delta+2n+|m|)}
\sqrt{
\frac{(\Delta)_{n+m_{+}}(\Delta)_{n+m_{-}}}
{(n+m_{+})!\,(n+m_{-})!}
}.
\end{equation}
This is the precise AdS$_3$/CFT$_2$ version of the \cite{Berenstein:2019} wave packet written in the same $(n,m)$ basis used for the bulk mode expansion \eqref{f}.

For fixed $m$, the probability distribution in the radial descendant number $n$ is
\begin{equation}\label{eq:fixed_m_probability_distribution}
|g_{\epsilon,\phi}(n,m)|^{2}
\propto
e^{-2\epsilon(\Delta+2n+|m|)}
\frac{(\Delta)_{n+m_{+}}(\Delta)_{n+m_{-}}}
{(n+m_{+})!\,(n+m_{-})!}.
\end{equation}
In the semiclassical regime $\Delta\gg 1$ and $\epsilon\ll 1$, this distribution is sharply peaked. For large $n$, the saddle point $n_{\ast}(m)$ is determined by
\begin{equation}\label{eq:n_star_fixed_m}
\frac{1}{n_{\ast}(m)}
+
\frac{1}{n_{\ast}(m)+|m|}
\;\simeq\;
\frac{4\epsilon}{\Delta-1},
\end{equation}
where we have further assumed that $n \gg \Delta$ to write a simple formula.
This is the AdS$_3$ specialization of the \cite{Berenstein:2019} saddle-point condition for fixed angular momentum. Importantly, the peak depends on $m$. Therefore, once one superposes different $m$'s, there is in general no single universal $n_*(m)$ independent of $m$.

The width of the radial distribution is obtained from the second derivative of $\log |g|^{2}$ at the saddle:
\begin{equation}\label{eq:n_width_fixed_m}
\sigma_{n}(m)^{-2}
\;\simeq\;
(\Delta-1)\left(
\frac{1}{n_{\ast}(m)^{2}}
+
\frac{1}{(n_{\ast}(m)+|m|)^{2}}
\right).
\end{equation}
Thus, for each fixed $m$, the Euclidean-regularized operator produces an approximately Gaussian packet in $n$-space. The corresponding mean energy is
\begin{equation}\label{eq:CFT_mean_energy_general}
\langle \hat{H}\rangle_{\epsilon}
=
\sum_{n,m}(\Delta+2n+|m|)\,|g_{\epsilon,\phi}(n,m)|^{2},
\end{equation}
and for a packet narrow in both $n$ and $m$ this becomes
\begin{equation}\label{eq:CFT_mean_energy_narrow_packet}
\langle \hat{H}\rangle_{\epsilon}
\;\simeq\;
\Delta+2n_{\ast}(m_{\rm peak})+|m_{\rm peak}|.
\end{equation}
This is the correct sharp-packet version of the \cite{Berenstein:2019} energy formula in our notation.

\subsection{Wave Packets Peaked in Angle and Spin}\label{subsec:CFT_angularly_localized_packet}

The state \eqref{eq:exact_descendant_expansion_nm}, equivalently the coefficients \eqref{eq:g_epsilon_theta_exact}, is localized at a boundary angle $\phi$, but it is not sharply localized in the angular mode label $m$: rather, it contains all values of $m$. To construct a packet peaked around a chosen mode label and a chosen angular location, let
\begin{equation}\label{eq:CFT_smearing_operator_definition}
\mathcal{O}_{F,\epsilon,\mu_{0},\varphi_{0}}
\equiv
\int_{0}^{2\pi}d\phi\,
F(\phi-\varphi_{0})\,e^{+i\mu_{0}(\phi-\varphi_0)}\,
\mathcal{O}_{\Delta,\epsilon}(\phi,0),
\end{equation}
where $F(\theta)$ is a real profile of compact support, or in practice, a narrow wrapped Gaussian centered at $\theta=0$ (as in our earlier discussions).

Defining $\theta\equiv \phi-\varphi_{0}$, the Fourier coefficients of $F$ are
\begin{equation}\label{eq:CFT_Fourier_of_F}
\widetilde{F}_{r}
\equiv
\int_{0}^{2\pi}d\theta\,F(\theta)e^{-ir\theta}.
\end{equation}
Then the normalized CFT state
\begin{equation}
|\Psi_{\rm CFT}(0)\rangle
\equiv
\mathcal{N}_{F,\epsilon,\mu_{0}}\,
\mathcal{O}_{F,\epsilon,\mu_{0},\varphi_{0}}|0\rangle
\end{equation}
has the expansion
\begin{equation}\label{eq:CFT_packet_general_nm_expansion}
|\Psi_{\rm CFT}(0)\rangle
=
\sum_{n,m}g_{\rm CFT}(n,m;0)\,|n,m\rangle_{\rm CFT},
\end{equation}
with
\begin{equation}\label{eq:CFT_packet_general_coefficients}
g_{\rm CFT}(n,m;0)
=
\mathcal{N}_{F,\epsilon,\mu_{0}}\,
\widetilde{F}_{m+\mu_{0}}\,
e^{\,i(m+\mu_{0})\varphi_{0}}\,
e^{-\epsilon(\Delta+2n+|m|)}
\sqrt{
\frac{(\Delta)_{n+m_{+}}(\Delta)_{n+m_{-}}}
{(n+m_{+})!\,(n+m_{-})!}
}.
\end{equation}
An overall phase $e^{-i\mu_{0}\varphi_{0}}$ has been absorbed into the normalization for convenience, and we have assumed that $F(\theta)$ is an even function of $\theta \equiv \phi-\varphi_0$.

A particularly useful choice is a narrow Gaussian
\begin{equation}\label{eq:CFT_boundary_gaussian_profile}
F(\theta)\propto e^{-\theta^{2}/(4\sigma_{\theta}^{2})},
\qquad \sigma_{\theta}\ll 1,
\end{equation}
for which
\begin{equation}\label{eq:CFT_Fourier_gaussian}
\widetilde{F}_{m+\mu_{0}}
\propto
e^{-\sigma_{\theta}^{2}(m+\mu_{0})^{2}}.
\end{equation}
Therefore the packet is approximately Gaussian in $m$-space, centered at $m=-\mu_{0}$ with variance
\begin{equation}\label{eq:CFT_m_variance_gaussian}
(\Delta m)^{2}
\;\simeq\;
\frac{1}{4\sigma_{\theta}^{2}},
\end{equation}
up to the mild extra $m$-dependence coming from the descendant normalization and the factor $e^{-\epsilon|m|}$.

The CFT interpretation of the angular data is now transparent:
\begin{itemize}
\item $\varphi_{0}$ is the center of the boundary angular smearing,
\item $\sigma_{\theta}$ controls the spread in the mode label $m$ as well as the angular smearing,
\item $-\mu_{0}$ is the mode-space center of the packet. (Note that the sign allows us to match $\mu_0=m_0$ in our earlier Gaussian conventions.)
\end{itemize}
Through the bulk/CFT identification of the one-particle basis, these are the CFT counterparts of the bulk angular location, angular width, and angular momentum data.

The parameter $\epsilon$ controls the radial descendant distribution and hence the mean
energy. To interpret this in terms of a classical bulk orbit, one should work in the
narrow-packet, large-$\Delta$ semiclassical regime. In that regime, the quantity that
plays the role of the worldline rest mass is $\Delta/\mathcal{R}$ (equivalently $\Delta$ in units
$\mathcal{R}=1$), whereas the bulk Klein--Gordon mass satisfies $M^2\mathcal{R}^2=\Delta(\Delta-2)$.
Here $M$ is the Klein--Gordon mass parameter appearing in the bulk wave equation, while
$\Delta/\mathcal{R}$ is the natural rest-energy scale seen directly in the CFT one-particle spectrum.
The two agree only asymptotically in the large-$\Delta$ semiclassical limit, since
$M\mathcal{R}=\sqrt{\Delta(\Delta-2)}=\Delta-1+O(\Delta^{-1})$.

For a timelike geodesic in the metric (2.1), the conserved energy and angular momentum
then satisfy (we set $\mathcal{R}=1$)
\begin{equation}
\left(\frac{E_{\rm gd}}{\Delta}\right)^2
=
\sec^2 \rho_* + \frac{L_{\rm gd}^2}{\Delta^2 \sin^2 \rho_*},
\qquad (\dot{\rho}=0).
\end{equation}
Equivalently, the two radial turning points are
\begin{equation}
2\sec^2 \rho_\pm
=
\widetilde E^{\,2}-\widetilde L^{\,2}+1
\pm
\sqrt{\left(\widetilde E^{\,2}-\widetilde L^{\,2}+1\right)^2-4\widetilde E^{\,2}},
\qquad
\widetilde E \equiv \frac{E_{\rm gd}}{\Delta},
\quad
\widetilde L \equiv \frac{L_{\rm gd}}{\Delta}.
\end{equation}
For the Euclidean-regularized packet, the natural semiclassical bulk interpretation is that
the packet is initially centered near the outer turning point $\rho_+$ (the aphelion). In
the special case of vanishing angular momentum, $L_{\rm gd}=0$, this reduces to
\begin{equation}
\sec \rho_* = \frac{E_{\rm gd}}{\Delta}.
\end{equation}
Thus, in the narrow-packet, large-$\Delta$ regime where the CFT packet behaves
semiclassically, $\epsilon$ fixes the mean energy and hence the radial turning point of the
associated bulk trajectory, up to the usual wave-packet spread corrections.
\subsection{Wave Packets with Momentum from Lorentzian Evolution}\label{subsec:CFT_Lorentzian_launch}

The Euclidean-regularized state constructed above should be viewed, in the narrow-packet,
large-$\Delta$ semiclassical regime, as a packet whose peak is initially located near the
outer turning point $\rho_+$ and whose mean radial velocity vanishes at the time of
insertion. This is not an exact statement about arbitrary states, but the semiclassical
interpretation of the Euclidean-regularized family when the packet is sufficiently narrow in
the common $(n,m)$ basis. To obtain a packet launched from a generic radial position
$\rho_0$ at $t=0$, with non-zero radial momentum, the correct procedure is simply
Lorentzian time evolution. This is a small extension of the cases considered in \cite{Berenstein:2019}. It
introduces a new parameter into the discussion that is a logical possibility in light of the
general Gaussian wave packets we discussed in earlier sections.

Define
\begin{equation}\label{eq:CFT_Lorentzian_shifted_state}
|\Psi_{\rm CFT}(t_{0})\rangle
\equiv
e^{-iHt_{0}}\,|\Psi_{\rm CFT}(0)\rangle.
\end{equation}
Since each basis state $|n,m\rangle_{\rm CFT}$ is an energy eigenstate with eigenvalue $\omega_{nm}=\Delta+2n+|m|$, we get
\begin{equation}\label{eq:CFT_time_evolved_coefficients}
g_{\rm CFT}(n,m;t_{0})
=
e^{-i\omega_{nm}t_{0}}\,
g_{\rm CFT}(n,m;0).
\end{equation}
Using \eqref{eq:CFT_packet_general_coefficients}, this becomes
\begin{equation}\label{eq:CFT_time_evolved_coefficients_explicit}
g_{\rm CFT}(n,m;t_{0})
=
\mathcal{N}_{F,\epsilon,\mu_{0}}\,
\widetilde{F}_{m+\mu_{0}}\,
e^{\,i(m+\mu_{0})\varphi_{0}}\,
e^{-i(\Delta+2n+|m|)t_{0}}\,
e^{-\epsilon(\Delta+2n+|m|)}
\sqrt{
\frac{(\Delta)_{n+m_{+}}(\Delta)_{n+m_{-}}}
{(n+m_{+})!\,(n+m_{-})!}
}.
\end{equation}
The full Lorentzian phase is $e^{-i(\Delta+2n+|m|)t_{0}}$. However, for
\emph{radial} observables the $\Delta$-piece is an overall phase and the $|m|$-dependent
piece cancels between bra and ket because those observables are diagonal in $m$. In
particular, in expressions such as $\langle \hat{\rho}\rangle$ that we will make explicit below, the relevant interference is
between different $n$-sectors at fixed $m$, so the only surviving nontrivial relative phase is
\begin{equation}\label{eq:CFT_radial_phase_from_time_evolution}
e^{-2 i n t_{0}}.
\end{equation}
In this precise sense, ordinary Lorentzian evolution provides the CFT origin of the bulk
``radial momentum'' datum: it is encoded in the
relative phase among neighboring radial descendants. By contrast, angular observables are
sensitive to coherence between different $m$-sectors, so the $|m|$-dependence is relevant
there even though the angular momentum itself remains conserved.

To summarize: once the angular profile $F$ is fixed, the parameters $\epsilon$ and $\mu_{0}$ determine the mean conserved charges and hence the corresponding semiclassical orbit. In a narrow-$m$ regime this may be viewed as saying that $\epsilon$ and $\mu_{0}$ fix the orbit. The parameter $t_{0}$ then selects where the packet sits on that orbit at the time slice that we choose to call $t=0$. Therefore a state with prescribed classical data $(\rho_{0},\phi_{0},\dot{\rho}_{0},\dot{\phi}_{0})$ is obtained as follows:
\begin{enumerate}
\item choose $\epsilon$ and $\mu_{0}$ so that the mean conserved charges $\langle \hat{H}\rangle$ and $\langle \hat{m}\rangle$ match the desired orbit;
\item choose $\phi_{0}$ through the center of the angular smearing;
\item choose $t_{0}$ so that the evolved packet has the desired radial position and radial velocity at $t=0$.
\end{enumerate}
Equivalently, in the semiclassical regime one may solve for $t_{0}$ by demanding
\begin{equation}\label{eq:CFT_match_initial_rho_and_rhodot}
\langle \hat{\rho}\rangle_{\rm CFT}(0)=\rho_{0},
\qquad
\left.\frac{d}{dt}\langle \hat{\rho}\rangle_{\rm CFT}(t)\right|_{t=0}
=
\dot{\rho}_{0},
\end{equation}
together with the analogous condition for $\phi$ if desired.

\subsection{Comparison with our Gaussian Wave Packets}\label{subsec:CFT_compare_with_bulk_gaussian}

We can now compare the CFT packet to the bulk Gaussian packet introduced earlier. The most efficient way to do this is to compare the two states in the common $(n,m)$ basis. The bulk Gaussian packet is specified by coefficients $g_{\rm bulk}(n,m)$ through
\begin{equation}
|\Psi_{\rm bulk}\rangle=\sum_{n,m}g_{\rm bulk}(n,m)\,|n,m\rangle,
\end{equation}
while the CFT packet is specified by \eqref{eq:CFT_time_evolved_coefficients_explicit}. Since the basis states are the same one-particle states, all observables computed earlier in the paper apply immediately to either construction after the replacement
\begin{equation}\label{eq:CFT_replace_g_by_gCFT}
g(n,m)\quad\longrightarrow\quad g_{\rm CFT}(n,m;t_{0}).
\end{equation}

In particular, the expectation values
\begin{equation}\label{eq:CFT_H_m_n_expectation_values}
\langle \hat{H}\rangle
=
\sum_{n,m}\omega_{nm}|g(n,m)|^{2},
\qquad
\langle \hat{m}\rangle
=
\sum_{n,m}m\,|g(n,m)|^{2},
\qquad
\langle \hat{n}\rangle
=
\sum_{n,m}n\,|g(n,m)|^{2},
\end{equation}
which are exactly the quantities already introduced in \eqref{tot_ang_mom}, \eqref{ang_modes}, and \eqref{rad_modes}. Likewise, the position-space profile is reconstructed from
\begin{equation}\label{eq:CFT_position_space_profile_again}
f_{\rm CFT}(\rho,\phi,t)
=
\frac{1}{\sqrt{2\pi}}
\sum_{n,m}
g_{\rm CFT}(n,m;t_{0})\,e^{-i\omega_{nm}t}R_{nm}(\rho)e^{im\phi},
\end{equation}
and therefore the expectation values $\langle \hat{\rho}\rangle$, $\langle \hat{\phi}\rangle$, $\langle \widehat{\cos(2\rho)}\rangle$, and $\langle \widehat{\mathcal{Z}}\rangle$ are obtained from the formulas of Sections \ref{sec:EV_of_rho_phi_Ttt} and \ref{sec:Q-C_corres} without any modification.

The radial mean and variance are thus
\begin{equation}\label{eq:CFT_rho_mean_variance}
\langle \hat{\rho}\rangle_{\rm CFT}(t)
=
\sum_{n,n',m}
I(n,n',m)\,
g_{\rm CFT}^{*}(n',m;t_{0})\,g_{\rm CFT}(n,m;t_{0})\,
e^{2i(n'-n)t},
\end{equation}
\begin{equation}\label{eq:CFT_rho2_overlap}
\langle \hat{\rho}^{2}\rangle_{\rm CFT}(t)
=
\sum_{n,n',m}
I^{(2)}(n,n',m)\,
g_{\rm CFT}^{*}(n',m;t_{0})\,g_{\rm CFT}(n,m;t_{0})\,
e^{2i(n'-n)t},
\end{equation}
where
\begin{equation}\label{eq:CFT_I2_definition}
I^{(2)}(n,n',m)
\equiv
\int d\rho\,\rho^{2}\tan\rho\,R^{*}_{n'm}(\rho)R_{nm}(\rho),
\end{equation}
and
\begin{equation}\label{eq:CFT_Delta_rho}
(\Delta \rho)_{\rm CFT}^{2}(t)
=
\langle \hat{\rho}^{2}\rangle_{\rm CFT}(t)
-
\langle \hat{\rho}\rangle_{\rm CFT}(t)^{2}.
\end{equation}
Analogous expressions define the angular mean and variance.

This makes the comparison to the bulk Gaussian packet completely concrete. The two packets should be identified semiclassically by matching the conserved charges, the initial position data, and the initial velocity data:
\begin{equation}\label{eq:CFT_matching_conditions}
\langle \hat{H}\rangle_{\rm CFT}\simeq \langle \hat{H}\rangle_{\rm bulk},
\qquad
\langle \hat{m}\rangle_{\rm CFT}\simeq \langle \hat{m}\rangle_{\rm bulk},
\qquad
\langle \hat{\rho}\rangle_{\rm CFT}(0)\simeq \rho_{0},
\qquad
\langle \hat{\phi}\rangle_{\rm CFT}(0)\simeq \phi_{0},
\end{equation}
together with
\begin{equation}\label{eq:CFT_matching_velocities_and_variances}
\left.\frac{d}{dt}\langle \hat{\rho}\rangle_{\rm CFT}(t)\right|_{t=0}\simeq \dot{\rho}_{0},
\qquad
\left.\frac{d}{dt}\langle \hat{\phi}\rangle_{\rm CFT}(t)\right|_{t=0}\simeq \dot{\phi}_{0},
\end{equation}
and the variances
\begin{equation}\label{eq:CFT_matching_variances}
(\Delta \rho)_{\rm CFT}(0)\simeq (\Delta \rho)_{\rm bulk}(0),
\qquad
(\Delta \phi)_{\rm CFT}(0)\simeq (\Delta \phi)_{\rm bulk}(0),
\qquad
(\Delta m)_{\rm CFT}\simeq (\Delta m)_{\rm bulk}.
\end{equation}
Operationally, this means:
\begin{align}\label{eq:CFT_bulk_parameter_dictionary}
\phi_{0} &\longleftrightarrow \text{center of the boundary angular smearing}, \nonumber\\
\sigma_{2} &\longleftrightarrow \text{width of }F(\theta)\text{ (equivalently the inverse }m\text{-space width)}, \nonumber\\
m_{0} &\longleftrightarrow \text{center of the angular mode-number smearing}, \nonumber\\
\epsilon &\longleftrightarrow \text{mean energy and (semi-classical) outer radial turning point}, \nonumber\\
n_{0} &\longleftrightarrow \text{the phase generated by }e^{-iHt_{0}}, \nonumber\\
\sigma_{1} &\longleftrightarrow \text{the width of the radial }n\text{-distribution around }n_{\ast}(m).
\end{align}

These correspondences are not meant as literal equalities between microscopic parameters.
Rather, they are a statement about how the same semiclassical information is controlled by matching the common $(n,m)$
coefficients and the resulting expectation values, velocities, and variances.

A final conceptual point. The CFT packet \eqref{eq:CFT_time_evolved_coefficients_explicit} need not be exactly Gaussian in the bulk radial coordinate $\rho$. What is approximately Gaussian, in the semiclassical regime, is the distribution in descendant number $n$ around the saddle $n_{\ast}(m)$. The claim is therefore {\em not} that the \cite{Berenstein:2019} packet and the bulk Gaussian profile are literally identical functions of $\rho$, but rather that they are two nearby descriptions of the same semiclassical one-particle state, and that the comparison should be carried out through the common mode coefficients and the observables built from them.

\section{Acknowledgments} We thank Vishal Gayari, Pradipta Pathak and Nirmal Raj for discussions and Gemini for assistance with speeding up our code. V.B. thanks Vishal Rao for discussions on the numerical implementation.

\appendix

\section{Geodesic Equation from Stress Tensor}\label{appendix_geo_from_stress_tensor}

The goal of this appendix is to demonstrate the emergence of the geodesic equation from our stress tensor operator approach. This has close connections to the work of Mathisson, Papapetrou, Dixon and others \cite{Papapetrou:1951,Mathisson1937,Dixon1970I,Dixon1970II,Dixon1974III} in classical general relativity. The key difference here is that instead of working with classical stress tensor configurations, we will be working with expectation values of stress tensor operators of quantum field theories in curved spacetime. This leads to some technical differences, but the broad outlines are similar -- so we will not review the work of \cite{Papapetrou:1951}.

We start with the background spacetime metric in ADM form:
\begin{align}
ds^{2} = -N_{\Sigma}^{2}\,dt^{2}+\sigma_{ab}(dx^{a}+N_{\Sigma}^{a}\,dt)(dx^{b}+N_{\Sigma}^{b}\,dt)
\end{align}
We define the mass associated to the stress tensor of a field living in this spacetime as (note that in general spacetimes this quantity may not be a conserved quantity):
\begin{align}\label{mass}
M = \int_{\Sigma} dV \sqrt{\sigma}N_{\Sigma}n_{\mu}n_{\nu}T^{\mu\nu}
\end{align}
Here $dV$ is the spatial volume element $d^{d-1}x$, $\Sigma$ is a constant time slice hypersurface, $\sigma$ is the determinant of the induced metric on $\Sigma$, and $n^{\mu}$ is the timelike unit normal to $\Sigma$. 

We define the center of mass of $T^{\mu\nu}$ as:
\begin{align}\label{definition of center of mass}
\bar{x}^{\sigma} = \frac{\int_{\Sigma}dV \sqrt{\sigma}N_{\Sigma}\,x^{\sigma}\,n_{\mu}\langle T^{\mu\nu}\rangle n_{\nu}}{\int_{\Sigma}dV\sqrt{\sigma}N_{\Sigma}n_{\mu}\langle T^{\mu\nu}\rangle n_{\nu}}
\end{align}
where the expectation of $T^{\mu\nu}$ is taken in a well-defined Hilbert space state. Using $\sqrt{\sigma}N_{\Sigma}=\sqrt{-g}$ and the normalized $n_{\mu} = (-N_{\Sigma},0,0,\dots)$, we define $\mathcal{T}^{tt} = \sqrt{-g}\,\,T^{tt}$. Equation \eqref{definition of center of mass} becomes:
\begin{align}\label{the final form of xbar}
\bar{x}^{\sigma} = \frac{\int_{\Sigma}dV x^{\sigma}\langle \mathcal{T}^{tt}(x)\rangle N_{\Sigma}^{2}(x)}{\int_{\Sigma}dV\langle \mathcal{T}^{tt}(x)\rangle N_{\Sigma}^{2}(x)}
\end{align}

\subsection{Definitions and Expansions}

From the ADM metric, the invariant volume element is $\sqrt{-g} = N_{\Sigma} \sqrt{\sigma}$. The normal co-vector to the spatial slice is $n_{\mu} = (-N_{\Sigma}, 0,0,0)$. This means the energy density simplifies exactly to $n_{\mu} n_{\nu} T^{\mu\nu} = N_{\Sigma}^2 T^{00}$. Defining the tensor density $\langle\mathcal{T}^{\mu\nu}\rangle = \sqrt{-g}\, \langle T^{\mu\nu}\rangle$, we have:
\begin{align}
M = \int d^3x \, N_{\Sigma}^2 \langle \mathcal{T}^{00}\rangle, \quad \bar{x}^i = \frac{1}{M} \int d^3x \, N_{\Sigma}^2 \langle \mathcal{T}^{00}\rangle x^i
\end{align}

Every time we evaluate a background field over the integral, we split it into its value at the center of mass $\bar{x}$ and its exact deviation $\delta$:
\begin{itemize}
    \item \textbf{Coordinates:} Let $x^i = \bar{x}^i + X^i$, where $X^i$ is the exact coordinate distance from the center of mass.
    \item \textbf{Connection:} Let $\Gamma^{\mu}_{\alpha\beta}(x) = \bar{\Gamma}^{\mu}_{\alpha\beta} + \delta\Gamma^{\mu}_{\alpha\beta}(x)$, where $\bar{\Gamma}^{\mu}_{\alpha\beta} = \Gamma^{\mu}_{\alpha\beta}(\bar{x})$.
    \item \textbf{Lapse Function:} Let $N_{\Sigma}^2(x) = \bar{N}_{\Sigma}^2 + \delta(N_{\Sigma}^2)(x)$, where $\bar{N}_{\Sigma}^2 = N_{\Sigma}^2(\bar{x})$.
\end{itemize}
We define the macroscopic moments as $M^{\mu\nu} = \int \langle \mathcal{T}^{\mu\nu}\rangle d^3x$. Note that the 00 component of $M^{\mu\nu}$ is not quite the mass $M$ defined in Eq. \eqref{mass}. The inner moments of the deviations are:
\begin{itemize}
    \item $D^{i\alpha\beta} \equiv \int X^i \, \langle \mathcal{T}^{\alpha\beta}\rangle d^3x$ (The dipole moments)
    \item $\delta\Gamma^\mu \equiv \int \delta\Gamma^{\mu}_{\alpha\beta} \, \langle \mathcal{T}^{\alpha\beta}\rangle d^3x$ (Connection deviation integral)
    \item $\delta\Gamma^{i\mu} \equiv \int X^i \, \delta\Gamma^{\mu}_{\alpha\beta} \, \langle\mathcal{T}^{\alpha\beta}\rangle d^3x$ (Dipole connection deviation)
\end{itemize}

\subsection{The Exact Center of Mass Dipole}

From the definition of $\bar{x}^i$, the integral of the displacement $X^i$ weighted by $N_{\Sigma}^2 \langle\mathcal{T}^{00}\rangle$ is strictly zero:
\begin{align}
\int X^i \, N_{\Sigma}^2\, \langle\mathcal{T}^{00}\rangle d^3x = 0
\end{align}
Substituting the lapse expansion $N_{\Sigma}^2(x) = \bar{N}_{\Sigma}^2 + \delta(N_{\Sigma}^2)(x)$:
\begin{align}
\int X^i \left( \bar{N}_{\Sigma}^2 + \delta(N_{\Sigma}^2) \right) \langle\mathcal{T}^{00}\rangle d^3x = 0 \nonumber \\
\bar{N}_{\Sigma}^2 \int X^i \langle\mathcal{T}^{00}\rangle d^3x + \int X^i \delta(N_{\Sigma}^2) \langle\mathcal{T}^{00}\rangle d^3x = 0 \nonumber \\
D^{i00} = -\frac{1}{\bar{N}_{\Sigma}^2} \int X^i \, \delta(N_{\Sigma}^2) \, \langle\mathcal{T}^{00}\rangle d^3x
\end{align}
$D^{i00}$ directly represents the coupling of the body's internal mass distribution to the gradient of the lapse.

\subsection{Exact Integration of Energy Conservation ($\mu=0$)}

The local conservation law is $\partial_0 \langle\mathcal{T}^{\mu0}\rangle + \partial_k \langle\mathcal{T}^{\mu k}\rangle + \Gamma^{\mu}_{\alpha\beta} \langle\mathcal{T}^{\alpha\beta}\rangle = 0$. For $\mu=0$:
\begin{align}
\partial_0 \langle\mathcal{T}^{00}\rangle + \partial_k \langle\mathcal{T}^{0k}\rangle + \bar{\Gamma}^{0}_{\alpha\beta} \langle\mathcal{T}^{\alpha\beta}\rangle + \delta\Gamma^{0}_{\alpha\beta} \langle\mathcal{T}^{\alpha\beta}\rangle = 0
\end{align}
Integrating over the spatial volume (boundary terms vanish):
\begin{align}
&\frac{d}{dt} \int \langle\mathcal{T}^{00}\rangle d^3x + \bar{\Gamma}^{0}_{\alpha\beta} \int \langle\mathcal{T}^{\alpha\beta}\rangle d^3x + \int \delta\Gamma^{0}_{\alpha\beta} \langle\mathcal{T}^{\alpha\beta}\rangle d^3x = 0 \nonumber \\
&\dot{M}^{00} = -\bar{\Gamma}^{0}_{\alpha\beta} M^{\alpha\beta} - \delta\Gamma^0
\end{align}

\subsection{Exact Integration for the Velocity Moment ($M^{0i}$)}

Multiply the $\mu=0$ conservation law by $x^i$ and integrate. Using integration by parts ($\int x^i \partial_k \langle\mathcal{T}^{0k}\rangle d^3x = -M^{0i}$):
\begin{align}
\frac{d}{dt} \int x^i \langle\mathcal{T}^{00}\rangle d^3x - M^{0i} + \int x^i \left( \bar{\Gamma}^{0}_{\alpha\beta} + \delta\Gamma^{0}_{\alpha\beta} \right) \langle\mathcal{T}^{\alpha\beta}\rangle d^3x = 0
\end{align}
Substituting $x^i = \bar{x}^i + X^i$:
\begin{align}
\frac{d}{dt} \left( \bar{x}^i M^{00} + D^{i00} \right) - M^{0i} + \bar{\Gamma}^{0}_{\alpha\beta} \left( \bar{x}^i M^{\alpha\beta} + D^{i\alpha\beta} \right) + \bar{x}^i \delta\Gamma^0 + \delta\Gamma^{i0} = 0 \nonumber \\
\dot{\bar{x}}^i M^{00} + \bar{x}^i \dot{M}^{00} + \dot{D}^{i00} - M^{0i} + \bar{x}^i \bar{\Gamma}^{0}_{\alpha\beta} M^{\alpha\beta} + \bar{\Gamma}^{0}_{\alpha\beta} D^{i\alpha\beta} + \bar{x}^i \delta\Gamma^0 + \delta\Gamma^{i0} = 0
\end{align}
Substituting $\dot{M}^{00}$ from Step 3:
\begin{align}
\dot{\bar{x}}^i M^{00} + \bar{x}^i ( -\bar{\Gamma}^{0}_{\alpha\beta} M^{\alpha\beta} - \delta\Gamma^0 ) + \dot{D}^{i00} - M^{0i} + \bar{x}^i \bar{\Gamma}^{0}_{\alpha\beta} M^{\alpha\beta} + \bar{\Gamma}^{0}_{\alpha\beta} D^{i\alpha\beta} + \bar{x}^i \delta\Gamma^0 + \delta\Gamma^{i0} = 0
\end{align}
After cancellations, we find:
\begin{align}
M^{0i} = \dot{\bar{x}}^i M^{00} + \epsilon^i, \quad \text{where } \epsilon^i \equiv \dot{D}^{i00} + \bar{\Gamma}^{0}_{\alpha\beta} D^{i\alpha\beta} + \delta\Gamma^{i0}
\end{align}

\subsection{Exact Integration for the Momentum Flux ($M^{ij}$)}

Following the same process for $\mu=j$, multiplied by $x^k$:
\begin{align}
\frac{d}{dt} \int x^k \langle\mathcal{T}^{j0}\rangle\, d^3x - M^{jk} + \int x^k \left( \bar{\Gamma}^{j}_{\alpha\beta} + \delta\Gamma^{j}_{\alpha\beta} \right) \langle\mathcal{T}^{\alpha\beta}\rangle\, d^3x = 0
\end{align}
Substituting $x^k = \bar{x}^k + X^k$, $\dot{M}^{j0} = -\bar{\Gamma}^{j}_{\alpha\beta} M^{\alpha\beta} - \delta\Gamma^j$, and $M^{j0} = \dot{\bar{x}}^j M^{00} + \epsilon^j$:
\begin{align}
M^{jk} = M^{00} \dot{\bar{x}}^j \dot{\bar{x}}^k + \epsilon^{jk}
\end{align}
Where $\epsilon^{jk} = \dot{\bar{x}}^k \epsilon^j + \dot{D}^{kj0} + \bar{\Gamma}^{j}_{\alpha\beta} D^{k\alpha\beta} + \delta\Gamma^{kj}$. It may not look symmetric but it is because both $M^{jk}$ and $M^{00} \dot{\bar{x}}^j \dot{\bar{x}}^k$ are symmetric in $j$ and $k$ \footnote{The reason for this is because we used $\mu=j$ and multiplied by $x^{k}$, but we could have set $\mu=k$ and multiplied by $x^{j}$ as well. }. Substituting these into $\dot{M}^{i0} = -\bar{\Gamma}^{i}_{\alpha\beta} M^{\alpha\beta} - \delta\Gamma^i$:
\begin{align}
\frac{d}{dt} \left( M^{00} \dot{\bar{x}}^i + \epsilon^i \right) = -\bar{\Gamma}^{i}_{\alpha\beta} \left( M^{00} \dot{\bar{x}}^\alpha \dot{\bar{x}}^\beta + \epsilon^{\alpha\beta} \right) - \delta\Gamma^i \nonumber \\
\dot{M}^{00} \dot{\bar{x}}^i + M^{00} \ddot{\bar{x}}^i + \dot{\epsilon}^i = -M^{00} \bar{\Gamma}^{i}_{\alpha\beta} \dot{\bar{x}}^\alpha \dot{\bar{x}}^\beta - \bar{\Gamma}^{i}_{\alpha\beta} \epsilon^{\alpha\beta} - \delta\Gamma^i
\end{align}
Substituting the exact $\dot{M}^{00}$ and dividing by $M^{00}$:
\begin{align}
\ddot{\bar{x}}^i + \bar{\Gamma}^{i}_{\alpha\beta} \dot{\bar{x}}^\alpha \dot{\bar{x}}^\beta - \bar{\Gamma}^{0}_{\alpha\beta} \dot{\bar{x}}^\alpha \dot{\bar{x}}^\beta \dot{\bar{x}}^i = \frac{1}{M^{00}} \left[ \dot{\bar{x}}^i \left( \bar{\Gamma}^{0}_{\alpha\beta} \epsilon^{\alpha\beta} + \delta\Gamma^0 \right) - \dot{\epsilon}^i - \bar{\Gamma}^{i}_{\alpha\beta} \epsilon^{\alpha\beta} - \delta\Gamma^i \right]
\end{align}
This result shows that the geodesic is followed by the center of mass $\bar{x}$ up to leading order in any metric and for any (free) field.

In Ad$S_3$ for a free scalar field, the above expression looks like:
\begin{itemize}
    \item \textbf{Radial Equation ($i=\rho$):}
\end{itemize}
\begin{align}\label{when you substitute i=rho in the final equation}
\ddot{\bar{\rho}} + \tan(\bar{\rho})(1 - \dot{\bar{\rho}}^2 - \dot{\bar{\phi}}^2) = \frac{1}{M^{00}} \left[ \dot{\bar{\rho}} \left( \bar{\Gamma}^{0}_{\alpha\beta} \epsilon^{\alpha\beta} + \delta\Gamma^0 \right) - \dot{\epsilon}^\rho - \bar{\Gamma}^{\rho}_{\alpha\beta} \epsilon^{\alpha\beta} - \delta\Gamma^\rho \right]
\end{align}

\begin{itemize}
    \item \textbf{Angular Equation ($i=\phi$):}
\end{itemize}
\begin{align}\label{when you substitute i=phi in the final equation}
\ddot{\bar{\phi}} + 2\dot{\bar{\rho}}\dot{\bar{\phi}} \cot\bar{\rho} = \frac{1}{M^{00}} \left[ \dot{\bar{\phi}} \left( \bar{\Gamma}^{0}_{\alpha\beta} \epsilon^{\alpha\beta} + \delta\Gamma^0 \right) - \dot{\epsilon}^\phi - \bar{\Gamma}^{\phi}_{\alpha\beta} \epsilon^{\alpha\beta} - \delta\Gamma^\phi \right]
\end{align}

\section{Position Operators in Static Spacetimes}\label{appendix_recipe_for_positiooperators_static}
Let us consider a static spacetime, which means we have a hypersurface-orthogonal timelike Killing vector $\xi_{\mu}$. In this case\footnote{What we will also need is ``something like'' global hyperbolicity: the static region together with the chosen boundary conditions should define a closed, unitary Klein–Gordon problem. We will not demand global hyperbolicity because technically AdS does not satisfy this, but it still is a good geometry for our purposes with reflecting boundary conditions at infinity.} we can choose coordinates where the metric components are independent of the coordinate $t$ and there are no spacetime cross terms \cite{Carroll:2004st}
\bea
\de_{t}g_{\mu\nu}=0, \,\,\,\,\,\,\,\,\,\,\,\,\,\,\,\,\,\,\,\,g_{ti}=0
\eea
The metric takes the form:
\bea
ds^{2} = g_{tt}\left(x^{i}\right)\,dt^{2}+\gamma_{ij}\left(x^{i}\right)dx^{i}dx^{j}
\eea
Since the metric is independent of time, the solution to the KG equation has the form:
\bea
\phi_{n}(x^{i},t)=\frac{1}{\sqrt{2\omega_{n}}}u_{n}\left(x^{i}\right)e^{-i\omega_{n}t}
\eea
where $u_{n}\left(x^{i}\right)$ is rest of the solution with no time dependence. The index $n$ includes all quantum numbers: in AdS$_3$, it takes the form $u_{nm}\left(x^{i}\right)=R_{nm}(\rho)\frac{e^{im\phi}}{\sqrt{2\pi}}$. The KG inner product is:
\bea
\langle\phi_{n},\phi_{m}\rangle=-i\int_{\Sigma}\left(\phi_{n}\nabla_{\mu}\phi^{*}_{m}-\phi^{*}_{m}\nabla_{\mu}\phi_{n}\right)n^{\mu}\sqrt{\gamma}\,d^{n-1}\boldsymbol{x}
\eea
Orthonormality of the modes guarantee that
\bea\label{The inner product integral}
\delta_{nm}
=\int d^{n-1}\boldsymbol{x}\,\frac{\sqrt{\gamma}}{\sqrt{g_{tt}}}\,u^{*}_{m}(\vec{x})u_{n}(\vec{x})
\eea
and completeness gives
\bea
\sum_{n}u^{*}_{n}(\boldsymbol{y})\,u_{n}(\boldsymbol{x})=\frac{\delta^{n-1}(\boldsymbol{x}-\boldsymbol{y})}{\frac{\sqrt{\gamma}}{\sqrt{g_{tt}}}}
\eea
The field mode expansion is:
\bea
\Phi(t,\boldsymbol{x}) = \sum_{n}\frac{1}{\sqrt{2\,\omega_{n}}}\Big[u_{n}e^{-i\omega_{n}t}a_{n}+u_{n}^{*}e^{i\omega_{n}t}a^{\dagger}_{n}\Big]
\eea
Now we define $a^{\dag}(\boldsymbol{x})|0\rangle=|\boldsymbol{x}\rangle$ as:
\bea
a^{\dag}(\boldsymbol{x})= \sum_{n}u_{n}^{*}(\boldsymbol{x})a^{\dagger}_{n}
\eea
The non trivial commutation relation is:
\bea
[a(\boldsymbol{x}),a^{\dagger}(\boldsymbol{y})]=\frac{\sqrt{g_{tt}}}{\sqrt{\gamma}}\delta^{n-1}(\boldsymbol{x}-\boldsymbol{y})
\eea
The general position operator in this class of spacetimes is:
\bea
\hat{\boldsymbol{x}} = \int d^{n-1}\boldsymbol{x}\,\frac{\sqrt{\gamma}}{\sqrt{g_{tt}}}\, a^{\dagger}(\boldsymbol{x})\,\boldsymbol{x}\,a(\boldsymbol{x})
\eea
which satisfies,
\bea
\hat{\boldsymbol{x}}|\boldsymbol{y}\rangle=\boldsymbol{y}|\boldsymbol{y}\rangle.
\eea

\section{Scalar Field in AdS$_3$}\label{appendix_scalar_field_ads3}

We work with the following AdS$_{3}$ metric ($\mathcal{R}=1$):
\bea\label{metric}
ds^{2} = \sec^{2}(\rho)dt^{2}-\sec^{2}(\rho)d\rho^{2}-\tan^{2}(\rho)d\phi^{2}
\eea

\subsection{Charges}\label{appendix_writing_hamiltonian_and _other_operators_ads3}
The Lagrangian density for the massive scalar field $\Phi$ is given by:
\bea\label{lagrangian}
\mathcal{L}= \frac{1}{2}\sqrt{g}\left(g^{\mu\nu}\partial_{\mu}\Phi\partial_{\nu}\Phi-M^{2}\Phi^{2}\right)
\eea
where the mode expansion of $\Phi$ is written as:
\bea\label{mode expansion1}
\Phi(t,\rho,\phi) = \sum_{m,n}\Big[\mathcal{U}_{mn}a_{mn}+\mathcal{U}_{mn}^{*}a^{\dagger}_{mn}\Big]
\eea
Here, we define $\mathcal{U}_{mn} = \frac{1}{\sqrt{2\omega_{mn}}}e^{-i\omega_{mn}t}\frac{e^{im\phi}}{\sqrt{2\pi}}R_{mn}(\rho)$. From the equation of motion for the above metric \eqref{metric}, the solution for the radial part of the field $\Phi$ becomes,
\bea \label{Form of Rnmrho that is used}
R_{mn}(\rho)= \frac{1}{\mathcal{N}_{mn}}(\sin\rho)^{|m|}(\cos\rho)^{\Delta}\hspace{1mm}_{2}F_{1}(-n,\Delta+|m|+n,|m|+1,\sin^{2}\rho)
\eea
Where $\mathcal{N}_{mn}$ is the normalization constant. The explicit form of the normalization is (see \cite{Kaplan_AdSCFT}):\footnote{There is additional factor of $ 2(2n+|m|+\Delta)$ because we choose our field expansion with an additional $\frac{1}{\sqrt{2\omega_{nm}}}$ factor and we want the commutation relation $\left[\Phi(\boldsymbol{x},t),\Pi(\boldsymbol{y},t)\right]=i\delta^{2}(\boldsymbol{x}-\boldsymbol{y})$ to hold (\cite{Kaplan_AdSCFT} has different).}
\bea
\mathcal{N}_{mn}= (-1)^{n}\sqrt{\frac{n!\Gamma^{2}(|m|+1)\Gamma(\Delta+n)}{\Gamma(n+|m|+1)\Gamma(\Delta+n+|m|) 2 (2n+|m|+\Delta)}}
\eea
We write the Hamiltonian density as:
\bea
\mathcal{H} = \Pi \Dot{\Phi} -\mathcal{L}
\eea
where $\Pi$ is the conjugate variable to $\Phi$. Now the Hamiltonian of a massive scalar field in AdS$_{3}$ becomes:
\bea\label{Ham_op_in_AdS3}
\hat{H} = \int d\rho\,d\phi\,\mathcal{\hat{H}} = \sum_{n,m}\omega_{nm}\,a^{\dagger}_{nm}a_{nm}
\eea
The above Hamiltonian is normal ordered and $\omega_{nm} = \Delta +2n+|m|$. See \cite{Nastase2015} for explicit calculation of $\omega_{nm}$ is AdS$_{3}$.

The expectation value of \eqref{Ham_op_in_AdS3} in the state \eqref{generic state} gives us the total energy:
\bea
\langle \hat{H}\rangle = \sum_{n,m}\omega_{nm}|g(n,m)|^{2}
\eea
Along with this, we can also define the following operators and their corresponding expectation values:
\begin{align} 
\hat{m}&=\sum_{n,m}\,m\,a^{\dagger}_{nm}a_{nm}\,\,\,\,\,\,\,\,;\,\,\,\,\hspace{1cm}\langle \hat{m}\rangle = \sum_{n,m}m|g(n,m)|^{2}\label{tot_ang_mom} \\
\hat{|m|}&=\sum_{n,m}\,|m|\,a^{\dagger}_{nm}a_{nm}\,\,\,\,\,;\hspace{1cm}\langle |\hat{m}| \rangle = \sum_{n,m}|m| |g(n,m)|^{2}\label{ang_modes}\\
\hat{n}&=\sum_{n,m}\,n\,a^{\dagger}_{nm}a_{nm}\,\,\,\,\,\,\,\,\,\,\,;\,\,\,\,\hspace{1cm}\langle \hat{n} \rangle = \sum_{n,m}n|g(n,m)|^{2}\label{rad_modes}
\end{align}
The expectation values $\langle \hat{H}\rangle,\langle \hat{m}\rangle$ come out to be the conserved charges $E_{T}$ and $L$, respectively, of the wave packet. We can see this as follows:

In the metric \eqref{metric}, the timelike Killing vector $\xi_{\nu}$ and the normal to the constant time surface $n_{\mu}$ are $\xi_{\nu}=\left(g_{tt},0,0\right)$ and $n_{\mu}=\left(\sqrt{g_{tt}},0,0\right)$ respectively. 
So the conserved charge is (see \cite{Carroll:2004st} for details of the method): 
$E_{T}=\int T^{\mu \nu} \xi_{\nu} n_{\mu} \sqrt{h}\, d\rho \, d\phi$. 
Now using Eq. \eqref{Stress Tensor for free scalar field} and the equation of motion, doing integration by parts, and finally promoting $\Phi$ to $\hat{\Phi}$ we get:
\begin{equation}\label{the conserved energy operator E_{T} written as integral over stress tensor}
\hat{E}_{T}=\frac{1}{2} \int d\rho\, d\phi\, \sqrt{g}\, g^{tt} \left(\dot{\hat{\Phi}}^2-\hat{\Phi}\,\de^{2}_{t}\hat{\Phi}\right) 
\end{equation}
Using the field expansion given in equation \eqref{mode expansion1} and applying Eq. \eqref{orthogonality relation} and \eqref{completeness relation}, this becomes
$\hat{E_{T}}=\frac{1}{2}\sum_{n,m} \omega_{nm}\left(a_{nm}a\dag_{nm}+a\dag_{nm}a_{nm}\right)$
which is precisely the Hamiltonian operator \eqref{Ham_op_in_AdS3} up to normal ordering.
Note that Eq. \eqref{the conserved energy operator E_{T} written as integral over stress tensor} (after normal ordering) was used to compute the wave packet's energy while doing the numerics in the stress tensor approach while Eq. \eqref{Ham_op_in_AdS3} was used in the position operator approach.

Now let us look at the conserved charge associated with the Killing vector $\partial_{\phi}$. This time we have the Killing vector $\xi_{\nu}=\left(0,0,g_{\phi \phi}\right)$.
With this we have the conserved charge as 
$L=\int T^{\mu \nu} \xi_{\nu} n_{\mu} \sqrt{h} d\rho\, d\phi$.
Upon simplification and promoting $\Phi$ to $\hat{\Phi}$, this integral becomes:
\begin{equation}\label{Angular Momentum Integral operator as integral}
\hat{L}=\int d\rho \, d\phi\, \sqrt{g}\, g^{tt}\dot{\hat{\Phi}}\,\de_{\phi}\hat{\Phi}
\end{equation}

Taking the expectation value of \eqref{Angular Momentum Integral operator as integral} in the single particle state \eqref{generic single particle state at t=0}, we get (after removing an additive constant):
$\langle\hat{L}\rangle= -\sum_{n,m}\,m\, |g(n,m)|^{2}$.
This is the same as (up to sign) the expectation value of the angular momentum operator $\hat{m}$ on a generic single particle state (see Eq. \eqref{tot_ang_mom}). Eq. \eqref{Angular Momentum Integral operator as integral} was used to compute the angular momentum of the wave packet in the stress tensor approach while Eq. \eqref{tot_ang_mom} was used in the position operator approach.

In the above discussion, $\langle|\hat{m}|\rangle,\langle \hat{n}\rangle$ are the total contribution of the angular and radially excited modes (descendants in the dual CFT, as we discuss in the main text) respectively, to the total energy $\langle \hat{H}\rangle$ of the wave packet. It is important to emphasize here that $\langle H\rangle$ and $\langle \hat{m}\rangle$ are conserved charges associated with the wave packet, not with the geodesic.

\subsection{Stress-Tensor}\label{appendix_free_scalar_field_stresstensor_details}

We will need mode-sum expressions as well as expectation values for the stress tensor, which we compute in this subsection.
For the scalar field $\Phi$ the stress-energy tensor is given by:
\bea\label{scalar_field_T}
\hat{T}^{\mu\nu} = \partial^{\mu}\Phi\partial^{\nu}\Phi - \frac{1}{2}g^{\mu\nu}(\partial^{\lambda}\Phi\partial_{\lambda}\Phi - M^{2}\Phi^{2} )
\eea
Using the above metric \eqref{metric} the $00^{th}$ component of \eqref{scalar_field_T} becomes:
\bea\label{T00}
\hat{T}^{00} = \frac{1}{2}(g^{tt})^{2}(\partial_{t}\Phi){^2}-\frac{1}{2}g^{tt}g^{\rho\rho}(\partial_{\rho}\Phi)^{2}-\frac{1}{2}g^{tt}g^{\phi\phi}(\partial_{\phi}\Phi)^{2}+\frac{1}{2}g^{tt}M^{2}\Phi^{2}
\eea
From \eqref{mode expansion1} and \eqref{Form of Rnmrho that is used} the above equation leads to:\footnote{Here we are only keeping the terms that have an equal number of creation and annihilation operators. Because these are the terms that survive when we take the expectation value of $\hat{T}^{00}$ in the state $|\Psi\rangle$, all other terms vanish.}
\bea\label{T00 simplified 1}
\hat{T}^{00} = \hspace{-4mm}\sum_{n_{1},m_{1},n_{2},m_{2}}\hspace{-2mm}\frac{A(n_{1},m_{1};n_{2},m_{2})}{4\pi\sqrt{\omega_{n_{1}m_{1}}\omega_{n_{2}m_{2}}}} e^{-i(\omega_{n_{1}m_{1}}-\omega_{n_{2}m_{2}})t}e^{i(m_{1}-m_{2})\phi}\Big(a_{n_{1}m_{1}}a^{\dagger}_{n_{2}m_{2}} + a^{\dagger}_{n_{2}m_{2}}a_{n_{1}m_{1}} \Big)
\eea
where,
\begin{align}\label{A1 App}
A = B^{*} &= \frac{1}{2}(g^{tt})^{2}R_{n_{1}m_{1}}(\rho)R^{*}_{n_{2}m_{2}}(\rho)\omega_{n_{1}m_{1}}\omega_{n_{2}m_{2}} - \frac{1}{2}g^{tt}g^{\rho\rho}\partial_{\rho}R^{}_{n_{1}m_{1}}(\rho)\partial_{\rho}R^{*}_{n_{2}m_{2}}(\rho) -  \nonumber\\
&\hspace{2cm}-\frac{1}{2}g^{tt}g^{\phi\phi}R_{n_{1}m_{1}}(\rho)R^{*}_{n_{2}m_{2}}(\rho)m_{1}m_{2} 
+ \frac{1}{2}M^{2}g^{tt}R_{n_{1}m_{1}}(\rho)R^{*}_{n_{2}m_{2}}(\rho)
\end{align}
The expectation value of $\hat{T}^{00}$ in the state \eqref{generic state} becomes:
\bea\label{T00_EV_intermediate}
\hspace{-4mm}\langle\hat{T}^{00}\rangle\hspace{-1mm} = \hspace{-7mm}\sum_{n_{1},m_{1},n_{2},m_{2}}\hspace{-3mm}\frac{A(n_{1},m_{1};n_{2},m_{2}) }{4\pi\sqrt{\omega_{n_{1}m_{1}}\omega_{n_{2}m_{2}}}} e^{-i(\omega_{n_{1}m_{1}}-\omega_{n_{2}m_{2}})t}e^{i(m_{1}-m_{2})\phi}\hspace{-1mm}\Big(\hspace{-1mm}\langle a_{n_{1}m_{1}}a^{\dagger}_{n_{2}m_{2}}\rangle\hspace{-1mm} + \hspace{-1mm}\langle a^{\dagger}_{n_{2}m_{2}}a_{n_{1}m_{1}}\rangle\Big)
\eea
To further simplify the above expression, we use the following identities:
\begin{align}
\langle a_{n_{1}m_{1}}a^{\dagger}_{n_{2}m_{2}}\rangle &= \langle a^{\dagger}_{n_{2}m_{2}}a_{n_{1}m_{1}}\rangle + \langle\delta_{n_{1}n_{2}}\delta_{m_{1}m_{2}}\rangle\\
\langle a^{\dagger}_{n_{2}m_{2}}a_{n_{1}m_{1}}\rangle &= g^{*}(n_{2},m_{2})g(n_{1},m_{1}) \\
\langle\delta_{n_{1}n_{2}}\delta_{m_{1}m_{2}}\rangle & = \delta_{n_{1}n_{2}}\delta_{m_{1}m_{2}}
\end{align}
In the last identity, we used $\langle\Psi|\Psi\rangle =1$. With this \eqref{T00_EV_intermediate} becomes:
\begin{align}\label{expression for T^tt in terms of A and gnm}
\hspace{-3mm}\langle\hat{T}^{00}\rangle &= \sum_{n_{1},m_{1},n_{2},m_{2}}\frac{A(n_{1},m_{1};n_{2},m_{2}) }{2\pi\sqrt{\omega_{n_{1}m_{1}}\omega_{n_{2}m_{2}}}}g^{*}(n_{2},m_{2})g(n_{1},m_{1}) e^{-i(\omega_{n_{1}m_{1}}-\omega_{n_{2}m_{2}})t}e^{i(m_{1}-m_{2})\phi} + \nonumber\\
&\hspace{8cm}+ \sum_{n_{1},m_{1}}\frac{A(n_{1},m_{1};n_{1},m_{1})}{4\pi\omega_{n_{1}m_{1}}}
\end{align}
The second term in the above expression is time-independent, so it will not affect the overall behavior of $\langle\hat{T}^{00}\rangle$ with respect to time. So we can remove it as a zero-point term. Substituting the expression of $A(n_{1},m_{1};n_{2},m_{2})$ from \eqref{A1 App} into \eqref{T00_EV_intermediate} gives us:
\bea
\langle\hat{T}^{00}\rangle = I + II + III + IV
\eea
where,
\bea
I = (g^{tt})^{2}|S|^{2},\hspace{5mm} II = -g^{tt}g^{\rho\rho}|T|^{2},\hspace{5mm} III = - g^{tt}g^{\phi\phi}|U|^{2},\hspace{5mm} IV = g^{tt}|V|^{2}
\eea
and,
\begin{align}
S &= \int d\rho' d\phi'\tan(\rho') G_{t}(\rho,\rho',\phi,\phi',t)f(\rho',\phi') \label{S}\\
T &= \int d\rho' d\phi'\tan(\rho') G_{\rho}(\rho,\rho',\phi,\phi',t)f(\rho',\phi')  \label{T}\\
U &= \int d\rho' d\phi'\tan(\rho') G_{p}(\rho,\rho',\phi,\phi',t)f(\rho',\phi')  \label{U}\\
V &= \int d\rho' d\phi'\tan(\rho') G_{M}(\rho,\rho',\phi,\phi',t)f(\rho',\phi') \label{V}
\end{align}
Here $G_{t}, G_{\rho}, G_{p}$ and $G_{M}$ are defined as:
\begin{align}\label{equation for the 4 propagators to calculate T^00}
G_{t}(\rho,\rho',\phi,\phi',t) &= \sum_{n,m}\frac{-i\omega_{nm}}{2\pi}\frac{R_{nm}(\rho)}{\sqrt{2\omega_{mn}}}R_{nm}^{*}(\rho')e^{im(\phi-\phi')}e^{-i\omega_{nm}t} \nonumber \\
G_{\rho}(\rho,\rho',\phi,\phi',t) &= \sum_{n,m}\frac{1}{2\pi}\frac{\partial_{\rho}R_{nm}(\rho)}{\sqrt{2\omega_{mn}}}R_{nm}^{*}(\rho')e^{im(\phi-\phi')}e^{-i\omega_{nm}t} \nonumber \\
G_{p}(\rho,\rho',\phi,\phi',t) &=\sum_{n,m}\frac{im}{2\pi}\frac{R_{nm}(\rho)}{\sqrt{2\omega_{mn}}}R_{nm}^{*}(\rho')e^{im(\phi-\phi')}e^{-i\omega_{nm}t} \nonumber\\
G_{M}(\rho,\rho',\phi,\phi',t) &=\sum_{n,m}\frac{M}{2\pi}\frac{R_{nm}(\rho)}{\sqrt{2\omega_{mn}}}R_{nm}^{*}(\rho')e^{im(\phi-\phi')}e^{-i\omega_{nm}t} 
\end{align}

\section{Why is \texorpdfstring{$\langle\hat{L}\rangle=-\langle\hat{m}\rangle=m_{0}$}{m0}?}\label{appendix_why_is_the_expval_m_0?}

The single particle initial state we used was 
\bea
f(\rho,\phi) = \mathcal{N}_{\rho}e^{-\frac{(\rho-\rho_{0})^{2}}{4\sigma_{1}^{2}}}e^{-in_0(\rho-\rho_0)}\mathcal{N}_{\phi}e^{-\frac{(\phi-\phi_{0})^{2}}{4\sigma_{2}^{2}}}e^{-im_0(\phi-\phi_0)} \label{Gauss-App}
\eea
For this profile, we got the conserved $\langle\hat{L}\rangle=-\langle\hat{m}\rangle=m_{0}$. Here we explain why this is the case. 

We start with equation \eqref{gnm} with $t=0$.
\bea
g(n,m) =\frac{1}{\sqrt{2\pi}}\int d\rho\,d\phi\,\tan(\rho)\,R^{*}_{nm}(\rho)f(\rho,\phi)e^{-im\phi}
\eea
We can invert this using the equations \eqref{orthogonality relation} and \eqref{completeness relation} to get
\bea
f(\rho,\phi) = \frac{1}{\sqrt{2\pi}}\sum_{n,m}g(n,m)R_{nm}(\rho)e^{im\phi}
\eea
Differentiating this w.r.t. $\phi$ and multiplying by $-i$ we get
\bea
-i\frac{d}{d\phi}f(\rho,\phi)=\frac{1}{\sqrt{2\pi}}\sum_{n,m}g(n,m)\,m\,R_{nm}(\rho)e^{im\phi}
\eea
Again we can invert this to get
\bea
m\,g(n,m) =\frac{1}{\sqrt{2\pi}}\int d\rho\,d\phi\,\tan(\rho)\,R^{*}_{nm}(\rho)\left(-i\frac{d}{d\phi}f(\rho,\phi)\right)e^{-im\phi}
\eea
Now we multiply both sides by $-g^{*}(n,m)$ and sum over $n,m$. Then integrating out over one of the set of variables $\rho,\phi$ we get
\bea\label{Angular momentum operator as in quantum mechanics}
-\sum_{n,m}m\, |g(n,m)|^{2}=\langle\hat{L}\rangle=-\int d\rho d\phi \tan(\rho)\left(-i\frac{d}{d\phi}f(\rho,\phi)\right)f^{*}(\rho,\phi).
\eea

We can now work out this integral explicitly for the Gaussian wave packet \eqref{Gauss-App} (or more correctly, \eqref{general choice of packet}). But it turns out that the result we are after, only depends on general properties of the packet, so   we will consider a more general form:
\bea
f(\rho,\phi) = h(\rho)\,g(\phi) e^{-im_{0}(\phi-\phi_{0})}.
\eea
The function $h(\rho)$ is complex, $g(\phi)$ is real, and they are both normalized. We will take $g(\phi)$ to be periodic in $\phi$: the image sum in \eqref{general choice of packet} guarantees periodicity\footnote{Even if we drop the image sum and approximate the wave packet as in \eqref{Gauss-App}, because of the decay of the angular Gaussian, the relevant $g(\phi)$ is still approximately periodic.}. 
Differentiating both sides w.r.t. $\phi$ and multiplying by $-if^{*}(\rho,\phi)$,
\bea
-if^{*}(\rho,\phi)\frac{d}{d\phi}f(\rho,\phi)=-i|h(\rho)|^{2}g^{*}(\phi)\frac{d}{d\phi}g(\phi)-m_{0}|f(\rho,\phi)|^{2}
\eea
Integrating both sides, we have
\bea
-i\int d\rho \tan(\rho)d\phi f^{*}(\rho,\phi)\frac{d}{d\phi}f(\rho,\phi)=-i\int d\rho \tan(\rho)d\phi|h(\rho)|^{2}g^{*}(\phi)\frac{d}{d\phi}g(\phi)\nonumber\\-\int d\rho \tan(\rho)d\phi\, m_{0}|f(\rho,\phi)|^{2}
\eea
Given that $f(\rho,\phi)$ and $h(\rho)$ are normalized, we get
\bea
-i\int d\rho \tan(\rho)d\phi f^{*}(\rho,\phi)\frac{d}{d\phi}f(\rho,\phi)=
-i\int d\phi g^{*}(\phi)\frac{d}{d\phi}g(\phi)-m_{0}
\eea
Since we have assumed that $g(\phi)$ is real, the first integral on RHS becomes,
\bea
\frac{i}{2}\int d\phi \frac{d}{d\phi}g^{2}(\phi)=\frac{i}{2}\left(g^{2}(2\pi)-g^{2}(0)\right)
\eea
The RHS vanishes because of periodicity (or approximate periodicity) and we have
\bea
-\int d\rho \tan(\rho)d\phi f^{*}(\rho,\phi)\Big(-i\frac{d}{d\phi}f(\rho,\phi)\Big)=
m_{0}
\eea
Using this back in \eqref{Angular momentum operator as in quantum mechanics}, we get
\bea
\langle\hat{L}\rangle=-\sum_{n,m}m\, |g(n,m)|^{2}=m_{0}
\eea
This demonstrates explicitly that the phase factor in the packet profile corresponds  to the expectation value of the angular momentum operator.

\section{Why is $n_{0}$ a Proxy for Initial Radial Momentum?}\label{appendix_why_is_n_0_initial_radial_mom?}

In Appendix \ref{appendix_why_is_the_expval_m_0?} we saw that the wave packet parameter $m_{0}$ is nothing but the angular momentum that we provided to the wave packet. In this Appendix, we will see why the parameter $n_{0}$ in the choice of wave packet can be thought of as initial radial momentum. It is clear from the numerics that when we put $n_{0}=0,$ the extrema of packet motion occurs at $t=0$ and $t=\frac{N\pi}{2}$ where $N\in\mathbb{Z}$. But when $n_{0}\neq0$, extrema will not occur at $t=0$ and $t=\frac{N\pi}{2}$.  We show this presently for $\langle\hat{\rho}\rangle$. 

The expectation value of $\hat{\rho}$ in a generic single particle state is (see Eq. \ref{expression for rhohat's expectation value in terms of g and I}): 
\bea
\langle\hat{\rho}\rangle = \sum_{n,n',m}I(n,n',m)g^{*}(n',m)g(n,m)e^{2i(n'-n)t}
\eea
where
\bea
I(n,n',m) = \int d\rho\,\rho\,\tan(\rho)\,R^{*}_{n'm}(\rho)R_{nm}(\rho)
\eea
We can write 
\begin{align}
\langle\hat{\rho}\rangle &= \sum_{n>n',m}I(n,n',m)g^{*}(n',m)g(n,m)e^{2i(n'-n)t} + \sum_{n<n',m}I(n,n',m)g^{*}(n',m)g(n,m)e^{2i(n'-n)t}+ \nonumber\\
&\hspace{10cm}+\text{time independent piece} 
\end{align}
We can ignore the time independent piece, because it will vanish when we will take time derivative of $\langle\hat{\rho}\rangle$. Now in the second term in the above equation we make the replacement $n\leftrightarrow n'$. With this the above equation becomes
\bea
\langle\hat{\rho}\rangle = \sum_{n>n',m}I(n,n',m)(g^{*}(n',m)g(n,m)e^{2i(n'-n)t} + g^{*}(n,m)g(n',m)e^{-2i(n'-n)t})
\eea
We know that
\bea
g(n,m)=\frac{1}{\sqrt{2\pi}}\int d\rho\,d\phi\,\tan(\rho)\,R^{*}_{nm}(\rho)f(\rho,\phi)e^{-im\phi}
\eea
Now we consider 
\bea
f(\rho,\phi)  = f_{1}(\rho)\,e^{-in_{0}(\rho-\rho_{0})}f_{2}(\phi)
\eea
with $f_{1}(\rho)$  real and $f_{2}(\phi)$ complex. This is a form that is general enough to contain our wave packets. With this the above expression of $g(n,m)$ becomes 
\bea
g(n,m) = \Big(\frac{1}{\sqrt{2\pi}}\int d\phi\,f_{2}(\phi)e^{-im\phi}\Big)\Big(\int d\rho\,R_{nm}(\rho)f_{1}(\rho)e^{-in_{0}(\rho-\rho_{0})}\Big) = P_{m}Q_{nm}
\eea
where we have defined 
\bea\label{additionally defined parameters for proving n0 is radial moemnta}
P_{m} = \frac{1}{\sqrt{2\pi}}\int d\phi\,f_{2}(\phi)e^{-im\phi} \,\, , \hspace{1cm}Q_{nm} = \int d\rho\,R_{nm}(\rho)f_{1}(\rho)e^{-in_{0}(\rho-\rho_{0})}      
\eea
In general $P_{m}\,, Q_{nm}\in \mathbb{C}$.  
With these, the expression for $\langle\hat{\rho}\rangle$ becomes,
\bea
\langle\hat{\rho}\rangle  = 2\sum_{n>n',m}I(n,n',m)|P_{m}|^{2}|Q_{n'm}||Q_{nm}|\cos\Big(\alpha(n,n',m,n_{0})-2(n-n')t\Big)
\eea
where we defined $Q_{nm} = |Q_{nm}|e^{i\theta(n,m,n_0)}$ and $\alpha(n,n',m,n_{0})=\theta(n,m,n_{0})-\theta(n',m,n_{0})$. Now its time derivative becomes
\bea \label{Final equation for drho/dt}
\frac{d\langle\hat{\rho}\rangle}{dt}= 4\hspace{-2mm}\sum_{n>n',m}(n-n')I(n,n',m)|P_{m}|^{2}|Q_{n'm}||Q_{nm}|\sin\Big(\alpha(n,n',m,n_{0})-2(n-n')t\Big)
\eea
Clearly, \eqref{Final equation for drho/dt} is zero at $t=0$ as well as any $t=\frac{Z\pi}{2}$ when $n_{0}=0$ (as $Q_{nm}$ is real in that case, and therefore the $\theta$'s vanish, leading to $\alpha=0)$. This is the natural result for $n_{0}=0$, if $n_0$ is the proxy for initial radial momentum.
 
But this is not sufficient to claim that $n_{0}$ can be thought of as the initial radial momentum. We also need to show that $\frac{d\langle\hat{\rho}\rangle}{dt}$ has directionality associated with $n_{0}$. This is basically a fancy way of saying that:
\bea
\frac{d\langle\hat{\rho}\rangle}{dt}(n_{0})\Big|_{t=0}=-\frac{d\langle\hat{\rho}\rangle}{dt}(-n_{0})\Big|_{t=0}.
\eea
That is $\frac{d\langle\hat{\rho}\rangle}{dt}$ is an odd function of $n_{0}$ at $t=0$. The fact that $\frac{d\langle\hat{\rho}\rangle}{dt}\Big|_{t=0}$ when $n_{0}=0$ trivially follows from this.
To see this, let's look at the second equation in \eqref{additionally defined parameters for proving n0 is radial moemnta}.
\bea
Q_{nm}(n_{0}) = \int d\rho\,R_{nm}(\rho)f_{1}(\rho)e^{-in_{0}(\rho-\rho_{0})}
\eea
Since $f_{1}(\rho)$ and $R_{nm}(\rho)$ are real we get:
\bea
Q_{nm}(-n_{0})= Q_{nm}^{*}(n_{0})
\eea
Now given that $Q_{nm}(n_{0}) = |Q_{nm}(n_{0})|e^{i\theta(n,m,n_0)}$, we have:
\bea
|Q_{nm}(-n_{0})|=|Q_{nm}(n_{0})| \,\,\,\, ,\,\,\,\, \theta(n,m,-n_0)=-\theta(n,m,n_0)
\eea
Also since $\alpha(n,n',m,n_{0})=\theta(n,m,n_{0})-\theta(n',m,n_{0})$ we get:
\bea
\alpha(n,n',m,-n_{0})=-\alpha(n,n',m,n_{0})
\eea
Using this result in equation \eqref{Final equation for drho/dt} with t=0, we get 
\begin{align} 
\frac{d\langle\hat{\rho}\rangle}{dt}(n_{0})\Big|_{t=0}&= 4\sum_{n>n',m}(n-n')I(n,n',m)|P_{m}|^{2}|Q_{n'm}(n_{0})||Q_{nm}(n_{0})|\sin\Big(\alpha(n,n',m,n_{0})\Big)\nonumber\\
&=4\sum_{n>n',m}(n-n')I(n,n',m)|P_{m}|^{2}|Q_{n'm}(-n_{0})||Q_{nm}(-n_{0})|\sin\Big(-\alpha(n,n',m,-n_{0})\Big)\nonumber\\
&=-4\sum_{n>n',m}(n-n')I(n,n',m)|P_{m}|^{2}|Q_{n'm}(-n_{0})||Q_{nm}(-n_{0})|\sin\Big(\alpha(n,n',m,-n_{0})\Big)\nonumber\\
&= -\frac{d\langle\hat{\rho}\rangle}{dt}(-n_{0})\Big|_{t=0}
\end{align}
This says that $n_{0}$ is an analogue for the initial radial momentum of the wave packet. 

\section{2D Plots: Stress Tensor Approach}\label{appendix_2d_plots_com}

In this section, we present the 2D plots for the cases examined in section \ref{sec:AdS3_implementation_com} along with other examples to see that the center of the packet $\bar{\rho}(t)$ and $\bar{\phi}(t)$ follow the classical geodesic trajectory.

\subsection{Massive Case: Elliptical-like}\label{appendix_2d_plots_com_massive_elliptical}

\noindent\textbf{Fix $m_{0}$, Vary $M$}

\begin{figure}[H]
    \centering

    \begin{subfigure}{0.48\textwidth}
        \includegraphics[width=0.7\linewidth]{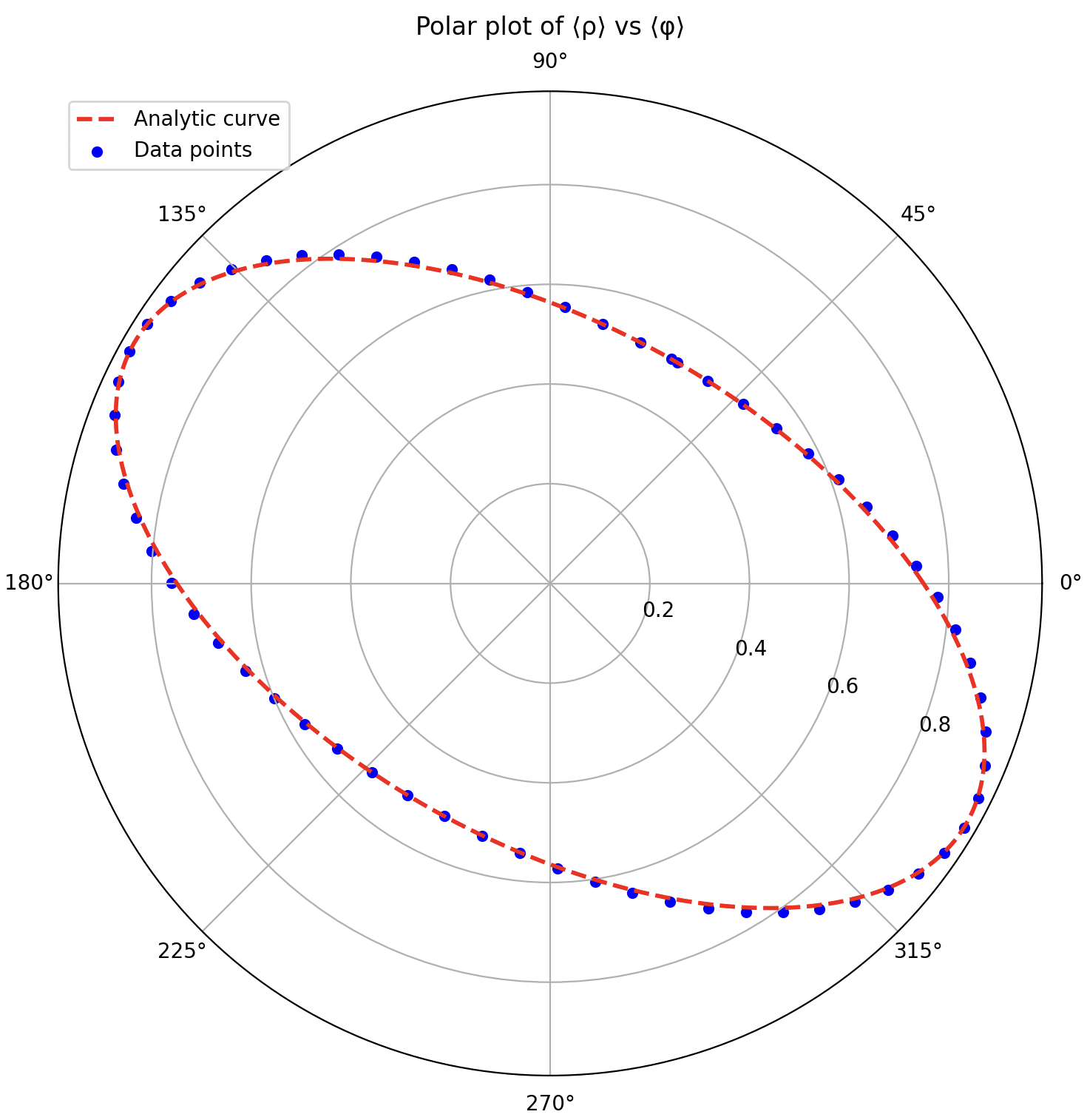}
        \caption{$m_{0}=-20, M=25,E\approx 53,n_{max}=m_{max}=75$}
    \end{subfigure}\hfill
    \begin{subfigure}{0.48\textwidth}
        \includegraphics[width=0.7\linewidth]{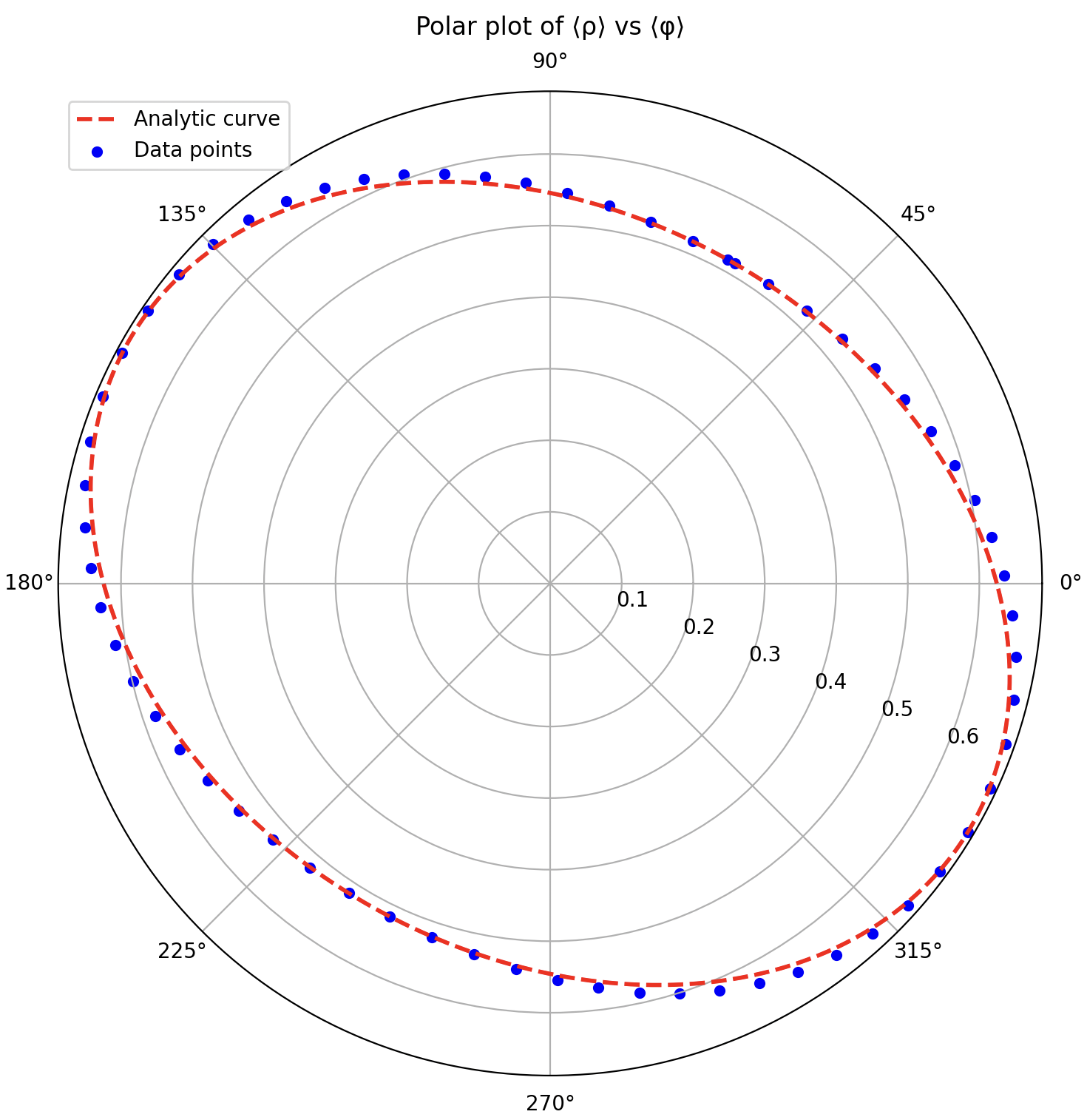}
        \caption{$m_{0}=-20, M=45, E\approx 69.3,n_{max}=m_{max}=75$}
    \end{subfigure}\hfill

    \begin{subfigure}{0.48\textwidth}
        \includegraphics[width=0.7\linewidth]{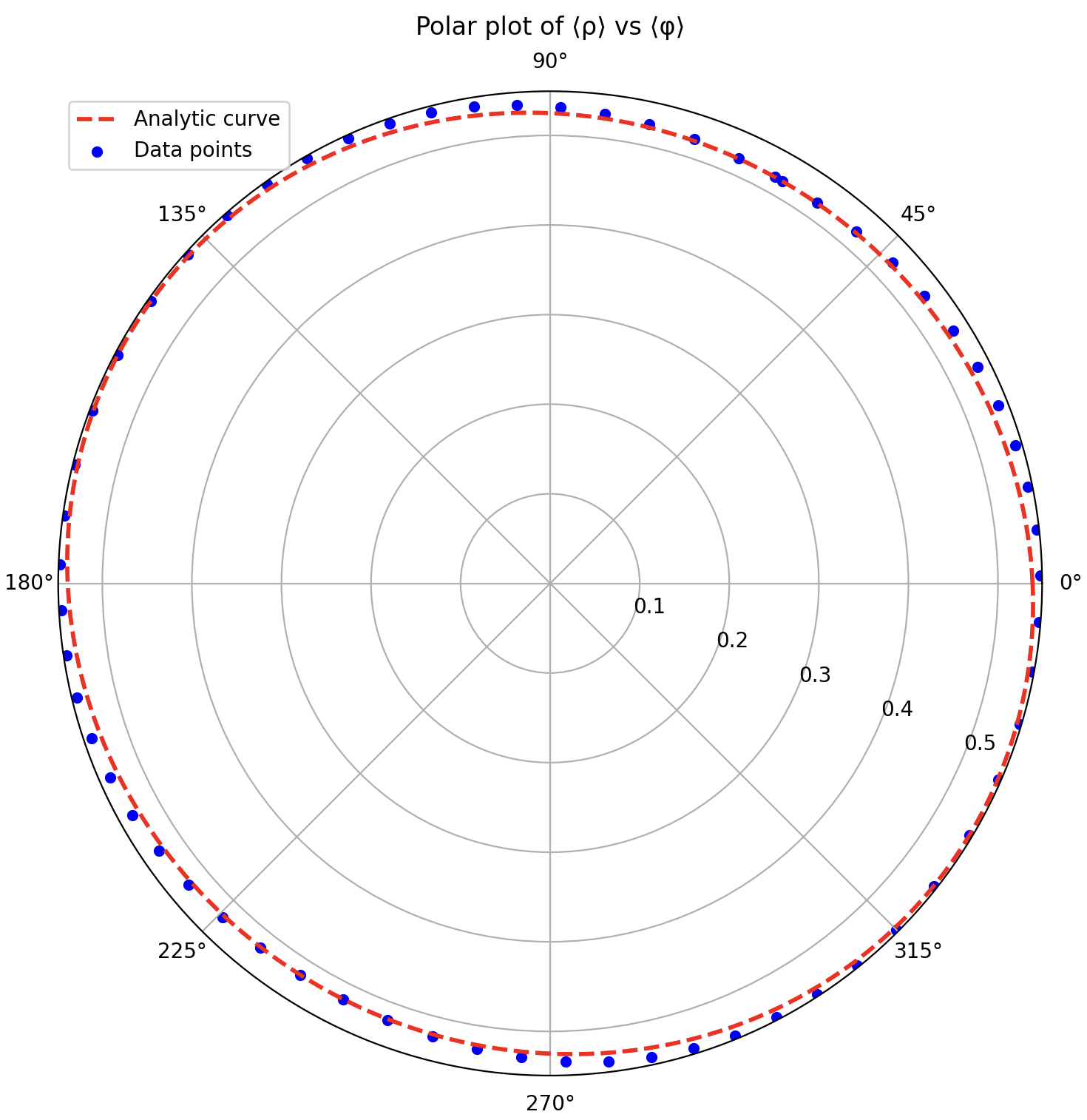}
        \caption{$m_{0}=-20, M=60, E\approx 83.5,n_{max}=m_{max}=75$}
    \end{subfigure}\hfill
    \begin{subfigure}{0.48\textwidth}
        \includegraphics[width=0.7\linewidth]{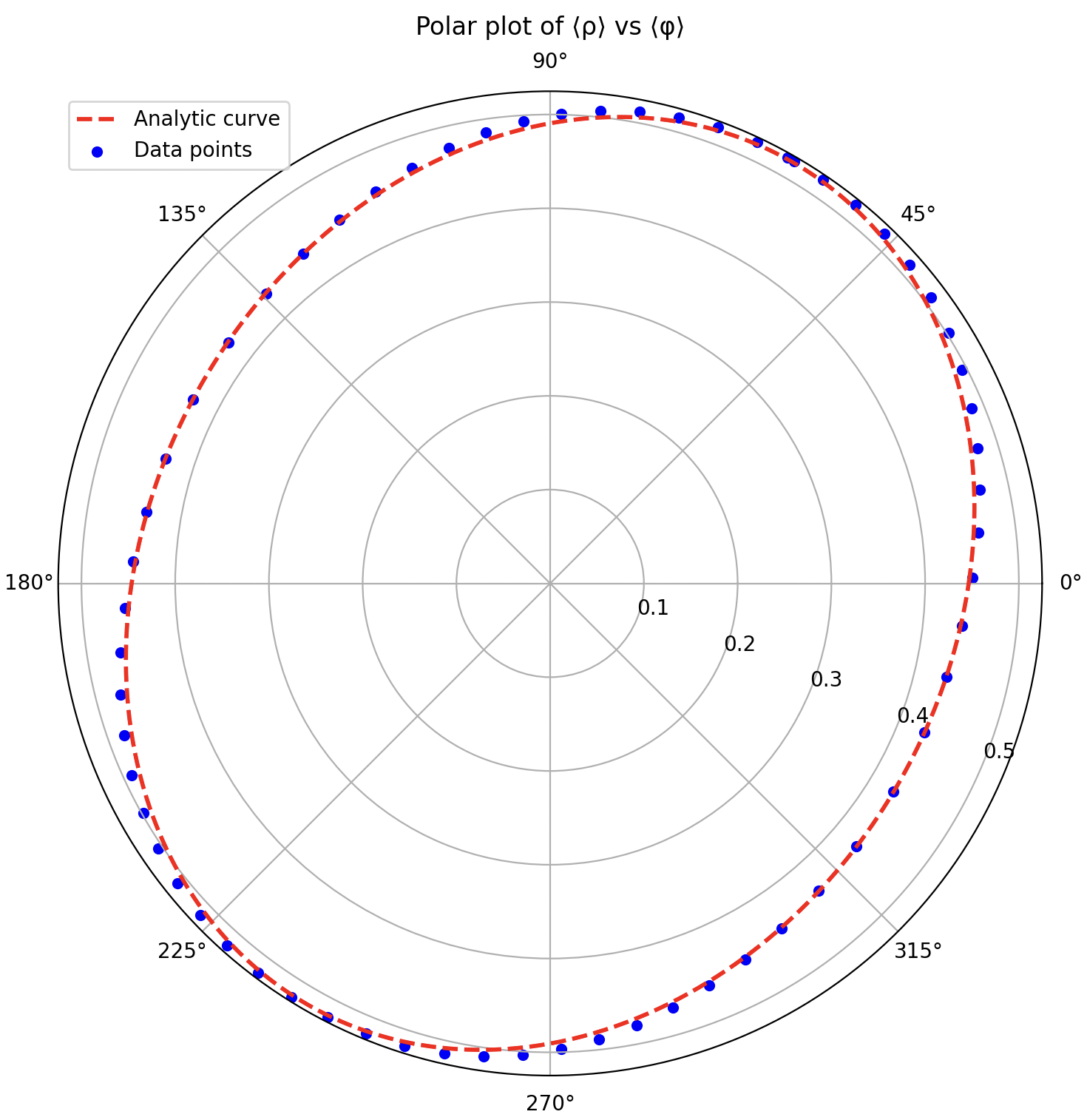}
        \caption{$m_{0}=-20, M=80, E\approx 103.8,n_{max}=m_{max}=75$}
    \end{subfigure}\hfill

    \caption{Parameters: $\rho_{0}=0.5,\phi_{0}=\pi/3,\sigma_{1}=0.09,\sigma_{2}=0.05$. In plots (a)--(d), the numerically determined wave packet coordinates $(\bar{\rho}(t),\bar{\phi}(t))$ (blue dots) are compared directly against the analytical solution of the geodesic equation (red dashed curve).}
    \label{fig:fig:com_massive_fix_m0_vary_M}
    
\end{figure}

Fig. \ref{fig:fig:com_massive_fix_m0_vary_M} presents the resulting 2D parametric plots in the $(\rho,\phi)$ plane for fix angular momentum $m_{0}=-20$ and mass $M=25,\,45,\,60,\,80$. 

As illustrated in the figures above, varying the mass $M$ produces a range of ellipse-like trajectories, including circular orbits.

\noindent\textbf{Fix $M$, Vary $m_{0}$}

\begin{figure}[H]
    \centering

    \begin{subfigure}{0.48\textwidth}
        \includegraphics[width=0.7\linewidth]{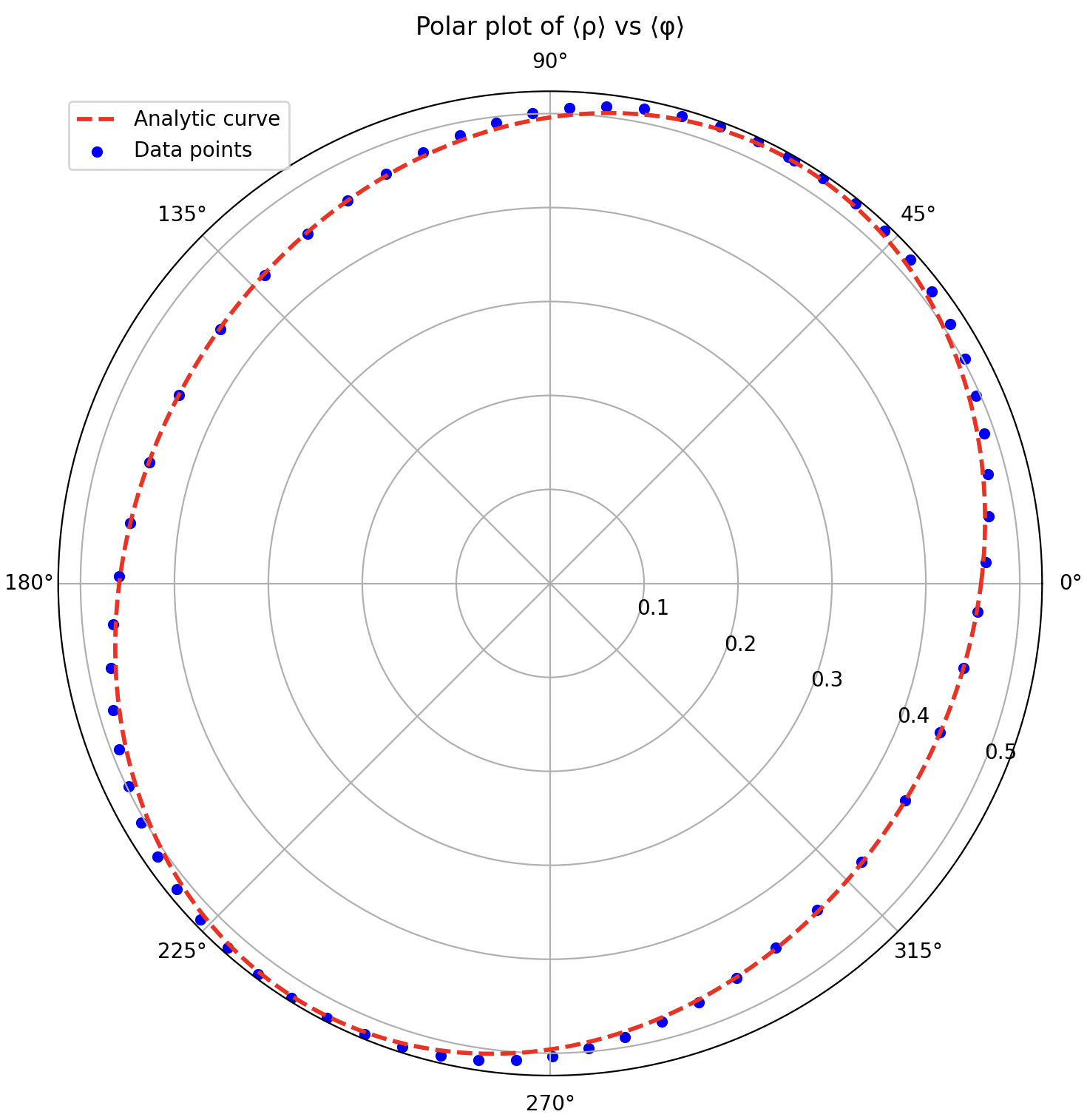}
        \caption{$M=60, m_{0}=-15,E\approx79,n_{max}=m_{max}=70$}
    \end{subfigure}\hfill
    \begin{subfigure}{0.48\textwidth}
        \includegraphics[width=0.7\linewidth]{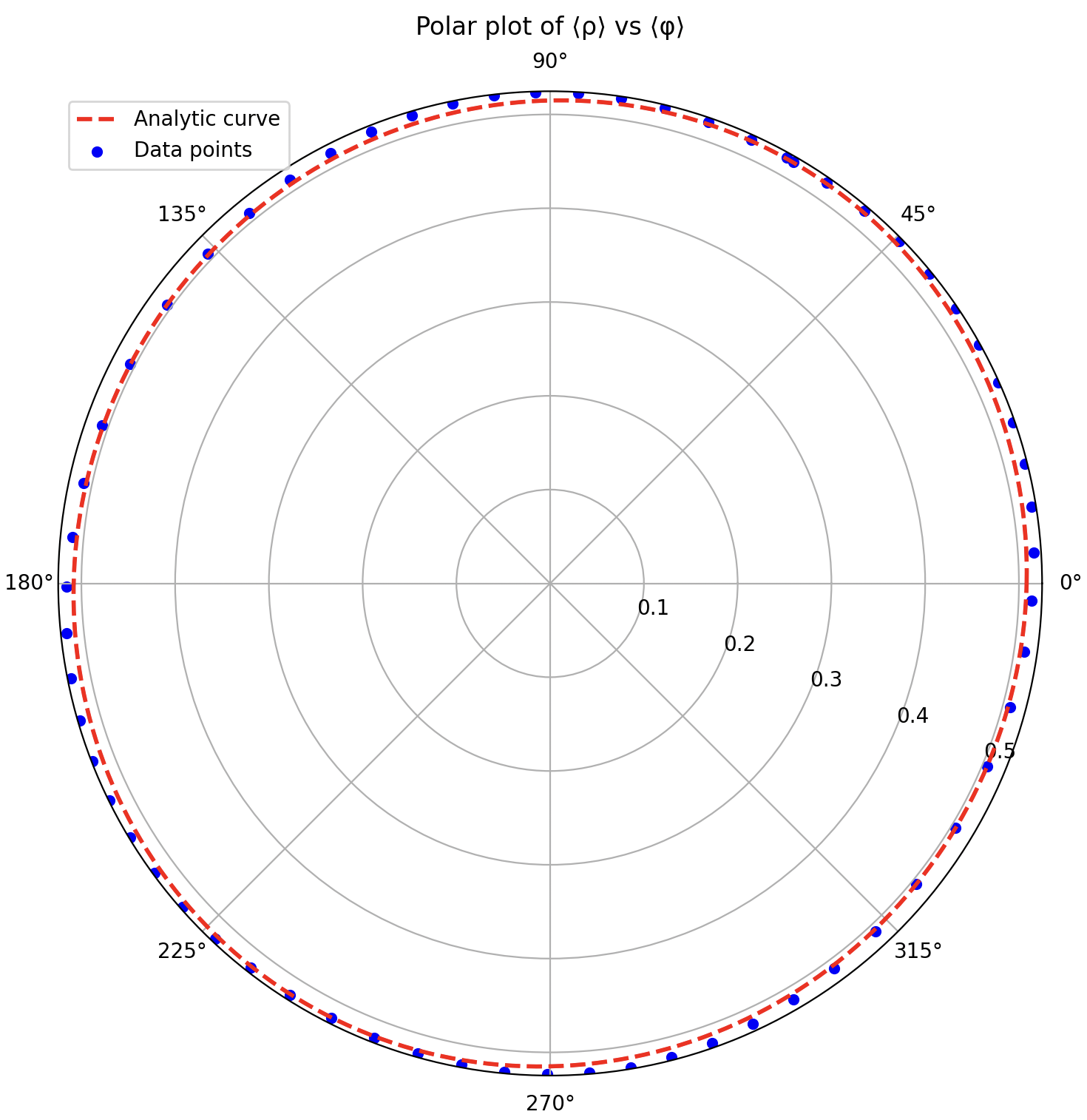}
        \caption{$M=60, m_{0}=-18,E\approx81.6,n_{max}=m_{max}=70$}
    \end{subfigure}\hfill

    \begin{subfigure}{0.48\textwidth}
        \centering
        \includegraphics[width=0.7\linewidth]{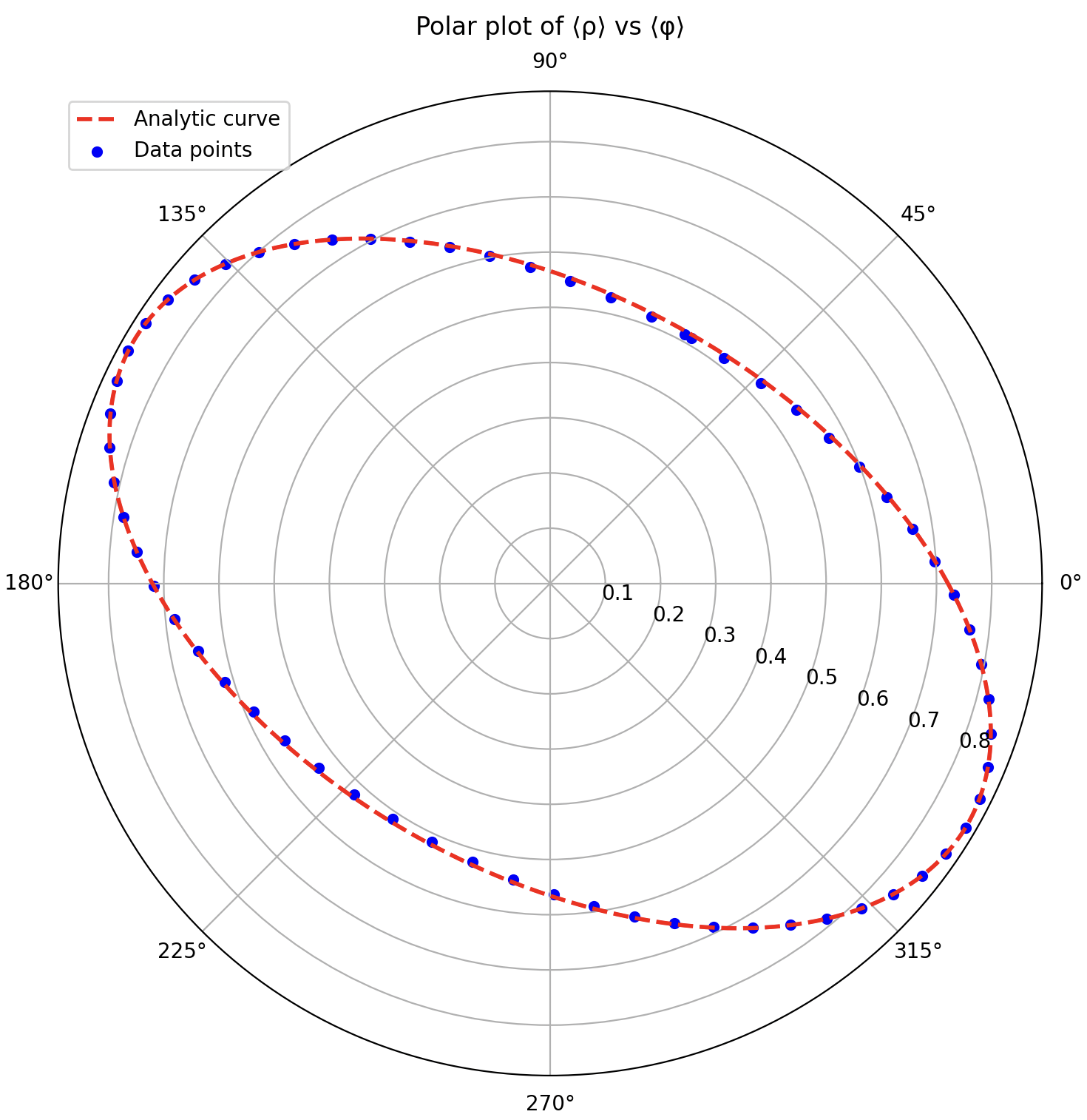}
        \caption{$M=60, m_{0}=-40, E\approx109.7, n_{max}=m_{max}=70$}
    \end{subfigure}\hfill
    \begin{subfigure}{0.48\textwidth}
        \centering
        \includegraphics[width=0.7\linewidth]{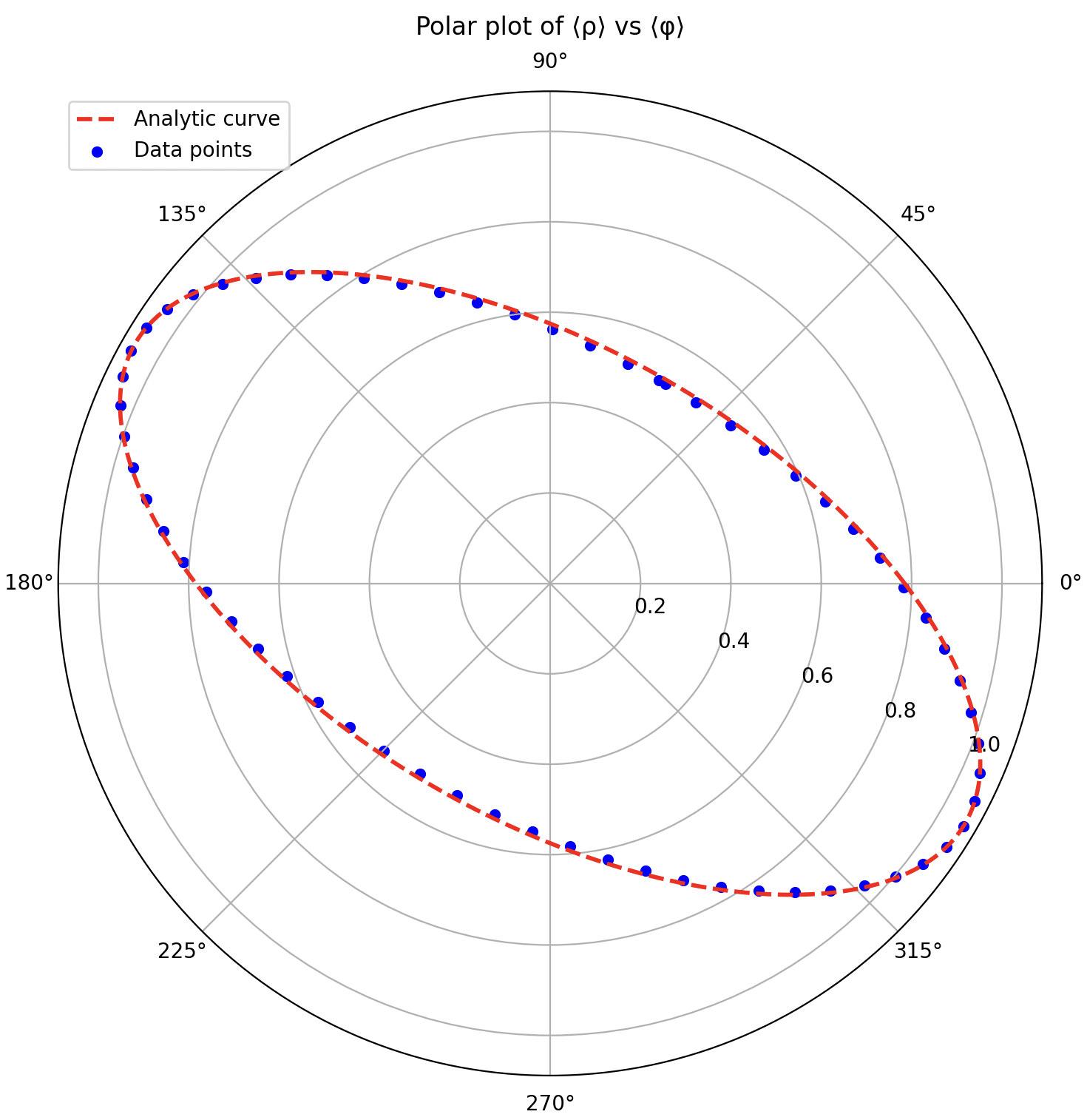}
        \caption{$M=60, m_{0}=-60, E\approx 143.7,n_{max}=m_{max}=100$}
    \end{subfigure}\hfill
    
    \caption{Parameters: $\rho_{0}=0.5,\phi_{0}=\pi/3,\sigma_{1}=0.09,\sigma_{2}=0.05$. In plots (a)--(d), the numerically determined wave packet coordinates $(\bar{\rho}(t),\bar{\phi}(t))$ (blue dots) are compared directly against the analytical solution of the geodesic equation (red dashed curve).}
    \label{fig:fig:com_massive_fix_M_vary_m0}

\end{figure}

Fig. \ref{fig:fig:com_massive_fix_M_vary_m0} presents the resulting 2D parametric plots in the $(\rho,\phi)$ plane for fix mass $M=60$ and angular momentum $m_{0}=-15,-18,-40,-60$.

We clearly see similar behavior as seen in the previous case with $m_{0}=\text{constant}$. 

\subsection{Massive Case: Radial Infall}\label{appendix_2d_plots_com_massive_radial_infall}

\begin{figure}[H]
    \centering

    \begin{subfigure}{0.48\textwidth}
        \includegraphics[width=0.7\linewidth]{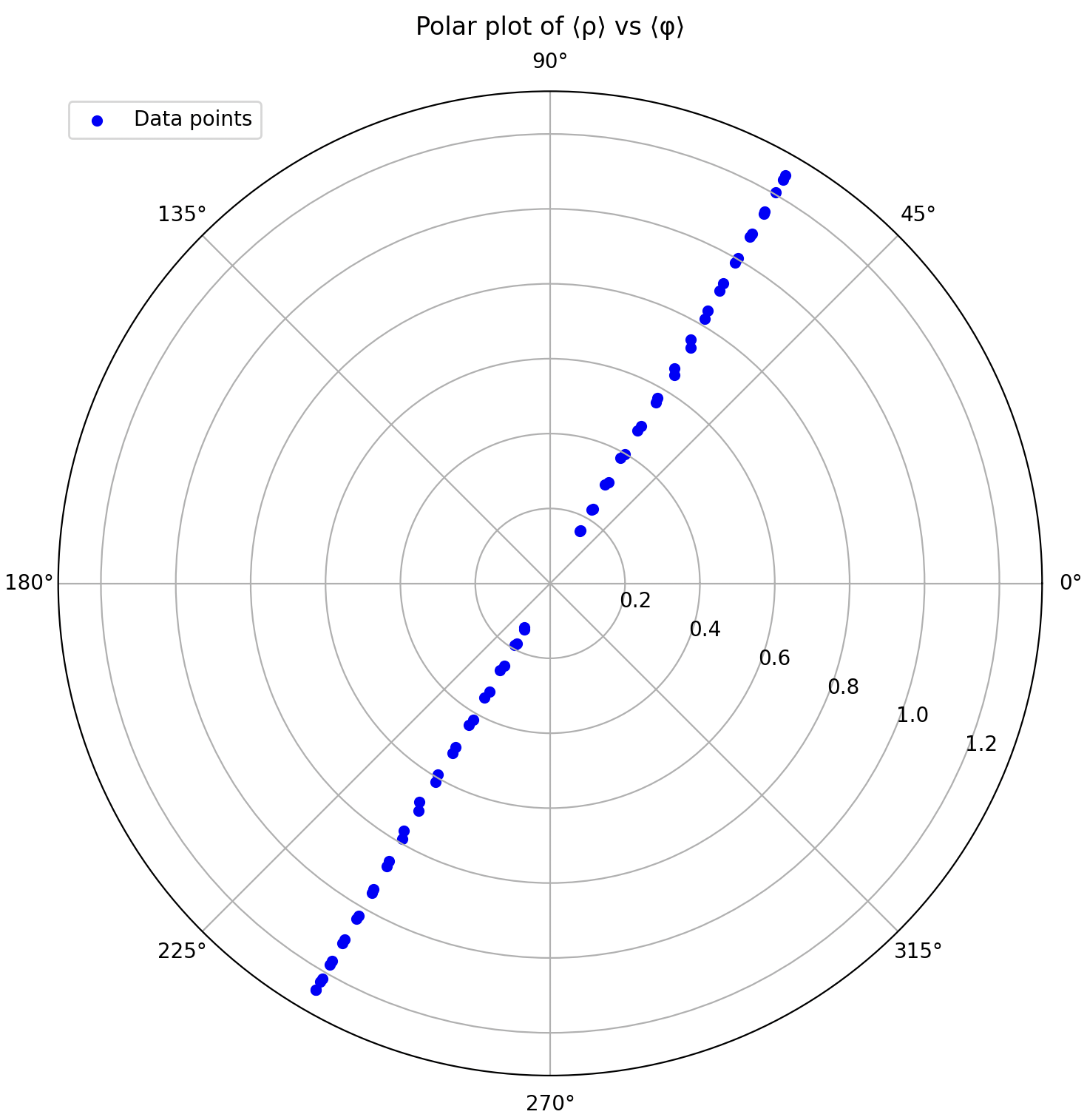}
        \caption{$M=20, m_{0}=0, E\approx 66.7,n_{max}=m_{max}=200$}
    \end{subfigure}\hfill
    \begin{subfigure}{0.48\textwidth}
        \includegraphics[width=0.7\linewidth]{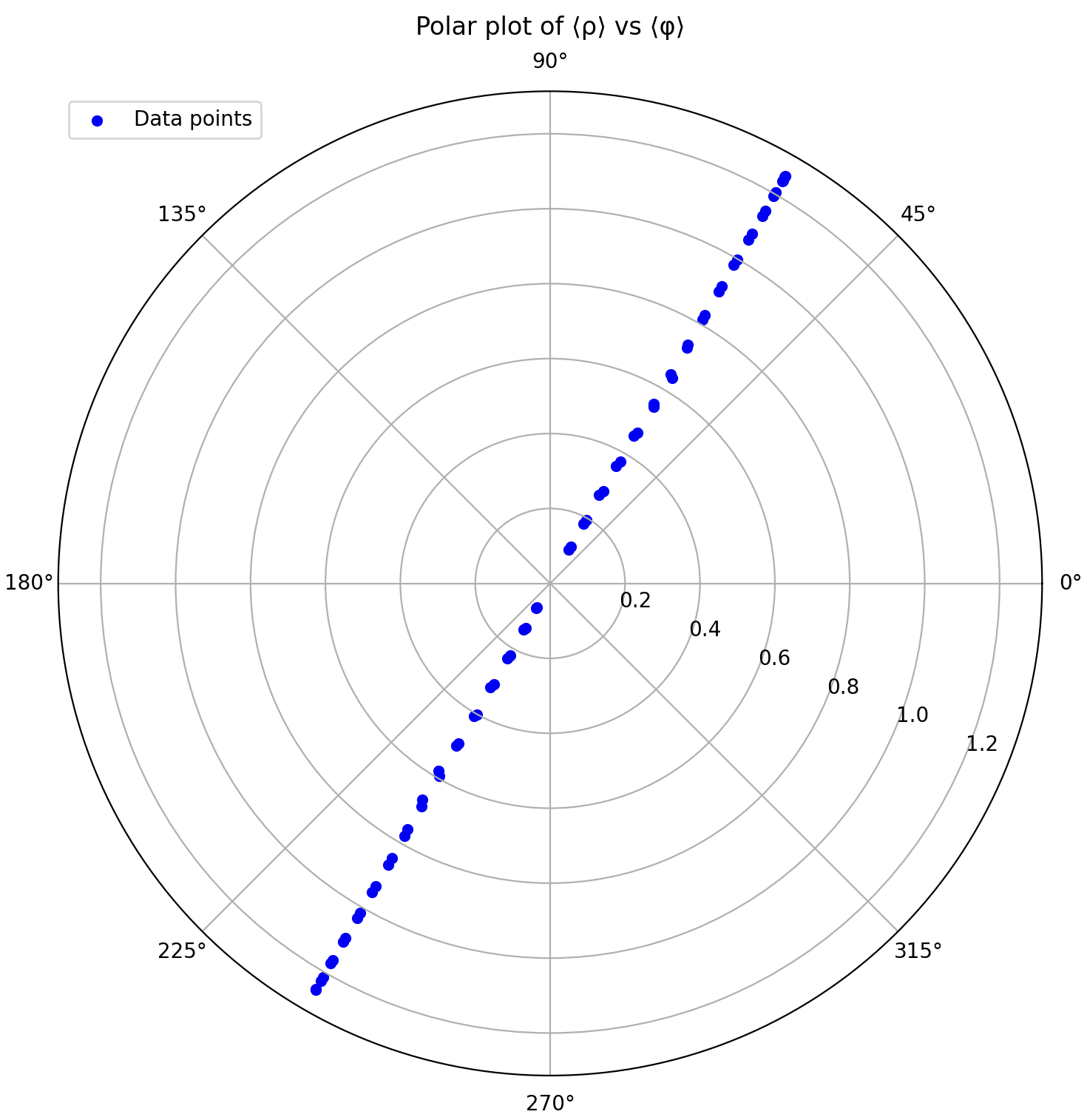}
        \caption{$M=40, m_{0}=0, E\approx 128,n_{max}=m_{max}=200$}
    \end{subfigure}\hfill

    \begin{subfigure}{0.48\textwidth}
        \includegraphics[width=0.7\linewidth]{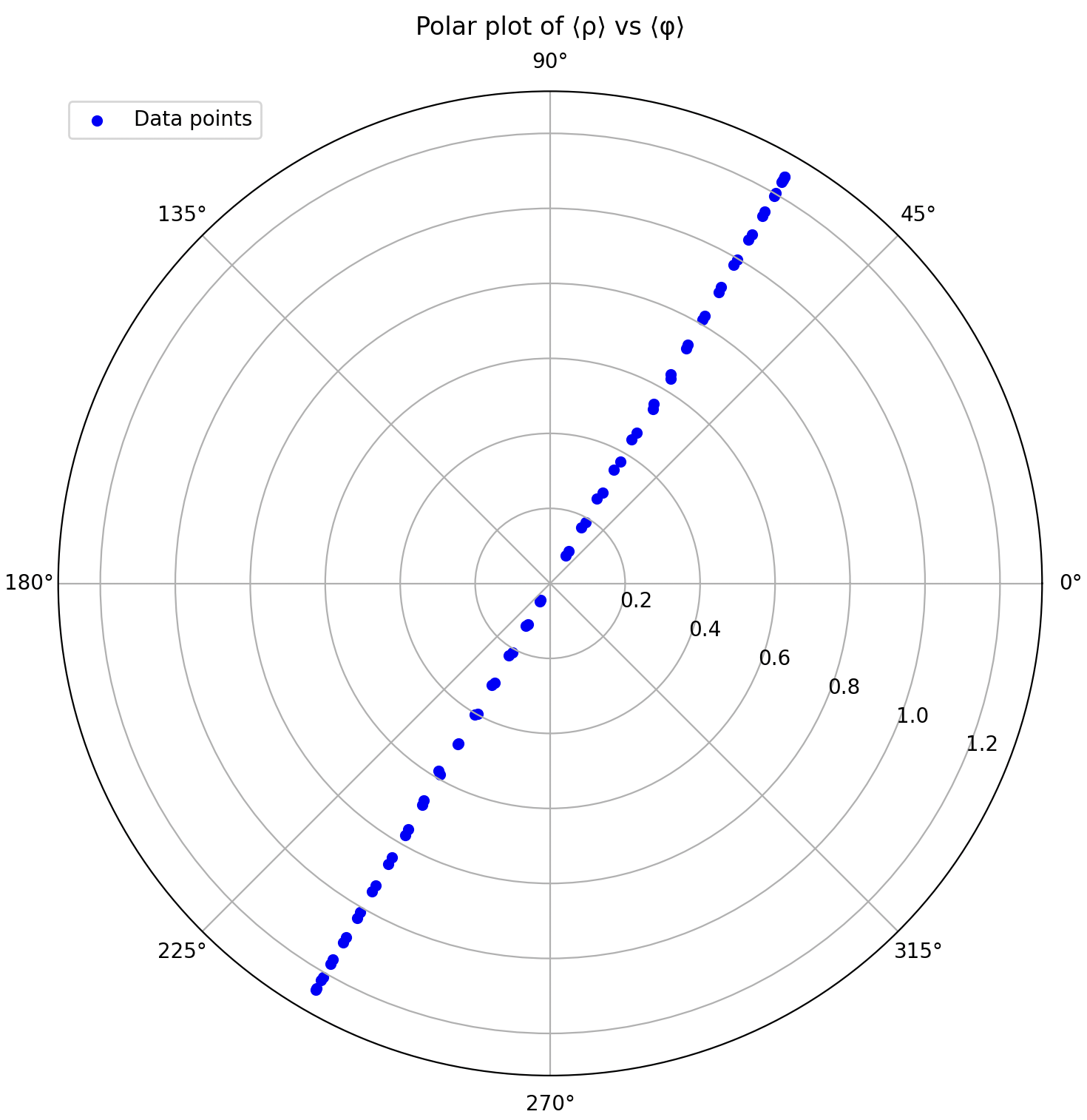}
        \caption{$M=60, m_{0}=0,E\approx 193,n_{max}=m_{max}=250$}
    \end{subfigure}\hfill
    \begin{subfigure}{0.48\textwidth}
        \includegraphics[width=0.7\linewidth]{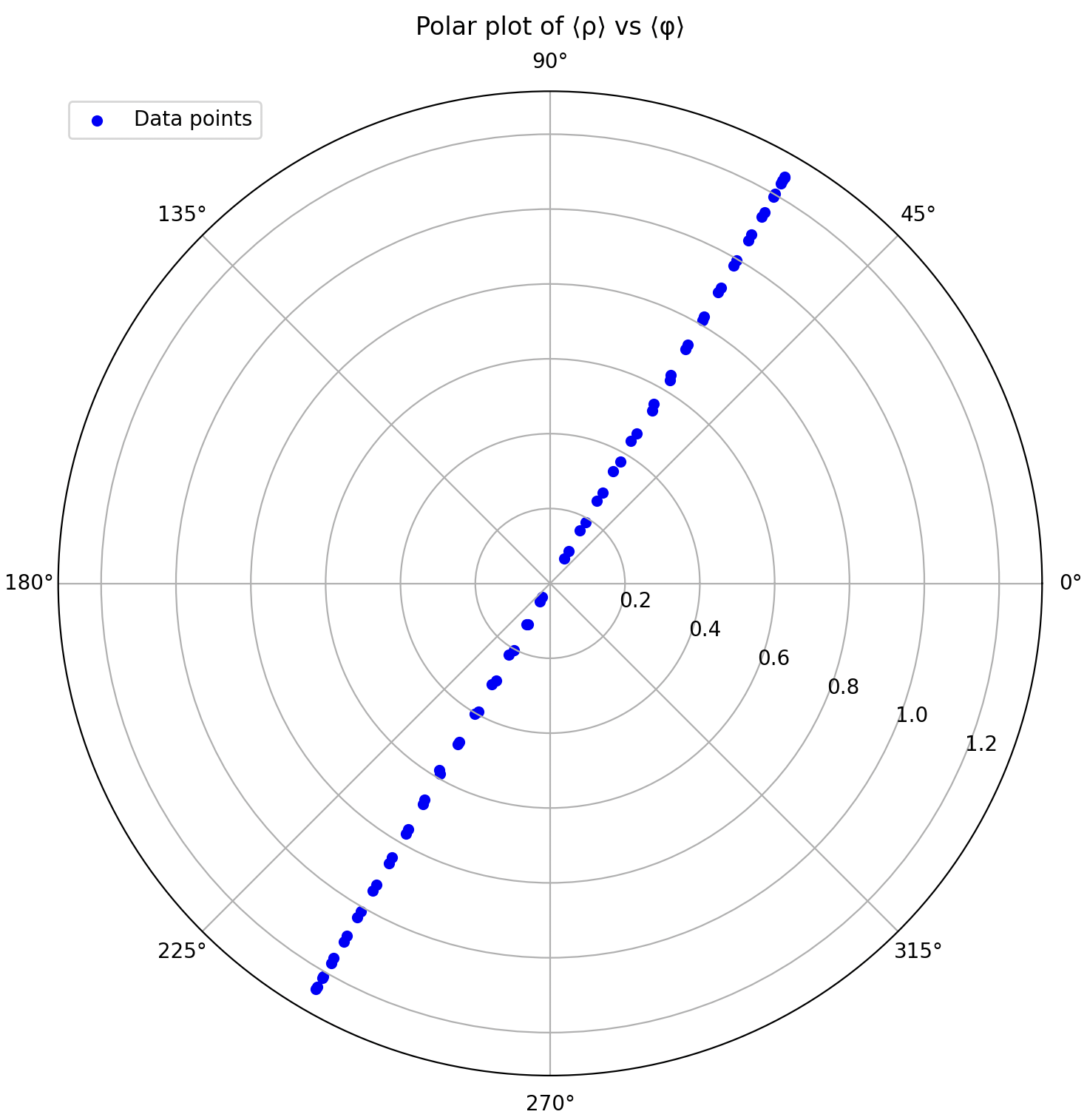}
        \caption{$M=80, m_{0}=0,E\approx 256,n_{max}=m_{max}=300$}
    \end{subfigure}\hfill
    
    \caption{Parameters: $\rho_{0}=1.2,\phi_{0}=\pi/3,\sigma_{1}=0.09,\sigma_{2}=0.05$.}
    \label{fig:fig:com_massive_radial_2D}

\end{figure}

Fig. \ref{fig:fig:com_massive_radial_2D} illustrate the resulting 2D trajectories in the $(\rho, \phi)$ plane (blue dots) for distinct parameter sets.

\subsection{$\bar{\rho}(t)$ vs $t$: Massive Case}\label{appendix_2d_plots_com_rho_vs_t}

To further characterize the dynamics of these trajectories, we now examine the time evolution of the radial coordinate's centroid $\bar{\rho}(t)$ as a function of the coordinate time $t$ for each of the scenarios discussed above in Fig. \ref{fig:fig:com_massive_radial_2D}.
\begin{figure}[H]
    \centering

    \begin{subfigure}{0.48\textwidth}
        \includegraphics[width=1.0\linewidth]{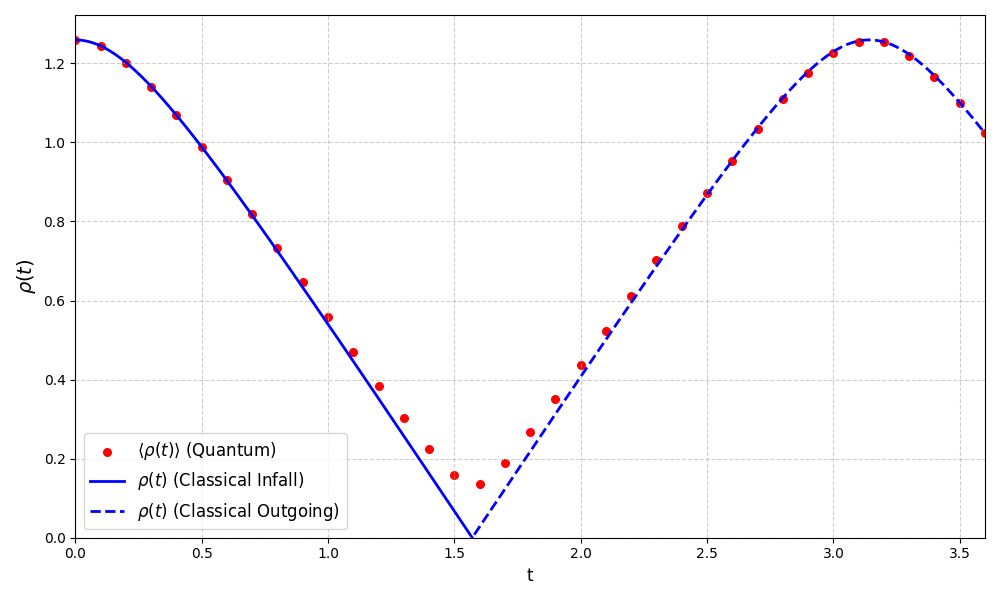}
        \caption{$M=20, m_{0}=0,n_{max}=m_{max}=200$}
    \end{subfigure}\hfill
    \begin{subfigure}{0.48\textwidth}
        \includegraphics[width=1.0\linewidth]{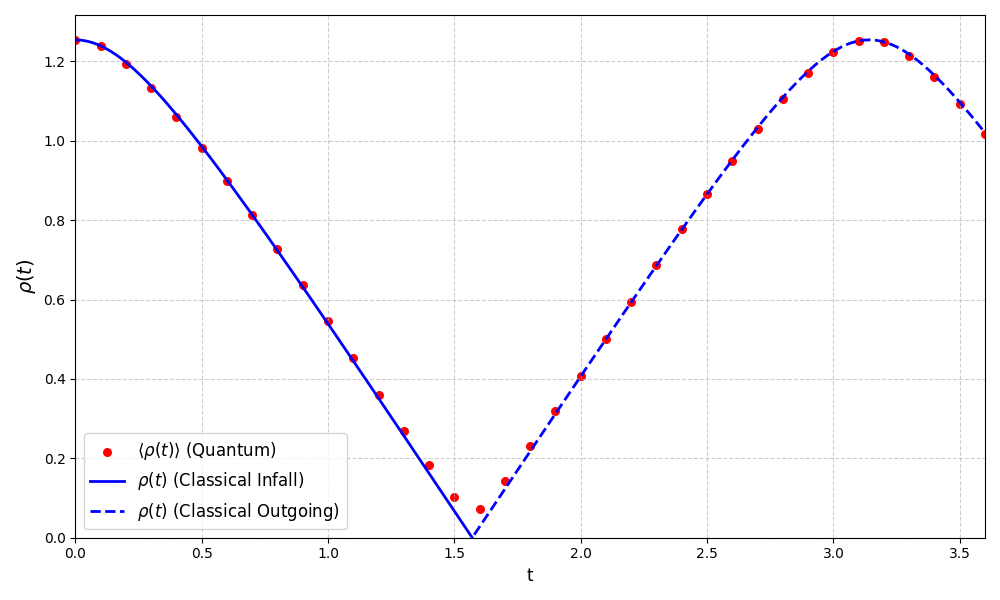}
        \caption{$M=40, m_{0}=0, n_{max}=m_{max}=200$}
    \end{subfigure}\hfill

    \begin{subfigure}{0.48\textwidth}
        \includegraphics[width=1.0\linewidth]{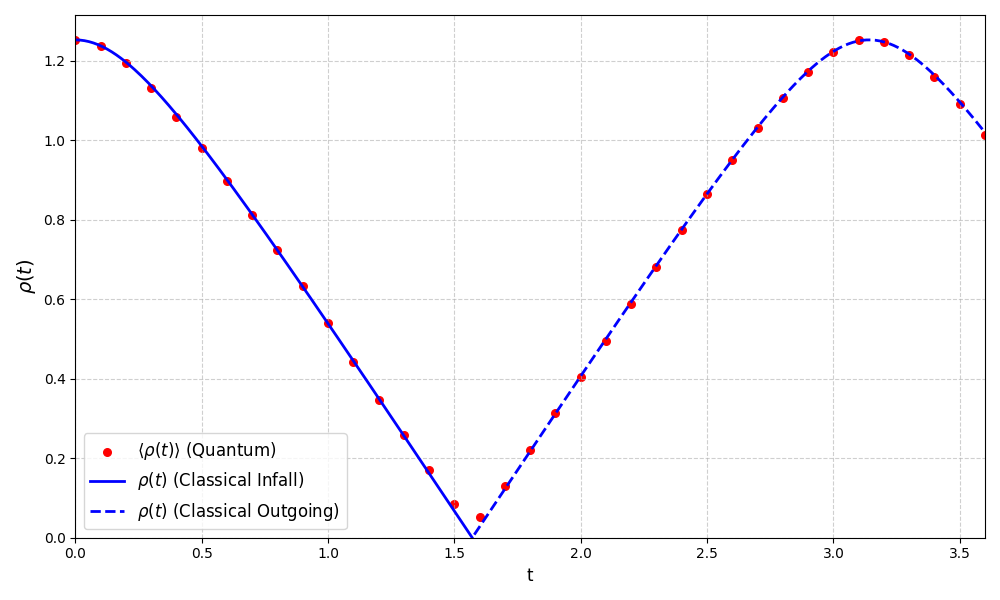}
        \caption{$M=60, m_{0}=0,n_{max}=m_{max}=250$}
    \end{subfigure}\hfill
    \begin{subfigure}{0.48\textwidth}
        \includegraphics[width=1.0\linewidth]{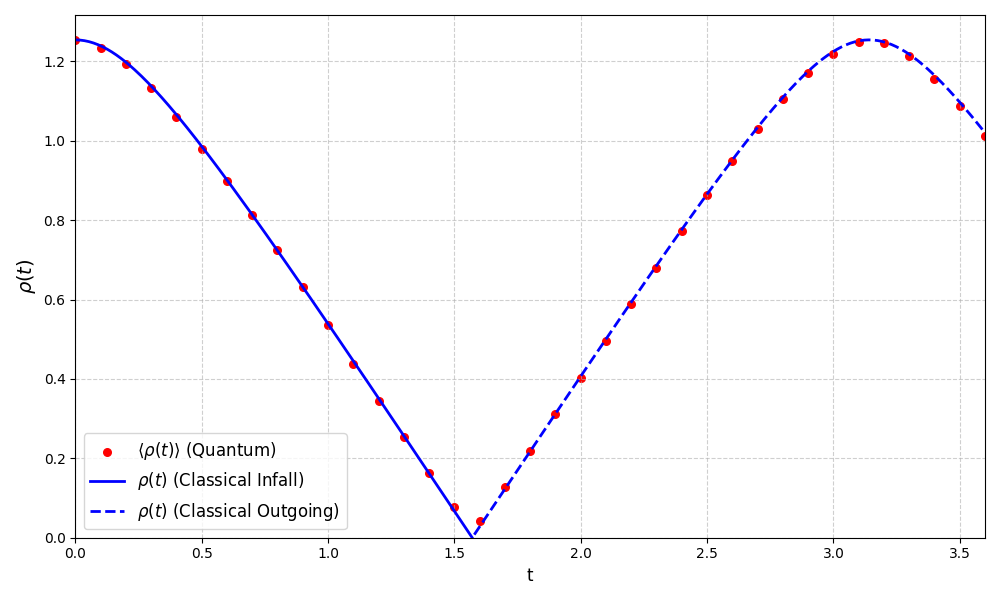}
        \caption{$M=80, m_{0}=0, n_{max}=m_{max}=300$}
    \end{subfigure}\hfill
    \caption{Parameters: $\rho_{0}=1.2,\phi_{0}=\pi/3,\sigma_{1}=0.09,\sigma_{2}=0.05$}
    \label{fig:fig:com_massive_radial_2D_rho_vs_t}
    
\end{figure}

As observed in these plots, the trajectory of the wave packet deviates slightly from the exact geodesic as it approaches the spatial origin ($\rho=0$). Notably, this deviation is systematically suppressed as the scalar mass $M$ is increased. The origin of this discrepancy is fundamentally a coordinate artifact: we have chosen the $\rho$ coordinate to define our operator\footnote{This statement is cleaner to see in the position operator approach, where the operator is literally constructed using the coordinate $\rho$.}, and when the location is the origin, contributions to the expectation value integral from {\em either} side of the location add constructively. (In other locations, the contributions are bigger on one side and lower on the other.) We emphasize that this is {\em not} a feature of the state, which is well-defined even at the origin. We can confirm this in flat space by working with radial and Cartesian coordinates: only the former exhibits the discrepancy. A larger mass/energy of the state yields a more tightly confined, more localized wave packet, hence reducing the sensitivity to this artifact.

\subsection{Null Case: Localized Wave Packet}\label{appendix_2d_plots_com_null_localized}

\begin{figure}[H]
    \centering

    \begin{minipage}{0.48\textwidth}
        \centering
        \includegraphics[width=0.7\linewidth]{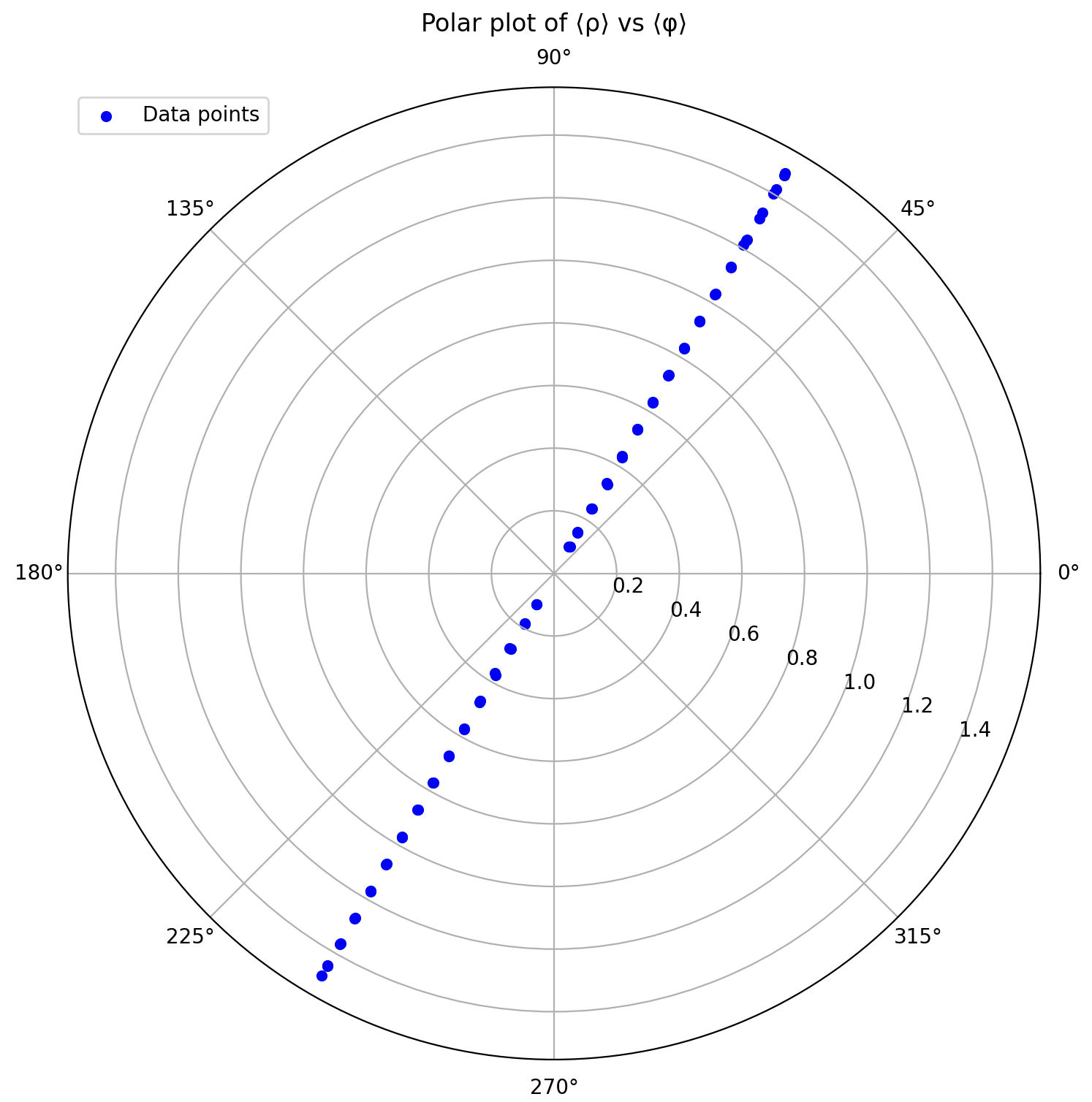}
    \end{minipage}\hfill
    \begin{minipage}{0.48\textwidth}
        \centering
        \includegraphics[width=0.7\linewidth]{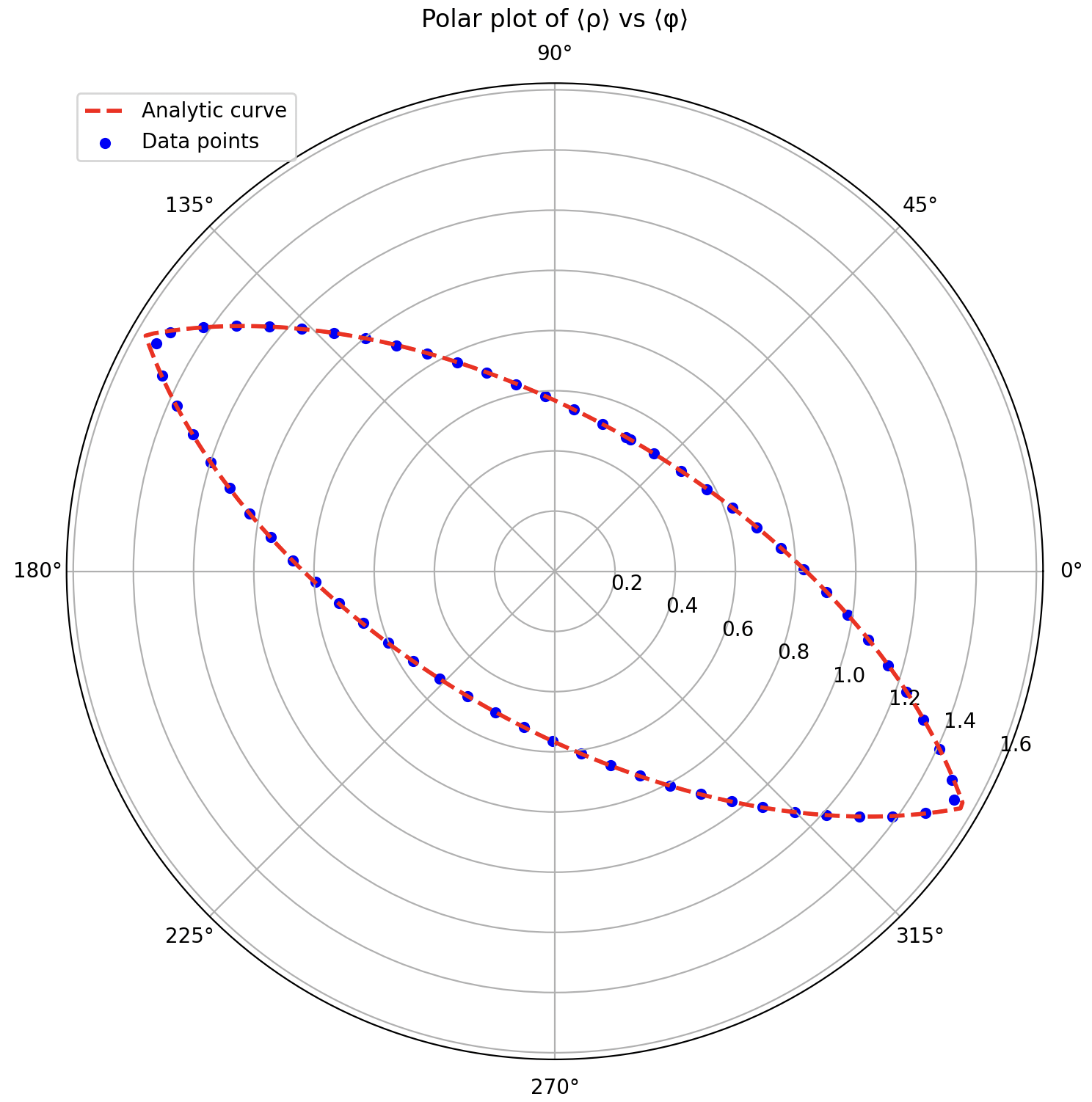}
    \end{minipage}
    \caption{Parameters: $\phi_{0}=\pi/3,\sigma_{1}=0.09,\sigma_{2}=0.05,n_{max}=m_{max}=300$, \textbf{Left:} $M=0, m_{0}=0, n_{0}=200,\rho_{0}=1.2, E\approx 200 $, \textbf{Right:} $M=0, m_{0}=-40,n_{0}=0,$ $\rho_{0}=0.5, E\approx 83$}
    \label{fig:fig:com_null_rad_and_ellip_2D}

\end{figure}

\begin{figure}[H]
    \centering
    \includegraphics[width=\textwidth, keepaspectratio]{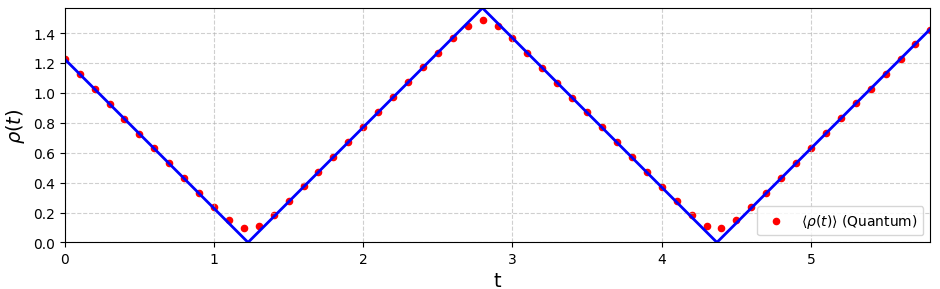}
    \caption{Parameters: $\rho_{0}=1.2,\phi_{0}=\pi/3,\sigma_{1}=0.09,\sigma_{2}=0.05,n_{max}=m_{max}=300$,$M=0, m_{0}=0, n_{0}=200$}
    \label{fig:fig:com_null_rad_2D_rho_vs_t}
\end{figure}

Fig. \ref{fig:fig:com_null_rad_and_ellip_2D} and \ref{fig:fig:com_null_rad_2D_rho_vs_t} illustrate the resulting 2D trajectories in the $(\rho,\phi)$ plane for various parameter configurations in the null regime.

\subsection{$\bar{\rho}(t)$ vs $t$: Null 
Case}\label{appendix_2d_plots_com_rho_vs_t_only_radial}

The following plot illustrates the time evolution of $\bar{\rho}(t)$ for a massless scalar field undergoing radial infall with large $n_{0}$ and hence large $E$. The solid blue line, characterized by a unit slope ($\pm 1$), represents the classical null trajectory of $\rho(t)$ in these coordinates. It is important to note that this light-like behavior is not exclusive to the $M=0$ regime; an analogous trajectory can be recovered for a massive particle by initializing the wave packet with a sufficiently large radial momentum $n_{0}$, effectively reaching the ultra-relativistic limit.
As before, we observe the coordinate offsets near the edges of the geometry. And as before, if the momentum parameter (here $n_{0}$) is increased, the state becomes more sharply localized, which systematically suppresses this boundary discrepancy.

\subsection{Circular Geodesics in \texorpdfstring{AdS$_{3}$}{AdS3}}\label{appendix_2d_plots_com_circular_orbits}

\begin{figure}[H]
    \centering
    \begin{minipage}{0.48\textwidth}
        \centering
        \includegraphics[width=1\linewidth]{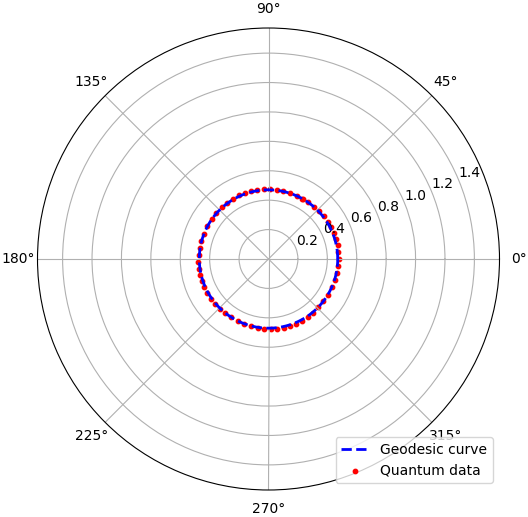}
        \caption{Parameters: $\rho_{0}=0.45, \phi_{0}=\pi/3, \sigma_{1}=0.09, \sigma_{2}=0.05, n_{\text{max}}=m_{\text{max}}=200, M=120, m_{0}=-30, n_{0}=0, E\approx154$.}
        \label{fig:fig51}
    \end{minipage}\hfill
    \begin{minipage}{0.48\textwidth}
        \centering
        \includegraphics[width=1\linewidth]{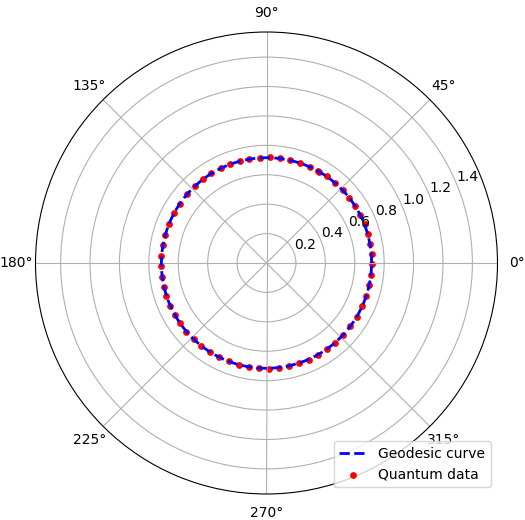}
        \caption{Parameters: $\rho_{0}=0.70, \phi_{0}=\pi/3, \sigma_{1}=0.09, \sigma_{2}=0.05, n_{\text{max}}=m_{\text{max}}=200, M=40, m_{0}=-30, n_{0}=0, E\approx72$.}
        \label{fig:fig52}
    \end{minipage}
\end{figure}

\begin{figure}[H]
    \centering
    \begin{minipage}{0.48\textwidth}
        \centering
        \includegraphics[width=1\linewidth]{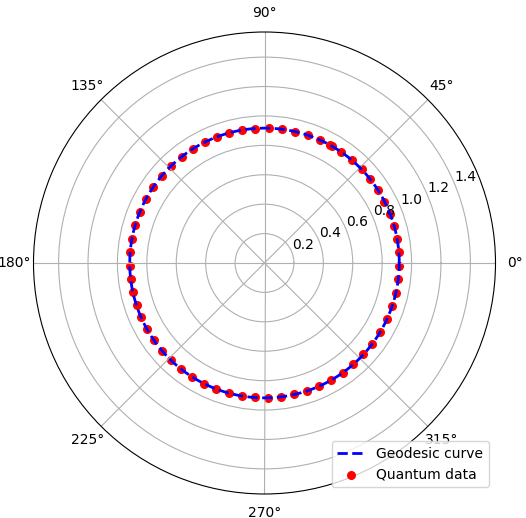}
        \caption{Parameters: $\rho_{0}=0.90, \phi_{0}=\pi/3, \sigma_{1}=0.09, \sigma_{2}=0.05, n_{\text{max}}=m_{\text{max}}=200, M=17.3, m_{0}=-30, n_{0}=0, E\approx49$.}
        \label{fig:fig53}
    \end{minipage}\hfill
    \begin{minipage}{0.48\textwidth}
        \centering
        \includegraphics[width=1\linewidth]{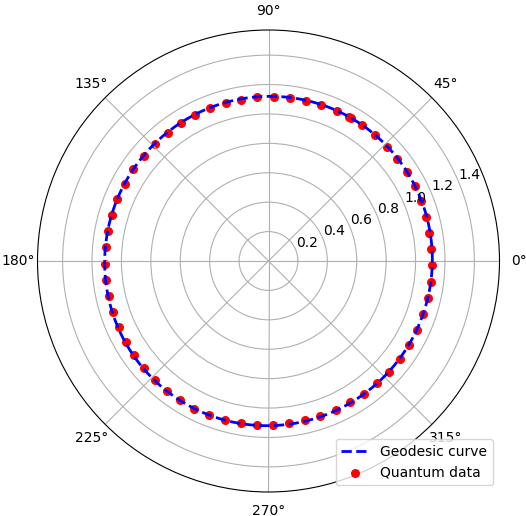}
        \caption{Parameters: $\rho_{0}=1.10, \phi_{0}=\pi/3, \sigma_{1}=0.09, \sigma_{2}=0.05, n_{\text{max}}=m_{\text{max}}=250, M=6.5, m_{0}=-30, n_{0}=0, E\approx38$.}
        \label{fig:fig54}
    \end{minipage}
\end{figure}

\begin{figure}[H]
    \centering

    \begin{minipage}{0.48\textwidth}
        \centering
        \includegraphics[width=1\linewidth]{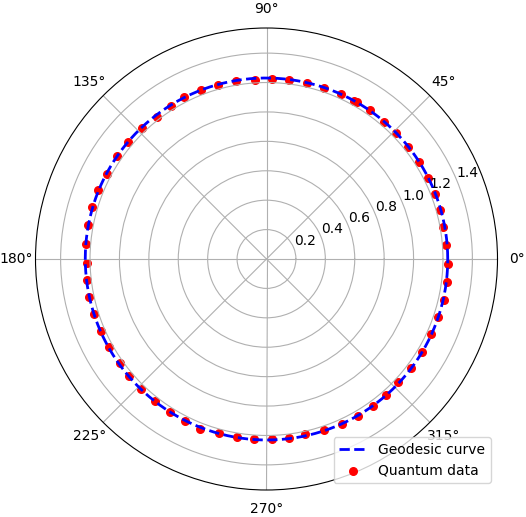}
    \end{minipage}\hfill
    \caption{Parameters: $\rho_{0}=1.20,\phi_{0}=\pi/3,\sigma_{1}=0.09,\sigma_{2}=0.05,n_{max}=m_{max}=350$,  $M=2.75, m_{0}=-30,n_{0}=0, E\approx34.5$}
    \label{fig:fig55}
\end{figure}

\section{2D Plots: Position Operator Approach}\label{appendix_2d_plots_pos_ope}
In this section, we present the 2D plots for the cases examined in section \ref{sec:pos} along with other examples to see that the center of the packet $\langle\hat{\rho}\rangle(t)$ and $\langle\hat{\phi}\rangle(t)$ follow the classical geodesic trajectory.

\subsection{Massive Case: Elliptical-like}\label{appendix_2d_plots_pos_ope_massive_elliptical}

\noindent\textbf{Fix $m_{0}$, Vary $M$}

Fig. \ref{fig:fig:pos_massive_fix_m0_vary_M_2D} below display 2D plots of $\langle\hat{\rho}\rangle$ and $\langle\hat{\phi}\rangle$ for fixed angular momentum $m_{0}=-20$ and mass $M=25,\,45,\,60,\,80$ and compares them with the corresponding geodesic solutions for various parameter choices.

\begin{figure}[H]
    \centering

    \begin{subfigure}{0.48\textwidth}
        \centering
        \includegraphics[width=0.7\linewidth]{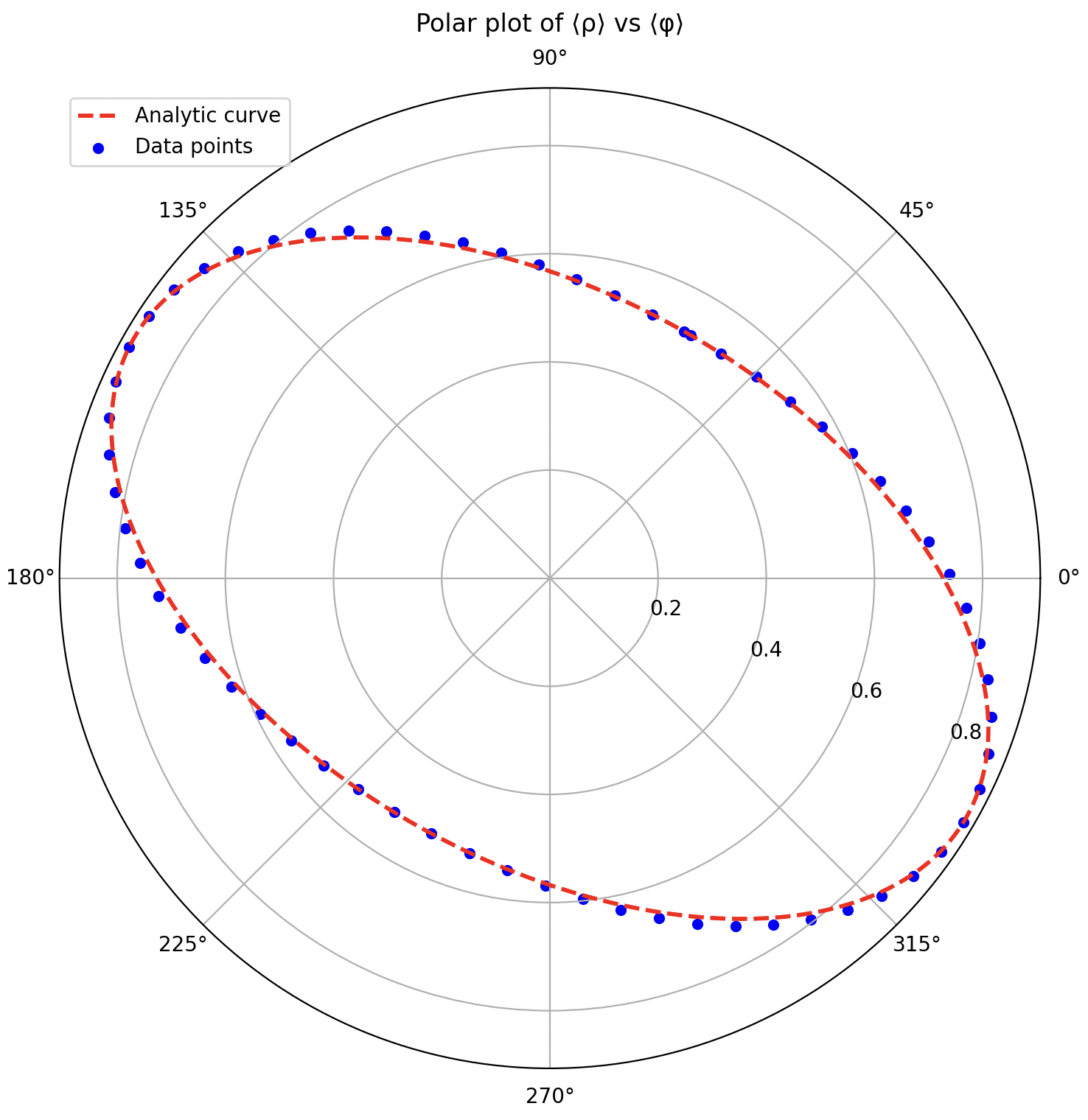}
        \caption{$m_{0}=-20$ , $M=25$ , $E\approx53,n_{max}=m_{max}=75$}
    \end{subfigure}\hfill
    \begin{subfigure}{0.48\textwidth}
        \centering
        \includegraphics[width=0.7\linewidth]{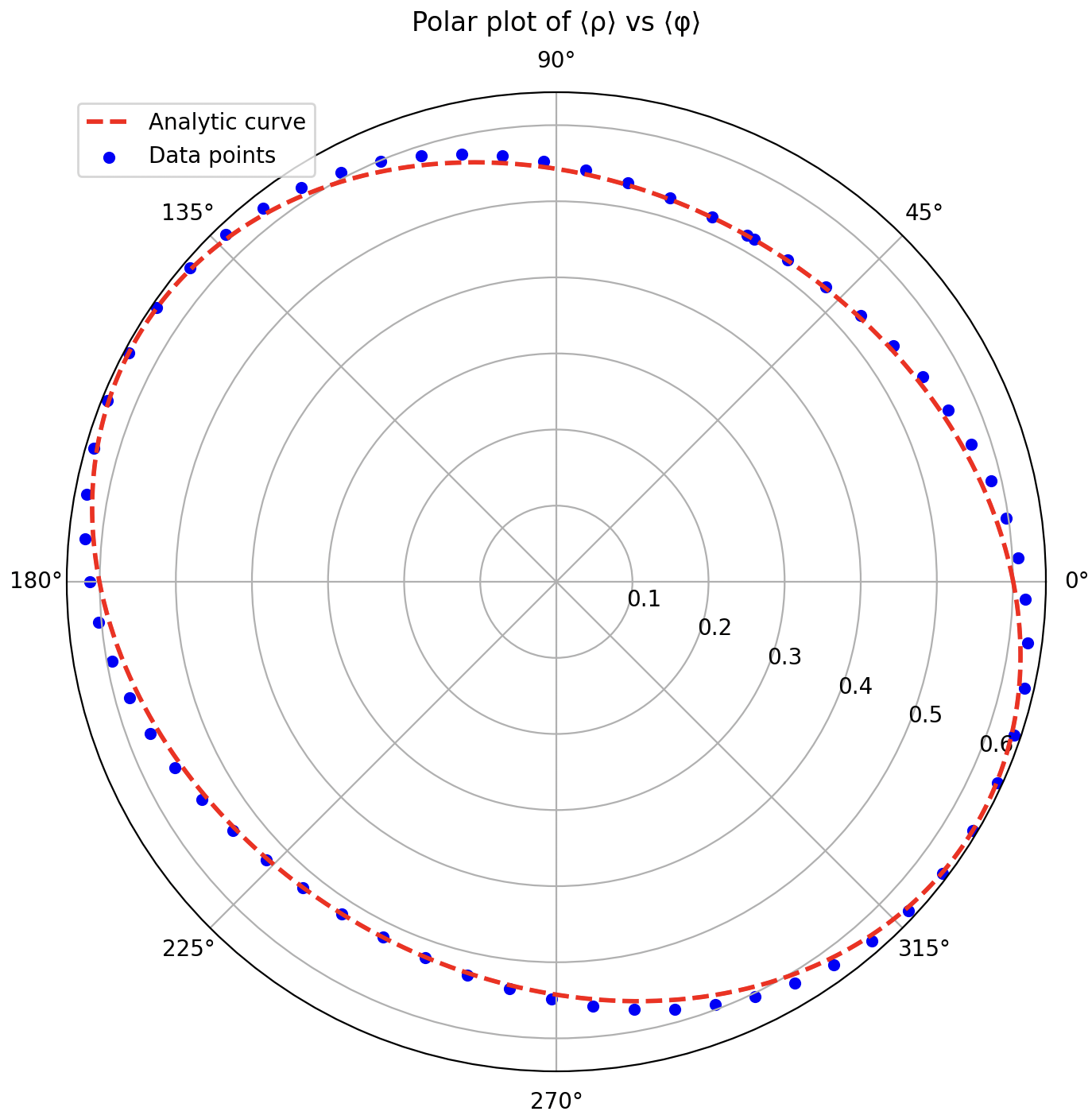}
        \caption{$m_{0}=-20$ , $M=45$, $E\approx69,n_{max}=m_{max}=75$}
    \end{subfigure}\hfill

    \begin{subfigure}{0.48\textwidth}
        \centering
        \includegraphics[width=0.7\linewidth]{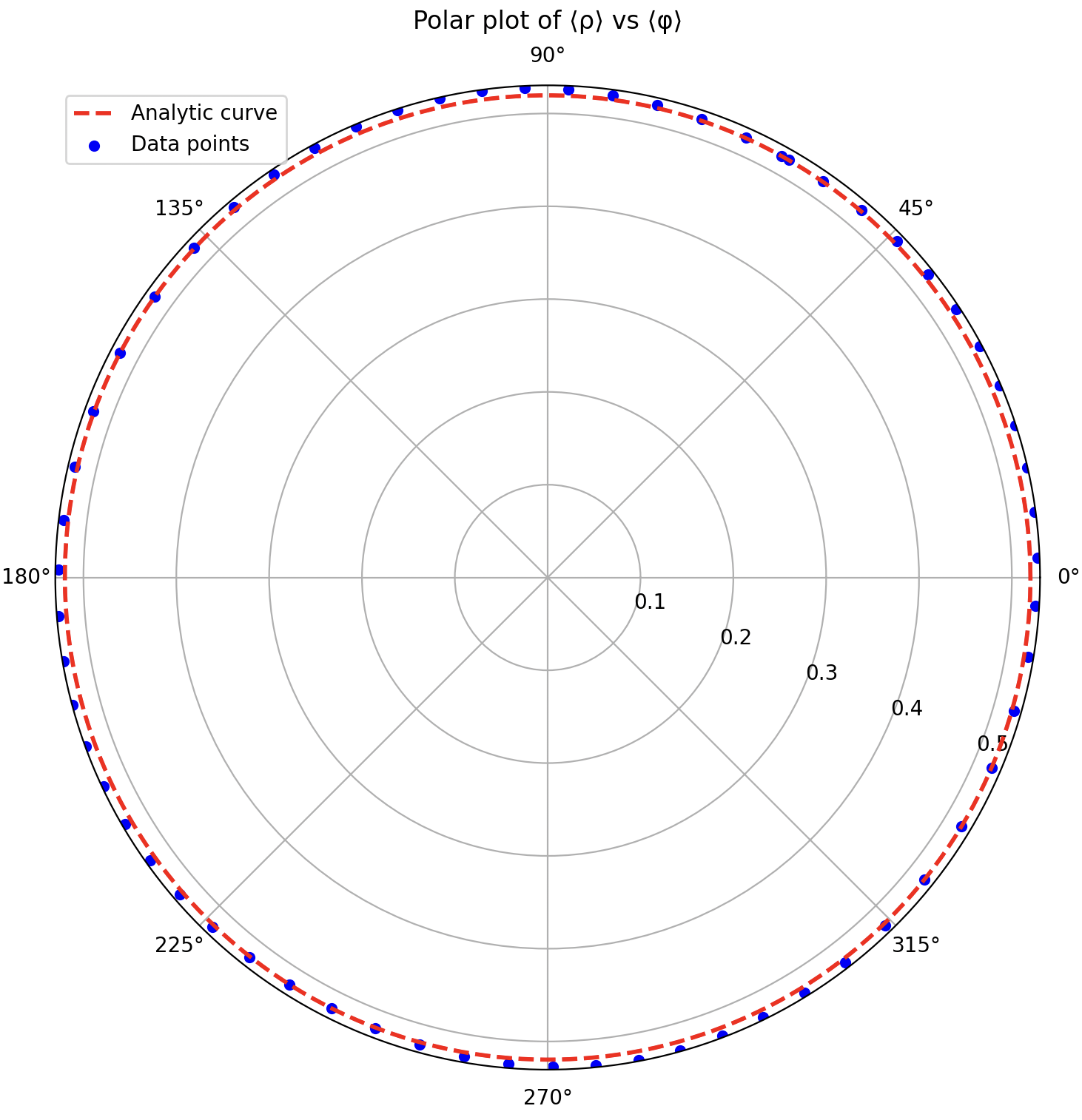}
        \caption{$m_{0}=-20, M=60 , E\approx83.5,n_{max}=m_{max}=75$}
    \end{subfigure}\hfill
    \begin{subfigure}{0.48\textwidth}
        \centering
        \includegraphics[width=0.7\linewidth]{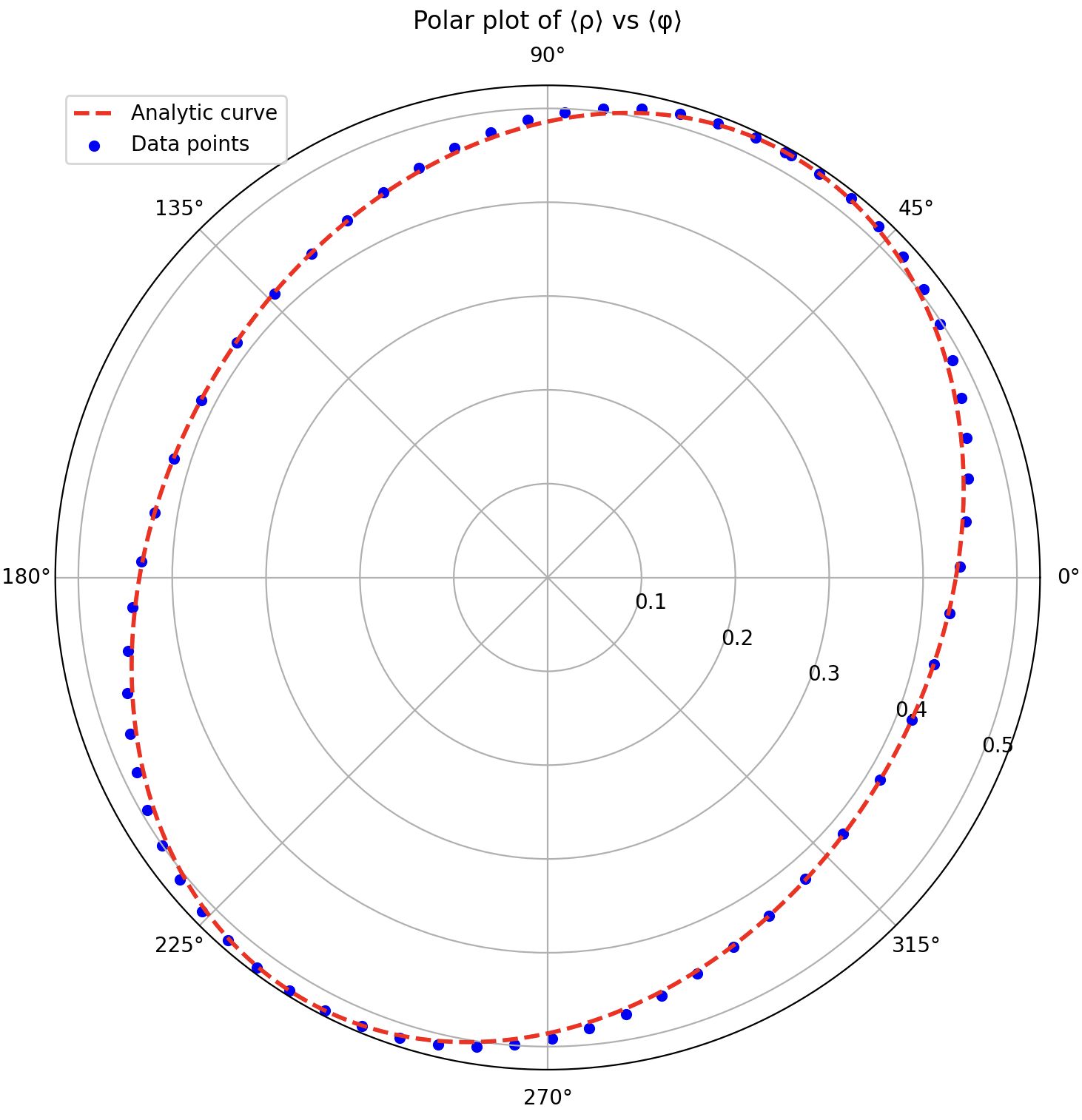}
        \caption{$m_{0}=-20, M=80, E\approx103.8,n_{max}=m_{max}=75$}
    \end{subfigure}\hfill
    
    \caption{Parameters: $\rho_{0}=0.5,\phi_{0}=\pi/3,\sigma_{1}=0.09,\sigma_{2}=0.05$}
    \label{fig:fig:pos_massive_fix_m0_vary_M_2D}

\end{figure}

As illustrated in the preceding figures, keeping the angular momentum parameter $m_{0}$ and other initial conditions fixed while progressively increasing the mass $M$ significantly alters the wave packet's trajectory. Specifically, the orbit transitions from an elliptical like shape, circularizes at a critical mass value, and subsequently elongates back again into an elliptical like path. 

\noindent\textbf{Fix $M$, Vary $m_{0}$}

\begin{figure}[H]
    \centering

    \begin{subfigure}{0.48\textwidth}
        \centering
        \includegraphics[width=0.7\linewidth]{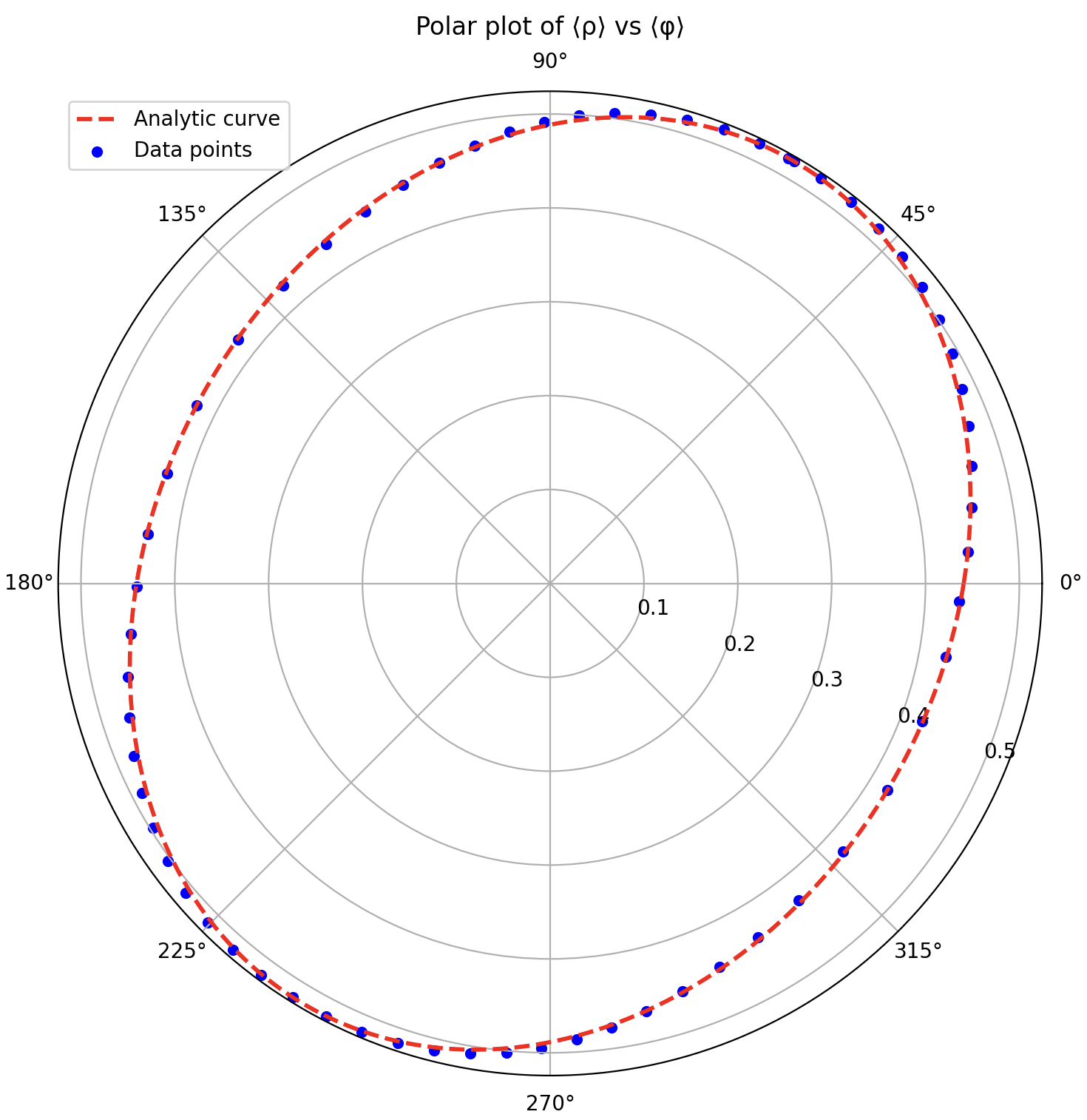}
        \caption{$M=60, m_{0}=-15, E\approx79,n_{max}=m_{max}=70$}
    \end{subfigure}\hfill
    \begin{subfigure}{0.48\textwidth}
        \centering
        \includegraphics[width=0.7\linewidth]{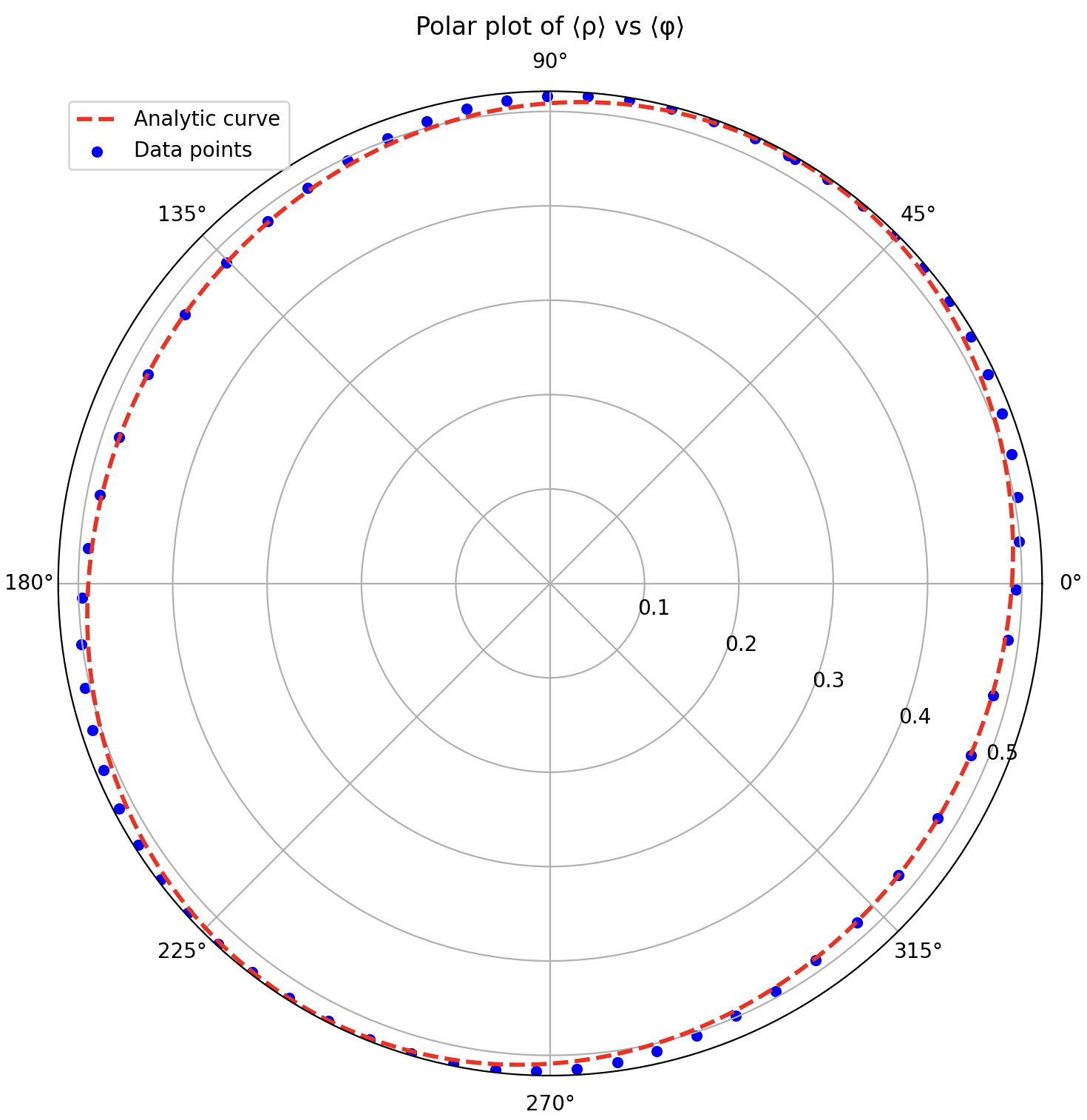}
        \caption{$M=60, m_{0}=-18, E\approx82,n_{max}=m_{max}=70$}
    \end{subfigure}

    \begin{subfigure}{0.48\textwidth}
        \centering
        \includegraphics[width=0.7\linewidth]{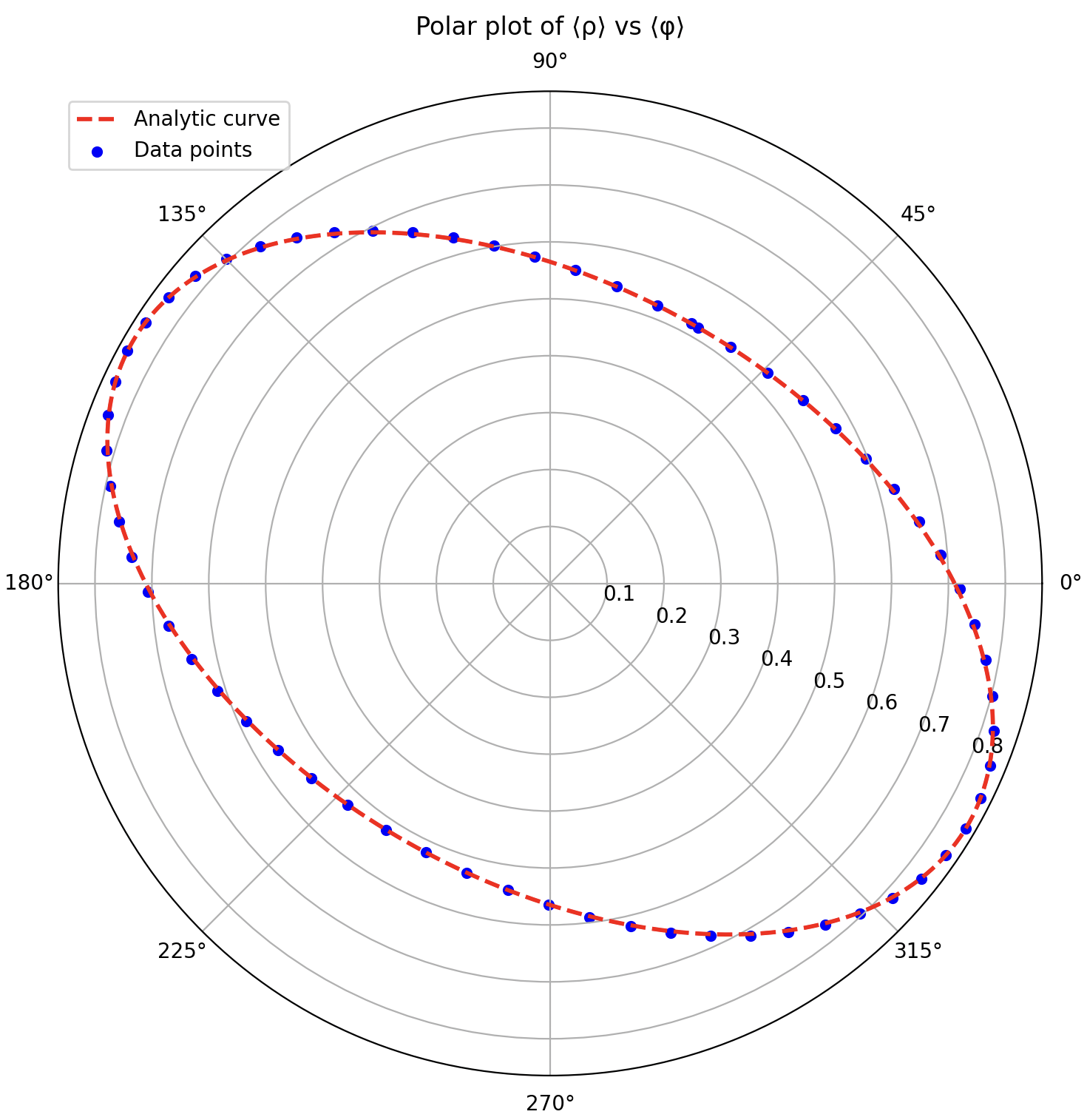}
        \caption{$M=60, m_{0}=-40,E\approx109.8,n_{max}=m_{max}=70$}
    \end{subfigure}\hfill
    \begin{subfigure}{0.48\textwidth}
        \centering
        \includegraphics[width=0.7\linewidth]{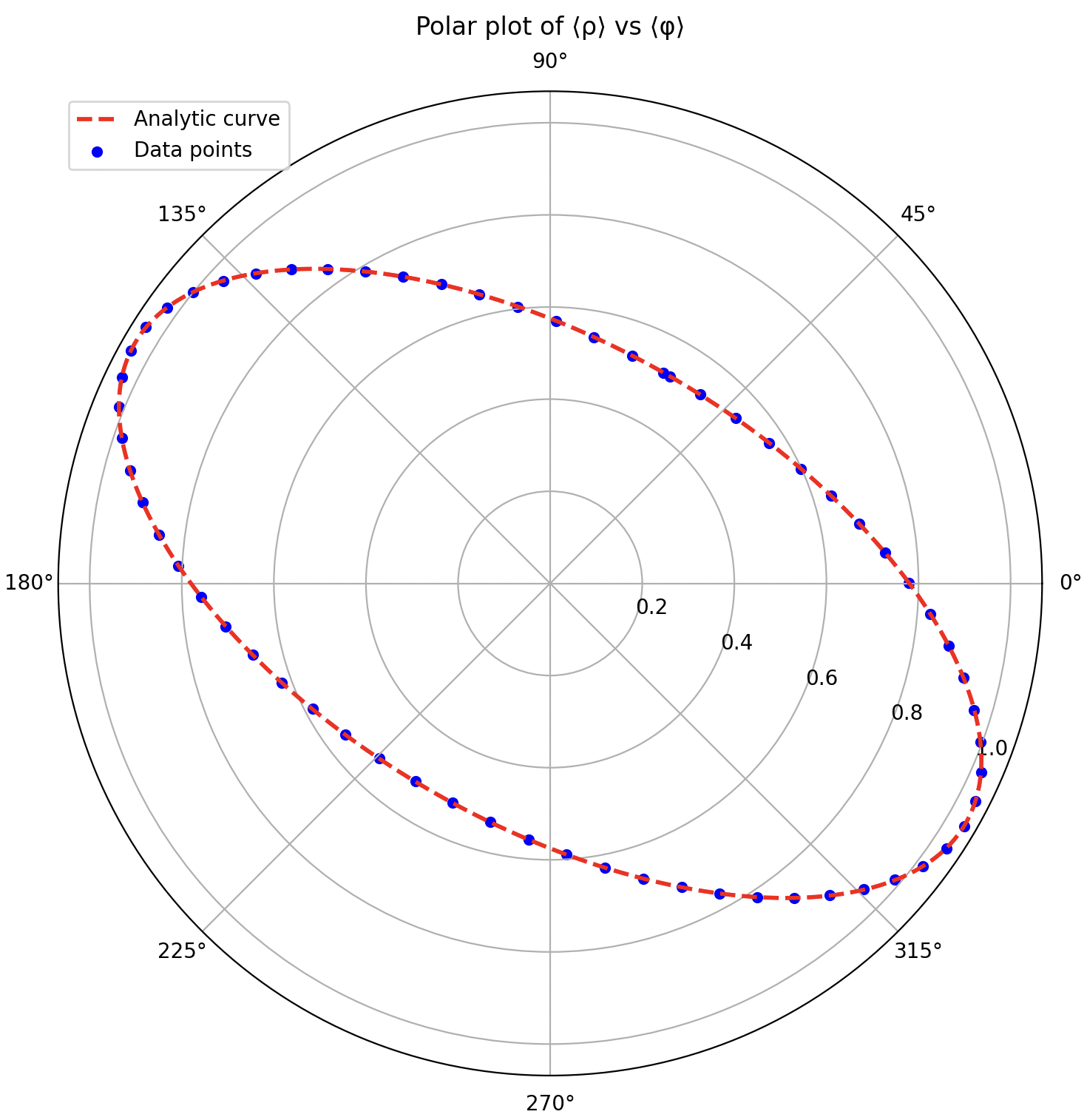}
        \caption{$M=60, m_{0}=-60,E\approx143.7,n_{max}=m_{max}=70,$}
    \end{subfigure}
    
    \caption{Parameters: $\rho_{0}=0.5,\phi_{0}=\pi/3,\sigma_{1}=0.09,\sigma_{2}=0.05$}
    \label{fig:fig:pos_massive_fix_M_vary_m0_2D}

\end{figure}

In the subsequent analysis, we hold the mass strictly constant at $M=60$ and systematically vary the angular momentum parameter $m_{0}=-15,-18,-40,-60$. See Fig. \ref{fig:fig:pos_massive_fix_M_vary_m0_2D}.

As expected, keeping the mass parameter $M$ and other initial conditions fixed while progressively increasing the angular momentum parameter $m_{0}$ significantly alters the wave packet's trajectory. Specifically, the orbit transitions from an elliptical like shape, circularizes at a critical angular momentum $m_{0}$, and subsequently elongates back again into an elliptical like path. 

In the above cases, tuning the scalar mass $M$ or angular momentum $m_{0}$ or both directly modulates the overall energy of the wave packet, thereby driving this orbital transition.

\subsection{Massive Case: Radial Infall}\label{appendix_2d_plots_pos_ope_massive_rad_infall}

\begin{figure}[H]
    \centering

    \begin{subfigure}{0.48\textwidth}
        \centering
        \includegraphics[width=0.7\linewidth]{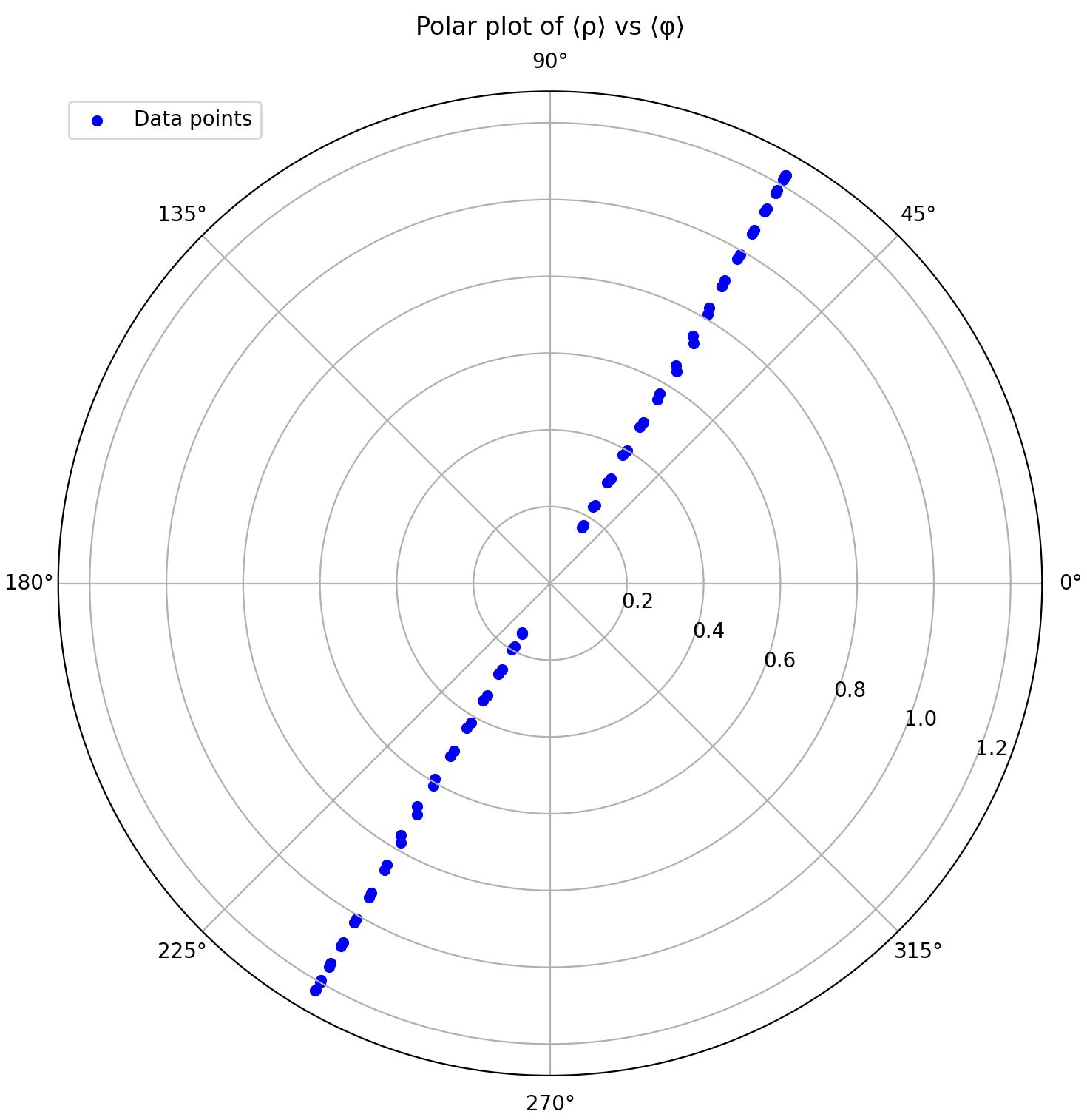}
        \caption{$M=20, m_{0}=0, E\approx66.7,n_{max}=m_{max}=200$}
    \end{subfigure}\hfill
    \begin{subfigure}{0.48\textwidth}
        \centering
        \includegraphics[width=0.7\linewidth]{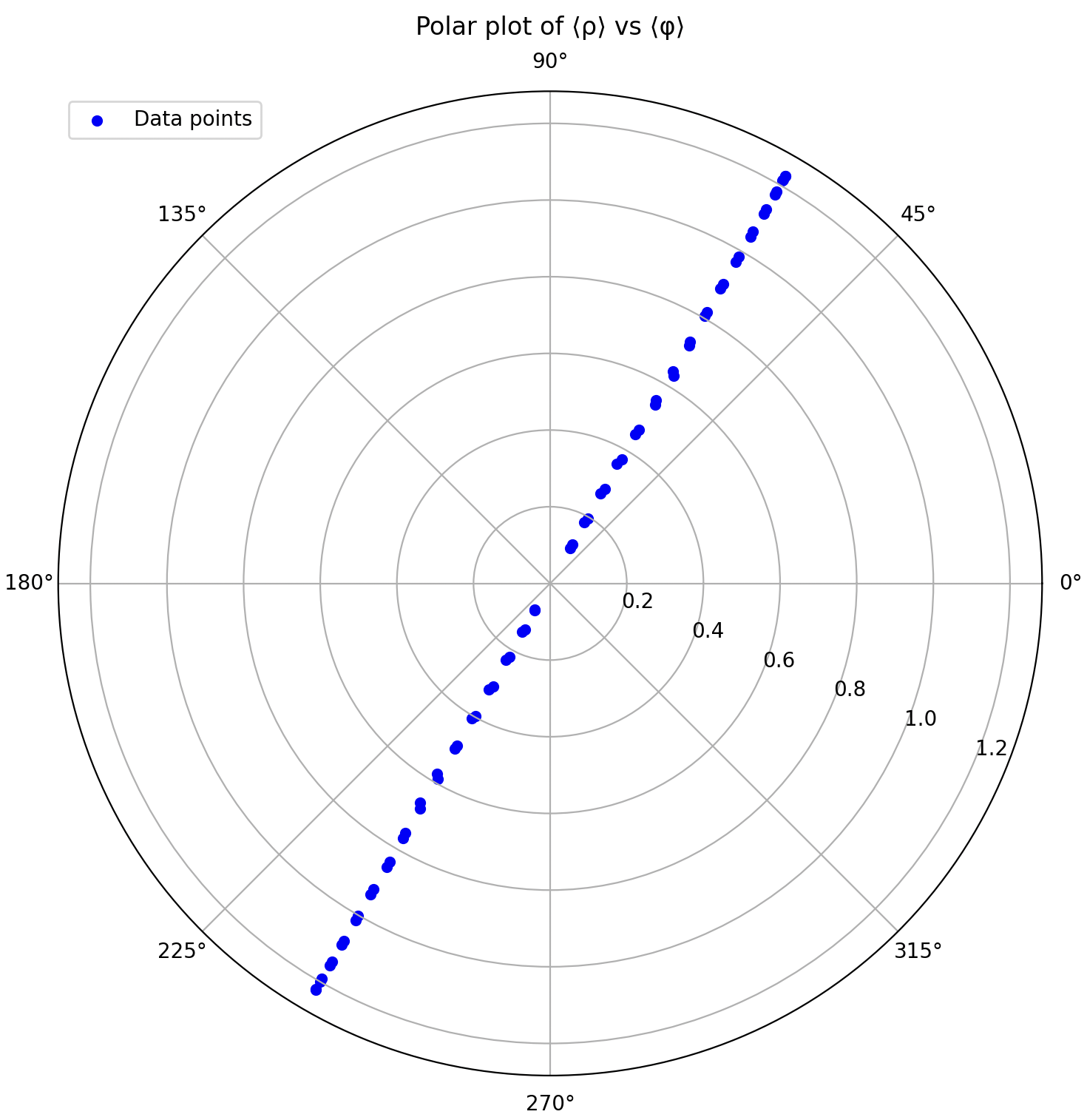}
        \caption{$M=40, m_{0}=0, E\approx128,n_{max}=m_{max}=200$}
    \end{subfigure}

    \begin{subfigure}{0.48\textwidth}
        \centering
        \includegraphics[width=0.7\linewidth]{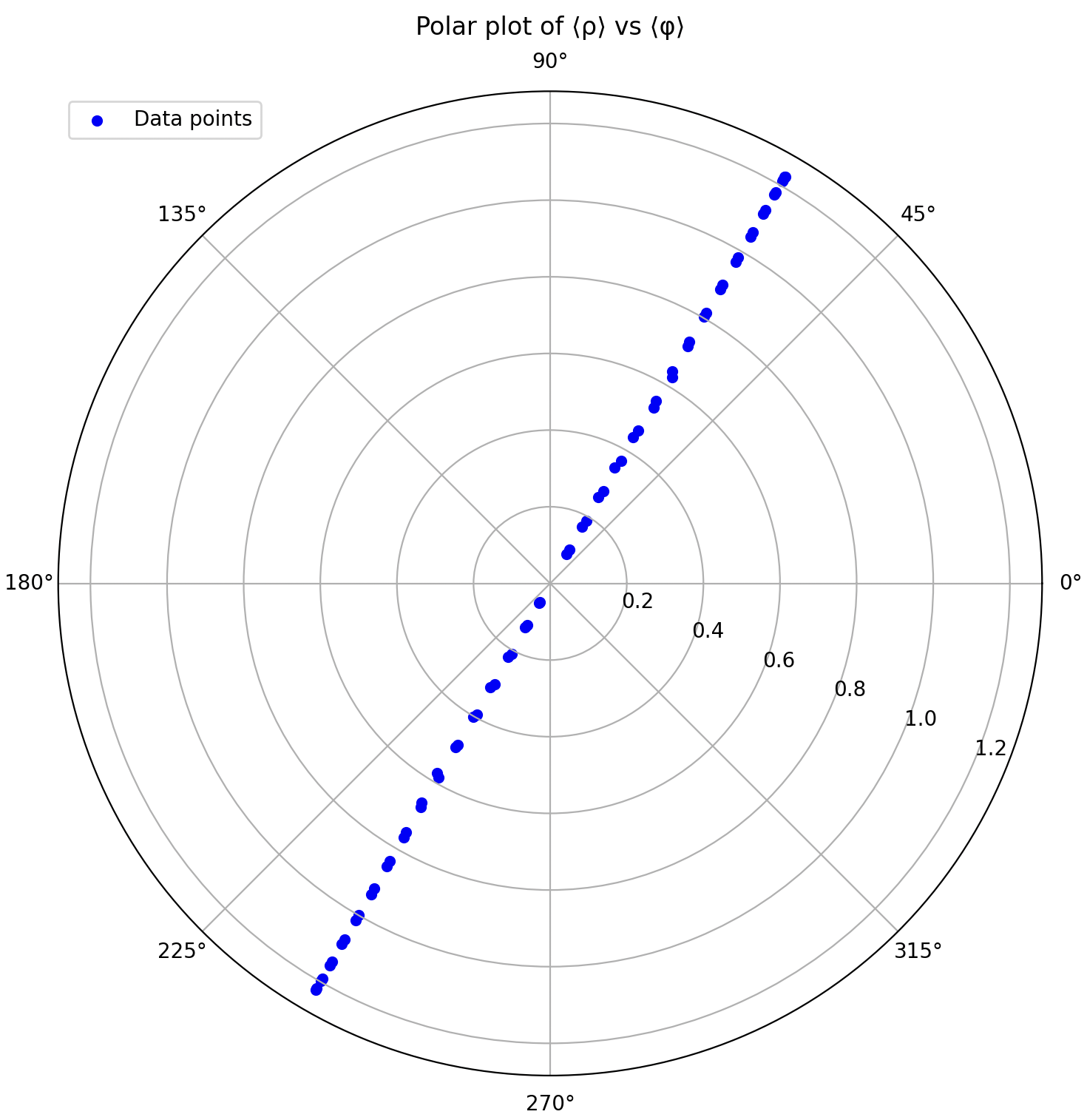}
        \caption{$M=60, m_{0}=0,n_{max}=m_{max}=250, E\approx193$}
    \end{subfigure}\hfill
    \begin{subfigure}{0.48\textwidth}
        \centering
        \includegraphics[width=0.7\linewidth]{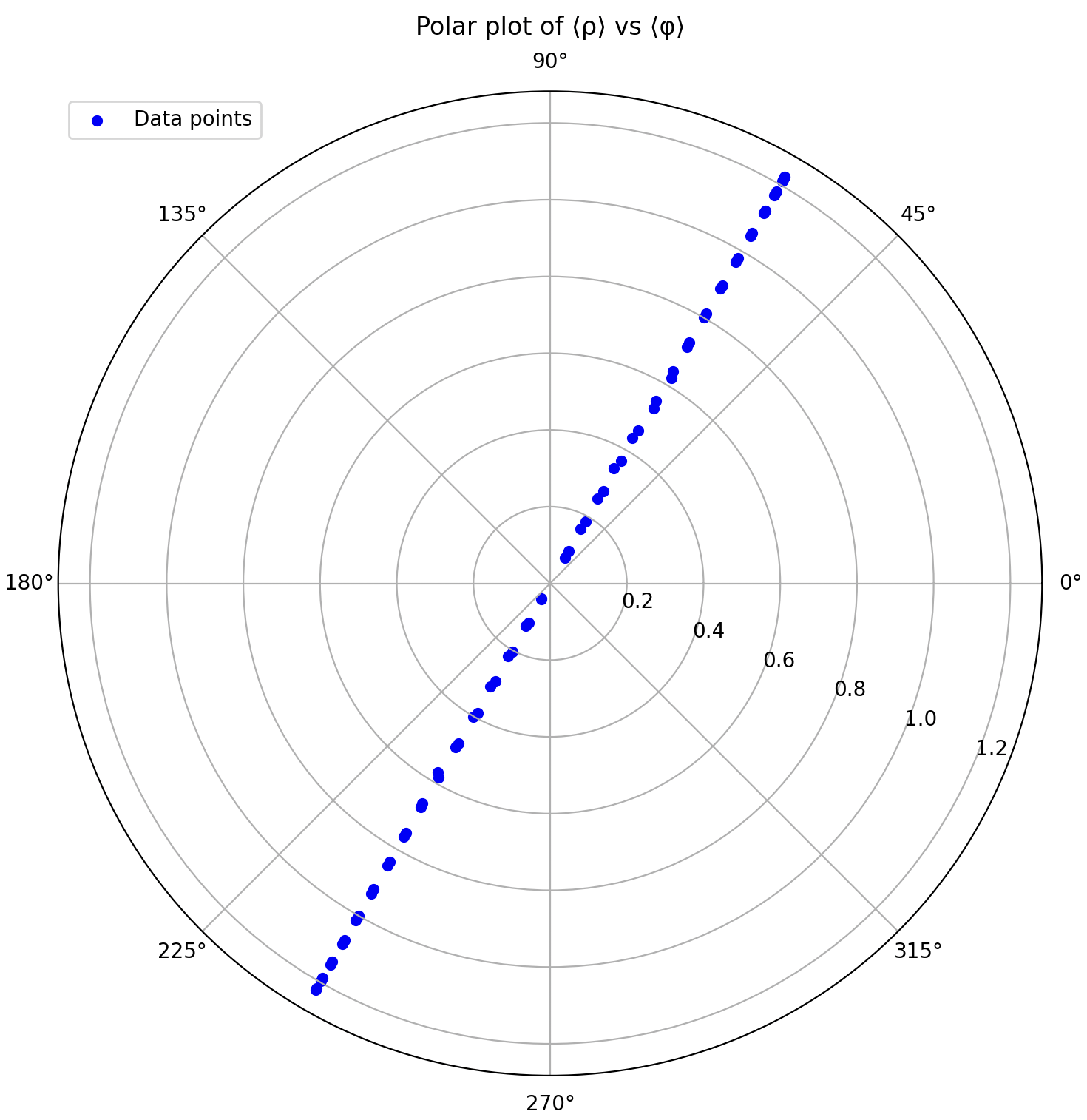}
        \caption{$M=80, m_{0}=0, n_{max}=m_{max}=300, E\approx256$}
    \end{subfigure}
    \caption{Parameters: $\rho_{0}=1.2,\phi_{0}=\pi/3,\sigma_{1}=0.09,\sigma_{2}=0.05$}
    \label{fig:fig:pos_massive_radial_2D}

\end{figure}

For radial infall, the resulting 2D parametric plots tracking the geometric coordinates $\langle\hat{\rho}\rangle$ versus $\langle\hat{\phi}\rangle$ are presented below (Fig. \ref{fig:fig:pos_massive_radial_2D}).

\subsection{$\langle\hat{\rho}\rangle$ vs $t$: Massive Case}\label{appendix_2d_plots_pos_ope_massive_rad_infall_rho_vs_t}

\begin{figure}[H]
    \centering

    \begin{subfigure}{0.48\textwidth}
        \centering
        \includegraphics[width=1.0\linewidth]{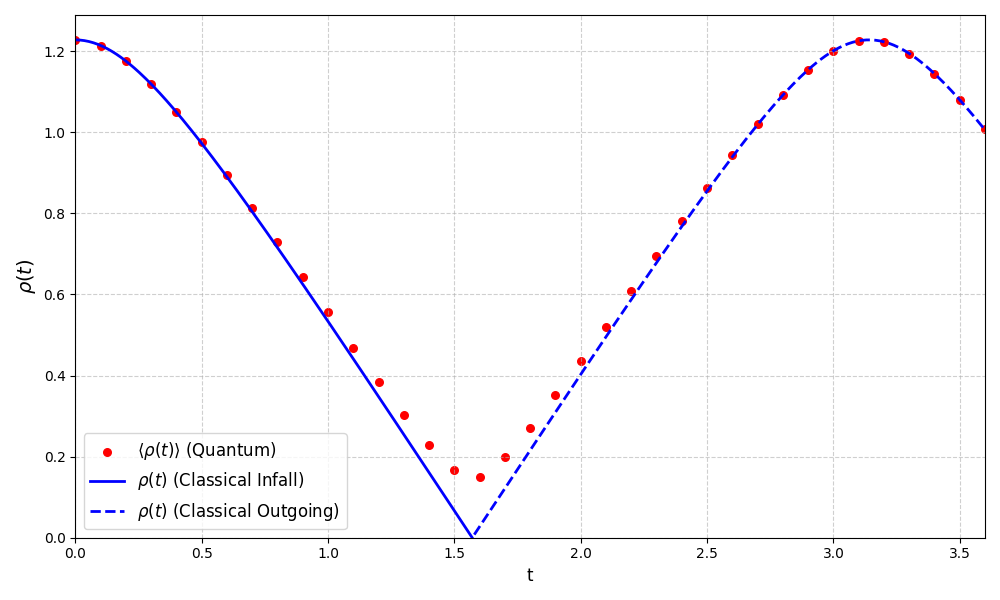}
        \caption{$M=20, m_{0}=0,n_{max}=m_{max}=200$}
    \end{subfigure}\hfill
    \begin{subfigure}{0.48\textwidth}
        \centering
        \includegraphics[width=1.0\linewidth]{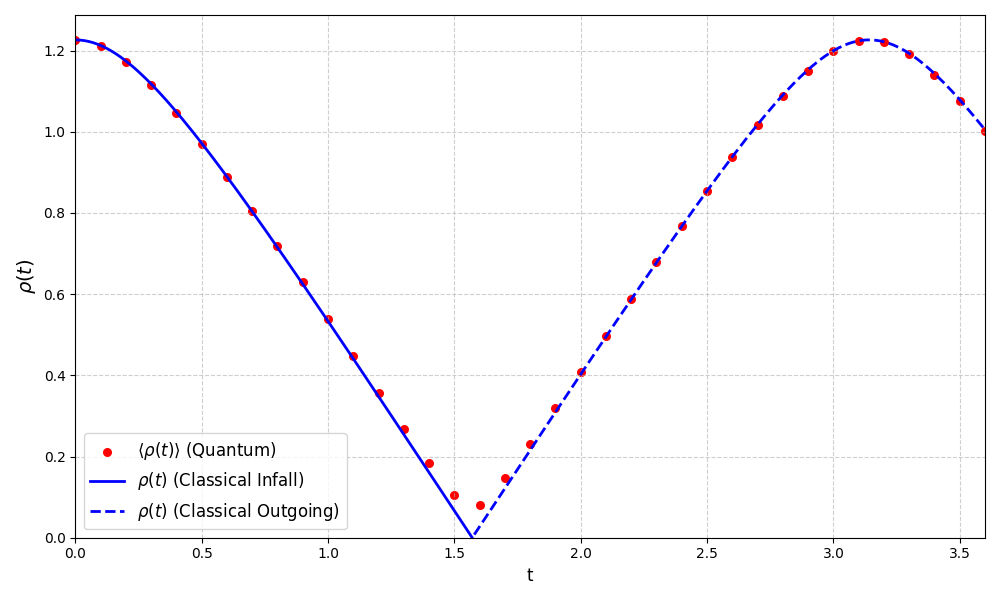}
        \caption{$M=40, m_{0}=0, n_{max}=m_{max}=200$}
    \end{subfigure}

    \begin{subfigure}{0.48\textwidth}
        \centering
        \includegraphics[width=1.0\linewidth]{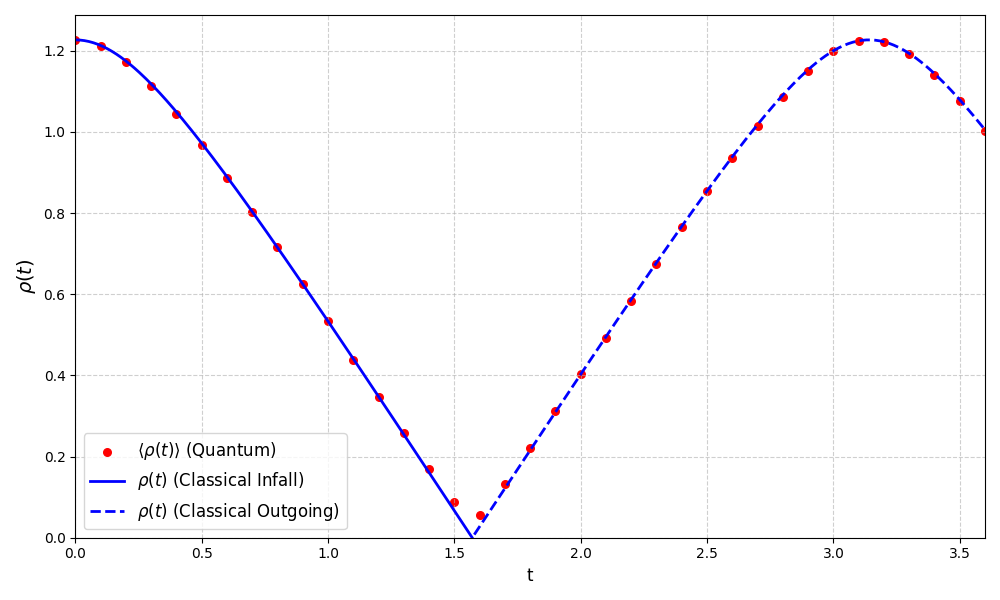}
        \caption{$M=60, m_{0}=0,n_{max}=m_{max}=250$}
    \end{subfigure}\hfill
    \begin{subfigure}{0.48\textwidth}
        \centering
        \includegraphics[width=1.0\linewidth]{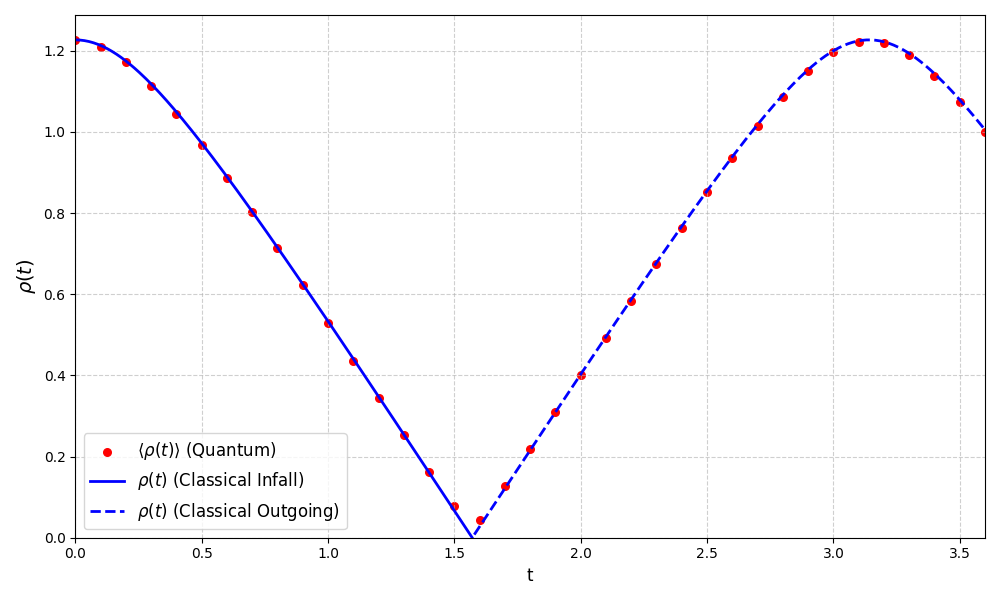}
        \caption{$M=80, m_{0}=0, n_{max}=m_{max}=300$}
    \end{subfigure}
    \caption{Parameters: $\rho_{0}=1.2,\phi_{0}=\pi/3,\sigma_{1}=0.09,\sigma_{2}=0.05$}
    \label{fig:fig:pos_massive_radial_rho_vs_t}
    
\end{figure}

We now examine the evolution of $\langle\hat{\rho}\rangle{(t)}$ as a function of the coordinate time $t$ for each of the scenarios discussed above. See Fig. \ref{fig:fig:pos_massive_radial_rho_vs_t}.

Consistent with the behavior observed for $\bar{\rho}(t)$, a deviation from the exact classical geodesic (of $\rho$) manifests as the wave packet traverses the spatial origin. This discrepancy originates from the same physical mechanism discussed previously: loosely, when $\langle \rho \rangle$ is close to zero, the fluctuations are significant. 

\subsection{Null Case: Localized Wave Packet}\label{appendix_2d_plots_pos_ope_null_localized_2d_plots}
\begin{figure}[H]
    \centering
    \begin{minipage}{0.48\textwidth}
        \centering
        \includegraphics[width=0.7\linewidth]{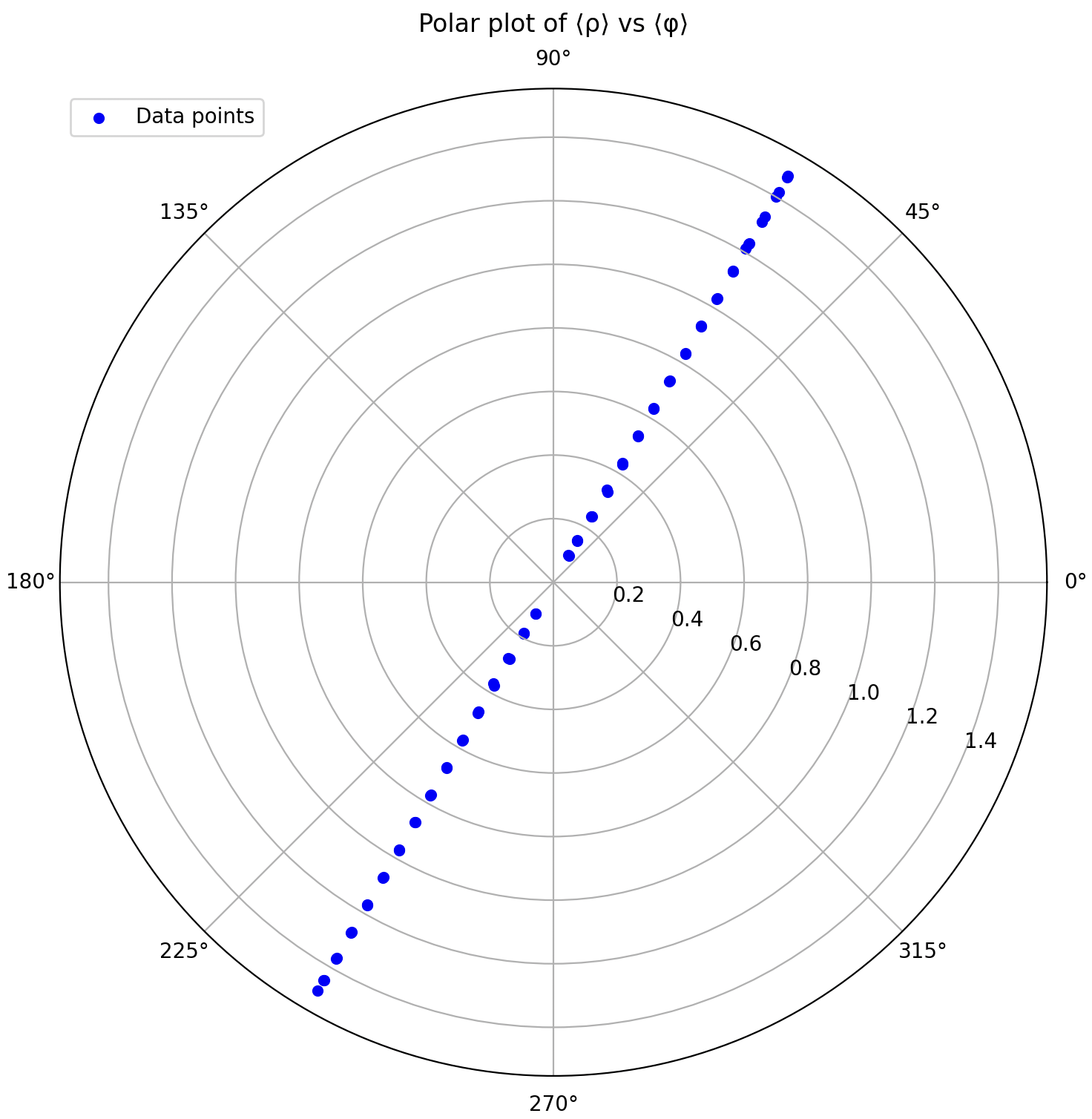}
    \end{minipage}\hfill
    \begin{minipage}{0.48\textwidth}
        \centering
        \includegraphics[width=0.7\linewidth]{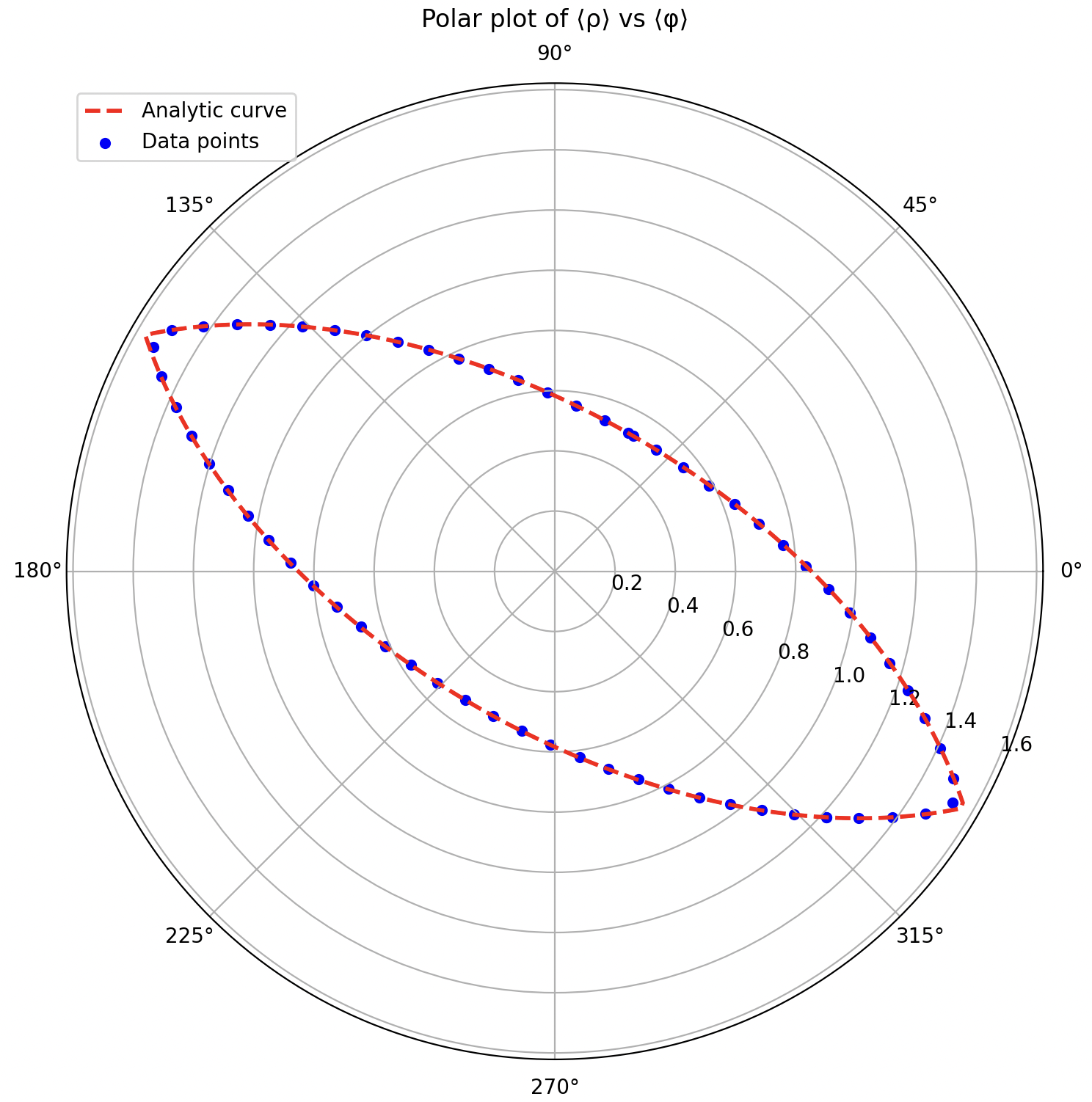}
    \end{minipage}
    \caption{Parameters: $\phi_{0}=\pi/3,\sigma_{1}=0.09,\sigma_{2}=0.05,n_{max}=m_{max}=300$ \textbf{Left:} $M=0, m_{0}=0,n_{0}=200,\rho_{0}=1.2, E\approx200$, \textbf{Right:} $M=0, m_{0}=-40,n_{0}=0,\rho_{0} = 0.5, E\approx83.5$}
    \label{fig:fig:pos_null_radial_and_ellip_2D}
\end{figure}

Here, we set $M=0$ and choose parameters such that it gives localized motion. Below we present the 2D plots between $\langle\hat{\rho}\rangle$ and $\langle\hat{\phi}\rangle$(Fig. \ref{fig:fig:pos_null_radial_and_ellip_2D}) for two cases.

\subsection{$\langle\hat{\rho}\rangle$ vs $t$: Null Case}\label{appendix_2d_plots_pos_ope_null_localized_radial_infall}

We now examine the evolution of $\langle\hat{\rho}\rangle$ with $t$ for a radially infalling massless scalar with large energy $E$. The solid blue line is the line with slope $\pm1$, which is the classical path.
\begin{figure}[H]
    \centering
    \includegraphics[width=\textwidth, keepaspectratio]{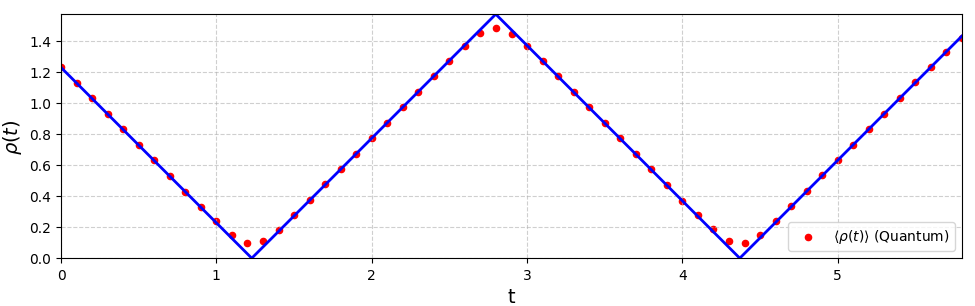}
    \caption{Parameters: $\rho_{0}=1.2,\phi_{0}=\pi/3,\sigma_{1}=0.09,\sigma_{2}=0.05,n_{max}=m_{max}=300$,$M=0, m_{0}=0, n_{0}=200$}
    \label{fig:fig:pos_null_radial_rho_vs_t_2D}
\end{figure}
Now we also see that there is slight offset at near the boundary $\rho=\frac{\pi}{2}$. Operationally this is because the finite spatial extent of the wave packet causes its leading tail to interact with the AdS boundary before its geometric center arrives.  This boundary interaction induces reflection, which accounts for the slight positional offset observed near the edges of the spacetime. However, as the momentum $n_{0}$ is increased, the wave packet becomes more sharply localized, systematically suppressing this offset.

\subsection{Circular Geodesics in \texorpdfstring{AdS$_{3}$}{AdS3}}\label{appendix_2d_plots_pos_ope_circular_orbits}

\begin{figure}[H]
    \centering
    \begin{minipage}{0.48\textwidth}
        \centering
        \includegraphics[width=1\linewidth]{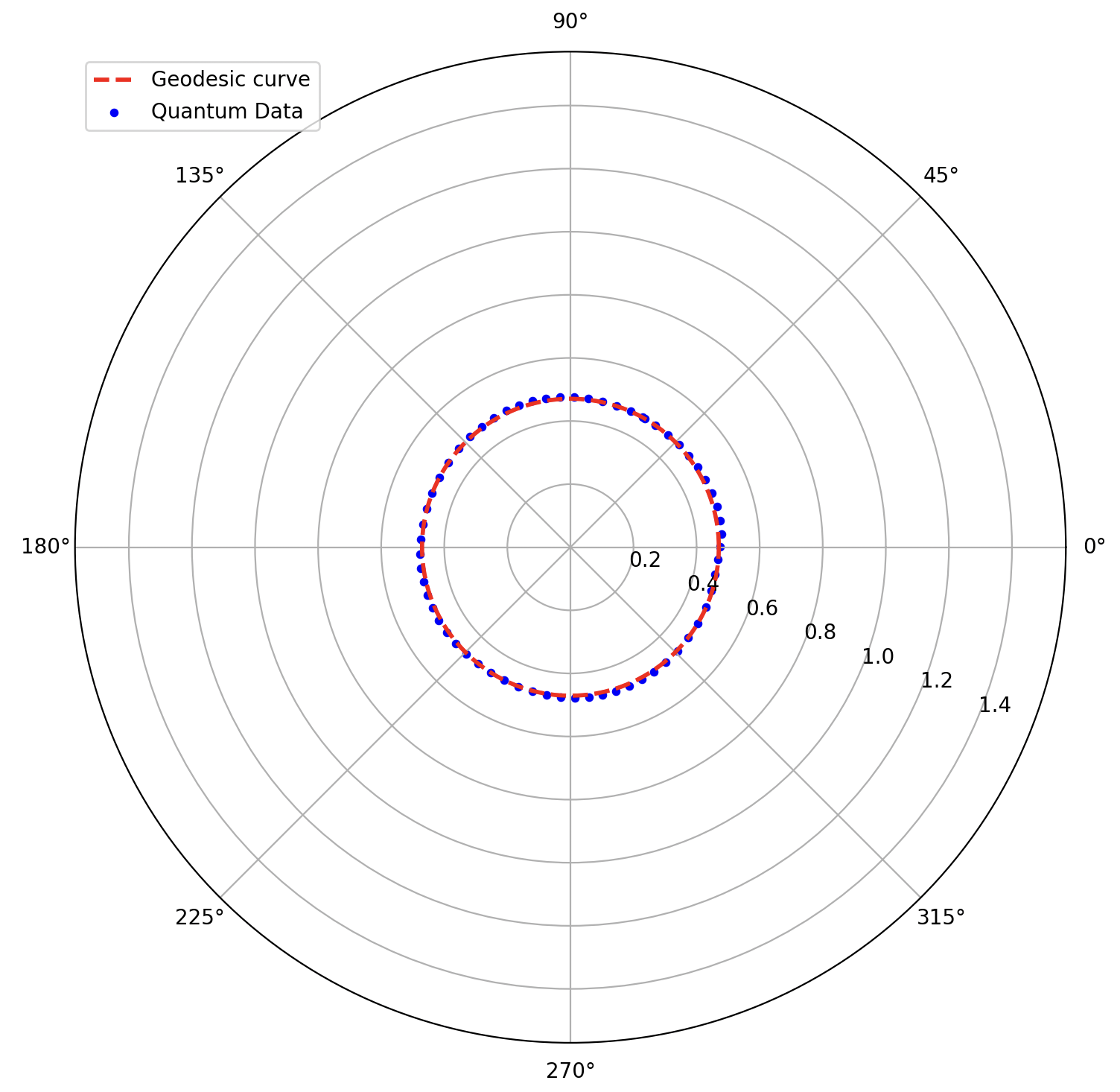}
        \caption{Parameters: $\rho_{0}=0.45, \phi_{0}=\pi/3, \sigma_{1}=0.09, \sigma_{2}=0.05, n_{\text{max}}=m_{\text{max}}=200, M=117, m_{0}=-30, n_{0}=0, E\approx151$.}
        \label{fig:fig66}
    \end{minipage}\hfill
    \begin{minipage}{0.48\textwidth}
        \centering
        \includegraphics[width=1\linewidth]{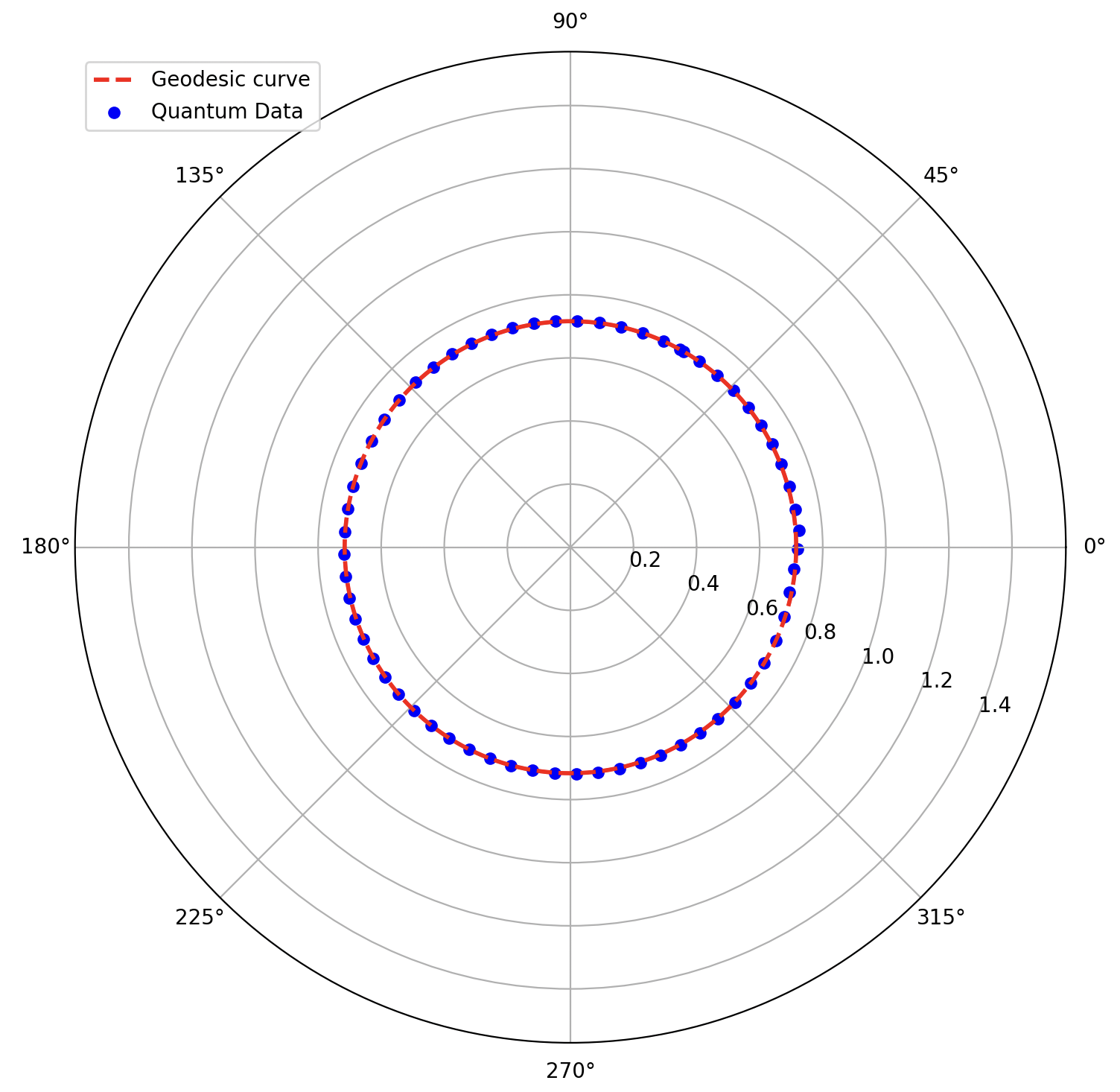}
        \caption{Parameters: $\rho_{0}=0.70, \phi_{0}=\pi/3, \sigma_{1}=0.09, \sigma_{2}=0.05, n_{\text{max}}=m_{\text{max}}=200, M=38, m_{0}=-30, n_{0}=0, E\approx70.3$.}
        \label{fig:fig67}
    \end{minipage}
\end{figure}

\begin{figure}[H]
    \centering
    \begin{minipage}{0.48\textwidth}
        \centering
        \includegraphics[width=1\linewidth]{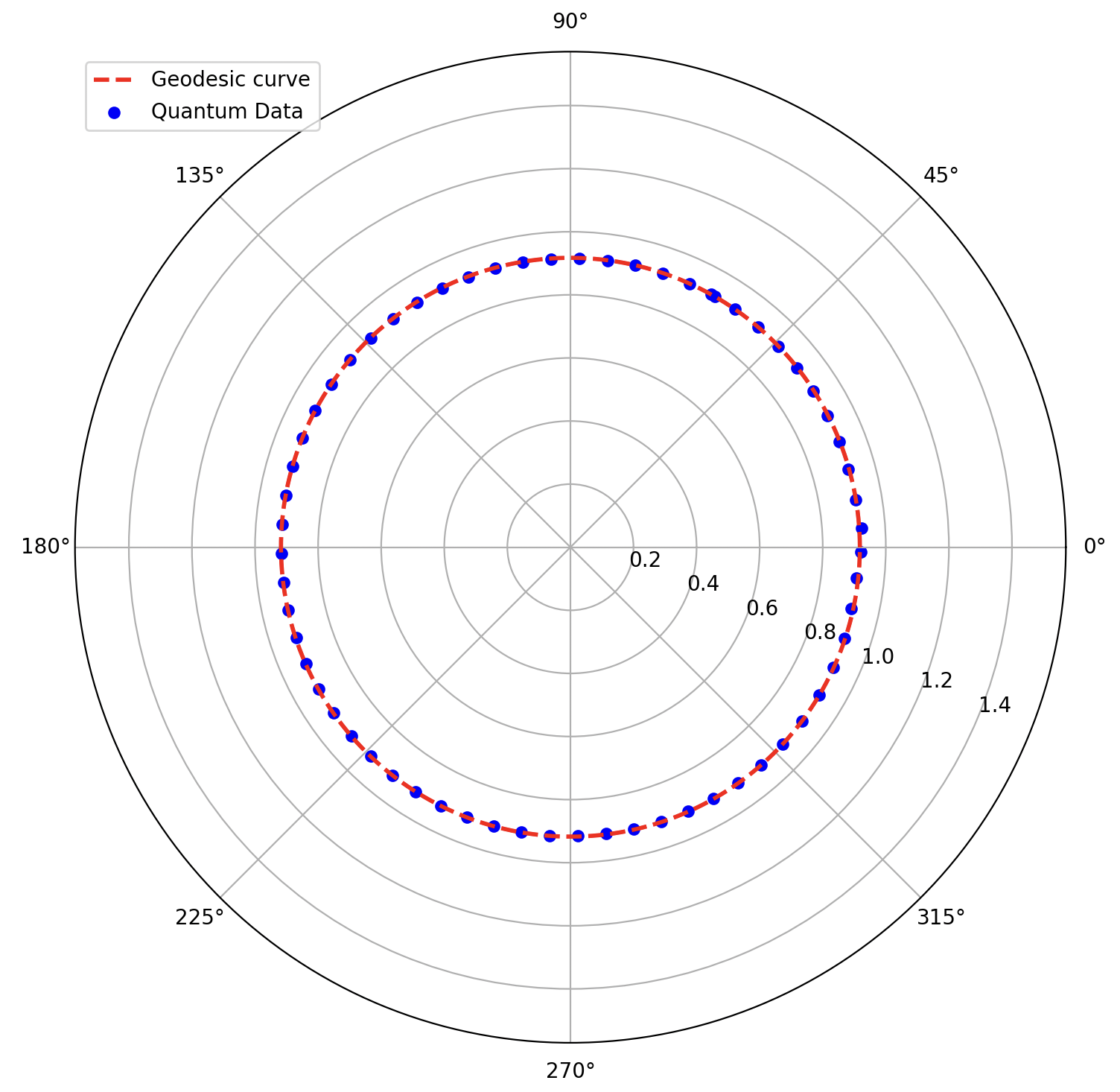}
        \caption{Parameters: $\rho_{0}=0.90, \phi_{0}=\pi/3, \sigma_{1}=0.09, \sigma_{2}=0.05, n_{\text{max}}=m_{\text{max}}=200, M=16, m_{0}=-30, n_{0}=0, E\approx47.8$.}
        \label{fig:fig68}
    \end{minipage}\hfill
    \begin{minipage}{0.48\textwidth}
        \centering
        \includegraphics[width=1\linewidth]{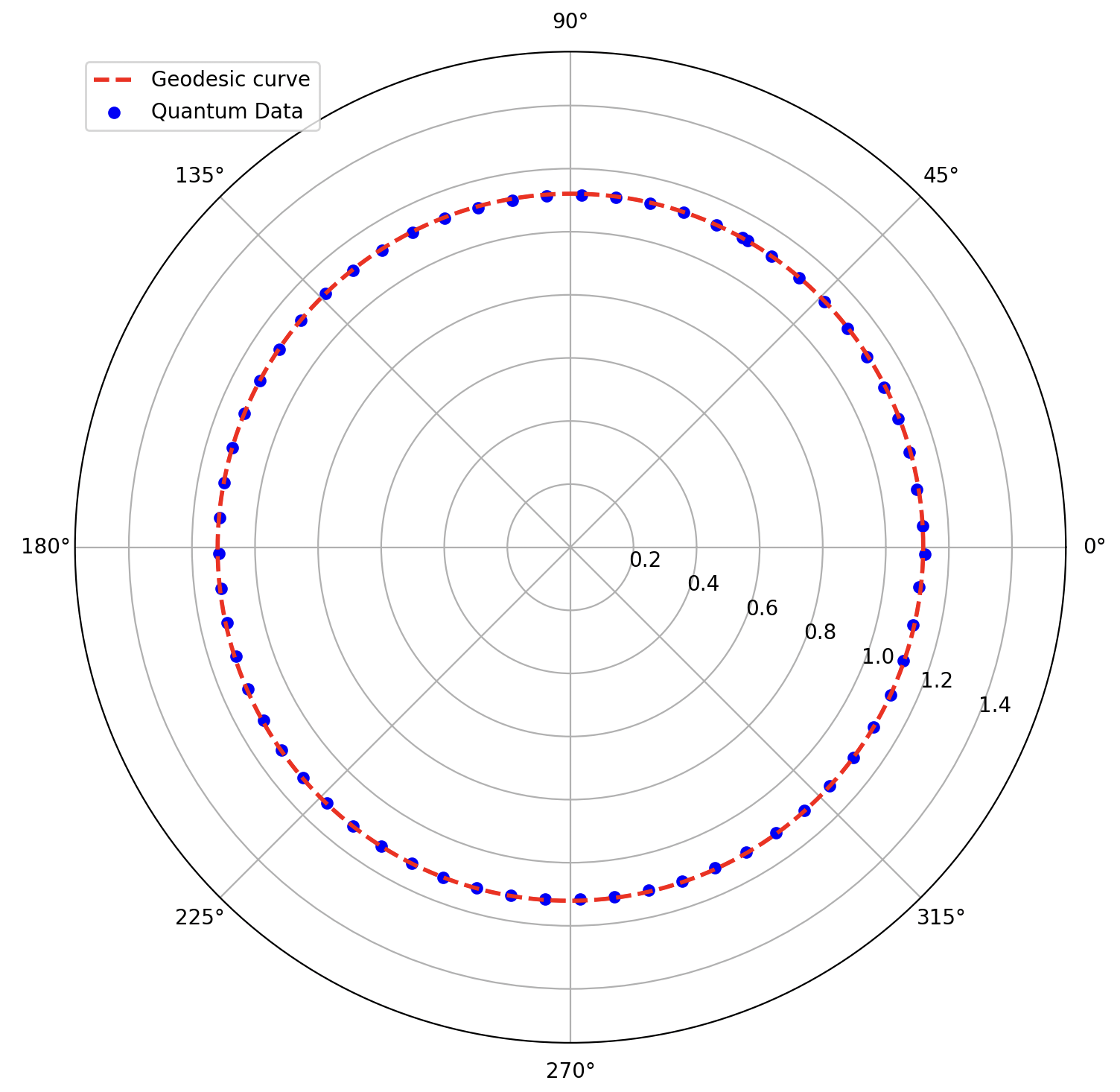}
        \caption{Parameters: $\rho_{0}=1.10, \phi_{0}=\pi/3, \sigma_{1}=0.09, \sigma_{2}=0.05, n_{\text{max}}=m_{\text{max}}=250, M=5.8, m_{0}=-30, n_{0}=0, E\approx37.3$.}
        \label{fig:fig69}
    \end{minipage}
\end{figure}

\begin{figure}[H]
    \centering

    \begin{minipage}{0.48\textwidth}
        \centering
        \includegraphics[width=1\linewidth]{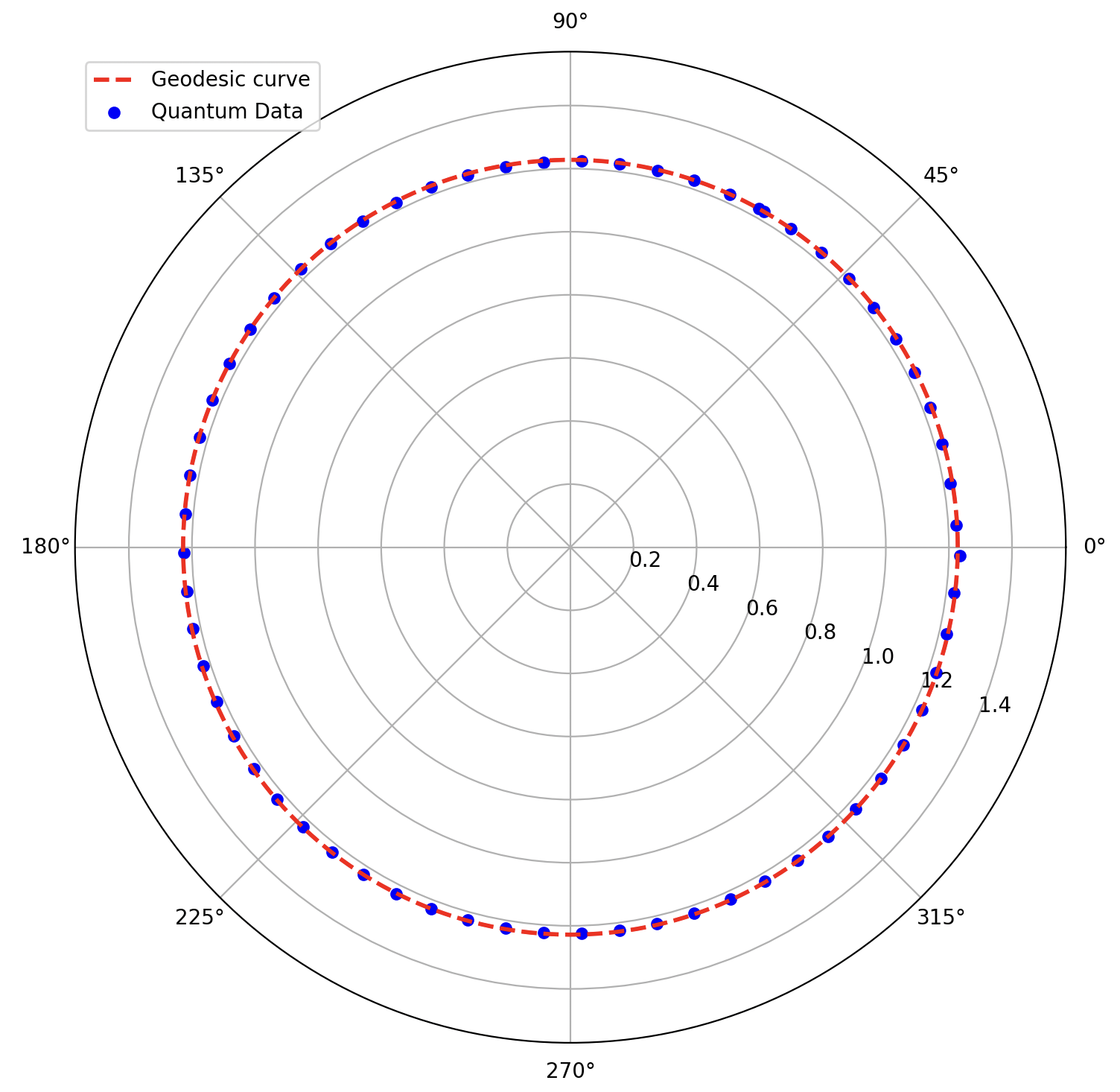}
    \end{minipage}\hfill
    \caption{Parameters: $\rho_{0}=1.20,\phi_{0}=\pi/3,\sigma_{1}=0.09,\sigma_{2}=0.05,n_{max}=m_{max}=350$,  $M=2.5, m_{0}=-30,n_{0}=0 , E\approx 34.3$, }
    \label{fig:fig70}
\end{figure}

\section{Flat Space Geodesics from Wave Packets}\label{appendix where we show that the definitions give straight line}
\subsection{Stress Tensor}\label{appendix_minkowski_com}
In this section, we shall show that the center of mass follows the geodesic equation in flat space (i.e straight line) for any choice of packet. In flat space (Minkowski metric), equation \eqref{definition of center of mass} becomes,
\bea\label{Flat space definition for center of mass}
\bar{x}^{\sigma} = \frac{\int dV x^{\sigma}\langle T^{tt}\rangle }{\int dV \langle T^{tt}\rangle }
\eea
As $\nabla_{\mu} \langle T^{\mu t}\rangle= \partial_{\mu} \langle T^{\mu t}\rangle=0$. Applying Stokes' theorem, we get:
\bea
\frac{d} {dt}\left (\int dV \langle T^{t t}\rangle\right)=0
\eea
So, let us label the denominator as $E$ which is a constant. We need to use one more identity. Multiplying both sides of the conservation equation by $x^{\sigma}$, applying chain rule and finally Stokes' theorem we get:
\bea
\int dV \langle T^{\sigma t}\rangle= \frac{d} {dt}\left (\int dV x^{\sigma} \langle T^{t t}\rangle\right)
\eea
We differentiate both sides of the equation \eqref{Flat space definition for center of mass} and use the above results to get,
\bea\label{Flat space differential form}
\frac{d \bar{x}^{\sigma}}{dt} = \frac{\int dV \langle T^{\sigma t}\rangle}{E}
\eea
Note that this equation is identity when $\sigma= t$. The numerator of the RHS is also a constant because of the conservation equation $\nabla_{\mu} \langle T^{\mu \sigma}\rangle= \partial_{\mu} \langle T^{\mu \sigma}\rangle=0$. So, the entire RHS is constant. Let us call the constant $\frac{\int dV \langle T^{i t}\rangle}{E}=\bar{v}^{i}$. So after integration we get,
\bea
\bar{x}^{i}(t)=\bar{x}^{i}(t_{0}) + \bar{v}^{i}(t-t_{0})
\eea
These are of course the geodesics in flat space. We get similar relation using the position operators in flat space as well, but the constant velocity we get will be different for both cases for the same packet choice. This will be evident when we write $\bar{v}^{i}$ in terms of packet profile in momentum space.

In flat space, a generic single particle state is given as:
\bea\label{Single particle state in flat space}
|\Psi(t) \rangle = \int d^{n}\mathbf{k}\hspace{0.5 mm} g(t,\mathbf{k})
  a^{\dagger} (\mathbf{k})|0\rangle
\eea
where,
\bea
g(t,\mathbf{k})=\tilde{g}(\mathbf{k}) e^{-i\omega(\mathbf{k})t}
\eea
where $\tilde{g}(\mathbf{k})$ is the packet profile in momentum space at t=0 and $\omega(\mathbf{k})=\sqrt{M^{2}+\mathbf{k}^{2}}$.
Let us define two operators $\hat{P}^{i}=\int dV \hat{T}^{it}$ and $\hat{E}=\int dV \hat{ T}^{t t}$. Each of these depend on the field expansion operator $\Phi(\mathbf{x},t)$.
Now, we take the expectation value of these two operators for the generic single particle state (equation \eqref{Single particle state in flat space}) and substitute in equation \eqref{Flat space differential form} to get,
\bea\label{flat space velocity}
\frac{d \bar{x}^{i}}{dt}=\bar{v}^{i}
= \frac{ \int d^{n}\boldsymbol{k}\, \boldsymbol{k}^{i}\tilde{g}(\boldsymbol{k})\tilde{g}^{*}(\boldsymbol{k})}
{ \int d^{n}\boldsymbol{k}\, \omega(\boldsymbol{k})\tilde{g}(\boldsymbol{k}) \tilde{g}^{*}(\boldsymbol{k})}
\eea
But the velocity in the operator formalism (we will see this in the next subsection) is \eqref{velocity for position operator flat space}
\bea
\langle \hat{\boldsymbol{v}}^{i}\rangle= {\int d^{n}\boldsymbol{k} \frac{\boldsymbol{k}^{i}}{\omega(\boldsymbol{k})}\tilde{g}(\boldsymbol{k}) \tilde{g}^{*}(\boldsymbol{k})}
\eea
which is clearly different from equation \eqref{flat space velocity}. This shows that the two formalisms are genuinely distinct: they do {\em not} produce identical auxiliary non-geometric quantities, even though they produce the same geodesics.

\subsection{Position Operator}\label{appendix_minkowski_pos_ope}
The field is of the following form,
\bea
   \phi (x) =  \int \frac{d^{n} \boldsymbol{k}}{\sqrt{(2 \pi)^{n} 2 \omega_{\boldsymbol{k} }}} \left ( {a}(\boldsymbol{k}) e^{-i kx} + 
  {a}^{\dagger} (\boldsymbol{k})e^{i kx}\right)
  \eea
We define two Fourier transformed quantities:
\bea\label{Fourier transform of ap in flat space}
  a(\boldsymbol{x}) = \frac{1}{\sqrt{(2 \pi)^{n}}} \int d^{n} \boldsymbol{k} \, a (\boldsymbol{k})\,e^{i \boldsymbol{k} \boldsymbol{x}} 
\eea
\bea\label{f in terms of g in flat space}
  f(t,\boldsymbol{x}) = \frac{1}{\sqrt{(2 \pi)^{n}}} \int d^{n} \boldsymbol{k} \, g(t,\boldsymbol{k})\,e^{i \boldsymbol{k} \boldsymbol{x}} 
\eea
where the creation and annihilation operators satisfy:
\bea
[{a}(\boldsymbol{k}),{a}^{\dagger}(\boldsymbol{k}')]=\delta(\boldsymbol{k}-\boldsymbol{k}'),
\eea
\bea
[{a}(\boldsymbol{k}),{a}(\boldsymbol{k}')]=0\hspace{1cm}[{a}^{\dagger}(\boldsymbol{k}),{a}^{\dagger}(\boldsymbol{k}')]=0
\eea
The infinite energy term $E_0$ is:  
\bea
 E_0= \frac{1}{2}\int d^3 \boldsymbol{k} \hspace{0.5mm}\delta^3(0) \hspace{0.5mm}\omega_{\boldsymbol{k}}
 ,\hspace{1cm}\omega_{\boldsymbol{k}} =\sqrt{\boldsymbol{k}^2+m^2}
\eea
So, the position operator takes the form:
\bea
   \hat{\boldsymbol{x}} = \int d^{n}\boldsymbol{x}  \, a^{\dagger} (\boldsymbol{x}) \boldsymbol{x} \, a(\boldsymbol{x})
\eea
This can be rewritten using equation \eqref{Fourier transform of ap in flat space} as,
\bea
  \hat{\boldsymbol{x}} = \int d^{n}\boldsymbol{p} \, a^{\dagger} (\boldsymbol{p}) i \nabla_{\boldsymbol{p}}\,( a(\boldsymbol{p})) 
\eea
Using this and equation \eqref{Single particle state in flat space} we get,
\bea\label{action of x hat in momentum basis}
  \langle \boldsymbol{p} |\hat{\boldsymbol{x}} |\Psi \rangle = i \nabla_{\boldsymbol{p}} g(t,\boldsymbol{p}) ,
\eea
The way we have written the field expansion, the basis ket is 
\bea
 |\boldsymbol{p} \rangle = a^{\dagger} (\boldsymbol{p}) |0\rangle
\eea
and hence,
\bea
g(t,\boldsymbol{p}) = \langle {\boldsymbol{p}}|\Psi \rangle
\eea
Using the above two equations in equation \eqref{action of x hat in momentum basis} we get,
\bea
\langle \Psi|\hat{\boldsymbol{x}}|\Psi\rangle=\int d^{n}\boldsymbol{p}\, g^{*}(t,\boldsymbol{p})\,i\nabla_{\boldsymbol{p}} \,\left( g(t,\boldsymbol{p})\right)
\eea
Finally using this equation,
\begin{align}
\langle\boldsymbol{x}\rangle_t&=
\langle \Psi|\hat{\boldsymbol{x}}|\Psi\rangle=
\int d^{n} \boldsymbol{p}\, \tilde{g}^{*}(\boldsymbol{p})e^{i( \omega_{\boldsymbol{p}} +E_{0}) t}i(-i g(t,\boldsymbol{p})(\nabla_{\boldsymbol{p}}\omega_{\boldsymbol{p}})t+ e^{-i( \omega_{\boldsymbol{p}} +E_0) t}(\nabla_{\boldsymbol{p}} \tilde{g}(\boldsymbol{p})))\nonumber\\
&=\int d^{n} \boldsymbol{p}\, \tilde{g}^{*}(\boldsymbol{p})e^{i( \omega_{\boldsymbol{p}} +E_0) t}( g(t,\boldsymbol{p})(\nabla_{\boldsymbol{p}}\omega_{\boldsymbol{p}})t+ie^{-i( \omega_{\boldsymbol{p}} +E_0) t}(\nabla_{\boldsymbol{p}} \tilde{g}(\boldsymbol{p}))\nonumber\\
&=\int d^{n} \boldsymbol{p}\, \tilde{g}^{*}(\boldsymbol{p})\tilde{g}(\boldsymbol{p})(\nabla_{\boldsymbol{p}}\omega_{\boldsymbol{p}}))t+\int d^{n} \boldsymbol{p}\,\tilde{g}^{*}(\boldsymbol{p})i(\nabla_{\boldsymbol{p}} \tilde{g}(\boldsymbol{p}))\nonumber\\
&=t\,\langle\Psi|\nabla_{\boldsymbol{p}}\, \hat{\omega}|\Psi\rangle_0+(\langle\Psi|\hat{\boldsymbol{x}}|\Psi\rangle)_0\nonumber\\
&=\langle \hat{\boldsymbol{v}}\rangle_0\,t +\langle \boldsymbol{x}\rangle_0
\end{align}
where the expectation value of velocity in a state $\tilde{g}(\boldsymbol{p})$ is
\bea\label{velocity for position operator flat space}
\langle \hat{\boldsymbol{v}}\rangle=\int d^{n} \boldsymbol{p}\, \tilde{g}^{*}(\boldsymbol{p})\tilde{g}(\boldsymbol{p})\frac{\boldsymbol{p}}{\sqrt{\boldsymbol{p}^{2}+M^{2}}}
\eea
Now let us go for the momentum,
\bea
  \hat{\boldsymbol{p}} = \int d^{n} \boldsymbol{p} \,a^{\dagger} (\boldsymbol{p}) \, \boldsymbol{p}\, a(\boldsymbol{p}) 
\eea
This satisfies,
\bea
\hat{\boldsymbol{p}}|\boldsymbol{k}\rangle=\boldsymbol{k}|\boldsymbol{k}\rangle
\eea
And from this,
\bea
  \langle \boldsymbol{p} |\hat{\boldsymbol{p}} |\Psi \rangle=\boldsymbol{p}\, \langle\boldsymbol{p}|\Psi\rangle = \boldsymbol{p} \,g(t,\boldsymbol{p})  ,
\eea
Likewise for the evolution of expectation value of momentum,
$$
\langle\boldsymbol{p}\rangle_{t}=
\langle \Psi|\hat{\boldsymbol{p}}|\Psi\rangle = \int d^{n} \boldsymbol{p} \langle\Psi|\boldsymbol{p} \rangle\langle\boldsymbol{p}|\hat{\boldsymbol{p}}|\Psi\rangle\\=\int d^{n}\boldsymbol{p}\,g^{*}(t,\boldsymbol{p})\boldsymbol{p} \,g(t,\boldsymbol{p})
$$
$$=\int d^{n}\boldsymbol{p}~ \tilde{g}^{*}(\boldsymbol{p})\boldsymbol{p} \,\tilde{g}(\boldsymbol{p})=\langle\boldsymbol{p}\rangle_{0}$$
So we have the evolution of expectation values as:
\bea
\langle\boldsymbol{p}\rangle_{t}=\langle\boldsymbol{p}\rangle_{0}
\eea
\bea\label{final form of flat space's expectation value of xhat}
\langle\boldsymbol{x}\rangle_{t}=\langle\boldsymbol{x}\rangle_{0}+\langle \nabla_{\boldsymbol{p}}\,\hat{\boldsymbol{\omega}}\rangle_{0}\,t=\langle\boldsymbol{x}\rangle_{0}+\langle \hat{\boldsymbol{v}}\rangle_{0}\,t
\eea
So we see that any normalizable and differentiable packet $g(t,\boldsymbol{p})$ follows a straight line trajectory.

\subsection{Flat Space in Plane Polar Coordinates}\label{appendi_flat_plane_polar}
Let us analyze what happens in flat space when we do similar things but now in plane polar coordinates. The metric is 
\bea
ds^{2}= dt^{2}-dr^{2}-r^{2}d\phi^{2}
\eea
The metric components are :
\bea
g_{tt}=1\,\,\,g_{rr}=-1\,\,\,g_{\phi\,\phi}=-r^{2}
\eea
So the non-vanishing Christoffel connections are:
\bea
\Gamma^{r}_{\phi\phi}=-r \,\,\,\,\Gamma^{\phi}_{r \phi}=\Gamma^{\phi}_{\phi r}=\frac{1}{r}
\eea
Hence the equations of motion are:
\begin{align}
&\ddot{r}-r\dot{\phi}^{2}=0\\
&\ddot{\phi}+\frac{2\dot{r}}{r}\dot{\phi}=0
\end{align}
The dots denote the derivatives w.r.t. the coordinate time $t$, not the proper time $\tau$.
Using this one can show that the equation of motion for $R=r^{2}$ becomes
\bea
\ddot{R}(t)=2v^{2}(t)
\eea
where $v^{2}(t)=\dot{r}^2(t)+(r(t)\dot{\phi}(t))^{2}$. Using $\epsilon=g_{\mu\nu}\frac{dx^{\mu}}{d\tau}\frac{dx^{\nu}}{d\tau}$ one can show that $v^{2}(t)$ is a constant along the geodesic. This means that the solution for $R(t)=r^{2}(t)$ is:
\bea\label{equation of motion for r^{2}}
r^{2}(t)= A\,+\,B\,t\,+\,C\,t^{2}\,
\eea
where $A,B,C$ are real parameters determined by the initial condition.
Like-wise consider the quantity $\mathcal{Z}(t)=r(t)e^{i\phi(t)}$. Using the equations of motion for $\ddot{r}$ and $\ddot{\phi}$, one can show:
\begin{align}\label{equation of motion for Z in pplane polar coordinates}
\frac{d^{2}\mathcal{Z}}{dt^2}&=0\nonumber\\
\implies \mathcal{Z}(t)&= \tilde{A}\,+\,\tilde{B}\,t
\end{align}
As before, $\tilde{A}$ and $\tilde{B}$ are determined by the initial condition. Note that these are in general complex numbers. Therefore, we need $4$ parameters to specify the motion. We see that $r^{2}(t)=\mathcal{Z}(t)\mathcal{Z}^{*}(t)$ where $\mathcal{Z}^{*}(t)$ is the complex conjugate of $\mathcal{Z}(t)$. Using this we see that $A=|\tilde{A}|^{2}$, $B=2\,\text{Re}(\tilde{A}\,\tilde{B}^{*})$ and $B=|\tilde{B}|^{2}$. So, this means the constants of $r^{2}(t)$ are related to the constants of $\mathcal{Z}(t)$ and $\mathcal{Z}(t)$ by itself is sufficient to describe the motion of the particle. One can easily show--by splitting Eq. \eqref{equation of motion for Z in pplane polar coordinates} 
into real and imaginary parts--that the coordinates $x(t)$ and $y(t)$ follow straight lines.

Now we will show that these equations (\eqref{equation of motion for r^{2}} and \eqref{equation of motion for Z in pplane polar coordinates}) hold true at the level of general expectation values for single particle states of a free scalar field. It is clear that the $\mathcal{Z}$ equation above turns into two real equations for the two Cartesian coordinates, and therefore one may suspect that they have a chance of holding. The non-trivial observation is that this is still true in {\em polar} coordinates: the technical reason is that the mode functions in plane polar coordinates are Bessel functions times $e^{i m \phi}$. When one acts with $r e^{i \phi}$ on them the angular part shifts to $e^{i(m+1)\phi}$ (another eigenmode) and the Bessel part can be re-written using a recurrence relation. 

We start with the KG equation:
\bea
\left(\square +  M^{2}\right)\Phi=0
\eea
Separation of variable gives the solution of the form $\Phi=R_{km}(r)\frac{e^{im\phi}}{\sqrt{2\pi}}e^{-i\omega(k)t}$ with $\omega(k)=\sqrt{k^{2}+M^{2}}$. The radial equation of motion is :
\bea
\frac{d^2\,R_{km}}{dr^2}+\frac{1}{r}\frac{d\,R_{km}}{dr}+\left(k^{2}-\frac{m^{2}}{r^{2}}\right)\,R_{km}=0
\eea
The solution to this ODE is basically Bessel function of first kind $J_{m}(kr)$ (see \cite{Arfken2012}) . But we write the solution as 
\bea
R_{mk}(r)= \sqrt{k}\,J_{m}(kr)
\eea
This change is made to get the desired commutation relation which we shall see shortly.
The solution is then:
\bea
\Phi(r, \phi, t) = \sum_{m=-\infty}^{\infty} \int_{0}^{\infty} dk \frac{1}{\sqrt{4\pi \omega(k)}} \left[ \sqrt{k}\, J_{m}(k\,r) \frac{e^{im\phi}}{\sqrt{2\pi}} e^{-i\omega(k)t} a_{km} + \text{h.c.} \right]
\eea
With this mode expansion, the Fourier transformed operator becomes:
\bea
a(r,\phi)=\sum_{m}\int\, dk\,\sqrt{k}\, J_{m}(k\,r)\frac{1}{\sqrt{2\pi}}e^{im\phi} a_{km}
\eea
We desire $[a_{km},a^{\dag}_{k'm'}]= \delta(k-k')\,\delta_{mm'}$ because this is the choice with which we can satisfy the canonical commutation relation for the $\Phi$ as above. With this the commutation relation for $a(r,\phi)$ becomes:
\begin{align}
&[a(r,\phi),a^{\dagger}(r',\phi')]= \frac{\delta(r-r^{'})\,\delta(\phi-\phi^{'})}{r}
\end{align}
Now the operator for $O(r,\phi)=O_{1}(r)O_{2}(\phi)$ becomes:
\bea\label{defining an oparator for function O(r,phi)}
\widehat{O(r,\phi)}=\int dr\,d\phi\,r\, a^{\dagger}(r,\phi)\,O(r,\phi)\,a(r,\phi)
\eea
Note that the above equation is linear in functions: 
\begin{equation}\label{linearity of the expectation value}
\begin{aligned}
\text{If } S(\rho, \phi) &= P_{1}(\rho, \phi) + P_{2}(\rho, \phi), \\
\text{then } \widehat{S(\rho, \phi)} &= \widehat{P_{1}(\rho, \phi)} + \widehat{P_{2}(\rho, \phi)}
\end{aligned}
\end{equation}
The time evolved generic single particle state (which we assume is normalizable) is:
\bea
|\Psi(t)\rangle=\sum_{m}\int dk\, g(k,m)\, e^{-i\omega(k)t} \,a^{\dagger}_{km}\, |0\rangle
\eea
Taking the expectation value of $\widehat{O(r,\phi)}$ we get:
\begin{align}
\langle \widehat{O(r,\phi)} \rangle_t &= \sum_{m_1, m_2} \int dk_1 dk_2 \, g^*(k_1, m_1) g(k_2, m_2) e^{i(\omega(k_1) - \omega(k_2))t}\times \nonumber \\
&\quad \times \left[ \frac{1}{2\pi} \int_0^{2\pi} d\phi \, e^{-i m_1 \phi} O_2(\phi) e^{i m_2 \phi} \right]\left[ \int_0^\infty dr \, r R_{k_1 m_1}^*(r) O_1(r) R_{k_2 m_2}(r) \right]
\end{align}
Let us first evaluate this for the complex position operator $\widehat{\mathcal{Z}} = \widehat{r e^{i\phi}}$, which tracks the straight-line classical geodesic. Here, $O_1(r) = r$ and $O_2(\phi) = e^{i\phi}$. The angular integral enforces the selection rule $\delta_{m_1, m_2+1}$. Setting $m_2 = m$ and $m_1 = m+1$, the radial integral becomes:
\bea
\mathcal{I}_{\mathcal{Z}} = \int_0^\infty dr \, r^2 \sqrt{k_1 k_2} J_{m+1}(k_1 r) J_m(k_2 r)
\eea
To evaluate this, let us derive an important identity. We start from the standard recurrence relations for Bessel functions of argument $x$, relating $J_m(x)$ and $J_{m+2}(x)$ to $J_{m+1}(x)$:
\begin{align}
J_m(x) + J_{m+2}(x) &= \frac{2(m+1)}{x} J_{m+1}(x) \nonumber \\
J_m(x) - J_{m+2}(x) &= 2 J_{m+1}'(x)
\end{align}
Adding these two relations and multiplying by $x/2$ yields:
\bea
x J_m(x) = (m+1) J_{m+1}(x) + x J_{m+1}'(x)
\eea
Now, setting $x = k_2 r$, the partial derivative with respect to $k_2$ is $\frac{\partial}{\partial k_2} J_{m+1}(k_2 r) = r J_{m+1}'(k_2 r)$. Substituting this back into our recurrence result, we can write it in terms of partial derivatives of $k_2$:
\begin{align}
k_2 r J_m(k_2 r) &= (m+1) J_{m+1}(k_2 r) + k_2 \frac{\partial}{\partial k_2} J_{m+1}(k_2 r) \nonumber \\
\implies r J_m(k_2 r) &= \left( \frac{\partial}{\partial k_2} + \frac{m+1}{k_2} \right) J_{m+1}(k_2 r)
\end{align}
To properly use this inside the radial integral $\mathcal{I}_{\mathcal{Z}}$, we must apply it to the normalized mode $R_{k_2 m}(r) = \sqrt{k_2} J_m(k_2 r)$. Using the product rule $\frac{\partial}{\partial k_2} [\sqrt{k_2} J_{m+1}(k_2 r)] = \frac{1}{2\sqrt{k_2}} J_{m+1}(k_2 r) + \sqrt{k_2} \frac{\partial}{\partial k_2} J_{m+1}(k_2 r)$, we rewrite our identity as:
\begin{align}
\sqrt{k_2} r J_m(k_2 r) &= \left( \frac{\partial}{\partial k_2} - \frac{1}{2k_2} + \frac{m+1}{k_2} \right) \left[ \sqrt{k_2} J_{m+1}(k_2 r) \right] \nonumber \\
&= \left( \frac{\partial}{\partial k_2} + \frac{m + 1/2}{k_2} \right) \left[ \sqrt{k_2} J_{m+1}(k_2 r) \right]
\end{align}
This neatly shifts the factor from $m+1$ to $m+1/2$. Substituting this into $\mathcal{I}_{\mathcal{Z}}$ and pulling the $k_2$ derivative outside the $r$ integration, we obtain:
\bea
\mathcal{I}_{\mathcal{Z}} = \left( \frac{\partial}{\partial k_2} + \frac{m + 1/2}{k_2} \right) \int_0^\infty dr \, r \left(\sqrt{k_1} J_{m+1}(k_1 r)\right) \left(\sqrt{k_2} J_{m+1}(k_2 r)\right)
\eea
Applying the orthogonality relation $\int r J_{m+1}(k_1 r) J_{m+1}(k_2 r) dr = \frac{1}{k_1} \delta(k_1 - k_2)$, the integral simply becomes $\delta(k_1 - k_2)$. Substituting this back into the full expectation value and integrating by parts with respect to $k_2$, the operator $( \partial_{k_2} + \frac{m+1/2}{k_2} )$ becomes $( -\partial_{k_2} + \frac{m+1/2}{k_2} )$ acting on the time-dependent state:
\begin{align}
\langle \widehat{\mathcal{Z}} \rangle_t &= \sum_{m} \int dk_1 dk_2 \, g^*(k_1, m+1) e^{i\omega(k_1)t} \delta(k_1 - k_2) \left[ -\frac{\partial}{\partial k_2} + \frac{m + 1/2}{k_2} \right] \left( g(k_2, m) e^{-i\omega(k_2)t} \right) \nonumber \\
&= \sum_m \int dk \, g^*(k, m+1) \left[ i t \frac{\partial \omega(k)}{\partial k} g(k, m) - \frac{\partial g(k,m)}{\partial k} + \frac{m + 1/2}{k} g(k,m) \right]
\end{align}
Grouping the time-independent terms as $A$, we see the expectation value perfectly separates into a constant term and a term that is linear in time:
\bea\label{quantum expectation value equation for Z(t) polar coordinates}
\langle \widehat{\mathcal{Z}} \rangle_t = \tilde{A} + \tilde{B}\,t
\eea
This is of the same form as its classical equation \eqref{equation of motion for Z in pplane polar coordinates}. 

We can use the above equation and the linearity condition  \eqref{linearity of the expectation value} to see that both $\langle\hat{x}\rangle(t)$ and $\langle\hat{y}\rangle(t)$ follow straight lines, where $\hat{x}(t)\equiv \widehat{r\cos{\phi}}(t)$ and $\hat{y}(t)\equiv \widehat{r\sin{\phi}}(t)$.  

Next, we evaluate the expectation value for the squared radial distance operator, $O(r,\phi) = r^2$. Here $O_1(r) = r^2$ and $O_2(\phi) = 1$. The angular integral enforces $m_1 = m_2 = m$. The radial integral becomes:
\bea
\mathcal{I}_{r^2} = \int_0^\infty dr \, r^3 R_{k_1 m}^*(r) R_{k_2 m}(r) = \int_0^\infty dr \, r \sqrt{k_1} J_m(k_1 r) \left[ r^2 \sqrt{k_2} J_m(k_2 r) \right]
\eea
Let us derive another important identity for the $r^2$ term, starting directly from the differential equation for the Bessel function $J_m(x)$:
\bea
x^2 J_m''(x) + x J_m'(x) + (x^2 - m^2) J_m(x) = 0
\eea
Setting $x = kr$, we use the partial derivatives $\frac{\partial}{\partial k} J_m(kr) = r J_m'(kr)$ and $\frac{\partial^2}{\partial k^2} J_m(kr) = r^2 J_m''(kr)$. Substituting these into the ODE yields:
\bea
k^2 r^2 J_m''(kr) + k r J_m'(kr) + (k^2 r^2 - m^2) J_m(kr) = 0 \nonumber \\
\implies r^2 J_m(kr) = \left[ -\frac{\partial^2}{\partial k^2} - \frac{1}{k} \frac{\partial}{\partial k} + \frac{m^2}{k^2} \right] J_m(kr)
\eea
We must evaluate the action of this operator on the normalized radial mode $R_{km}(r) = \sqrt{k} J_m(kr)$. Let $F(k) = \sqrt{k} J_m(kr)$. Operating on $k^{-1/2} F(k)$, we find:
\begin{align}
\left[ -\frac{\partial^2}{\partial k^2} - \frac{1}{k} \frac{\partial}{\partial k} + \frac{m^2}{k^2} \right] \left( k^{-1/2} F \right) &= - \frac{\partial}{\partial k} \left( k^{-1/2} F' - \frac{1}{2} k^{-3/2} F \right)- \nonumber \\
&- \frac{1}{k} \left( k^{-1/2} F' - \frac{1}{2} k^{-3/2} F \right) + \frac{m^2}{k^2} k^{-1/2} F \nonumber \\
&= k^{-1/2} \left[ -F'' + \frac{m^2 - 1/4}{k^2} F \right]
\end{align}
Multiplying by $\sqrt{k}$, the first derivative terms beautifully cancel out, leaving a remarkably simple, manifestly Hermitian operator:
\bea
r^2 R_{km}(r) = \left[ -\frac{\partial^2}{\partial k^2} + \frac{m^2 - 1/4}{k^2} \right] R_{km}(r) \equiv \widetilde{\mathcal{D}}_k R_{km}(r)
\eea
This resolves any ambiguities with integration by parts, as $\widetilde{\mathcal{D}}_k$ is self-adjoint. The radial integral becomes:
\bea
\mathcal{I}_{r^2} = \int_0^\infty dr \, r R_{k_1 m}(r) \widetilde{\mathcal{D}}_{k_2} R_{k_2 m}(r) = \widetilde{\mathcal{D}}_{k_2} \delta(k_1 - k_2)
\eea
Substituting this into the expectation value and integrating by parts twice with respect to $k_2$ transfers $\widetilde{\mathcal{D}}_{k_2}$ onto the state $g(k_2, m) e^{-i\omega(k_2)t}$:
\begin{align}
\langle \widehat{r^2} \rangle_t &= \sum_m \int dk \, g^*(k, m) e^{i\omega(k)t} \left[ -\frac{\partial^2}{\partial k^2} + \frac{m^2 - 1/4}{k^2} \right] \left( g(k,m) e^{-i\omega(k)t} \right)
\end{align}
The second-order derivative acts on the time-dependent phase twice, pulling down both linear and quadratic terms in time.

Grouping the terms by powers of $t$, we obtain $\langle \widehat{r^2} \rangle_t$ to be of the form:
\bea
\langle \widehat{r^2} \rangle_t = A\,+\,B\,t\,+\,C\,t^{2}
\eea
This equation also has the same form as its classical equation \eqref{equation of motion for r^{2}}. Note that these constants $A,\,B,\,C$ are not related to the constants $\tilde{A}$ and $\tilde{B}$ in the quantum equation \eqref{quantum expectation value equation for Z(t) polar coordinates} the same way they relate classically. This is because $\langle \widehat{\mathcal{Z}} \widehat{\mathcal{Z}}^{*} \rangle_t \neq \langle \widehat{\mathcal{Z}} \rangle_t \langle \widehat{\mathcal{Z}}^{*} \rangle_t$ (see Eq. \eqref{defining an oparator for function O(r,phi)}).
This is not a contradiction because in the expectation value formulas, the constants $A, B, C, \tilde A$ and $\tilde B$ are state-dependent quantities.

\section{Quantum-Classical Correspondence: Stress Tensor Operator}\label{appendix_Q-C_corresp_com}

Using \eqref{the final form of xbar}, the stress-tensor centroids of $\cos(2\rho)$ and
$\sin(\rho)e^{i\phi}$ in AdS$_3$ are
\begin{align}\label{centroid of cos2rho and sinrhoexpiphi}
\overline{\cos(2\rho)}_{t}
&=
\frac{\int d\rho\,d\phi\,\cos(2\rho)\,\langle\mathcal{T}^{tt}\rangle \sec^{2}\rho}
{\int d\rho\,d\phi\,\langle\mathcal{T}^{tt}\rangle \sec^{2}\rho},
\nonumber\\[2mm]
\overline{\sin(\rho)e^{i\phi}}_{t}
&=
\frac{\int d\rho\,d\phi\,\sin(\rho)e^{i\phi}\,\langle\mathcal{T}^{tt}\rangle \sec^{2}\rho}
{\int d\rho\,d\phi\,\langle\mathcal{T}^{tt}\rangle \sec^{2}\rho}.
\end{align}
Here $\mathcal{T}^{tt}=\sqrt{g}\,T^{tt}$ and $\sqrt{g}=\sin\rho/\cos^{3}\rho$.
Using \eqref{expression for T^tt in terms of A and gnm}, and dropping the state-independent
constant term, we have
\begin{equation}
\langle \hat T^{tt}\rangle
=
\sum_{n_1,m_1,n_2,m_2}
\frac{A(n_1,m_1;n_2,m_2,\rho)}
{2\pi\sqrt{\omega_{n_1m_1}\omega_{n_2m_2}}}
\,g^*(n_2,m_2)g(n_1,m_1)\,
e^{-i(\omega_{n_1m_1}-\omega_{n_2m_2})t}e^{i(m_1-m_2)\phi},
\end{equation}
with
\begin{align}
A(n_1,m_1;n_2,m_2,\rho)=\frac{1}{2}\cos^4\rho
\Biggl[
&R_{n_1m_1}R_{n_2m_2}\,\omega_{n_1m_1}\omega_{n_2m_2}
+\partial_\rho R_{n_1m_1}\,\partial_\rho R_{n_2m_2}
\nonumber\\
&\qquad\qquad
+\left(\frac{m_1m_2}{\sin^2\rho}+\frac{M^2}{\cos^2\rho}\right)
R_{n_1m_1}R_{n_2m_2}
\Biggr].
\end{align}
Since the denominator in \eqref{centroid of cos2rho and sinrhoexpiphi} is just the conserved
energy $E$, the numerator for a general insertion $O_1(\rho)O_2(\phi)$ can be written as
\begin{align}
N(t)
=
\sum_{n_1,m_1,n_2,m_2}
\frac{g^*(n_2,m_2)g(n_1,m_1)\,e^{-i(\omega_{n_1m_1}-\omega_{n_2m_2})t}}
{\sqrt{\omega_{n_1m_1}\omega_{n_2m_2}}}
\left[\int d\rho\,O_1(\rho)\frac{\sin\rho}{\cos^5\rho}A\right] \times \hspace{0.5in} \nonumber \\
\hspace{3.5in} \times \left[\frac{1}{2\pi}\int d\phi\,e^{i(m_1-m_2)\phi}O_2(\phi)\right].
\end{align}

We first take $O_1(\rho)=\cos(2\rho)$ and $O_2(\phi)=1$. The $\phi$ integral gives
$\delta_{m_1m_2}$, so after summing over $m_2$ the radial integral becomes
\begin{equation}
I(n_1,n_2,m_1)=I_1+I_2+I_3,
\end{equation}
where
\begin{align}
I_1(n_1,n_2,m_1)
&=
\frac{\omega_{n_1m_1}\omega_{n_2m_1}}{2}
\int d\rho\,\cos(2\rho)\tan\rho\,R_{n_1m_1}R_{n_2m_1},
\\
I_2(n_1,n_2,m_1)
&=
\frac{1}{2}\int d\rho\,\cos(2\rho)\tan\rho\,
\partial_\rho R_{n_1m_1}\partial_\rho R_{n_2m_1},
\\
I_3(n_1,n_2,m_1)
&=
\frac{1}{2}\int d\rho\,\cos(2\rho)\tan\rho
\left(\frac{m_1^2}{\sin^2\rho}+\frac{M^2}{\cos^2\rho}\right)
R_{n_1m_1}R_{n_2m_1}.
\end{align}
The first term is just the overlap $X^{m_1}_{n_2,n_1}$ of \eqref{X}, up to the harmless
$\omega$-factors, so it is non-zero only for $n_2=n_1,n_1\pm1$.
The nontrivial point is that $I_2+I_3$ has exactly the same support; we show this in the
next subsection. Therefore $I(n_1,n_2,m_1)$ also vanishes unless $n_2=n_1,n_1\pm1$.

Since $\omega_{n+1,m}-\omega_{n,m}=2$, the surviving terms in
$\overline{\cos(2\rho)}_t$ contain only frequencies $0,\pm2$, and hence
\begin{equation}
\overline{\cos(2\rho)}_{t}
=
A + C e^{2it} + C^* e^{-2it}
=
B + A' \cos(2t+\delta).
\end{equation}
Equivalently,
\bea \label{exact centroid of cos2rho}
\overline{\cos(2\rho)}_{t}= A + B\cos(2t+\delta),
\eea
which is of the same form as the classical relation \eqref{exact relation 1}. As before, this
can be rewritten as
\bea \label{the final rhobar with the variance like corretion}
\bar\rho_t
=
\frac{1}{2}\cos^{-1}\!\Big(A+B\cos(2t+\delta)+\tilde\Delta_1(t)\Big),
\eea
where
\bea
\tilde\Delta_1(t)\equiv \cos(2\bar\rho_t)-\overline{\cos(2\rho)}_t.
\eea
Thus $\bar\rho_t$ obeys the classical equation up to the variance-like correction
$\tilde\Delta_1(t)$.

Next take $O_1(\rho)=\sin\rho$ and $O_2(\phi)=e^{i\phi}$. The angular integral gives
\bea
\frac{1}{2\pi}\int d\phi\,e^{i(m_1+1-m_2)\phi}=\delta_{m_2,m_1+1},
\eea
so the radial integral becomes
\bea
\tilde I(n_1,n_2,m_1)=\tilde I_1+\tilde I_2+\tilde I_3,
\eea
with
\begin{align}
\tilde I_1(n_1,n_2,m_1)
&=
\frac{\omega_{n_1m_1}\omega_{n_2,m_1+1}}{2}
\int d\rho\,\sin\rho\,\tan\rho\,R_{n_1m_1}R_{n_2,m_1+1},
\\
\tilde I_2(n_1,n_2,m_1)
&=
\frac{1}{2}\int d\rho\,\sin\rho\,\tan\rho\,
\partial_\rho R_{n_1m_1}\partial_\rho R_{n_2,m_1+1},
\\
\tilde I_3(n_1,n_2,m_1)
&=
\frac{1}{2}\int d\rho\,\sin\rho\,\tan\rho
\left(\frac{m_1(m_1+1)}{\sin^2\rho}+\frac{M^2}{\cos^2\rho}\right)
R_{n_1m_1}R_{n_2,m_1+1}.
\end{align}
Again, $\tilde I_1$ is the same overlap as $Y^{m_1}_{n_2,n_1}$ in \eqref{Y}, up to
$\omega$-factors, so it obeys the same selection rules. The combination
$\tilde I_2+\tilde I_3$ is shown in subsection I.2 to be proportional to $\tilde I_1$, and
therefore has the same support.

Using the equally spaced spectrum once more, the surviving terms carry only frequencies
$\pm1$, so we obtain
\bea\label{exact centroid of sinrhoexpiphi}
\overline{\sin(\rho)e^{i\phi}}_{t}=A\,e^{it}+B\,e^{-it},
\eea
in precise agreement with the classical equation for $\sin\rho\,e^{i\phi}$.
Hence
\bea\label{the final argexpiphi with the variance like corretion}
\arg\!\left(\overline{e^{i\phi}}_{t}\right)
=
\arg\!\left(A\,e^{it}+B\,e^{-it}+\tilde\Delta_2(t)\right),
\eea
where
\bea
\tilde\Delta_2(t)
\equiv
\overline{\sin(\rho)}_t\,\overline{e^{i\phi}}_t-\overline{\sin(\rho)e^{i\phi}}_t.
\eea

\subsection{Exact Equations}

The key input is the radial equation
\bea\label{diff equation}
-\frac{d}{d\rho}\Big[\tan\rho\,\frac{dR_{nm}}{d\rho}\Big]
+\tan\rho\Big[\frac{m^2}{\sin^2\rho}+\frac{M^2}{\cos^2\rho}\Big]R_{nm}
=
\omega_{nm}^2\tan\rho\,R_{nm}.
\eea
Equivalently,
\begin{equation}\label{substituting the radial solution}
\partial_\rho\!\left(\tan\rho\,\partial_\rho R_{nm}\right)
=
\tan\rho\left(\frac{m^2}{\sin^2\rho}+\frac{M^2}{\cos^2\rho}-\omega_{nm}^2\right)R_{nm}.
\end{equation}

Starting from $I_2$, integrating by parts, and then using
\eqref{substituting the radial solution} for $R_{n_1m_1}$ gives
\begin{equation}
I_2+I_3
=
\int d\rho\,\tan\rho\,R_{n_2m_1}
\left[
\frac{\omega_{n_1m_1}^2}{2}\cos(2\rho)\,R_{n_1m_1}
+\sin(2\rho)\,\partial_\rho R_{n_1m_1}
\right].
\end{equation}
Repeating the same step with $n_1\leftrightarrow n_2$ and averaging the two expressions,
\begin{align}
I_2+I_3
&=
\frac{\omega_{n_1m_1}^2+\omega_{n_2m_1}^2}{4}
\int d\rho\,\tan\rho\,\cos(2\rho)\,R_{n_1m_1}R_{n_2m_1}
\nonumber\\
&\qquad
+\frac{1}{2}\int d\rho\,\tan\rho\,\sin(2\rho)\,
\partial_\rho\!\left(R_{n_1m_1}R_{n_2m_1}\right).
\end{align}
Now use $\tan\rho\,\sin(2\rho)=1-\cos(2\rho)$ and integrate by parts once more:
\begin{align}
\frac{1}{2}\int d\rho\,(1-\cos2\rho)\,\partial_\rho(R_{n_1m_1}R_{n_2m_1})
&=
-\int d\rho\,\sin(2\rho)\,R_{n_1m_1}R_{n_2m_1}
\nonumber\\
&=
-\int d\rho\,\tan\rho\,(1+\cos2\rho)\,R_{n_1m_1}R_{n_2m_1}.
\end{align}
Therefore
\begin{align}
I_2+I_3
=
\left(\frac{\omega_{n_1m_1}^2+\omega_{n_2m_1}^2}{4}-1\right)
\int d\rho\,\tan\rho\,\cos(2\rho)\,R_{n_1m_1}R_{n_2m_1}
-
\int d\rho\,\tan\rho\,R_{n_1m_1}R_{n_2m_1}.
\end{align}
The second term is proportional to $\delta_{n_1n_2}$ by orthogonality, while the first term
is again the $X$-type overlap and therefore vanishes unless $n_2=n_1,n_1\pm1$.
Hence $I_2+I_3$ has the same support as $I_1$, which is the basic reason for our observation about $\overline{\cos(2\rho)}_t$.

The derivation for $\overline{\sin(\rho)e^{i\phi}}_t$ is parallel. Let
\bea
R_1\equiv R_{n_1m_1}(\rho),\qquad
R_2\equiv R_{n_2,m_1+1}(\rho),\qquad
\omega_1\equiv \omega_{n_1m_1},\qquad
\omega_2\equiv \omega_{n_2,m_1+1}.
\eea
Starting from $\tilde I_2+\tilde I_3$, integrate by parts once and use
\eqref{substituting the radial solution} for $R_2$:
\begin{align}
\tilde I_2+\tilde I_3
&=
\frac{\omega_2^2}{\omega_1\omega_2}\,\tilde I_1
+\frac{1}{2}\int d\rho
\left[
-\frac{m_1+1}{\cos\rho}\,R_1R_2
-\sin\rho\,R_1\partial_\rho R_2
\right].
\end{align}
Doing the same with the derivative moved the other way gives
\begin{equation}
\tilde I_2+\tilde I_3
=
\frac{\omega_1^2}{\omega_1\omega_2}\,\tilde I_1
+\frac{1}{2}\int d\rho
\left[
\frac{m_1}{\cos\rho}\,R_1R_2
-\sin\rho\,R_2\partial_\rho R_1
\right].
\end{equation}
Adding the two forms eliminates the cross-derivative terms and yields
\begin{equation}
\tilde I_2+\tilde I_3
=
\left(\frac{\omega_1^2+\omega_2^2-1}{2\omega_1\omega_2}\right)\tilde I_1.
\end{equation}
Hence the full integral is simply
\begin{equation}
\tilde I
=
\tilde I_1+\tilde I_2+\tilde I_3
=
\left(\frac{(\omega_1+\omega_2)^2-1}{2\omega_1\omega_2}\right)\tilde I_1.
\end{equation}
So $\tilde I$ is proportional to the same $Y$-type overlap that already appeared in the
position-operator analysis, and therefore inherits exactly the same selection rules. This is
why only the frequencies $\pm1$ survive in
\eqref{exact centroid of sinrhoexpiphi}.

Finally, recall the equations
\eqref{when you substitute i=rho in the final equation} and
\eqref{when you substitute i=phi in the final equation}, which contain the error terms on
their right-hand sides. For the free scalar in AdS$_3$, the net effect of those terms is simply
to reorganize into the variance-like corrections $\tilde\Delta_1(t)$ and $\tilde\Delta_2(t)$
appearing in \eqref{the final rhobar with the variance like corretion} and
\eqref{the final argexpiphi with the variance like corretion}.

\section{Origin of the Selection Rules}\label{appendix_sec:symmetry_origin}

In this Appendix, we trace the origin of the selection rules noticed in \ref{appendix_Q-C_corresp_com} to the algebraic structure of the radial mode functions as orthogonal polynomials and the linearity of AdS modes. We also contrast the situation with flat Minkowski space in 2+1 dimensions.

\subsection{Jacobi Polynomial: Three-term Recurrence}\label{sec:jacobi_recurrence}

The selection rules for the overlap integrals have a direct algebraic origin: the radial
AdS$_3$ mode functions are Jacobi polynomials in the variable
\bea
u \equiv \sin^2\rho, \qquad x \equiv 1-2u=\cos(2\rho).
\eea
Recalling \eqref{Form of Rnmrho that is used}, we may write
\bea
R_{nm}(\rho)
=\frac{1}{\mathcal N_{nm}}(\sin\rho)^{|m|}(\cos\rho)^{\Delta}\,
{}_2F_1(-n,\Delta+|m|+n,|m|+1,\sin^2\rho).
\eea
Since the first argument of the hypergeometric function is $-n$, this is a polynomial of
degree $n$ in $u$. More precisely,
\bea\label{Jacobi identification}
{}_2F_1(-n,\Delta+|m|+n,|m|+1,u)\propto
P_n^{(|m|,\Delta-1)}(1-2u),
\eea
so that, up to normalization,
\bea
R_{nm}(\rho)\propto
u^{|m|/2}(1-u)^{\Delta/2}\,
P_n^{(|m|,\Delta-1)}(1-2u).
\eea
Thus the radial Hilbert space at fixed $m$ is organized by Jacobi polynomials with
parameters
\bea
\alpha=|m|,\qquad \beta=\Delta-1.
\eea

\paragraph{The \texorpdfstring{$X^m_{n_2,n_1}$}{X} overlap.}
Consider first the overlap appearing in the $\widehat{\cos(2\rho)}$ analysis:
\bea
X^{m}_{n_2,n_1}=\int d\rho\,\tan\rho\,\cos(2\rho)\,R_{n_1m}(\rho)R_{n_2m}(\rho).
\eea
Using $d\rho\,\tan\rho=\frac{du}{2(1-u)}$, this becomes
\bea
X^{m}_{n_2,n_1}\propto
\int_0^1 du\;
u^{\alpha}(1-u)^{\beta}\,x\,
P_{n_1}^{(\alpha,\beta)}(x)\,
P_{n_2}^{(\alpha,\beta)}(x),
\qquad x=1-2u.
\eea
Now Jacobi polynomials satisfy the standard three-term recurrence relation
\bea\label{three-term recurrence}
x\,P_n^{(\alpha,\beta)}(x)
=
A_n\,P_{n+1}^{(\alpha,\beta)}(x)
+
B_n\,P_n^{(\alpha,\beta)}(x)
+
C_n\,P_{n-1}^{(\alpha,\beta)}(x)
\eea
for suitable coefficients $A_n,B_n,C_n$. Therefore the factor of $x=\cos(2\rho)$ maps
$P_{n_1}^{(\alpha,\beta)}$ into a linear combination involving only
$n_1+1$, $n_1$, and $n_1-1$. Orthogonality with respect to the Jacobi weight
$u^\alpha(1-u)^\beta$ then implies
\bea
X^m_{n_2,n_1}=0
\qquad \text{unless} \qquad
n_2\in\{n_1-1,n_1,n_1+1\}.
\eea
This is precisely the selection rule used in the discussion of
$\langle \widehat{\cos(2\rho)}\rangle_t$.

\paragraph{The \texorpdfstring{$Y^{m_1}_{n_2,n_1}$}{Y} overlap.}
The overlap relevant for $\widehat{\mathcal Z}=\widehat{\sin\rho\,e^{i\phi}}$ is
\bea
Y^{m_1}_{n_2,n_1}
=
\int d\rho\,\tan\rho\,\sin\rho\,
R_{n_1m_1}(\rho)\,R_{n_2,m_1+1}(\rho).
\eea
The angular factor $e^{i\phi}$ gives the Fourier selection rule
$m_2=m_1+1$. The radial analysis is slightly subtler than for $X$, because the two mode
functions now belong to \emph{different} Jacobi families: the parameter $\alpha=|m|$ shifts
by one unit. Thus the three-term recurrence by itself is not enough; one also needs the
standard one-step \emph{connection relations} between Jacobi polynomials with neighboring
values of $\alpha$.

When $m_1\ge 0$, we have $|m_1+1|=m_1+1$, and using
$d\rho\,\tan\rho\,\sin\rho=\frac12 du\,u^{1/2}(1-u)^{-1}$ one finds
\bea
Y^{m_1}_{n_2,n_1}
\propto
\int_0^1 du\;
u^{m_1+1}(1-u)^{\beta}\,
P_{n_1}^{(m_1,\beta)}(x)\,
P_{n_2}^{(m_1+1,\beta)}(x),
\qquad x=1-2u.
\eea
Now one uses the one-step connection relation
\bea
P_n^{(\alpha,\beta)}(x)
=
a_n\,P_n^{(\alpha+1,\beta)}(x)
+
b_n\,P_{n-1}^{(\alpha+1,\beta)}(x),
\eea
for suitable coefficients $a_n,b_n$. Rewriting
$P_{n_1}^{(m_1,\beta)}$ in the $(m_1+1,\beta)$ family and then using orthogonality with
weight $u^{m_1+1}(1-u)^\beta$ gives
\bea
Y^{m_1}_{n_2,n_1}=0
\qquad \text{unless} \qquad
n_2=n_1 \quad \text{or} \quad n_2=n_1-1,
\qquad (m_1\ge 0).
\eea

When $m_1<0$, write
\bea
m_1=-(\ell+1),\qquad \ell\ge 0,
\eea
so that $m_1+1=-\ell$, $|m_1|=\ell+1$, and $|m_1+1|=\ell$. Then
\bea
Y^{m_1}_{n_2,n_1}
\propto
\int_0^1 du\;
u^{\ell+1}(1-u)^{\beta}\,
P_{n_1}^{(\ell+1,\beta)}(x)\,
P_{n_2}^{(\ell,\beta)}(x).
\eea
In this case one uses the inverse one-step connection relation
\bea
P_n^{(\alpha-1,\beta)}(x)
=
c_n\,P_n^{(\alpha,\beta)}(x)
+
d_n\,P_{n-1}^{(\alpha,\beta)}(x),
\eea
for suitable coefficients $c_n,d_n$. Rewriting
$P_{n_2}^{(\ell,\beta)}$ in the $(\ell+1,\beta)$ family and then using orthogonality yields
\bea
Y^{m_1}_{n_2,n_1}=0
\qquad \text{unless} \qquad
n_2=n_1 \quad \text{or} \quad n_2=n_1+1,
\qquad (m_1<0).
\eea

So the overlap $Y^{m_1}_{n_2,n_1}$ obeys the weaker bound
$|n_1-n_2|\le 1$, but in fact its support is more restrictive and depends on the sign of
$m_1$:
\bea
m_1\ge 0:\qquad n_2=n_1 \ \text{or}\  n_1-1,
\eea
\bea
m_1<0:\qquad n_2=n_1 \ \text{or}\  n_1+1.
\eea
This sign-dependent pattern is exactly what was used implicitly in the discussion of
$\langle \widehat{\mathcal Z}\rangle_t$ in Section \ref{sec:Q-C_corres} and Appendix
\ref{appendix_Q-C_corresp_com}.

The basic mechanism is therefore clear. For $X$, multiplication by
$\cos(2\rho)=x$ stays within a single Jacobi family and the three-term recurrence gives the
selection rule directly. For $Y$, the factor $\sin\rho\,e^{i\phi}$ changes the angular
quantum number by one unit, so one must combine orthogonality with a one-step Jacobi
parameter shift. In both cases the result is finite support in neighboring radial levels.

\subsection{The Role of the Equally Spaced Spectrum}\label{sec:equally_spaced}

The selection rules by themselves are not enough to guarantee the exactness of
\eqref{exact relation 1} and \eqref{exact relation 2}. The second crucial ingredient is the
equally spaced AdS$_3$ spectrum
\bea
\omega_{nm}=\Delta+2n+|m|.
\eea

For $\langle \widehat{\cos(2\rho)}\rangle_t$, the angular selection rule enforces
$m_2=m_1$, while the $X$-overlap restricts $n_2$ to $n_1$, $n_1\pm1$. Therefore
\bea
\omega_{n_2m}-\omega_{n_1m}=2(n_2-n_1),
\eea
so the only non-constant frequencies that can appear are $\pm 2$. The diagonal terms
$n_2=n_1$ give the constant piece, while the off-diagonal terms $n_2=n_1\pm1$ all
oscillate with the same frequency. This is why the full answer collapses exactly to
\bea
\langle \widehat{\cos(2\rho)}\rangle_t = A + B\cos(2t+\delta).
\eea

For $\langle \widehat{\mathcal Z}\rangle_t$, the relevant frequency difference is
\bea
\Delta\omega
\equiv
\omega_{n_2,m_1+1}-\omega_{n_1,m_1}
=
2(n_2-n_1)+|m_1+1|-|m_1|.
\eea
Now the sign-dependent support derived above becomes essential.

If $m_1\ge 0$, then $|m_1+1|-|m_1|=+1$, while the allowed values are
$n_2=n_1$ or $n_2=n_1-1$. Hence
\bea
\Delta\omega=+1 \quad \text{or} \quad -1.
\eea

If $m_1<0$, then $|m_1+1|-|m_1|=-1$, while the allowed values are
$n_2=n_1$ or $n_2=n_1+1$. Again,
\bea
\Delta\omega=-1 \quad \text{or} \quad +1.
\eea

So every surviving term in $\langle \widehat{\mathcal Z}\rangle_t$ carries frequency
either $+1$ or $-1$, independent of $n_1$ and $m_1$. Therefore
\bea
\langle \widehat{\mathcal Z}\rangle_t = A\,e^{it}+B\,e^{-it}
\eea
exactly.

The Jacobi structure gives finite support in
neighboring radial levels, and the equally spaced AdS$_3$ spectrum makes all of the surviving
terms oscillate with the same frequency. Without the equal spacing, the same finite-support
selection rules would still leave an $n$-dependent set of frequencies, and the result would
generically exhibit beating or quasi-periodic motion rather than a single harmonic.

Equivalently, from the representation-theoretic point of view, the one-particle Hilbert
space is built from highest-weight modules of
$\mathfrak{sl}(2,\mathbb R)_L\oplus \mathfrak{sl}(2,\mathbb R)_R$, and the compact global
time generator has evenly spaced eigenvalues. The exact quantum equations are therefore a
consequence of \emph{both} facts: Jacobi-polynomial selection rules and the linear AdS$_3$
spectrum.

\subsection{Comparison with $2+1$ Flat Minkowski Space}

It is useful to compare the AdS$_3$ discussion with $2+1$ dimensional flat Minkowski
space. The exact AdS$_3$ relations of this appendix rely on a particular combination of
selection rules and spectrum, but flat space also has exact one-particle statements, realized
in a different way.

In flat $2+1$ Minkowski space with polar coordinates, the metric is
\begin{equation}
ds^2 = dt^2-dr^2-r^2 d\phi^2 .
\end{equation}
The natural analog of $Z=\sin\rho\,e^{i\phi}$ is
\begin{equation}
Z_{\rm flat}=r e^{i\phi}=x+iy .
\end{equation}
Classically, a free particle satisfies $\ddot x=\ddot y=0$, and hence
\begin{equation}
\frac{d^2 Z_{\rm flat}}{dt^2}=0 .
\end{equation}
Thus $Z_{\rm flat}(t)=Z_0+Vt$ simply describes straight-line geodesic motion. At the quantum
level, the same statement is immediate in Cartesian coordinates:
\begin{equation}
\hat x(t)=\hat x(0)+\hat v_x\, t,\qquad
\hat y(t)=\hat y(0)+\hat v_y\, t ,
\end{equation}
so that
\begin{equation}
\hat Z_{\rm flat}(t)=\hat Z_{\rm flat}(0)+\hat V\, t,
\qquad
\frac{d^2}{dt^2}\langle \hat Z_{\rm flat}\rangle_t=0
\end{equation}
for any normalizable single-particle state.

If one instead works in polar coordinates, the mode expansion involves Bessel functions,
\begin{equation}
\Phi \sim e^{-i\omega(k)t} e^{im\phi} J_m(kr),\qquad
\omega(k)=\sqrt{k^2+M^2},
\end{equation}
so the radial label $k$ is continuous. As discussed in Appendix \ref{appendi_flat_plane_polar},
the angular part of $\hat Z_{\rm flat}=\widehat{r e^{i\phi}}$ still imposes the simple selection
rule $m\to m+1$. The radial action, however, is represented through Bessel recursion
relations as
\begin{equation}
r\,R_{km}(r)
=
\left(\partial_k+\frac{m+\tfrac12}{k}\right) R_{k,m+1}(r),
\end{equation}
with $R_{km}(r)=\sqrt{k}\,J_m(kr)$, and similarly
\begin{equation}
r^2 R_{km}(r)
=
\left[-\partial_k^2+\frac{m^2-\tfrac14}{k^2}\right] R_{km}(r).
\end{equation}
So in the polar basis the relevant operators act on a continuous Bessel label rather than by
finite-band recurrences on a discrete Jacobi basis.

One may also ask whether a more AdS-like structure emerges if flat space is placed in a
finite circular cavity so that $k$ becomes discrete. Even then, however, the frequencies
\begin{equation}
\omega_{nm}=\sqrt{k_{nm}^2+M^2}
\end{equation}
are not equally spaced in $n$. Therefore the time dependence does not collapse in the same
way as in AdS$_3$.

To summarize: In flat space, exact geodesic motion is most
transparent in Cartesian coordinates, where free motion is linear. In AdS$_3$, the exact
relations derived in this appendix are naturally tied to the discrete normal-mode basis and
to the coexistence of finite-band selection rules with an equally spaced spectrum. We will
not attempt here to relate the two structures; the point of
the comparison is only to note that both settings admit exact statements.

\section{Geodesic from the Eikonal Limit of the Scalar Field}\label{app:eikonal}

In this appendix, we show how the classical geodesic equation emerges from the equation of motion of a massive scalar field in curved spacetime via the eikonal/WKB approximation. This is a claim that is more often heard than derived, so we present it explicitly -- we follow \cite{Post} who gives an analogous derivation for the (massless) Maxwell field. 

Consider a real scalar field $\phi$ of mass $M$, governed by the Lagrangian
\bea
\mathcal{L} = \frac{1}{2}\,\partial_{\mu}\phi\,\partial^{\mu}\phi - \frac{1}{2}\,M^{2}\phi^{2}\,.
\eea
The corresponding equation of motion in a curved background is
\bea\label{eq:KG}
\frac{1}{\sqrt{|g|}}\,\partial_{\mu}\!\Big(\sqrt{|g|}\,g^{\mu\nu}\partial_{\nu}\phi\Big) = -\frac{M^{2}}{\hbar^{2}}\,\phi\,.
\eea
Throughout this appendix, we set $c = 1$ but retain $\hbar$ explicitly, so as to identify the terms that dominate in the classical ($\hbar \to 0$) limit.

\subsection{WKB Ansatz and the Eikonal Equation}

We substitute the WKB ansatz\footnote{Note that when we do an eikonal limit, we are thinking of a particular classical field configuration and not simply the abstract classical field. Therefore this will connect with wave functions in the one particle sector, once we quantize the field theory. We discuss this in Appendix \ref{eknl_lim_get_geo_and_all}.}
\bea\label{eq:WKB}
\phi = A\, e^{iS/\hbar}\,,
\eea
where both $A$ and $S$ are real-valued functions of spacetime, into the equation of motion~\eqref{eq:KG}. Expanding the covariant d'Alembertian and grouping terms by powers of $\hbar$, one obtains
\begin{align}\label{eq:expanded}
  -A\,(\partial_{\mu}S)(\partial^{\mu}S) 
  + i\hbar\Big[A\,\Box S + 2\,(\partial_{\mu}A)(\partial^{\mu}S)\Big]
  + \hbar^{2}\,\Box A
  = -M^{2}A\,,
\end{align}
where $\Box \equiv \frac{1}{\sqrt{|g|}}\partial_{\mu}(\sqrt{|g|}\partial^{\mu})$. In the limit $\hbar \to 0$, only the leading-order (zeroth-order in $\hbar$) terms survive, yielding
\bea\label{eq:eikonal}
g^{\mu\nu}(\partial_{\mu}S)(\partial_{\nu}S) = M^{2}\,.
\eea
This is the \emph{eikonal equation}, which is recognized as the relativistic Hamilton--Jacobi equation for a point particle of mass $M$, with $S$ playing the role of Hamilton's principal function.

\subsection{From the Eikonal Equation to the Geodesic}

We now interpret \eqref{eq:eikonal} in the language of Hamiltonian mechanics. Defining the canonical momentum $k_{\mu} \equiv \partial_{\mu}S$, the eikonal equation takes the form $g^{\mu\nu}k_{\mu}k_{\nu} - M^{2} = 0$. We therefore introduce the Hamiltonian
\bea\label{eq:ham}
H = \frac{1}{2}\,g^{\mu\nu}k_{\mu}k_{\nu} - \frac{M^{2}}{2}\,,
\eea
which vanishes on shell as a consequence of the mass-shell constraint. Hamilton's equations give
\bea\label{eq:xdot}
\dot{x}^{\mu} = \frac{\partial H}{\partial k_{\mu}} = g^{\mu\nu}k_{\nu} = k^{\mu}\,,
\eea
where an overdot denotes differentiation with respect to the affine parameter\footnote{The Hamiltonian in \eqref{eq:ham} should not be confused with the Hilbert space Hamiltonian that evolves the quantum field in a chosen coordinate time. It is the reparameterization-invariant mass-shell Hamiltonian whose flow generates the characteristics of the Hamilton-Jacobi equation. The parameter along this Hamiltonian flow is an auxiliary worldline parameter, which we denote by $\tau$:  $\frac{dx^\mu}{d\tau} \equiv k^\mu$. With the normalization in \eqref{eq:ham} it is an affine parameter, because $k^\mu\nabla_\mu k_\nu=\frac{1}{2} \nabla_\nu(k^\mu k_\mu)=\frac{1}{2}\nabla_\mu (M^2)=0$. It is proportional (but not quite equal) to proper time when the scalar is massive. The reason is that  $g^{\mu\nu}k_{\mu}k_{\nu}$ is $M^{2}$ and not $1$. This can be changed by setting $\frac{dx^\mu}{d\tau}=\frac{k^\mu}{M}$ instead, but we will not do so here.} $\tau$. Performing a Legendre transformation, we obtain the corresponding worldline Lagrangian
\bea
L = k_{\mu}\,\dot{x}^{\mu} - H = \frac{1}{2}\,g_{\mu\nu}\,\dot{x}^{\mu}\dot{x}^{\nu} + \frac{M^{2}}{2}\,,
\eea
where in the second equality we have used \eqref{eq:xdot}. The constant term $M^{2}/2$ does not affect the equations of motion, so the dynamics is governed by the action
\bea\label{eq:action}
\mathcal{S} = \frac{1}{2}\int d\tau\; g_{\mu\nu}\,\dot{x}^{\mu}\dot{x}^{\nu}\,.
\eea
We now derive the equation of motion by demanding $\delta \mathcal{S} = 0$. The variation of \eqref{eq:action} gives
\bea
\delta \mathcal{S} = \int d\tau\;\bigg[g_{\mu\nu}\,\dot{x}^{\nu}\,\frac{d}{d\tau}(\delta x^{\mu}) + \frac{1}{2}\,\partial_{\sigma}g_{\mu\nu}\;\delta x^{\sigma}\,\dot{x}^{\mu}\dot{x}^{\nu}\bigg]\,.
\eea
Integrating the first term by parts and discarding the boundary contribution, we obtain
\bea\label{eq:deltaS}
\delta \mathcal{S} = \int d\tau\;\bigg[-\frac{d}{d\tau}\!\big(g_{\mu\nu}\,\dot{x}^{\nu}\big)\,\delta x^{\mu} + \frac{1}{2}\,\partial_{\sigma}g_{\mu\nu}\;\delta x^{\sigma}\,\dot{x}^{\mu}\dot{x}^{\nu}\bigg]\,.
\eea
Expanding the total $\tau$-derivative in the first term and relabelling the free index as $\sigma$, we find
\begin{align}
\frac{d}{d\tau}\!\big(g_{\sigma\nu}\,\dot{x}^{\nu}\big) 
&= g_{\sigma\nu}\,\ddot{x}^{\nu} + \partial_{\mu}g_{\sigma\nu}\;\dot{x}^{\mu}\dot{x}^{\nu} \nonumber\\
&= g_{\sigma\nu}\,\ddot{x}^{\nu} + \frac{1}{2}\big(\partial_{\mu}g_{\sigma\nu} + \partial_{\nu}g_{\sigma\mu}\big)\dot{x}^{\mu}\dot{x}^{\nu}\,,
\end{align}
where in the last step we have symmetrized in $\mu$ and $\nu$ by relabeling the dummy indices. Substituting back into \eqref{eq:deltaS} and setting $\delta \mathcal{S} = 0$ for arbitrary $\delta x^{\sigma}$, we arrive at
\bea
g_{\sigma\nu}\,\ddot{x}^{\nu} + \frac{1}{2}\Big(\partial_{\mu}g_{\sigma\nu} + \partial_{\nu}g_{\sigma\mu} - \partial_{\sigma}g_{\mu\nu}\Big)\dot{x}^{\mu}\dot{x}^{\nu} = 0\,.
\eea
Contracting with the inverse metric $g^{\lambda\sigma}$, this becomes
\bea
\ddot{x}^{\lambda} + \Gamma^{\lambda}{}_{\mu\nu}\;\dot{x}^{\mu}\dot{x}^{\nu} = 0\,,
\eea
where
\bea
\Gamma^{\lambda}{}_{\mu\nu} = \frac{1}{2}\,g^{\lambda\sigma}\Big(\partial_{\mu}g_{\sigma\nu} + \partial_{\nu}g_{\sigma\mu} - \partial_{\sigma}g_{\mu\nu}\Big)
\eea
is the Christoffel connection. This is the geodesic equation for a point particle propagating in the curved background, confirming that the classical particle limit of the massive scalar field theory reproduces the expected geodesic.

\section{Eikonal Limit: Classical Fields vs One-Particle Wave Functions}\label{eknl_lim_get_geo_and_all}

The position-operator construction gives a precise Hilbert-space meaning to the
coordinate-space wavefunction
\[
f_\Psi(x,t) = \langle x|\Psi(t)\rangle .
\]
The eikonal approximation is the local semiclassical approximation to this object, because classical field configurations map to wave functions in the one-particle Hilbert space. Writing
\[
f_\Psi(x,t) = A(x,t)e^{iS(x,t)/\hbar},
\]
the phase $S$ determines the local covariant momentum
\[
k_\mu = \partial_\mu S .
\]
The Klein--Gordon equation then implies, at leading order in the semiclassical
expansion, that $S$ obeys the Hamilton--Jacobi equation. Equivalently, the rays
generated by $k_\mu$ are geodesics. Thus the position expectation value
$\langle \hat x^i\rangle_t$ tracks one of these rays when the amplitude $A$ is
sufficiently localized and the phase gradient varies slowly across the support
of the packet.

Now let us do the explicit calculation to find $A(\rho,\phi,t)$ and $S(\rho,\phi,t)$ in terms of the parameters that go in the choice of the initial wave packet (see Eq. \eqref{approximate choice of packet}) in our AdS$_3$ discussion. Much of the discussion generalizes to more general spacetimes, as we comment along the way.

\subsection{Radial Equation in WKB Approximation}

We start by writing the WKB form of the radial equation \eqref{Diff equation}. Substituting $R_{nm}(\rho)=u_{nm}(\rho)/\sqrt{\tan\rho}$, we get the following equation:
\begin{equation}\label{diff_eqn_for_u}
 \frac{d^{2}u_{nm}(\rho)}{d \rho^{2}}+k^{2}_{nm}(\rho) u_{nm}(\rho)=0   
\end{equation}
where,
\begin{equation}\label{expression_for_k^2_nm_WKB}
k^{2}_{nm}(\rho)= \omega_{nm}^{2}-\frac{m^{2}-\frac{1}{4}}{\sin^{2}(\rho)} - \frac{M^{2}+\frac{3}{4}}{\cos^{2}(\rho)}   
\end{equation}
The solution of \eqref{diff_eqn_for_u} in the WKB approximation is \cite{Sakurai1994}:
\begin{equation}
u_{nm}(\rho)\approx \frac{C}{\sqrt{k_{\rho}}} e^{\pm i \int d \rho' \, k_{nm}(\rho')}    
\end{equation}
where $C$ is a complex number (i.e. $C=|C|\,e^{i\varphi_{0}}$). We know that $R_{nm}(\rho)$ should be a real valued function. So, we write the following form of $R_{nm}(\rho)$ to work with:
\begin{equation}\label{R_nm_WKB_form}
R_{nm}(\rho) \approx \frac{|C|}{\sqrt{\tan(\rho) k_{nm}(\rho)}}\left( e^{ i \int_{\rho_{t}}^{\rho} d \rho' \, k_{nm}(\rho')+i \varphi_{0}} + e^{ -i \int_{\rho_{t}}^{\rho} d \rho' \, k_{nm}(\rho')-i \varphi_{0}} \right)    
\end{equation}
In this equation, $\rho_t$ is an arbitrary reference point in the classically allowed region and we chose the positive root of \eqref{expression_for_k^2_nm_WKB} (i.e. $k_{nm}(\rho)>0$). We will work in a regime where $n,m,\Delta\,, n_{0}, m_{0}\gg 1$. We expect the WKB approximation to be valid in this regime \footnote{The WKB approximation is valid under the adiabaticity condition $|dk_{nm}/d\rho| \ll k^2_{nm}$. Given that $k^2_{nm}(\rho) = \omega_{nm}^2 - V_{\text{eff}}(\rho)$, where $V_{\text{eff}}(\rho) = \frac{m^2 - 1/4}{\sin^2 \rho} + \frac{M^2 + 3/4}{\cos^2 \rho}$, this requirement simplifies to $y\equiv R(\rho) = \frac{|V_{\text{eff}}'(\rho)|}{2 (k^2(\rho))^{3/2}} \ll 1$. By plotting $R(\rho)$ against the threshold line $y=1$, we can determine the parameter regimes that satisfy this adiabaticity condition. While increasing the radial quantum number ($n  \gg 1$) forces the ratio below the $y=1$ line across the entire domain from the origin to $\rho = \pi/2$, a large $n_0$ coupled with relatively small $m$ and $M$ leads to almost purely radial infalling trajectories. To support non-radial trajectories, the angular momentum must be comparable to the radial excitation ($m_0 \sim n_0$). However, if the mass $M$ remains small compared to these scales, the system stays within a quasi-null regime, yielding trajectories similar to the right-hand image in Fig.~\ref{fig:fig:pos_null_radial_and_ellip_2D}. Consequently, to recover the full diversity of massive trajectories within a valid WKB framework, we work in the limit $M \sim |m_{0}| \sim n_{0} \gg 1$ and consequently $n,\,|m|\gg 1$}.

\subsection{Initial Profile in Momentum Space}
We now use Eq. \eqref{gnm} with $t=0$ to get the initial profile in momentum space corresponding to \eqref{approximate choice of packet}: 
\begin{align}
&g(n,m) \nonumber \\
&=\frac{1}{\sqrt{2\pi}}\left(\int d\rho \tan\left(\rho\right)R^{*}_{nm}(\rho) \mathcal{N}_{\rho}\,e^{-\frac{(\rho-\rho_{0})^{2}}{4\sigma_{1}^{2}}}e^{-in_0(\rho-\rho_0)}\right)\left(\int d\phi\,e^{-im\phi} \mathcal{N}_{\phi}\,e^{-\frac{(\phi-\phi_{0})^{2}}{4\sigma_{2}^{2}}}e^{-im_0(\phi-\phi_0)}  \right) \label{firstexp_for_gnm_wkb}       
\end{align}
We assume that the packet is highly localized (i.e. $\sigma_{1}, \sigma_{2}\ll 1$). Under this assumption, the angular integral can be evaluated to be $\approx  2 \sqrt{\pi} \mathcal{ N_{\phi}}\, \sigma_{2}e^{-\sigma_{2}^{2}(m+m_{0})^{2}}e^{-im\phi_{0}}$. Small width does not cause the problem of delocalization over time, because the state has a large energy when we choose $n_{0}, m_{0}, M \gg 1$. 

We substitute the $R_{nm}(\rho)$ from \eqref{R_nm_WKB_form} in \eqref{firstexp_for_gnm_wkb}. Given that $n_{0}\gg 1$ and we have chosen the positive root of \eqref{expression_for_k^2_nm_WKB} as $k_{nm}(\rho)$ in \eqref{R_nm_WKB_form}, after taking the conjugate only the exponent with $+ i$ will contribute to the integral over $\rho$ in \eqref{R_nm_WKB_form}. The exponent with $-i$ is always negative and hence will oscillate very rapidly averaging its contribution to 0. Note that the other way round would have happened with a negative value of $n_{0}$, but still satisfying $-n_{0}\gg 1$. Likewise, the positive value of $m_{0}$ is also only a matter of choice. Only the magnitude of these quantities matter for the calculation.

Since $\sigma_{1}\ll 1$, we can Taylor expand the phase factor around $\rho_{0}$ as: $\int_{\rho_{t}}^{\rho}\,d\rho'k_{nm}(\rho')=\int_{\rho_{t}}^{\rho_{0}}\,d\rho'k_{nm}(\rho')+k_{nm}(\rho_{0})(\rho-\rho_{0})+...$. We pull the non-phase factors $\sqrt{\tan(\rho)}$ and $k_{nm}(\rho)$ outside the integral as $\sqrt{\tan(\rho_{0})}$ and $k_{nm}(\rho_{0})$ respectively. The final form of the radial integral is:
\begin{align}
&\approx \frac{|C|\,\mathcal{N}_{\rho}\sqrt{\tan\rho_{0}}}{\sqrt{k_{nm}(\rho_{0})}}\,e^{i \int_{\rho_{t}}^{\rho_{0}}\,d\rho'k_{nm}(\rho')+i \varphi_{0}} \int d\rho e^{-\frac{(\rho-\rho_{0})^{2}}{4\sigma_{1}^{2}}}e^{-i(n_0-k_{nm}(\rho_{0}))(\rho-\rho_0)} \nonumber \\   
&\approx \frac{|C|\,\mathcal{N}_{\rho}\sqrt{\tan\rho_{0}}}{\sqrt{k_{nm}(\rho_{0})}}\,e^{i \int_{\rho_{t}}^{\rho_{0}}\,d\rho'k_{nm}(\rho')+i \varphi_{0}} 2\sqrt{\pi} \sigma_{1}\, e^{-\sigma_{1}^{2}(k_{nm}(\rho_{0})-n_0)^{2}}
\end{align}
So the final form of $g(n,m)$ takes the form $g(n,m)=|g(n,m)|e^{i\Phi_{nm}}$ with:
\begin{align}
&|g(n,m)|\approx G_{0}\,e^{-\sigma_{1}^{2}(k_{nm}(\rho_{0})-n_0)^{2}}\,e^{-\sigma_{2}^{2}(m+m_{0})^{2}}\\
&\Phi_{nm}= \int_{\rho_{t}}^{\rho_{0}}\,d\rho'k_{nm}(\rho') -m\phi_{0}\,+ \varphi_{0}
\end{align}
where $G_{0}$ absorbs all the constants that appeared so far. We can see that $|g(n,m)|$ peaks at around $\bar{m}=-m_{0}$ and $\bar{n}$ such that $k_{\bar{n}\bar{m}}(\rho_{0})=n_{0}$. 

\subsection{Solving For The Wave Function}

Now to the most important bit. We substitute this $g(n,m)$ obtained after the WKB approximation into \eqref{f}. 
\begin{equation}\label{f_rho_phi_t_wkb_in_terms_of_gauss_n_m}
f(\rho,\phi,t)\approx\frac{G_{0}}{\sqrt{2\pi}}\sum_{n,m}  e^{-\sigma_{1}^{2}(k_{nm}(\rho_{0})-n_0)^{2}}\,e^{-\sigma_{2}^{2}(m+m_{0})^{2}}\,e^{i\int_{\rho_{t}}^{\rho_{0}}\,d\rho'k_{nm}(\rho')-im\phi_{0}+i \varphi_{0}}R_{nm}(\rho)e^{im\phi}\,e^{-i\omega_{nm}t}  
\end{equation}
Again, we see that only the exponent in \eqref{R_nm_WKB_form} with $-i$ contributes. This term also cancels the $+i \varphi_{0}$ phase appearing in \eqref{f_rho_phi_t_wkb_in_terms_of_gauss_n_m}. So using \eqref{R_nm_WKB_form} in \eqref{f_rho_phi_t_wkb_in_terms_of_gauss_n_m}, and after some simplification, we get:
\begin{equation}
f(\rho,\phi,t)\approx\frac{\mathcal{K}}{\sqrt{\tan(\rho)}}\sum_{n,m}\,\frac{1}{\sqrt{k_{nm}(\rho)}}\,e^{-\sigma_{1}^{2}(k_{nm}(\rho_{0})-n_0)^{2}}\,e^{-\sigma_{2}^{2}(m+m_{0})^{2}}\,e^{iS_{nm}(\rho,\phi,t)}    
\end{equation}
where $\mathcal{K}$ absorbs all the constants like $|C|,\,\sqrt{2\pi}$, etc. and:
\begin{equation}\label{the_phase_action_nm_S_nm_WKB}
S_{nm}(\rho,\phi,t)=-\omega_{nm}t+m(\phi-\phi_{0})-\int_{\rho_{0}}^{\rho} d\rho'\, k_{nm}(\rho')
\end{equation}
Since we are in the regime of large quantum numbers and mass, we can expand this equation and $k_{nm}(\rho_{0})$ around $\bar{n}$ and $\bar{m}$ (as these are the peaks) and replace the sums by integrals. We can also pull out the $\sqrt{k_{nm}(\rho)}$ in the numerator out of the sum writing $\sqrt{k_{\bar{n}\bar{m}}(\rho)}$. This gives:
\begin{equation}
f(\rho,\phi,t)\approx \frac{\mathcal{K}}{\sqrt{\tan(\rho)k_{\bar{n}\bar{m}}}} \mathcal{I}(\bar{n},\bar{m},\rho,\phi,t)e^{iS_{\bar{n}\bar{m}}(\rho,\phi,t)}    
\end{equation}
where,
\begin{align}
\mathcal{I}(\bar{n},\bar{m},\rho,\phi,t)&=\int_{0}^{\infty} dn\, \int_{-\infty}^{\infty} dm\, e^{-\sigma_{1}^{2}[p_{\bar{n}\bar{m}}(n-\bar{n})+q_{\bar{n}\bar{m}}(m-\bar{m})]^{2}}\,e^{-\sigma_{2}^{2}(m-\bar{m})^{2}}\,e^{i\left(D_{\bar{n}}(n-\bar{n})+D_{\bar{m}}(m-\bar{m})\right)}   \nonumber \\
&=\int_{-\bar{n}}^{\infty} dn'\, \int_{-\infty}^{\infty} dm'\, e^{-\sigma_{1}^{2}(p_{\bar{n}\bar{m}}n'+q_{\bar{n}\bar{m}}m')^{2}}\,e^{-\sigma_{2}^{2}m'^{2}}\,e^{i\left(D_{\bar{n}}n'+D_{\bar{m}}m'\right)}\label{the_int_to_evl_using_standard_gaussian}
\end{align}
where,
\begin{itemize}
    \item $p_{\bar{n}\bar{m}} = \left. \frac{\partial k_{nm}(\rho_{0})}{\partial n} \right|_{\bar{n},\bar{m}}$
    \item $q_{\bar{n}\bar{m}} = \left. \frac{\partial k_{nm}(\rho_{0})}{\partial m} \right|_{\bar{n},\bar{m}}$
    \item $S_{\bar{n}\bar{m}}(\rho,\phi,t) = - \omega_{\bar{n}\bar{m}}t +\bar{m}(\phi-\phi_0)- \int_{\rho_0}^{\rho} k_{\bar{n}\bar{m}}(\rho')\,d\rho'$ which is our desired action.
\end{itemize}
and,   
\begin{equation}\label{D_n_WKB}
D_{\bar{n}} = \left. \frac{\partial S}{\partial n} \right|_{\bar{n},\bar{m}} = -2t - \int_{\rho_0}^{\rho} \left. \frac{\partial k_{nm}(\rho')}{\partial n} \right|_{\bar{n},\bar{m}} d\rho'
\end{equation}

\begin{equation}\label{D_m_WKB}
D_{\bar{m}} = \left. \frac{\partial S}{\partial m} \right|_{\bar{n},\bar{m}} = (\phi-\phi_0) - \text{sgn}(\bar{m})t - \int_{\rho_0}^{\rho} \left. \frac{\partial k_{nm}(\rho')}{\partial m} \right|_{\bar{n},\bar{m}} d\rho'
\end{equation}
are obtained via the Taylor expansion of the action around $\bar{n},\bar{m}$.

Since, $\bar{n}\gg1$, we can effectively replace the range of $n'$ integration in \eqref{the_int_to_evl_using_standard_gaussian} to be from $-\infty$ to $\infty$. In this case, we can use the standard result $\int_{-\infty}^{\infty}d ^n X e^{-\frac{1}{2}X^{T}\,M\,X+i J^{T}X}=\sqrt{\frac{(2\pi)^{n}}{\text{det M}}}e^{-\frac{1}{2}J^{T}M^{-1}J}$ (see \cite{Schwartz2014}). In our case, $J^{T}=(D_{\bar{n}}\,\,\,D_{\bar{m}})$ and $X^{T}=(n'\,\,\,m')$ and the matrix $M$ is:
\begin{equation}
\label{the_matrix_required}
M= 2
\begin{bmatrix}
\sigma_{1}^{2}p^{2}_{\bar{n}\bar{m}} & \sigma^{2}_{1}p_{\bar{n}\bar{m}}q_{\bar{n}\bar{m}}\\
\sigma^{2}_{1}p_{\bar{n}\bar{m}}q_{\bar{n}\bar{m}} & \sigma^{2}_{1}q^{2}_{\bar{n}\bar{m}}+\sigma^{2}_{2}
\end{bmatrix}
\end{equation}
The determinant can be evaluated to be: $\text{det} M=4\sigma^{2}_{1}\sigma^{2}_{2}p^{2}_{\bar{n\bar{m}}}$, which is clearly non zero. So the inverse $M^{-1}$ exists and is given by:
\begin{equation}
\label{the_inverse_of_the_matrix}
M^{-1}= \frac{1}{2\sigma^{2}_{1}\sigma^{2}_{2}p^{2}_{\bar{n\bar{m}}}}
\begin{bmatrix}
\sigma^{2}_{1}q^{2}_{\bar{n}\bar{m}}+\sigma^{2}_{2}& -\sigma^{2}_{1}p_{\bar{n}\bar{m}}q_{\bar{n}\bar{m}}\\
-\sigma^{2}_{1}p_{\bar{n}\bar{m}}q_{\bar{n}\bar{m}} & \sigma_{1}^{2}p^{2}_{\bar{n}\bar{m}}
\end{bmatrix}
\end{equation}
After some simplification, we get $-\frac{1}{2}J^{T}M^{-1}J=-\frac{1}{4} \left[ \frac{(D_{\bar{m}} - \frac{q_{\bar{n}\bar{m}}}{p_{\bar{n}\bar{m}}} D_{\bar{n}})^2}{\sigma_2^2} + \frac{D_{\bar{n}}^2}{\sigma_1^2 p^2_{\bar{n}\bar{m}}} \right]$. So with this we get the desired results:
\begin{itemize}
    \item \textbf{The Amplitude}: \begin{equation}\label{final_form_of_A}
    A(\rho,\phi,t)\approx\frac{\pi \mathcal{K}}{\sigma_{1}\sigma_{2}|p_{\bar{n}\bar{m}}|\sqrt{\tan(\rho)k_{\bar{n}\bar{m}}(\rho)}}\text{exp}\left[-\frac{1}{4} \left( \frac{(D_{\bar{m}} - \frac{q_{\bar{n}\bar{m}}}{p_{\bar{n}\bar{m}}} D_{\bar{n}})^2}{\sigma_2^2} + \frac{D_{\bar{n}}^2}{\sigma_1^2 p^2_{\bar{n}\bar{m}}} \right)\right]    
    \end{equation}
    \item \textbf{The Phase:}
    \begin{equation}\label{final_form_of_S}
    S(\rho,\phi,t)= S_{\bar{n}\bar{m}}(\rho,\phi,t) = - \omega_{\bar{n}\bar{m}}t +\bar{m}(\phi-\phi_0)- \int_{\rho_0}^{\rho} k_{\bar{n}\bar{m}}(\rho')\,d\rho' 
    \end{equation}
\end{itemize}
We set $k_{\mu}=\partial_{\mu}S$, the momenta. Clearly, $\partial_{t}S=-\omega_{\bar{n}\bar{m}}$, $\partial_{\rho}S=-k_{\bar{n}\bar{m}}(\rho)$ and $\partial_{\phi}S=\bar{m}=-m_{0}$.  Likewise the expression for $A(\rho,\phi,t)$ (Eq. \eqref{final_form_of_A}) shows that the motion is peaked around the region where $D_{\bar{n}}=0$ and $D_{\bar{m}}=0$. This means the classical trajectory is the path obtained by solving \eqref{D_n_WKB} and \eqref{D_m_WKB}.

In generic static space times, the metric components $g_{\mu\nu}$ often depend on the azimuthal coordinate $\phi$, meaning the field expansion is not a simple $e^{im\phi}$ mode sum. This lack of axial symmetry prevents the field equations from separating into a neat radial ODE where a 1D WKB approximation can be blindly applied; instead, one would be dealing with a full PDE requiring more complex semiclassical treatments. However, in cases where the symmetry does allow for a radial reduction, any resulting second-order ODE can be mapped to the normal form $u''(\rho) + k^2(\rho)u(\rho) = 0$ through appropriate coordinate transformations. In this regime, the semiclassical solution $u(\rho) \sim k^{-1/2} \exp\left(\pm i \int p(\rho') d\rho'\right)$ recovers the phase information needed to track the classical trajectories, just as we did for Ad$S_{3}$.
 
\subsection{Wave Packet Peaks as Geodesics}

The classical path of the wave packet is determined by the stationary points of the amplitude $A(\rho, \phi, t)$, which occur when the phase derivatives $D_{\bar{n}}$ and $D_{\bar{m}}$ vanish. In this semiclassical regime where $\bar{n}, \bar{m}, M \gg 1$, we can write the local momentum $k_{\bar{n}\bar{m}}(\rho)$ by neglecting the $1/4$ and $3/4$ shifts in Eq. \eqref{expression_for_k^2_nm_WKB}:
\begin{equation}
k(\rho) = \sqrt{\omega^2 - \frac{m^2}{\sin^2\rho} - \frac{M^2}{\cos^2\rho}}
\end{equation}
where $\omega = \omega_{\bar{n}\bar{m}}$ and $m = \bar{m}$. We also note that for the $AdS_3$ spectrum, $\omega = 2n + |m| + \Delta$, which implies the derivative $\partial k / \partial n$ involves a factor of $2$ from the energy dependence.

Setting $D_{\bar{n}} = 0$ in Eq. \eqref{D_n_WKB} defines the trajectory $\rho(t)$. The condition $D_{\bar{n}} = 0$ leads to the integral:
\begin{equation}\label{time_as_fn_of_rho_WKB}
t = -\frac{1}{2} \int_{\rho_0}^{\rho} \frac{\partial k(\rho')}{\partial n} d\rho' = - \int_{\rho_0}^{\rho} \frac{\omega}{k(\rho')} d\rho'
\end{equation}
By performing the substitution $u = \sin^2\rho'$, this integral takes the standard form $\int \frac{du}{\sqrt{-au^2 + bu - c}}$, where the solution is given by $\frac{1}{\sqrt{a}}\sin^{-1}\left(\frac{2au-b}{\sqrt{b^2-4ac}}\right)$. Evaluating this integral and performing the algebraic simplification, we arrive at:
\begin{equation}\label{exact_rel_cos2rho_wkb}
\cos(2\rho(t)) = \mathcal{C}_1 + \mathcal{C}_2 \cos(2t + \delta)
\end{equation}
where $\mathcal{C}_1 = \frac{M^2 - m^2}{\omega^2}$, $\mathcal{C}_2 = \frac{\sqrt{(\omega^2 + m^2 - M^2)^2 - 4\omega^2 m^2}}{\omega^2}$, and $\delta$ can be obtained by putting $\rho=\rho_0$ with $t=0$ in \eqref{exact_rel_cos2rho_wkb}.

To relate this to the classical physics of a point particle, we recall that for a highly localized packet, the effective stress-energy tensor $T^{\mu\nu}$ is proportional to $u^\mu u^\nu$, where $u^\mu$ is the four-velocity. The conserved charges of the geodesic—energy $E$ and angular momentum $L$ per unit mass—are related to the field parameters via $E = \omega/M$ and $L = m/M$. (See footnote \ref{foot} for a closely related discussion.) Making these substitutions, we get: $\mathcal{C}_{1}=\frac{1 - L^2}{E^2},\, \mathcal{C}_{2}=\frac{\sqrt{(E^{2}+L^{2}-1)^{2}-4L^{2}E^{2}}}{E^{2}}$ which are same as the results obtained for massive particle in \eqref{exact relation 1} (with $\mathcal{R}=1$). Given $\mathcal{C}_{1}$ and $\mathcal{C}_{2}$ match with $A$ and $B$ in \eqref{exact relation 1}, and since $\delta$ is an initial condition choice, this establishes that \eqref{exact_rel_cos2rho_wkb} successfully reproduces \eqref{exact relation 1}.

Next, we determine the equation of orbit by setting $D_{\bar{m}} = 0$. Using Eq. \eqref{D_m_WKB}, and substituting for $t$ from \eqref{time_as_fn_of_rho_WKB} , we find:
\begin{equation}\label{dphi_by_drho_for_WKB}
\frac{d\phi}{d\rho} = -\frac{m}{k(\rho) \sin^2\rho}
\end{equation}
Substituting $m = LM$ and $\omega = EM$ in \eqref{dphi_by_drho_for_WKB}, we get:
\begin{equation}\label{d_rho_by_dt_WKB_appx}
\frac{d\rho}{d\phi} =- \frac{\sin^{2}\rho}{L}\sqrt{E^2 - \frac{L^2}{\sin^{2}\rho} - \frac{1}{\cos^{2}\rho}}
\end{equation}
This is same as \eqref{d_rho_by_dt_mmaassiivvee_case}, which is the expected geodesic equation for a massive particle in $AdS_3$ (with $\mathcal{R}=1$). 

\subsection{Geodesics as Position Expectation Values in the Eikonal Limit}

Now, let us use the wave functions in the eikonal limit to show that the position expectation values directly yield geodesic trajectories. Our single particle expectation value equations \eqref{EV rho} and \eqref{EV phi new} can also be rewritten as
\begin{align}
&\langle\hat{\rho}\rangle_{t} = \frac{\int d\rho\,d\phi\,\tan(\rho)\,\rho\,|f(\rho,\phi,t)|^{2}}{\int d\rho\,d\phi\,\tan(\rho)\,|f(\rho,\phi,t)|^{2}} \label{new_expect_rho_for_WKB_to_get_geo}\\
&\langle\widehat{e^{i\phi}}\rangle_{t} = \frac{\int d\rho\,d\phi\,\tan(\rho)\,e^{i\phi}\,|f(\rho,\phi,t)|^{2}}{\int d\rho\,d\phi\,\tan(\rho)\,|f(\rho,\phi,t)|^{2}}
\end{align}
since the denominator is normalized to $1$. We shall use our WKB form $f(\rho,\phi,t) \approx A(\rho,\phi,t)e^{iS(\rho,\phi,t)}$ to see how we get the expectation value of $\hat{\rho}$ to follow the classical path. The $\tan(\rho)$ factor in the integration measure \eqref{new_expect_rho_for_WKB_to_get_geo} is exactly canceled by the $1/\tan(\rho)$ factor coming from the square of the amplitude $|A(\rho,\phi,t)|^2$. Consequently, the expectation value simplifies to:
\begin{align}
\langle\hat{\rho}\rangle_{t} &= \frac{\int d\rho\,d\phi\,\frac{1}{k_{\bar{n}\bar{m}}(\rho)}\,\rho\,\exp\left[-\frac{1}{2} \left( \frac{(D_{\bar{m}} - \frac{q}{p} D_{\bar{n}})^2}{\sigma_2^2} + \frac{D_{\bar{n}}^2}{\sigma_1^2 p^2} \right)\right]}{\int d\rho\,d\phi\,\frac{1}{k_{\bar{n}\bar{m}}(\rho)}\,\exp\left[-\frac{1}{2} \left( \frac{(D_{\bar{m}} - \frac{q}{p} D_{\bar{n}})^2}{\sigma_2^2} + \frac{D_{\bar{n}}^2}{\sigma_1^2 p^2} \right)\right]}
\end{align}
where we set $p_{\bar{n}\bar{m}}=p$ and $q_{\bar{n}\bar{m}}=q$. As established earlier, the amplitude $A(\rho,\phi,t)$ is sharply peaked during all times at the coordinate location $(\rho_{geo}, \phi_{geo})$ corresponding to the classical geodesic path (where $D_{\bar{n}}=0$ and $D_{\bar{m}}=0$). Since the Gaussian envelope provides support only in a vanishingly small neighborhood around this trajectory, we may treat the coordinate $\rho$ as a constant and pull it outside the integral.

This results in:
\begin{align}
\langle\hat{\rho}\rangle_{t} &= \rho_{geo}(t) \frac{\int d\rho\,d\phi\,\frac{1}{k_{\bar{n}\bar{m}}(\rho_{geo})}\,|A_{env}|^2}{\int d\rho\,d\phi\,\frac{1}{k_{\bar{n}\bar{m}}(\rho_{geo})}\,|A_{env}|^2} \nonumber \\
&= \rho_{geo}(t)
\end{align}
By applying the same logic to the angular operator, we find:
\begin{align}
\langle\widehat{e^{i\phi}}\rangle_{t} &= e^{i\phi_{geo}(t)} \frac{\int d\rho\,d\phi\,\frac{1}{k_{\bar{n}\bar{m}}(\rho_{geo})}\,|A_{env}|^2}{\int d\rho\,d\phi\,\frac{1}{k_{\bar{n}\bar{m}}(\rho_{geo})}\,|A_{env}|^2} \nonumber \\
&= e^{i\phi_{geo}(t)}
\end{align}
In both cases, the integrals in the numerator and denominator cancel perfectly, demonstrating that the expectation values of the operators precisely track the classical geodesic path derived from the WKB phase. This confirms that the WKB construction provides a robust bridge between the quantum field theoretic description and the semiclassical particle limit.

\end{document}